\documentclass[a4paper,10pt]{article}
\usepackage{amssymb}
\usepackage{amsmath}
\usepackage{bm}
\usepackage{boxedminipage}
\usepackage[pdftex]{graphicx}
\usepackage{graphicx}
\@addtoreset{equation}{section} \makeatother
\renewcommand{\theequation}{\thesection.\arabic{equation}}
\newif\ifpdf \ifx\pdfoutput\undefined \pdffalse
\else \pdfoutput=1 \pdftrue \fi \ifpdf \else \fi
\input epsf

\begin{document}

\ifpdf\DeclareGraphicsExtensions{.pdf, .jpg, .tif} \else%
\DeclareGraphicsExtensions{.eps, .jpg} \fi
\begin{titlepage}

    \thispagestyle{empty}
    \begin{flushright}
        \hfill{CERN-PH-TH/2006-154}\\
        \hfill{UCLA/07/TEP/8}\\
    \end{flushright}

    \vspace{5pt}
    \begin{center}
        { \huge{\textbf{Mirror Fermat Calabi-Yau Threefolds} \\\vspace{12pt}
 \textbf{and} \\\vspace{20pt} \textbf{Landau-Ginzburg Black Hole Attractors}}}

        \vspace{28pt}

        {\bf Stefano Bellucci$^{\spadesuit}$, \ Sergio Ferrara$^{\diamondsuit,\maltese,\spadesuit}$, \ Alessio Marrani$^{\heartsuit,\spadesuit}$ and \ Armen Yeranyan$^{\clubsuit,\spadesuit}$}

        \vspace{15pt}

        {$\spadesuit$ \it INFN - Laboratori Nazionali di Frascati, \\
        Via Enrico Fermi 40,00044 Frascati, Italy\\
        \texttt{bellucci,marrani@lnf.infn.it}}

        \vspace{10pt}

        {$\diamondsuit$ \it Physics Department,Theory Unit, CERN, \\
        CH 1211, Geneva 23, Switzerland\\
        \texttt{sergio.ferrara@cern.ch}}

        \vspace{10pt}

        {$\maltese$ \it Department of Physics and Astronomy,\\
        University of California, Los Angeles, CA USA\\
        \texttt{ferrara@physics.ucla.edu}}

        \vspace{10pt}

        {$\heartsuit$ \it Museo Storico della Fisica e\\
        Centro Studi e Ricerche ``Enrico Fermi"\\
        Via Panisperna 89A, 00184 Roma, Italy}

        \vspace{10pt}

        {$\clubsuit$ \it Department of Physics, Yerevan State University, \\Alex  Manoogian St., 1, Yerevan,
        375025, Armenia\\
        \texttt{ayeran@ysu.am}}

        \vspace{15pt}

        \vspace{20pt}

        {ABSTRACT}
    \end{center}

    \vspace{5pt}
We study black hole attractor equations for one-(complex
structure)modulus Calabi-Yau spaces which are the mirror dual of
Fermat Calabi-Yau threefolds ($CY_{3}$s).

When exploring non-degenerate solutions near the Landau-Ginzburg
point of the moduli space of such 4-dimensional compactifications,
we always find two species of
extremal black hole attractors, depending on the choice of the $Sp\left( 4,%
\mathbb{Z}\right) $ symplectic charge vector, one $\frac{1}{2}$%
-BPS (which is always stable, according to general results of special K\"{a}%
hler geometry) and one non-BPS. The latter turns out to be stable
(local minimum of the ``effective black hole potential'' $V_{BH}$)
for non-vanishing central charge, whereas it is unstable (saddle
point of $V_{BH} $) for the case of vanishing central charge.

This is to be compared to the large volume limit of one-modulus $CY_{3}$%
-compactifications (of Type II A superstrings), in which the
homogeneous symmetric special K\"{a}hler geometry based on cubic
prepotential admits (beside the $\frac{1}{2}$-BPS ones) only non-BPS
extremal black hole attractors with non-vanishing central charge,
which are always stable.

    \vspace{150pt}

\end{titlepage}
\newpage \baselineskip6 mm \tableofcontents\newpage

\section{Introduction\label{Intro}}

Extremal black hole (BH) attractors
\cite{FKS}-\nocite{Strom,FK1}\cite{FK2} have been recently widely
investigated \cite{Sen-old1}-\nocite
{GIJT,Sen-old2,K1,TT,G,GJMT,Ebra1,K2,Ira1,BFM,AoB,Lust1,Sen1,FKlast,Ebra2,FG2,BFGM1,rotating-attr,K3,Misra1,Lust2,Morales,CYY,ADFT,MRS,CdWMa,DFT07-1,
BFM-SIGRAV06, ADFT-2} \cite{Saraikin-Vafa-1}, especially in
connection with new classes of solutions to the attractor equations
corresponding to non-BPS (Bogomol'ny-Prasad-Sommerfeld) horizon
geometries, supported by particular configurations of the BH
electric and magnetic charges. Such geometries are
\textit{non-degenerate}, \textit{i.e.} they have a finite,
non-vanishing horizon area, and their Bekenstein-Hawking entropy
\cite{BH1}\textbf{\ }is obtained by extremizing an ``effective BH
potential''.

In $\mathcal{N}=2$, $d=4$ Maxwell-Einstein supergravity theories (MESGTs),
non-degenerate attractor horizon geometries correspond to BH solitonic
states belonging to ``short massive multiplets'' (for the $\frac{1}{2}$-BPS
case, with $0<\left| Z\right| _{H}=M_{ADM,H}$) and to ``long massive
multiplets'', either with non-vanishing or vanishing central charge $Z$ not
saturating the BPS bound\footnote{%
Here and in what follows, the subscript ``$H$'' will denote values at the BH
event horizon.} \cite{BPS}
\begin{equation}
0\leq \left| Z\right| _{H}<M_{ADM,H}.
\end{equation}
The Arnowitt-Deser-Misner (ADM) mass \cite{ADM} at the BH horizon is
obtained by extremizing a positive-definite ``effective BH potential''%
\footnote{%
Here and below ``$\phi $'' denotes the set of real scalars relevant for
Attractor Mechanism, \textit{i.e.} the $2n_{V}$ ones coming from the $n_{V}$
vector supermultiplets coupled to the supergravity multiplet.} $V_{BH}\left(
\phi ,\widetilde{\Gamma }\right) $, where $\widetilde{\Gamma }$ is the $%
1\times \left( 2n_{V}+2\right) $ symplectic vector of the BH field-strength
fluxes , \textit{i.e.} of the magnetic and electric charges:
\begin{equation}
\widetilde{\Gamma }\equiv \left( p^{\Lambda },q_{\Lambda }\right)
,~~~~~p^{\Lambda }\equiv \frac{1}{4\pi }\int_{S_{\infty }^{2}}\mathcal{F}%
^{\Lambda },~~~q_{\Lambda }\equiv \frac{1}{4\pi }\int_{S_{\infty }^{2}}%
\mathcal{G}_{\Lambda },~~~\Lambda =0,1,...,n_{V},  \label{Gamma-tilde}
\end{equation}
where $n_{V}$ denotes the number of Abelian vector supermultiplets coupled
to the supergravity one (containing the graviphoton $A^{0}$); moreover, $%
\mathcal{F}^{\Lambda }=dA^{\Lambda }$, and $\mathcal{G}_{\Lambda }$ is the
``dual'' field-strength two-form \cite{CDF,CDFVP}.

The BH entropy $S_{BH}$ is given by the Bekenstein-Hawking entropy-area
formula \cite{BH1,FGK}
\begin{equation}
S_{BH}\left( \widetilde{\Gamma }\right) =\frac{A_{H}\left( \widetilde{\Gamma
}\right) }{4}=\pi \left. V_{BH}\left( \phi ,\widetilde{\Gamma }\right)
\right| _{\partial V_{BH}=0}=\pi V_{BH}\left( \phi _{H}\left( \widetilde{%
\Gamma }\right) \right) ,  \label{BHEA}
\end{equation}
where $A_{H}$ is the event horizon area, and the solution $\phi _{H}\left(
\widetilde{\Gamma }\right) $ to the criticality condition
\begin{equation}
\partial _{\phi }V_{BH}\left( \phi ,\widetilde{\Gamma }\right) =0
\label{crit-cond-gen}
\end{equation}
is properly named \textit{attractor} if the critical $\left( 2n_{V}+2\right)
\times \left( 2n_{V}+2\right) $ real symmetric Hessian matrix
\begin{equation}
\left. \frac{\partial ^{2}V_{BH}\left( \phi ,\widetilde{\Gamma }\right) }{%
\partial \phi \partial \phi }\right| _{\phi =\phi _{H}\left( \widetilde{%
\Gamma }\right) }  \label{crit-Hessian-gen}
\end{equation}
is a strictly positive-definite matrix\footnote{%
The opposite is in general not true, \textit{i.e.} there can be attractor
points corresponding to critical Hessian matrices with ``flat'' directions (%
\textit{i.e.} vanishing eigenvalues). In general, when a critical Hessian
matrix exhibits some vanishing eigenvalues, one has to look at higher-order
derivatives of $V_{BH}$ evaluated at the considered point, and study their
sign. Dependingly on the values of the supporting BH charges, one can
obtains stable or unstable critical points. Examples in \ literature of
investigations beyond the Hessian level can be found in \cite{TT,K3,Misra1}.}%
.

Although non-supersymmetric BH attractors exist also in $\mathcal{N}>2$, $%
d=4 $ and $d=5$ supergravities \cite{FG,FKlast}, the most interesting
examples arise in $\mathcal{N}=2$, $d=4$ MESGTs, where the scalar
fluctuations relevant for the BH Attractor Mechanism parametrize a special
K\"{a}hler (SK) manifold. Recently, the classification of ``attractor
solutions'' for extremal BHs has been performed in full generality for the
whole class of homogeneous symmetric SK geometries \cite{BFGM1}, and three
distinct classes of extremal BH attractors (namely $\frac{1}{2}$-BPS,
non-BPS $Z\neq 0$ and non-BPS $Z=0$ ones) were found as solutions to Eqs. (%
\ref{crit-cond-gen}). In such a framework, the non-BPS charge orbits have
been found to depend on whether the supporting charge vector $\widetilde{%
\Gamma }$ is such that the $\mathcal{N}=2$ central charge vanishes or not.
Moreover, the critical Hessian matrix (\ref{crit-Hessian-gen}) was usually
found to exhibit zero modes (\textit{i.e.} ``flat'' directions), whose
attractor nature seemingly further depends on additional conditions on the
charge vector $\widetilde{\Gamma }$, other than the ones given by the
extremality conditions (\ref{crit-cond-gen}) (see \textit{e.g.} \cite{TT}%
).\medskip

The aim of the present work is to study a particular class of (1-modulus) SK
geometries, namely the ones underlying the complex structure moduli space of
(mirror) Fermat Calabi-Yau threefolds ($CY_{3}$s) (classified by the \textit{%
Fermat parameter }$k=5,6,8,10$, and firstly found in \cite{Strom-Witten}).
The fourth order linear Picard-Fuchs (PF) ordinary differential equations
determining the holomorphic fundamental period $4\times 1$ vector for such a
class of 1-modulus $CY_{3}$s were found some time ago for $k=5$ in \cite
{CDLOGP1,CDLOGP2} (see in particular Eq. (3.9) of \cite{CDLOGP1}; see also
\cite{Cadavid-Ferrara}) and for $k=6,8,10$ in \cite{KT} (see also \cite{Font}%
).

In the present work we will make distinction between two different (but, as
we will explicitly check in the abovementioned context, completely
equivalent) approaches to extremal BH attractors, namely:

\textit{i}) the so-called \textit{``criticality condition'' }approach, based
on the extremization of a suitably defined, positive-definite ``effective BH
potential'' $V_{BH}\left( \phi ,\widetilde{\Gamma }\right) $, whose
criticality conditions (\ref{crit-cond-gen}) are called \textit{``Attractor
Equations''};

\textit{ii}) the so-called \textit{``SK geometrical identities'' }approach
(exploited in Sect. \ref{SKG-id-approach}), based on the evaluation of some
fundamental geometrical identities of SK geometry along the \textit{loci} of
the scalar manifold determined by some ``critical constraints''.\bigskip

The plan of the paper is as follows.

In Sect. \ref{SKG-gen} we sketchily recall the fundamentals of the local SK
geometry of the scalar manifolds of $\mathcal{N}=2$, $d=4$ MESGTs. After the
treatment of the general $n_{V}$-moduli case in Subsect. \ref{SKG-gen-n}, in
Subsect. \ref{SKG-gen-1} we focus on the $1$-modulus case.

Thence, Sect. \ref{Sect4} is devoted to the issue of the stability of the
critical points of $V_{BH}$. After a general discussion in Subsect. \ref
{Sect4-n} (including a treatment of the $\frac{1}{2}$-BPS attractors,
specialized for $n_{V}=1$ in Subsubsect. \ref{stab-BPS}), in Subsect. \ref
{Sect4-1} we determine (sufficient, but in general not necessary) conditions
for non-BPS $Z\neq 0$ (Subsubsect. \ref{stab-non-BPS-Z<>0}) and non-BPS $Z=0$
(Subsubsect. \ref{stab-non-BPS-Z=0}) (classes of) critical points of $V_{BH}$
to be (\textit{at least local}) minima of $V_{BH}$, and thus to be \textit{%
attractors} in a strict sense.

Sects. \ref{HG} and \ref{HG2} are devoted to present the holomorphic
geometry embedded in SK geometry. Such a geometry is relevant in order to
introduce the PF ordinary linear differential Eqs.. Once again, the $1$%
-modulus case is considered in detail in Subsects \ref{HG-1} and \ref{HG2-1}.

Then, in Sect. \ref{GA} the formalism of (mirror) Fermat $CY_{3}$s
(classified by the Fermat parameter $k=5,6,8,10$), in particular near the LG
point $\psi =0$ of their (complex structure deformation) moduli space, is
introduced. The general analysis of Sect. \ref{GA} is consequently
specialized to the study of non-degenerate extremal BH LG attractors in the
complex structure moduli space of the four mirror Fermat $CY_{3}$s,
corresponding to $k=5$ (Sect. \ref{quintic}), $k=6$ (Sect. \ref{sixtic}), $%
k=8$ (Sect. \ref{eightic}), and $k=10$ (Sect. \ref{tentic}).

The abovementioned \textit{``SK geometrical identities'' }approach is
exploited in Sect. \ref{SKG-id-approach}, in order to study the extremal BH
LG attractors for the above-mentioned class of $CY_{3}$s. After an
introduction to $n_{V}=1$ SK geometrical identities near the LG point in
Subsect. \ref{SKG-id-approach-gen}, in Subsect. \ref{SKG-id-approach-apply}
their evaluation along the \textit{loci} corresponding to the various
species of critical points of $V_{BH}$ is performed. By analyzing the $\frac{%
1}{2}$-BPS, non-BPS $Z\neq 0$ and non-BPS $Z=0$ classes respectively in
Subsubsects. \ref{SKG-id-approach-BPS}, \ref{SKG-id-approach-non-BPS-Z<>0}
and \ref{SKG-id-approach-non-BPS-Z=0}, we obtain results identical to the
ones we got in Sects. \ref{quintic}-\ref{tentic} by exploiting the
abovementioned \textit{``criticality condition'' }approach, corresponding to
solve near the LG point the 2 real \textit{criticality conditions} of $%
V_{BH} $, which in the $1$-modulus case are nothing but the real and
imaginary part of the so-called $\mathcal{N}=2$, $d=4$ supergravity AEs.

Sect. \ref{const-norm} deals with the Yukawa coupling function of $1$%
-modulus SK geometry. Subsect. \ref{const-norm-1} presents three different,
equivalent approaches to compute such a crucial quantity, and apply them to
study the LG limit of the moduli space of (mirror) Fermat $CY_{3}$s. In such
a framework, the $1$-modulus stability conditions for non-BPS $Z\neq 0$ and
non-BPS $Z=0$ critical points of $V_{BH}$ (respectively obtained in
Subsubsects. \ref{stab-non-BPS-Z<>0} and \ref{stab-non-BPS-Z=0}) are thence
explicitly checked in Subsect. \ref{const-norm-1-bis}, obtaining results
consistent with the ones of Sects. \ref{quintic}-\ref{tentic}. Finally,
Subsect. \ref{const-norm-2} deals with the global, exact expression of the
holomorphic Yukawa coupling function of the $1$-modulus SK geometry
underlying the moduli space of (mirror) Fermat $CY_{3}$s, and with its
limits near the three typologies of regular singular points of the PF
differential Eqs. introduced in Sects. \ref{HG} and \ref{HG2}.

Concluding remarks, summarizing observations and outlooking comments are the
contents of the final Sect. \ref{Conclusion}.

\section{\label{SKG-gen}Summary of Special K\"{a}hler Geometry}

\setcounter{equation}0
\def\theequation{2.\arabic{subsection}.\arabic{equation}}

\subsection{\label{SKG-gen-n}$n_{V}$-Moduli}

In $\mathcal{N}=2$, $d=4$ MESGT the following formula holds\footnote{%
Here and below we switch to the complex parametrization of the set of
scalars being considered:
\begin{equation*}
\left\{ \phi ^{a}\right\} _{a=1,...,2n_{V}}\longrightarrow \left\{ z^{i},%
\overline{z}^{\overline{i}}\right\} _{i,\overline{i}=1,...,n_{V}}.
\end{equation*}
The relation between such two equivalent parametrizations of the SK scalar
manifold is given by Eq. (4.2) of \cite{BFM}.} \cite{FK1,FK2,CDF}
\begin{equation}
V_{BH}\left( z,\overline{z};q,p\right) =\left| Z\right| ^{2}\left( z,%
\overline{z};q,p\right) +g^{j\overline{j}}\left( z,\overline{z}\right)
D_{j}Z\left( z,\overline{z};q,p\right) \overline{D}_{\overline{j}}\overline{Z%
}\left( z,\overline{z};q,p\right) .  \label{VBH1}
\end{equation}
Consequently, the criticality conditions (\ref{crit-cond-gen}) can be easily
shown to acquire the form \cite{FGK}
\begin{equation}
2\overline{Z}D_{i}Z+g^{j\overline{j}}\left( D_{i}D_{j}Z\right) \overline{D}_{%
\overline{j}}\overline{Z}=0;  \label{AEs1}
\end{equation}
this is what one should rigorously refer to as the $\mathcal{N}=2$, $d=4$
supergravity attractor equations (AEs). $g^{j\overline{j}}\left( z,\overline{%
z}\right) $ is the contravariant K\"{a}hler metric tensor, satisfying the
usual orthonormality condition:
\begin{equation}
g^{i\overline{j}}\left( z,\overline{z}\right) \partial _{i}\overline{%
\partial }_{\overline{k}}K\left( z,\overline{z}\right) =\delta _{\overline{k}%
}^{\overline{j}},
\end{equation}
where $K\left( z,\overline{z}\right) $ is the real K\"{a}hler potential. As
previously mentioned, $Z\left( z,\overline{z};q,p\right) $ is the $\mathcal{N%
}=2$ central charge function
\begin{equation}
Z\left( z,\overline{z};q,p\right) \equiv e^{\frac{1}{2}K\left( z,\overline{z}%
\right) }\widetilde{\Gamma }\Omega \Pi \left( z\right) =e^{\frac{1}{2}%
K\left( z,\overline{z}\right) }\left[ q_{\Lambda }X^{\Lambda }\left(
z\right) -p^{\Lambda }F_{\Lambda }\left( z\right) \right] \equiv e^{\frac{1}{%
2}K\left( z,\overline{z}\right) }W\left( z;q,p\right) ,  \label{Z}
\end{equation}
where $\Omega $ is the $\left( 2n_{V}+2\right) $-dim. symplectic metric
(subscripts denote dimensions)
\begin{equation}
\Omega \equiv \left(
\begin{array}{ccc}
0_{n_{V}+1} &  & -\mathbb{I}_{n_{V}+1} \\
&  &  \\
\mathbb{I}_{n_{V}+1} &  & 0_{n_{V}+1}
\end{array}
\right) ,  \label{Omega}
\end{equation}
and $\Pi \left( z\right) $ is the $\left( 2n_{V}+2\right) \times 1$
holomorphic period vector in symplectic basis
\begin{equation}
\Pi \left( z\right) \equiv \left(
\begin{array}{c}
X^{\Lambda }\left( z\right) \\
\\
F_{\Lambda }\left( z\right)
\end{array}
\right) ,  \label{PI-PI}
\end{equation}
with $X^{\Lambda }\left( z\right) $ and $F_{\Lambda }\left( z\right) $ being
the holomorphic sections of the $U(1)$ line (Hodge) bundle over the SK
manifold (clearly, due to holomorphicity they do not belong to the related $%
U(1)$ ring). Finally, $W\left( z;q,p\right) $ is the holomorphic $\mathcal{N}%
=2$ central charge function, also named $\mathcal{N}=2$ superpotential.

Let us here recall that $Z$ has K\"{a}hler weights $\left( p,\overline{p}%
\right) =\left( 1,-1\right) $; thus, its K\"{a}hler-covariant derivatives
read
\begin{equation}
\begin{array}{l}
D_{i}Z=\left( \partial _{i}+\frac{1}{2}\partial _{i}K\right) Z, \\
\\
\overline{D}_{\overline{i}}Z=\left( \overline{\partial }_{\overline{i}}-%
\frac{1}{2}\overline{\partial }_{\overline{i}}K\right) Z.
\end{array}
\end{equation}

The non-holomorphic basic, defining differential relations of SK geometry are%
\footnote{%
Actually, there are different (equivalent) defining approaches to SK
geometry. For subtleties and further elucidation concerning such an issue,
see \textit{e.g.} \cite{Craps1} and \cite{Craps2}.} (see \textit{e.g.} \cite
{CDF}):
\begin{equation}
\left\{
\begin{array}{l}
D_{i}Z=Z_{i}; \\
\\
D_{i}Z_{j}=iC_{ijk}g^{k\overline{k}}\overline{D}_{\overline{k}}\overline{Z}%
=iC_{ijk}g^{k\overline{k}}\overline{Z}_{\overline{k}}; \\
\\
D_{i}\overline{D}_{\overline{j}}\overline{Z}=D_{i}\overline{Z}_{\overline{j}%
}=g_{i\overline{j}}\overline{Z}; \\
\\
D_{i}\overline{Z}=0,
\end{array}
\right.  \label{SKG-rels1}
\end{equation}
where the first relation is nothing but the definition of the ``matter
charges'' $Z_{i}$s and the fourth relation expresses the
K\"{a}hler-covariant holomorphicity of $Z$. $C_{ijk}$ is the rank-3,
completely symmetric, covariantly holomorphic tensor of SK geometry (with
K\"{a}hler weights $\left( 2,-2\right) $) (see \textit{e.g.} \cite{CDF}-
\nocite{Castellani1}\cite{DFF}):
\begin{equation}
\begin{array}{l}
\left\{
\begin{array}{l}
C_{ijk}=C_{(ijk)}=\left\langle D_{i}D_{j}V,D_{k}V\right\rangle =e^{K}\left(
\partial _{i}\mathcal{N}_{\Lambda \Sigma }\right) D_{j}X^{\Lambda
}D_{k}X^{\Sigma }= \\
\\
=e^{K}\left( \partial _{i}X^{\Lambda }\right) \left( \partial _{j}X^{\Sigma
}\right) \left( \partial _{k}X^{\Xi }\right) \partial _{\Xi }\partial
_{\Sigma }F_{\Lambda }\left( X\right) \equiv e^{K}W_{ijk},\text{~~}\overline{%
\partial }_{\overline{l}}W_{ijk}=0; \\
\\
C_{ijk}=D_{i}D_{j}D_{k}\mathcal{S},~~\mathcal{S}\equiv -iL^{\Lambda
}L^{\Sigma }Im\left( F_{\Lambda \Sigma }\right) ,~~F_{\Lambda \Sigma }\equiv
\frac{\partial F_{\Lambda }}{\partial X^{\Sigma }},F_{\Lambda \Sigma }\equiv
F_{\left( \Lambda \Sigma \right) }~; \\
\\
C_{ijk}=-ig_{i\overline{l}}\overline{f}_{\Lambda }^{\overline{l}%
}D_{j}D_{k}L^{\Lambda },~~~\overline{f}_{\Lambda }^{\overline{l}}\left(
\overline{D}\overline{L}_{\overline{s}}^{\Lambda }\right) \equiv \delta _{%
\overline{s}}^{\overline{l}};
\end{array}
\right. \\
\\
\overline{D}_{\overline{i}}C_{jkl}=0\text{ (\textit{covariant holomorphicity}%
)}; \\
\\
R_{i\overline{j}k\overline{l}}=-g_{i\overline{j}}g_{k\overline{l}}-g_{i%
\overline{l}}g_{k\overline{j}}+C_{ikp}\overline{C}_{\overline{j}\overline{l}%
\overline{p}}g^{p\overline{p}}\text{ (usually named \textit{SKG constraints})%
}; \\
\\
D_{[i}C_{j]kl}=0,
\end{array}
\label{C}
\end{equation}
where round (square) brackets denote (anti)symmetrization with respect to
enclosed indices, and $R_{i\overline{j}k\overline{l}}$ is the
Riemann-Christoffel tensor\footnote{%
Notice that the third of Eqs. (\ref{C}) correctly defines the Riemann tensor
$R_{i\overline{j}k\overline{l}}$, and it is actual the opposite of the one
which may be found in a large part of existing literature (see \textit{e.g.}
\cite{Bagger-Witten}). Such a formulation of the so-called \textit{SKG
constraints} is well defined, because, as discussed \textit{e.g.} in \cite
{BFM-SIGRAV06}, it yields negative values of the constant scalar curvature
of ($n_{V}=1$-dim.) homogeneous symmetric compact SK manifolds.} of
K\"{a}hler geometry:
\begin{equation}
R_{i\overline{j}k\overline{l}}\equiv g^{m\overline{n}}\left( \overline{%
\partial }_{\overline{l}}\overline{\partial }_{\overline{j}}\partial
_{m}K\right) \partial _{i}\overline{\partial }_{\overline{n}}\partial _{k}K-%
\overline{\partial }_{\overline{l}}\partial _{i}\overline{\partial }_{%
\overline{j}}\partial _{k}K=g_{k\overline{n}}\partial _{i}\overline{\Gamma }%
_{\overline{l}\overline{j}}^{~~\overline{n}}=g_{n\overline{l}}\overline{%
\partial }_{\overline{j}}\Gamma _{ki}^{~~n},~~~~~~~~\overline{R_{i\overline{j%
}k\overline{l}}}=R_{j\overline{i}l\overline{k}};\   \label{Riemann}
\end{equation}
\begin{equation}
\Gamma _{ij}^{~~l}\equiv -g^{l\overline{l}}\partial _{i}g_{j\overline{l}%
}=-g^{l\overline{l}}\partial _{i}\overline{\partial }_{\overline{l}}\partial
_{j}K=\Gamma _{\left( ij\right) }^{~~l},  \label{Christoffel-n}
\end{equation}
$\Gamma _{ij}^{~~l}$ standing for the Christoffel symbols of the second kind
of the K\"{a}hler metric $g_{i\overline{j}}$. For later convenience, it is
here worth writing the expression for the holomorphic covariant derivative
of $C_{ijk}$:
\begin{equation}
D_{i}C_{jkl}=D_{(i}C_{j)kl}=\partial _{i}C_{jkl}+\left( \partial
_{i}K\right) C_{jkl}+\Gamma _{ij}^{~~m}C_{mkl}+\Gamma
_{ik}^{~~m}C_{mjl}+\Gamma _{il}^{~~m}C_{mjk},  \label{DC}
\end{equation}
where we used the last of Eqs. (\ref{C}), which is a consequence, through
the SKG constraints and the covariant holomorphicity of $C_{ijk}$, of the
Bianchi identities for the Riemann tensor $R_{i\overline{j}k\overline{l}}$
(see \textit{e.g.} \cite{Castellani1}).

By using the first two of relations (\ref{SKG-rels1}), the $\mathcal{N}=2$
AEs (\ref{AEs1}) can be recast as follows \cite{FGK}:
\begin{equation}
2\overline{Z}Z_{i}+iC_{ijk}g^{j\overline{j}}g^{k\overline{k}}\overline{Z}_{%
\overline{j}}\overline{Z}_{\overline{k}}=0.  \label{AEs2}
\end{equation}

It is now worth recalling some fundamental identities defining the geometric
structure of SK manifolds \cite{FBC,K1,K2,BFM,AoB,K3}
\begin{equation}
\widetilde{\Gamma }^{T}-i\Omega \mathcal{M}\left( \mathcal{N}\right)
\widetilde{\Gamma }^{T}=e^{K/2}\left[ -2iZ\overline{\Pi }-2ig^{j\overline{j}%
}\left( \overline{D}_{\overline{j}}\overline{Z}\right) D_{j}\Pi \right] ,
\label{SKG-identities1}
\end{equation}
where $\mathcal{M}\left( \mathcal{N}\right) $ denotes the $\left(
2n_{V}+2\right) \times \left( 2n_{V}+2\right) $ real symmetric matrix \cite
{CDF,FK1,FK2}
\begin{equation}
\mathcal{M}\left( \mathcal{N}\right) \equiv \left(
\begin{array}{cccc}
Im\left( \mathcal{N}\right) +Re\left( \mathcal{N}\right) \left( Im\left(
\mathcal{N}\right) \right) ^{-1}Re\left( \mathcal{N}\right) &  &  &
-Re\left( \mathcal{N}\right) \left( Im\left( \mathcal{N}\right) \right) ^{-1}
\\
&  &  &  \\
-\left( Im\left( \mathcal{N}\right) \right) ^{-1}Re\left( \mathcal{N}\right)
&  &  & \left( Im\left( \mathcal{N}\right) \right) ^{-1}
\end{array}
\right) ,
\end{equation}
where $\mathcal{N}_{\Lambda \Sigma }$ is a $\left( n_{V}+1\right) \times
\left( n_{V}+1\right) $ complex symmetric, moduli-dependent, K\"{a}hler
gauge-invariant matrix, the so-called \textit{kinetic matrix} of $\mathcal{N}%
=2$, $d=4$ MESGT (see \textit{e.g.} the report \cite{CDF}). Moreover, it
should be here reminded that
\begin{equation}
\begin{array}{l}
D_{i}\Pi =\left( \partial _{i}+\partial _{i}K\right) \Pi , \\
\\
\overline{D}_{\overline{i}}\Pi =\overline{\partial }_{\overline{i}}\Pi =0,
\end{array}
\end{equation}
since $\Pi $ is holomorphic with K\"{a}hler weights $\left( 2,0\right) $.

The $2n_{V}+2$ complex identities (\ref{SKG-identities1}) (whose real and
imaginary parts are related by a suitable ``rotation'' \cite{AoB}) express
nothing but a \textit{change of basis} of the BH charge configurations,
between the K\"{a}hler-invariant $1\times \left( 2n_{V}+2\right) $\
symplectic (magnetic/electric) basis vector $\widetilde{\Gamma }$ defined by
Eq. (\ref{Gamma-tilde}) and the complex, moduli-dependent $1\times \left(
n_{V}+1\right) $ \textit{matter charges} vector (with K\"{a}hler weights $%
\left( 1,-1\right) $)
\begin{equation}
\mathcal{Z}\left( z,\overline{z}\right) \equiv \left( Z\left( z,\overline{z}%
\right) ,Z_{i}\left( z,\overline{z}\right) \right) _{i=1,...,n_{V}}.
\label{Z-call}
\end{equation}
It should be recalled that the BH charges are conserved due to the overall $%
\left( U(1)\right) ^{n_{V}+1}$ gauge-invariance of the system under
consideration, and $\widetilde{\Gamma }$ and $\mathcal{Z}\left( z,\overline{z%
}\right) $ are two \textit{equivalent} basis for them. Their very
equivalence relations are given by the SKG identities (\ref{SKG-identities1}%
) themselves. By its very definition (\ref{Gamma-tilde}), $\widetilde{\Gamma
}$\ is \textit{moduli-independent} (at least in a stationary, spherically
symmetric and asymptotically flat extremal BH background, as it is the case
being treated here), whereas $Z$ is \textit{moduli-dependent}, since it
refers to the eigenstates of the $\mathcal{N}=2$, $d=4$ supergravity
multiplet and of the $n_{V}$\ Maxwell vector supermultiplets.

Moreover, it is important to stress that identities (\ref{SKG-identities1})
entail 2 redundant degrees of freedom, encoded in the homogeneity (of degree
1) of (\ref{SKG-identities1}) under complex rescalings of $\widetilde{\Gamma
}$. Indeed, by recalling the definition (\ref{Z}) it can be readily checked
that the right-hand side of (\ref{SKG-identities1}) acquires an overall
factor $\lambda $ under the rescaling
\begin{equation}
\widetilde{\Gamma }\longrightarrow \lambda \widetilde{\Gamma },~~~\lambda
\in \mathbb{C}.
\end{equation}
We will reconsider such a point in Sect. \ref{SKG-id-approach}, when
treating the $1$-modulus case more in detail.

It should also be noticed that the $\mathcal{N}=2$ ``effective BH
potential'' given by Eq. (\ref{VBH1}) can also be rewritten as \cite
{FK1,FK2,CDF}
\begin{equation}
V_{BH}\left( z,\overline{z};q,p\right) =-\frac{1}{2}\widetilde{\Gamma }%
\mathcal{M}\left( \mathcal{N}\right) \widetilde{\Gamma }^{T},  \label{VBH2}
\end{equation}
and therefore it can be identified with the first, positive-definite real
invariant of SK geometry (see \textit{e.g.} \cite{K3,CDF}). It is
interesting to remark that the result (\ref{VBH2}) can be elegantly obtained
from the SK geometry identities (\ref{SKG-identities1}) by making use of the
following relations \cite{FKlast}:
\begin{equation}
\mathcal{M}\left( \mathcal{N}\right) \Pi =i\Omega \Pi ,~~\mathcal{M}\left(
\mathcal{N}\right) \overline{D}_{\overline{j}}\overline{\Pi }=i\Omega
\overline{D}_{\overline{j}}\overline{\Pi },~~\forall \overline{j}.
\label{rela1}
\end{equation}
By introducing the $\left( 2n_{V}+2\right) \times \left( n_{V}+1\right) $
rectangular matrix ($I=1,...,n_{V},n_{V}+1$)
\begin{eqnarray}
\mathcal{V} &\equiv &\left( \overline{D}_{\overline{i}}\overline{\Pi },\Pi
\right) =\left(
\begin{array}{l}
\overline{D}_{\overline{i}}\overline{X}^{\Lambda },X^{\Lambda } \\
~ \\
\overline{D}_{\overline{i}}\overline{F}_{\Lambda },F_{\Lambda }
\end{array}
\right) \equiv e^{-K/2}\left(
\begin{array}{l}
\overline{D}_{\overline{i}}\overline{L}^{\Lambda },L^{\Lambda } \\
~ \\
\overline{D}_{\overline{i}}\overline{M}_{\Lambda },M_{\Lambda }
\end{array}
\right) =e^{-K/2}\left(
\begin{array}{l}
f_{I}^{\Lambda } \\
~ \\
h_{I\Lambda }
\end{array}
\right) ,~  \label{def-V} \\
&&  \notag \\
~f_{I}^{\Lambda } &\equiv &\left( \overline{D}_{\overline{i}}\overline{L}%
^{\Lambda },L^{\Lambda }\right) ,\text{ \ }h_{I\Lambda }\equiv \left(
\overline{D}_{\overline{i}}\overline{M}_{\Lambda },M_{\Lambda }\right) ,
\label{def-f-h}
\end{eqnarray}
the relations (\ref{rela1}) can be recast as\footnote{%
Up to an exchange of order of $L^{\Lambda },M_{\Lambda }$ and $\overline{D}_{%
\overline{i}}\overline{L}^{\Lambda },\overline{D}_{\overline{i}}\overline{M}%
_{\Lambda }$ in the definitions (\ref{def-f-h}), Eq. (\ref{Eqq1}) matches
the $\mathcal{N}\geqslant 2$ generalization of SKG (and its application to
the case $\mathcal{N}=2$) performed in Sect. 2 of \cite{FKlast}.}
\begin{equation}
\mathcal{M}\left( \mathcal{N}\right) \mathcal{V}=i\Omega \mathcal{V}%
\Leftrightarrow \frac{1}{2}\left( \mathcal{M}\left( \mathcal{N}\right)
+i\Omega \right) \mathcal{V}=i\Omega \mathcal{V}.  \label{Eqq1}
\end{equation}
It is also here worth recalling that the fundamental \textit{Ans\"{a}tze}
\cite{CDF} of SKG (solving the \textit{SKG constraints} given by the third
of Eqs. (\ref{C}))
\begin{equation}
M_{\Lambda }=\mathcal{N}_{\Lambda \Sigma }L^{\Sigma },~~D_{i}M_{\Lambda }=%
\overline{\mathcal{N}}_{\Lambda \Sigma }D_{i}L^{\Sigma }.  \label{Ans1}
\end{equation}
univoquely determine $\mathcal{N}_{\Lambda \Sigma }\left( z,\overline{z}%
\right) $ as
\begin{equation}
\mathcal{N}_{\Lambda \Sigma }\left( z,\overline{z}\right) =h_{I\Lambda
}\left( z,\overline{z}\right) \circ \left( f^{-1}\right) _{\Sigma
}^{I}\left( z,\overline{z}\right) ,
\end{equation}
where $\circ $ denotes the usual matrix product, $\left( f^{-1}\right)
_{\Sigma }^{I}f_{I}^{\Lambda }=\delta _{\Sigma }^{\Lambda }$, $\left(
f^{-1}\right) _{\Lambda }^{I}f_{J}^{\Lambda }=\delta _{J}^{I}$, and $%
f_{I}^{\Lambda }$ and $h_{I\Lambda }$ are $\left( n_{V}+1\right)
\times \left( n_{V}+1\right) $ complex matrices defined by Eqs.
(\ref{def-f-h}). \setcounter{equation}0
\def\theequation{2.\arabic{subsection}.\arabic{equation}}

\subsection{\label{SKG-gen-1}$1$-Modulus}

In the 1-modulus case ($n_{V}=1$) major simplifications occur in the
formalism of SKG.

The potential $V_{BH}$ (\ref{VBH1}) and the AEs (given by Eqs. (\ref{AEs1})
and (\ref{AEs2})) respectively reduce to\footnote{%
In the present paper $C$ will be used to denote two different entities in
different contexts:
\par
1) the Yukawa coupling function in $1$-dim. SK geometry, which is nothing
but the tensor $C_{ijk}$ (defined by Eqs. (\ref{C})) for $n_{V}=1$ (see also
Sect. \ref{const-norm});
\par
2) some ($k$-parameterized) constants in the general formalism of $1$-dim.
SK geometry arising from compactifications on (mirror) Fermat $CY_{3}$s, as
given by the first of definitions (\ref{omega0-2}).
\par
Attention should be paid to correctly identify case-by-case the meaning of $%
C $, which however can be effortlessly understood in the various contexts.} (%
$z^{1}\equiv \psi $, $D_{1}\equiv D_{\psi }\equiv D$, $g_{\psi \overline{%
\psi }}\equiv g$, $C_{111}\equiv C_{\psi \psi \psi }\equiv C$)
\begin{eqnarray}
V_{BH}\left( \psi ,\overline{\psi };q,p\right) &\equiv &\left| Z\right|
^{2}\left( \psi ,\overline{\psi };q,p\right) +g^{-1}\left( \psi ,\overline{%
\psi }\right) \left| DZ\right| ^{2}\left( \psi ,\overline{\psi };q,p\right) =
\notag \\
&=&e^{K\left( \psi ,\overline{\psi }\right) }\left[ \left| W\right|
^{2}\left( \psi ,\overline{\psi };q,p\right) +g^{-1}\left( \psi ,\overline{%
\psi }\right) \left| DW\right| ^{2}\left( \psi ,\overline{\psi };q,p\right)
\right] ;  \label{VBH1-1-modulus}
\end{eqnarray}
\begin{equation}
2\overline{W}DW+g^{-1}\left( D^{2}W\right) \overline{D}\overline{W}=2%
\overline{W}DW+iCg^{-2}\left( \overline{D}\overline{W}\right) ^{2}=0,
\label{AEs-1-modulus-W}
\end{equation}
where the SK geometry has been assumed to be \textit{regular }(\textit{i.e.}
with $\left| K_{k}\left( \psi ,\overline{\psi }\right) \right| <\infty $
globally holding), and the (holomorphic) K\"{a}hler weight $2$ of the
holomorphic superpotential $W$ has been exploited:
\begin{equation}
DW=\partial W+\left( \partial K\right) W=e^{-K/2}DZ=e^{-K/2}D\left(
e^{K/2}W\right) .  \label{DW}
\end{equation}
By specializing\footnote{%
In the present paper $\Gamma $ will be used to denote different entities in
different contexts:
\par
1) the Christoffel affine connection of ($1$-dim.) K\"{a}hler geometry, as
defined by Eq. (\ref{Christoffel-n}) ((\ref{Christoffel-1}));
\par
2) (one of the forms of) the symplectic charge vector, as given by the first
of definitions (\ref{Gamma-Fermat});
\par
3) the Euler Gamma-function, as defined by Eq. (\ref{Euler-Gamma}).
\par
Attention should be paid to carefully identify the meaning of $\Gamma $,
which however can be understood at a glance in the various frameworks. It
should also be recalled that $\widetilde{\Gamma }$ denotes the symplectic
charge vector, as defined by Eq. (\ref{Gamma-tilde}), and $\widehat{\Gamma }$
denotes the (holomorphic) Christoffel connection of ($1$-dim.) holomorphic
geometry, as defined by Eqs. (\ref{Gamma-hat})-(\ref{Gamma-hat2}) (the first
of definitions (\ref{jazz2})).} the definition (\ref{Christoffel-n}) for $%
n_{V}=1$:
\begin{equation}
\Gamma _{11}^{~~~~1}\equiv \Gamma _{\psi \psi }^{~~~~\psi }\equiv \Gamma
=-g^{-1}\partial g=-\partial ln\left( g\right) ,  \label{Christoffel-1}
\end{equation}
one can compute
\begin{equation}
\begin{array}{l}
D^{2}W=\partial DW+\left( \partial K\right) DW+\Gamma DW= \\
\\
~=\partial \left[ \partial W+\left( \partial K\right) W\right] +\left(
\partial K\right) \left[ \partial W+\left( \partial K\right) W\right] -\left[
\partial ln\left( g\right) \right] DW= \\
\\
=\left\{ \partial ^{2}+\partial ^{2}K+2\partial K\partial +\left( \partial
K\right) ^{2}-\left[ \partial ln\left( \partial \overline{\partial }K\right)
\right] \left( \partial +\partial K\right) \right\} W.
\end{array}
\label{DDW}
\end{equation}

Thus, the three different classes of non-degenerate, critical points of $%
V_{BH}$ are in order:

$\frac{1}{2}$\textbf{-BPS class}, \textit{i.e.} the geometrical \textit{locus%
} defined by the constraints
\begin{equation}
Z\neq 0,~DZ=0\Longleftrightarrow W\neq 0,~DW=0,  \label{BPS-constr}
\end{equation}
which directly solve the AEs (\ref{AEs-1-modulus-W}).

\textbf{Non-BPS, }$Z\neq 0$\textbf{\ class}, \textit{i.e.} the geometrical
\textit{locus} defined by
\begin{equation}
Z\neq 0,~DZ\neq 0\Longleftrightarrow W\neq 0,~DW\neq 0,  \label{rain1}
\end{equation}
further constrained to solve AEs (\ref{AEs-1-modulus-W}), and thus yielding
the following relations \cite{BFM} ($k\in \mathbb{Z}$)
\begin{eqnarray}
&&
\begin{array}{c}
D^{2}W=-2g\overline{W}\frac{DW}{\overline{D}\overline{W}} \\
\Updownarrow \\
\left| D^{2}W\right| =2g\left| W\right| ,~~2\varphi _{DW}=\varphi
_{D^{2}W}+\varphi _{W}+\left( 1+2k\right) \pi ;
\end{array}
\\
&&  \notag \\
&&  \notag \\
&&
\begin{array}{c}
DW=-\frac{i}{2}Cg^{-2}\frac{\left( \overline{D}\overline{W}\right) ^{2}}{%
\overline{W}}=-\frac{i}{2}C\overline{W}g^{-2}\overline{Dln\left( W\right) }%
^{2} \\
\Updownarrow \\
\left| DW\right| =2g^{2}\frac{\left| W\right| }{\left| C\right| },~~\varphi
_{DW}=\varphi _{C}+\varphi _{W}+\left( \frac{3}{2}+2k\right) \pi ,
\end{array}
\notag \\
&&  \label{DZ-non-BPS-Z<>0}
\end{eqnarray}
where the $\varphi $s stand for phases. Notice that Eqs. (\ref
{AEs-1-modulus-W}) and (\ref{rain1}) necessarily yield that
\begin{equation}
C_{non-BPS,Z\neq 0}\neq 0.  \label{rain2}
\end{equation}
For this class, one can also define the \textit{supersymmetry-breaking order
parameter} as follows:
\begin{eqnarray}
\mathcal{O}_{non-BPS,Z\neq 0} &\equiv &\left[ \frac{g^{-1}|DZ|^{2}}{|Z|^{2}}%
\right] _{non-BPS,Z\neq 0}=\left[ \frac{g^{-1}|DW|^{2}}{|W|^{2}}\right]
_{non-BPS,Z\neq 0}=  \label{SUSY-breaking} \\
&=&4\left[ \frac{g^{3}}{\left| C\right| ^{2}}\right] _{non-BPS,Z\neq 0},
\label{SUSY-breaking2}
\end{eqnarray}
where in the second line we used Eq. (\ref{DZ-non-BPS-Z<>0}). It is worth
noticing that for a cubic prepotential (in a suitable special projective
coordinate, with K\"{a}hler gauge fixed such that $X^{0}=1$) $\mathcal{F}%
\left( z\right) =\varrho z^{3}$ it holds that $\mathcal{O}_{non-BPS,Z\neq
0}=3$ $\forall \varrho \in \mathbb{C}$ \cite{BFGM1}; such a result actually
holds for cubic prepotentials in generic $n_{V}$-moduli SK geometries, such
as the ones arising in the large volume limit of $CY_{3}$-compactifications
of Type II A superstring theory (see Eq. (111) of \cite{TT}).

\textbf{Non-BPS, }$Z=0$\textbf{\ class}, \textit{i.e.} the geometrical
\textit{locus} defined by
\begin{equation}
Z=0,~DZ\neq 0\Longleftrightarrow W=0,~DW\neq 0,  \label{non-BPS-Z=0-2}
\end{equation}
further constrained to solve AEs (\ref{AEs-1-modulus-W}), and thus yielding
\begin{equation}
\left. D^{2}W\right| _{non-BPS,Z=0}=0\overset{\left. DW\right|
_{non-BPS,Z=0}\neq 0}{\Longleftrightarrow }C_{non-BPS,Z=0}=0.
\label{non-BPS-Z=0-1}
\end{equation}
\medskip

Along the \textit{loci} of its critical points, the ``BH effective
potential'' $V_{BH}$ respectively reads
\begin{eqnarray}
&&
\begin{array}{l}
V_{BH,\frac{1}{2}-BPS}=\left| Z\right| _{\frac{1}{2}-BPS}^{2};
\end{array}
\\
&&
\begin{array}{l}
V_{BH,non-BPS,Z\neq 0}=\left| Z\right| _{non-BPS,Z\neq 0}^{2}\left[ 1+4\frac{%
g^{3}}{\left| C\right| ^{2}}\right] _{non-BPS,Z\neq 0};
\end{array}
\\
&&
\begin{array}{l}
V_{BH,non-BPS,Z=0}=\left| DZ\right| _{non-BPS,Z=0}^{2}.
\end{array}
\end{eqnarray}

In the present paper we will investigate AE (\ref{AEs-1-modulus-W}) near one
of three typologies of regular singular points in the complex structure
moduli space of (mirror) Fermat $CY_{3}$s, namely near the so-called
Landau-Ginzburg (LG) point $\psi =0$. In such a framework, the identities (%
\ref{SKG-identities1}) of SK geometry, when considered in the 1-modulus case
and in correspondence of the various above-mentioned species of critical
points of $V_{BH}$, can be used to find the BH charge configurations
supporting the LG point $\psi =0$ to be an attractor point of the considered
kind. It will be shown that, in spite of the fact that identities (\ref
{SKG-identities1}) give 4 real Eqs. in the 1-modulus case, only 2 of them
are independent, and they are completely equivalent to the 2 real
rigorously-named $\mathcal{N}=2$, $d=4$ supergravity AEs (\ref
{AEs-1-modulus-W}), which are nothing but the criticality condition $%
\partial V_{BH}=0$.\medskip

\section{\label{Sect4}Stability of Critical Points of $V_{BH}$}

\setcounter{equation}0
\def\theequation{3.\arabic{subsection}.\arabic{equation}}

\subsection{\label{Sect4-n}$n_{V}$-Moduli}

In order to decide whether a critical point of $V_{BH}$ is an attractor in
strict sense, one has to consider the following condition:
\begin{equation}
H_{\mathbb{R}}^{V_{BH}}\equiv H_{ab}^{V_{BH}}\equiv D_{a}D_{b}V_{BH}>0~~~%
\text{at}~~~D_{c}V_{BH}=\frac{\partial V_{BH}}{\partial \phi ^{c}}=0\text{~~~%
}\forall c=1,...,2n_{V},  \label{stab}
\end{equation}
\textit{i.e.} the condition of (strict) positive-definiteness of the real $%
2n_{V}\times 2n_{V}$ Hessian matrix $H_{\mathbb{R}}^{V_{BH}}\equiv
H_{ab}^{V_{BH}}$ of $V_{BH}$ (which is nothing but the squared mass matrix
of the moduli) at the critical points of $V_{BH}$, expressed in the real
parameterization through the $\phi $-coordinates. Since $V_{BH}$ is
positive-definite, a stable critical point (namely, an attractor in strict
sense) is necessarily a(n at least local) minimum, and therefore it fulfills
the condition (\ref{stab}).

In general, $H_{\mathbb{R}}^{V_{BH}}$ may be block-decomposed in $%
n_{V}\times n_{V}$ real matrices:
\begin{equation}
H_{\mathbb{R}}^{V_{BH}}=\left(
\begin{array}{ccc}
\mathcal{A} &  & \mathcal{C} \\
&  &  \\
\mathcal{C}^{T} &  & \mathcal{B}
\end{array}
\right) ,  \label{Hessian-real}
\end{equation}
with $\mathcal{A}$ and $\mathcal{B}$ being $n_{V}\times n_{V}$ real
symmetric matrices:
\begin{equation}
\mathcal{A}^{T}=\mathcal{A},~\mathcal{B}^{T}=\mathcal{B}\Leftrightarrow
\left( H_{\mathbb{R}}^{V_{BH}}\right) ^{T}=H_{\mathbb{R}}^{V_{BH}}.
\end{equation}

In the local complex $\left( z,\overline{z}\right) $-parameterization, the $%
2n_{V}\times 2n_{V}$ Hessian matrix of $V_{BH}$ reads
\begin{equation}
H_{\mathbb{C}}^{V_{BH}}\equiv H_{\widehat{i}\widehat{j}}^{V_{BH}}\equiv
\left(
\begin{array}{ccc}
D_{i}D_{j}V_{BH} &  & D_{i}\overline{D}_{\overline{j}}V_{BH} \\
&  &  \\
D_{j}\overline{D}_{\overline{i}}V_{BH} &  & \overline{D}_{\overline{i}}%
\overline{D}_{\overline{j}}V_{BH}
\end{array}
\right) =\left(
\begin{array}{ccc}
\mathcal{M}_{ij} &  & \mathcal{N}_{i\overline{j}} \\
&  &  \\
\overline{\mathcal{N}_{i\overline{j}}} &  & \overline{\mathcal{M}_{ij}}
\end{array}
\right) ,  \label{Hessian-complex}
\end{equation}
where the hatted indices $\hat{\imath}$ and $\hat{\jmath}$ may be
holomorphic or antiholomorphic. $H_{\mathbb{C}}^{V_{BH}}$ is the matrix
actually computable in the SKG formalism presented in Sect. \ref{SKG-gen}
(see below, Eqs. (\ref{M}) and (\ref{N})).

In general, $\frac{1}{2}$-BPS critical points are (at least local) minima of
$V_{BH}$, and therefore they are stable; thus, they are \textit{attractors}
in strict sense. Indeed, the $2n_{V}\times 2n_{V}$ (covariant) Hessian
matrix $H_{\mathbb{C}}^{V_{BH}}$ evaluated at such points is strictly
positive-definite \cite{FGK} :
\begin{eqnarray}
&&
\begin{array}{l}
\left( D_{i}D_{j}V_{BH}\right) _{\frac{1}{2}-BPS}=\left( \partial
_{i}\partial _{j}V_{BH}\right) _{\frac{1}{2}-BPS}=0, \\
\\
\left( D_{i}\overline{D}_{\overline{j}}V_{BH}\right) _{\frac{1}{2}%
-BPS}=\left( \partial _{i}\overline{\partial }_{\overline{j}}V_{BH}\right) _{%
\frac{1}{2}-BPS}=2\left( g_{i\overline{j}}V_{BH}\right) _{\frac{1}{2}%
-BPS}=2\left. g_{i\overline{j}}\right| _{\frac{1}{2}-BPS}\left| Z\right| _{%
\frac{1}{2}-BPS}^{2}>0,
\end{array}
\notag \\
&&  \label{SUSY-crit}
\end{eqnarray}
where here and below the notation ``$>0$'' (``$<0$'') is understood as
strict positive-(negative-)definiteness. The Hermiticity and (strict)
positive-definiteness of the (covariant) Hessian matrix $H_{\mathbb{C}%
}^{V_{BH}}$ at the $\frac{1}{2}$-BPS critical points are due to the
Hermiticity and - assumed - (strict) positive-definiteness (actually holding
globally) of the metric $g_{i\overline{j}}$ of the SK scalar manifold being
considered.

On the other hand, non-BPS critical points of $V_{BH}$ does not
automatically fulfill the condition (\ref{stab}), and a more detailed
analysis \cite{BFGM1,AoB} is needed.

Using the properties of SKG, one obtains:
\begin{eqnarray}
&&
\begin{array}{l}
\mathcal{M}_{ij}\equiv D_{i}D_{j}V_{BH}=D_{j}D_{i}V_{BH}= \\
\\
=3\overline{Z}D_{i}D_{j}Z+g^{k\overline{k}}\left( D_{i}D_{j}D_{k}Z\right)
\overline{D}_{\overline{k}}\overline{Z}=e^{K}\left[ 3\overline{W}%
D_{i}D_{j}W+g^{k\overline{k}}\left( D_{i}D_{j}D_{k}W\right) \overline{D}_{%
\overline{k}}\overline{W}\right] \\
\\
=4i\overline{Z}C_{ijk}g^{k\overline{k}}\left( \overline{D}_{\overline{k}}%
\overline{Z}\right) +i\left( D_{j}C_{ikl}\right) g^{k\overline{k}}g^{l%
\overline{l}}\left( \overline{D}_{\overline{k}}\overline{Z}\right) \left(
\overline{D}_{\overline{l}}\overline{Z}\right) = \\
\\
=e^{K}\left[ 4i\overline{Z}C_{ijk}g^{k\overline{k}}\left( \overline{D}_{%
\overline{k}}\overline{W}\right) +i\left( D_{j}C_{ikl}\right) g^{k\overline{k%
}}g^{l\overline{l}}\left( \overline{D}_{\overline{k}}\overline{W}\right)
\left( \overline{D}_{\overline{l}}\overline{W}\right) \right] ;
\end{array}
\notag \\
&&  \label{M}
\end{eqnarray}
\begin{equation}
\begin{array}{l}
\mathcal{N}_{i\overline{j}}\equiv D_{i}\overline{D}_{\overline{j}}V_{BH}=%
\overline{D}_{\overline{j}}D_{i}V_{BH}= \\
\\
=2g_{i\overline{j}}\left| Z\right| ^{2}+2\left( D_{i}Z\right) \left(
\overline{D}_{\overline{j}}\overline{Z}\right) +g^{k\overline{k}}\left[
\left( \overline{D}_{\overline{j}}D_{i}D_{k}Z\right) \overline{D}_{\overline{%
k}}\overline{Z}+\left( D_{i}D_{k}Z\right) \overline{D}_{\overline{j}}%
\overline{D}_{\overline{k}}\overline{Z}\right] = \\
\\
=e^{K}\left\{ 2g_{i\overline{j}}\left| W\right| ^{2}+2\left( D_{i}W\right)
\left( \overline{D}_{\overline{j}}\overline{W}\right) +g^{k\overline{k}}%
\left[ \left( \overline{D}_{\overline{j}}D_{i}D_{k}W\right) \overline{D}_{%
\overline{k}}\overline{W}+\left( D_{i}D_{k}W\right) \overline{D}_{\overline{j%
}}\overline{D}_{\overline{k}}\overline{W}\right] \right\} = \\
\\
=2\left[ g_{i\overline{j}}\left| Z\right| ^{2}+\left( D_{i}Z\right) \left(
\overline{D}_{\overline{j}}\overline{Z}\right) +g^{l\overline{n}}C_{ikl}%
\overline{C}_{\overline{j}\overline{m}\overline{n}}g^{k\overline{k}}g^{m%
\overline{m}}\left( \overline{D}_{\overline{k}}\overline{Z}\right) \left(
D_{m}Z\right) \right] = \\
\\
=2e^{K}\left[ g_{i\overline{j}}\left| W\right| ^{2}+\left( D_{i}W\right)
\left( \overline{D}_{\overline{j}}\overline{W}\right) +g^{l\overline{n}%
}C_{ikl}\overline{C}_{\overline{j}\overline{m}\overline{n}}g^{k\overline{k}%
}g^{m\overline{m}}\left( \overline{D}_{\overline{k}}\overline{W}\right)
\left( D_{m}W\right) \right] ,
\end{array}
\label{N}
\end{equation}
with $D_{j}C_{ikl}$ given by Eq. (\ref{DC}). Clearly, evaluating Eqs. (\ref
{M}) and (\ref{N}) constrained by the $\frac{1}{2}$-BPS conditions $%
D_{i}Z=0,\forall i=1,...,n_{V}$, one reobtains the results (\ref{SUSY-crit}%
). Here we limit ourselves to point out that further noteworthy elaborations
of $\mathcal{M}_{ij}$ and $\mathcal{N}_{i\overline{j}}$ can be performed in
homogeneous symmetric SK manifolds, where $D_{j}C_{ikl}=0$ globally \cite
{BFGM1}, and that the K\"{a}hler-invariant $\left( 2,2\right) $-tensor $g^{l%
\overline{n}}C_{ikl}\overline{C}_{\overline{j}\overline{m}\overline{n}}$ can
be rewritten in terms of the Riemann-Christoffel tensor $R_{i\overline{j}k%
\overline{m}}$ by using the so-called ``SKG constraints'' (see the third of
Eqs. (\ref{C})) \cite{AoB}. Moreover, the differential Bianchi identities
for $R_{i\overline{j}k\overline{m}}$ imply $\mathcal{M}_{ij}$ to be
symmetric (see comment below Eqs. (\ref{C}) and (\ref{DC})).

Thus, one gets the following global properties:
\begin{equation}
\mathcal{M}^{T}=\mathcal{M},~~\mathcal{N}^{\dag }=\mathcal{N}\Leftrightarrow
\left( H_{\mathbb{C}}^{V_{BH}}\right) ^{T}=H_{\mathbb{C}}^{V_{BH}},
\end{equation}
implying that
\begin{equation}
\left( H_{\mathbb{C}}^{V_{BH}}\right) ^{\dag }=H_{\mathbb{C}%
}^{V_{BH}}\Leftrightarrow \mathcal{M}^{\dag }=\mathcal{M},~~\mathcal{N}^{T}=%
\mathcal{N}\Leftrightarrow \overline{\mathcal{M}}=\mathcal{M},~~\overline{%
\mathcal{N}}=\mathcal{N}.
\end{equation}
It should be stressed clearly that the symmetry but non-Hermiticity of $H_{%
\mathbb{C}}^{V_{BH}}$ actually does not matter, because what one is
ineterested in are the eigenvalues of the real form $H_{\mathbb{R}}^{V_{BH}}$%
, which is real and symmetric, and therefore admitting $2n_{V}$ \textit{real}
eigenvalues.

The relation between $H_{\mathbb{R}}^{V_{BH}}$ expressed by Eq. (\ref
{Hessian-real}) and $H_{\mathbb{C}}^{V_{BH}}$ given by Eq. (\ref
{Hessian-complex}) is expressed by the following relations between the $%
n_{V}\times n_{V}$ sub-blocks of $H_{\mathbb{R}}^{V_{BH}}$ and $H_{\mathbb{C}%
}^{V_{BH}}$ \cite{BFM,BFM-SIGRAV06}:
\begin{equation}
\left\{
\begin{array}{l}
\mathcal{M}=\frac{1}{2}\left( \mathcal{A}-\mathcal{B}\right) +\frac{i}{2}%
\left( \mathcal{C}+\mathcal{C}^{T}\right) ; \\
\\
\mathcal{N}=\frac{1}{2}\left( \mathcal{A}+\mathcal{B}\right) +\frac{i}{2}%
\left( \mathcal{C}^{T}-\mathcal{C}\right) ,
\end{array}
\right.  \label{4jan1}
\end{equation}
or its inverse
\begin{equation}
\left\{
\begin{array}{l}
\mathcal{A}=Re\mathcal{M}+Re\mathcal{N}; \\
\\
\mathcal{B}=Re\mathcal{N}-Re\mathcal{M}; \\
\\
\mathcal{C}=Im\mathcal{M}-Im\mathcal{N}.
\end{array}
\right.  \label{4jan2}
\end{equation}

The structure of the Hessian matrix gets simplified at the critical points
of $V_{BH}$, because the covariant derivatives may be substituted by the
flat ones; the critical Hessian matrices in complex
holomorphic/antiholomorphic and real local parameterizations respectively
read
\begin{eqnarray}
&&
\begin{array}{l}
\left. H_{\mathbb{C}}^{V_{BH}}\right| _{\partial V_{BH}=0}\equiv \left(
\begin{array}{ccc}
\partial _{i}\partial _{j}V_{BH} &  & \partial _{i}\overline{\partial }_{%
\overline{j}}V_{BH} \\
&  &  \\
\partial _{j}\overline{\partial }_{\overline{i}}V_{BH} &  & \overline{%
\partial }_{\overline{i}}\overline{\partial }_{\overline{j}}V_{BH}
\end{array}
\right) _{\partial V_{BH}=0}=\left(
\begin{array}{ccc}
\mathcal{M} &  & \mathcal{N} \\
&  &  \\
\overline{\mathcal{N}} &  & \overline{\mathcal{M}}
\end{array}
\right) _{\partial V_{BH}=0};
\end{array}
\label{Hessian-complex-crit} \\
&&  \notag \\
&&  \notag \\
&&
\begin{array}{l}
\left. H_{\mathbb{R}}^{V_{BH}}\right| _{\partial V_{BH}=0}=\left. \frac{%
\partial ^{2}V_{BH}}{\partial \phi ^{a}\partial \phi ^{b}}\right| _{\partial
V_{BH}=0}=\left(
\begin{array}{ccc}
\mathcal{A} &  & \mathcal{C} \\
&  &  \\
\mathcal{C}^{T} &  & \mathcal{B}
\end{array}
\right) _{\partial V_{BH}=0}.
\end{array}
\label{Hessian-real-crit}
\end{eqnarray}
Thus, the study of the condition (\ref{stab}) finally amounts to the study
of the \textit{eigenvalue problem} of the real symmetric $2n_{V}\times
2n_{V} $ critical Hessian matrix $\left. H_{\mathbb{R}}^{V_{BH}}\right|
_{\partial V_{BH}=0}$ given by Eq. (\ref{Hessian-real-crit}), which is the
Cayley-transformed of the complex (symmetric, but not necessarily Hermitian)
$2n_{V}\times 2n_{V}$ critical Hessian $\left. H_{\mathbb{C}%
}^{V_{BH}}\right| _{\partial V_{BH}=0}$ given by Eq. (\ref
{Hessian-complex-crit}). \setcounter{equation}0
\def\theequation{3.\arabic{subsection}.\arabic{equation}}

\subsection{\label{Sect4-1}$1$-Modulus}

Once again, the situation strongly simplifies in $n_{V}=1$ SKG.

Indeed, for $n_{V}=1$ the moduli-dependent matrices $\mathcal{A}$, $\mathcal{%
B}$, $\mathcal{C}$, $\mathcal{M}$ and $\mathcal{N}$ introduced above are
simply scalar functions. In particular, $\mathcal{N}$ is real, since $%
\mathcal{C}$ trivially satisfies $\mathcal{C}=\mathcal{C}^{T}$. The
stability condition (\ref{stab}) can thus be written as
\begin{equation}
H_{\mathbb{R}}^{V_{BH}}\equiv D_{a}D_{b}V_{BH}>0\text{,~}\left(
a,b=1,2\right) ~~~\text{at}~~~D_{c}V_{BH}=\frac{\partial V_{BH}}{\partial
\phi ^{c}}=0\text{~~~}\forall c=1,2,  \label{stabi-1}
\end{equation}
and Eqs. (\ref{M}) and (\ref{N}) respectively simplify to
\begin{eqnarray}
&&
\begin{array}{l}
\mathcal{M}\equiv D^{2}V_{BH}= \\
\\
=3\overline{Z}D^{2}Z+g^{-1}\left( D^{3}Z\right) \overline{D}_{\overline{k}}%
\overline{Z}=e^{K}\left[ 3\overline{W}D^{2}W+g^{-1}\left( D^{3}W\right)
\overline{D}\overline{W}\right] = \\
\\
=4i\overline{Z}Cg^{-1}\overline{D}\overline{Z}+i\left( DC\right)
g^{-2}\left( \overline{D}\overline{Z}\right) ^{2}=ie^{K}g^{-1}\left(
\overline{D}\overline{W}\right) \left[ 4\overline{W}C+\left( DC\right) g^{-1}%
\overline{D}\overline{W}\right] ;
\end{array}
\label{M-1} \\
&&  \notag \\
&&  \notag \\
&&
\begin{array}{l}
\mathcal{N}\equiv D\overline{D}V_{BH}=\overline{D}DV_{BH}= \\
\\
=2g\left| Z\right| ^{2}+2\left| DZ\right| ^{2}+g^{-1}\left[ \left( \overline{%
D}D^{2}Z\right) \overline{D}\overline{Z}+\left| D^{2}Z\right| ^{2}\right] =
\\
\\
=e^{K}\left\{ 2g\left| W\right| ^{2}+2\left| DW\right| ^{2}+g^{-1}\left[
\left( \overline{D}D^{2}W\right) \overline{D}\overline{W}+\left|
D^{2}W\right| ^{2}\right] \right\} = \\
\\
=2\left[ g\left| Z\right| ^{2}+\left| DZ\right| ^{2}+\left| C\right|
^{2}g^{-3}\left| DZ\right| ^{2}\right] =2e^{K}\left[ g\left| W\right|
^{2}+\left| DW\right| ^{2}+\left| C\right| ^{2}g^{-3}\left| DW\right| ^{2}%
\right] ,
\end{array}
\label{N-1}
\end{eqnarray}
where $DC$ is given by the case $n_{V}=1$ case of Eq. (\ref{DC}) (also
recall definition (\ref{Christoffel-1})):
\begin{equation}
DC=\partial C+\left[ \left( \partial K\right) +3\Gamma \right] C=\partial C+%
\left[ \left( \partial K\right) -3\partial ln\left( g\right) \right]
C=\left\{ \partial +\left[ \partial ln\left( \frac{e^{K}}{\left( \overline{%
\partial }\partial K\right) ^{3}}\right) \right] \right\} C.  \label{DC-1}
\end{equation}
It is easy to show that the stability condition (\ref{stabi-1}) for critical
points of $V_{BH}$ in $n_{V}=1$ SKG can be equivalently reformulated as the
strict bound
\begin{equation}
\left. \mathcal{N}\right| _{\partial V_{BH}=0}>\left| \mathcal{M}\right|
_{\partial V_{BH}=0}.  \label{stab-1}
\end{equation}
Let us now see how such a bound can be further elaborated for the
three possible classes of critical points of $V_{BH}$.
\setcounter{equation}0
\def\theequation{3.2.\arabic{subsubsection}.\arabic{equation}}
\subsubsection{\label{stab-BPS}$\frac{1}{2}$-BPS}

\begin{eqnarray}
&&
\begin{array}{l}
\mathcal{M}_{\frac{1}{2}-BPS}\equiv \left. D^{2}V_{BH}\right| _{\frac{1}{2}%
-BPS}= \\
\\
=\left\{ e^{K}\left[ 3\overline{W}D^{2}W+g^{-1}\left( D^{3}W\right)
\overline{D}\overline{W}\right] \right\} _{\frac{1}{2}-BPS}=0;
\end{array}
\label{M-1-BPS} \\
&&  \notag \\
&&  \notag \\
&&
\begin{array}{l}
\mathcal{N}_{\frac{1}{2}-BPS}\equiv \left. D\overline{D}V_{BH}\right| _{%
\frac{1}{2}-BPS}= \\
\\
=\left[ e^{K}\left\{ 2g\left| W\right| ^{2}+g^{-1}\left| D^{2}W\right|
^{2}\right\} \right] _{\frac{1}{2}-BPS}=2\left[ ge^{K}\left| W\right| ^{2}%
\right] _{\frac{1}{2}-BPS}.
\end{array}
\label{N-1-BPS}
\end{eqnarray}
Eqs. (\ref{M-1-BPS}) and (\ref{N-1-BPS}) are nothing but the 1-modulus case
of Eq. (\ref{SUSY-crit}), and they directly satisfy the bound (\ref{stab-1}%
). Thus, consistently with what stated above, the $\frac{1}{2}$-BPS
class of critical points of $V_{BH}$ actually is a class of
attractors in strict sense (\textit{at least} local minima of
$V_{BH}$). \setcounter{equation}0
\def\theequation{3.2.\arabic{subsubsection}.\arabic{equation}}

\subsubsection{\label{stab-non-BPS-Z<>0}Non-BPS, $Z\neq 0$}

\begin{eqnarray}
&&
\begin{array}{l}
\mathcal{M}_{non-BPS,Z\neq 0}\equiv \left. D^{2}V_{BH}\right|
_{non-BPS,Z\neq 0}= \\
\\
=\left\{ e^{K}\left[ 3\overline{W}D^{2}W+g^{-1}\left( D^{3}W\right)
\overline{D}\overline{W}\right] \right\} _{non-BPS,Z\neq 0}= \\
\\
=\left\{ e^{K}\left[ -6g\left( DW\right) \frac{\overline{W}^{2}}{\overline{D}%
\overline{W}}+g^{-1}\left( D^{3}W\right) \overline{D}\overline{W}\right]
\right\} _{non-BPS,Z\neq 0}= \\
\\
=\left[ e^{K}\left| DW\right| ^{2}\left( -6\frac{g}{\overline{Dln\left(
W\right) }^{2}}+g^{-1}\frac{D^{3}W}{DW}\right) \right] _{non-BPS,Z\neq 0}=
\\
\\
=i\left\{ e^{K}g^{-1}\left[ 4\overline{W}C\overline{D}\overline{W}+\left(
DC\right) g^{-1}\left( \overline{D}\overline{W}\right) ^{2}\right] \right\}
_{non-BPS,Z\neq 0}= \\
\\
=-2\left\{ e^{K}g^{-1}\overline{W}DW\left[ g^{-2}\left| C\right|
^{2}Dln\left( W\right) +gDln\left( C\right) \right] \right\} _{non-BPS,Z\neq
0}= \\
\\
=i\left\{ e^{K}Cg^{-3}\left( \overline{D}\overline{W}\right) ^{2}\left[
g^{-2}\left| C\right| ^{2}Dln\left( W\right) +gDln\left( C\right) \right]
\right\} _{non-BPS,Z\neq 0};
\end{array}
\notag \\
&&  \label{M-1-non-BPS-Z<>0}
\end{eqnarray}
\begin{eqnarray}
&&
\begin{array}{l}
\mathcal{N}_{non-BPS,Z\neq 0}\equiv \left. D\overline{D}V_{BH}\right|
_{non-BPS,Z\neq 0}=\left. \overline{D}DV_{BH}\right| _{non-BPS,Z\neq 0}= \\
\\
=\left\{ e^{K}\left[ 2g\left| W\right| ^{2}+2\left| DW\right|
^{2}+g^{-1}\left( \overline{D}D^{2}W\right) \overline{D}\overline{W}%
+g^{-1}\left| D^{2}W\right| ^{2}\right] \right\} _{non-BPS,Z\neq 0}= \\
\\
=\left\{ e^{K}\left| DW\right| ^{2}\left[ 2+6g\frac{\left| W\right| ^{2}}{%
\left| DW\right| ^{2}}+g^{-1}\frac{\overline{D}D^{2}W}{DW}\right] \right\}
_{non-BPS,Z\neq 0}= \\
\\
=\left\{ e^{K}\left| DW\right| ^{2}\left[ 2+\frac{3}{2}g^{-3}\left| C\right|
^{2}+g^{-1}\frac{\overline{D}D^{2}W}{DW}\right] \right\} _{non-BPS,Z\neq 0}=
\\
\\
=2\left\{ e^{K}\left| DW\right| ^{2}\left[ 1+\frac{5}{4}g^{-3}\left|
C\right| ^{2}\right] \right\} _{non-BPS,Z\neq 0},
\end{array}
\notag \\
&&  \label{N-1-non-BPS-Z<>0}
\end{eqnarray}
where $D^{2}W$ is given by Eq. (\ref{DDW}) and
\begin{eqnarray}
\overline{D}D^{2}W &=&\overline{\partial }D^{2}W=\overline{\partial }\left\{
\left[ \partial ^{2}+\partial ^{2}K+2\partial K\partial +\left( \partial
K\right) ^{2}-\left( \partial ln\left( \partial \overline{\partial }K\right)
\right) \left( \partial +\partial K\right) \right] W\right\} =  \notag \\
&=&g^{-2}\left| C\right| ^{2}DW\neq D\overline{D}DW=gDW;
\end{eqnarray}
\begin{eqnarray}
D^{3}W &=&\partial D^{2}W+\left( \partial K\right) D^{2}W+2\Gamma D^{2}W=
\notag \\
&=&\partial \left\{ \left[ \partial ^{2}+\partial ^{2}K+2\partial K\partial
+\left( \partial K\right) ^{2}-\left( \partial ln\left( \partial \overline{%
\partial }K\right) \right) \left( \partial +\partial K\right) \right]
W\right\} +  \notag \\
&&+\left[ \left( \partial K\right) -2\left( \partial ln\left( \partial
\overline{\partial }K\right) \right) \right] \left\{ \left[ \partial
^{2}+\partial ^{2}K+2\partial K\partial +\left( \partial K\right)
^{2}-\left( \partial ln\left( \partial \overline{\partial }K\right) \right)
\left( \partial +\partial K\right) \right] W\right\}
\end{eqnarray}
Eq. (\ref{M-1-non-BPS-Z<>0}) yields that
\begin{eqnarray}
\left| \mathcal{M}\right| _{non-BPS,Z\neq 0}^{2} &=&\left\{ e^{2K}\left|
DW\right| ^{4}\left[ 36\frac{g^{2}}{\left| Dln\left( W\right) \right| ^{4}}%
+g^{-2}\frac{\left| D^{3}W\right| ^{2}}{\left| DW\right| ^{2}}-12Re\left(
\frac{D^{3}W}{DW}\frac{1}{\overline{Dln\left( W\right) }^{2}}\right) \right]
\right\} _{non-BPS,Z\neq 0}=  \notag \\
&&  \notag \\
&=&\left\{ e^{2K}\left| DW\right| ^{2}\left[ 36g^{2}\frac{\left| W\right|
^{4}}{\left| DW\right| ^{2}}+g^{-2}\left| D^{3}W\right| ^{2}-12Re\left[
\overline{W}\left( D^{3}W\right) \overline{Dln\left( W\right) }\right]
\right] \right\} _{non-BPS,Z\neq 0}=  \notag \\
&&  \notag \\
&=&4\left\{ e^{2K}\left| DW\right| ^{4}\left[ \left| C\right| ^{4}g^{-6}+%
\frac{1}{4}g^{-4}\left| DC\right| ^{2}+2g^{-3}Re\left[ C\left( \overline{D}%
\overline{C}\right) Dln\left( W\right) \right] \right] \right\}
_{non-BPS,Z\neq 0}.  \notag \\
&&  \label{M-non-BPS-Z<>0}
\end{eqnarray}
By substituting Eqs. (\ref{N-1-non-BPS-Z<>0}) and (\ref{M-non-BPS-Z<>0})
into the strict inequality (\ref{stab-1}), one finally obtains the stability
condition for non-BPS, $Z\neq 0$ critical points of $V_{BH}$ in $n_{V}=1$
SKG \cite{BFM}:
\begin{gather}
\mathcal{N}_{non-BPS,Z\neq 0}>\left| \mathcal{M}\right| _{non-BPS,Z\neq 0};
\\
\Updownarrow  \notag \\
\left[ 2\left| DW\right| +6g\frac{\left| W\right| ^{2}}{\left| DW\right| }%
+g^{-1}\left( \overline{D}D^{2}W\right) \overline{D}\overline{W}\right]
_{non-BPS,Z\neq 0}>\sqrt{\left\{
\begin{array}{l}
36g^{2}\frac{\left| W\right| ^{4}}{\left| DW\right| ^{2}}+g^{-2}\left|
D^{3}W\right| ^{2}+ \\
\\
-12Re\left[ \overline{W}\left( D^{3}W\right) \overline{Dln\left( W\right) }%
\right]
\end{array}
\right\} }_{non-BPS,Z\neq 0} \\
\Updownarrow  \notag \\
\notag \\
1+\frac{5}{4}\left( \left| C\right| ^{2}g^{-3}\right) _{non-BPS,Z\neq 0}>%
\sqrt{\left[ \left| C\right| ^{4}g^{-6}+\frac{1}{4}g^{-4}\left| DC\right|
^{2}+2g^{-3}Re\left[ C\left( \overline{D}\overline{C}\right) \left(
\overline{D}ln\overline{Z}\right) \right] \right] _{non-BPS,Z\neq 0}}.
\label{stab-1-1}
\end{gather}
As it is seen from such a condition, in general $\left( DC\right)
_{non-BPS,Z\neq 0}$ is the fundamental geometrical quantity playing a key
role in determining the stability of non-BPS, $Z\neq 0$ critical points of $%
V_{BH}$ in $1$-modulus SK geometry. \setcounter{equation}0
\def\theequation{3.2.\arabic{subsubsection}.\arabic{equation}}

\subsubsection{\label{stab-non-BPS-Z=0}Non-BPS, $Z=0$}

\begin{eqnarray}
&&
\begin{array}{l}
\mathcal{M}_{non-BPS,Z=0}\equiv \left. D^{2}V_{BH}\right| _{non-BPS,Z=0}= \\
\\
=\left[ e^{K}g^{-1}\left( D^{3}W\right) \overline{D}\overline{W}\right]
_{non-BPS,Z=0}=\left[ e^{K}g^{-1}\left( D^{3}W\right) \overline{\partial }%
\overline{W}\right] _{non-BPS,Z=0}= \\
\\
=i\left[ e^{K}g^{-2}\left( DC\right) \left( \overline{D}\overline{W}\right)
^{2}\right] _{non-BPS,Z=0}=i\left[ e^{K}g^{-2}\left( \partial C\right)
\left( \overline{\partial }\overline{W}\right) ^{2}\right] _{non-BPS,Z=0};
\end{array}
\label{M-1-non-BPS-Z=0} \\
&&  \notag \\
&&  \notag \\
&&
\begin{array}{l}
\mathcal{N}_{non-BPS,Z=0}\equiv \left. D\overline{D}V_{BH}\right|
_{non-BPS,Z=0}= \\
\\
=\left\{ e^{K}\left[ 2\left| DW\right| ^{2}+g^{-1}\left( \overline{\partial }%
D^{2}W\right) \overline{D}\overline{W}\right] \right\}
_{non-BPS,Z=0}=\left\{ e^{K}\left[ 2\left| \partial W\right|
^{2}+g^{-1}\left( \overline{\partial }D^{2}W\right) \overline{\partial }%
\overline{W}\right] \right\} _{non-BPS,Z=0}= \\
\\
=2e^{K}\left| DW\right| _{non-BPS,Z=0}^{2}=2e^{K}\left| \partial W\right|
_{non-BPS,Z=0}^{2},
\end{array}
\notag \\
&&  \label{N-1-non-BPS-Z=0}
\end{eqnarray}
where Eqs. (\ref{non-BPS-Z=0-2}) and (\ref{non-BPS-Z=0-1}) have been used.
Eq. (\ref{M-1-non-BPS-Z=0}) yields that
\begin{equation}
\left| \mathcal{M}\right| _{non-BPS,Z=0}=\left[ e^{K}g^{-1}\left|
D^{3}W\right| \left| \partial W\right| \right] _{non-BPS,Z=0}=\left[
e^{K}g^{-2}\left| \partial C\right| \left| \partial W\right| ^{2}\right]
_{non-BPS,Z=0}.  \label{M-non-BPS-Z=0}
\end{equation}
By substituting Eqs. (\ref{N-1-non-BPS-Z=0}) and (\ref{M-non-BPS-Z=0}) into
the strict inequality (\ref{stab-1}), one finally obtains the stability
condition for non-BPS, $Z=0$ critical points of $V_{BH}$ in $n_{V}=1$ SKG:
\begin{gather}
\mathcal{N}_{non-BPS,Z=0}>\left| \mathcal{M}\right| _{non-BPS,Z=0}; \\
\Updownarrow  \notag \\
2g_{non-BPS,Z=0}+\left( \frac{\overline{\partial }D^{2}W}{\partial W}\right)
_{non-BPS,Z=0}>\left| \frac{D^{3}W}{\partial W}\right| _{non-BPS,Z=0}; \\
\Updownarrow  \notag \\
2g_{non-BPS,Z=0}^{2}>\left| \partial C\right| _{non-BPS,Z=0}.
\label{stab-1-2}
\end{gather}
Even though the stability condition (\ref{stab-1-2}) have been obtained by
correctly using Eqs. (\ref{non-BPS-Z=0-2}) and (\ref{non-BPS-Z=0-1}),
holding at the non-BPS, $Z=0$ critical points of $V_{BH}$, in some cases it
may happen that, in the limit of approaching the non-BPS, $Z=0$ critical
point of $V_{BH}$, in $DC$ (given by Eq. (\ref{DC-1})) the ``connection
term'' $\left[ \left( \partial K\right) +3\Gamma \right] C$ is not
necessarily sub-leading with respect to the ``differential term'' $\partial
C $. Thus, for later convenience, it is worth rewriting the condition (\ref
{stab-1-2}) as follows:
\begin{equation}
2\left( \overline{\partial }\partial K\right) _{non-BPS,Z=0}^{2}>\left|
\left\{ \partial +\left[ \left( \partial K\right) -3\partial ln\left(
g\right) \right] \right\} C\right| _{non-BPS,Z=0}=\left| \left\{ \partial +%
\left[ \partial ln\left( \frac{e^{K}}{\left( \overline{\partial }\partial
K\right) ^{3}}\right) \right] \right\} C\right| _{non-BPS,Z=0}.
\label{stab-1-3}
\end{equation}
\medskip

\subsection*{Remark}

Let us consider the $1$-modulus stability conditions (\ref{stab-1-1}) and (%
\ref{stab-1-2})-\ref{stab-1-3}). It is immediate to realize that they are
both satisfied when the function $C$ is globally covariantly constant:
\begin{equation}
DC=\partial C+\left[ \left( \partial K\right) +3\Gamma \right] C=0,
\label{DC=0-1}
\end{equation}
\textit{i.e.} for the so-called homogeneous symmetric ($dim_{\mathbb{C}%
}=n_{V}=1$) SK geometry \cite{CKV,CVP}, univoquely associated to the coset
manifold $\frac{SU(1,1)}{U(1)}$. Such a SK manifold can be twofold
characterized as:

\textit{i}) the $n=0$ element of the irreducible rank-$1$ infinite sequence $%
\frac{SU(1,1+n)}{U(1)\otimes SU(1+n)}$ (with $n_{V}=n+rank=n+1$), endowed
with a quadratic holomorphic prepotential function reading (in a suitable
projective special coordinate, with K\"{a}hler gauge fixed such that $X^{0}=1
$; see \cite{BFGM1} and Refs. therein)
\begin{equation}
\mathcal{F}(z)=\frac{i}{4}\left( z^{2}-1\right) .  \label{quadr-1}
\end{equation}
By recalling the first of Eqs. (\ref{C}), such a prepotential yields $C=0$
globally (and thus Eq. (\ref{DC=0-1})), and therefore, by using the SKG
constraints (i.e. the third of Eqs. (\ref{C})) it yields also the constant
scalar curvature
\begin{equation}
\mathcal{R}\equiv g^{-2}R=-2,  \label{curv-quadr-1}
\end{equation}
where $R\equiv R_{1\overline{1}1\overline{1}}$ denotes the unique
component of the Riemann tensor. As obtained in \cite{BFGM1},
quadratic (homogeneous symmetric) SK geometries only admit
$\frac{1}{2}$-BPS and non-BPS, $Z=0$ critical points of $V_{BH}$.
Thus, it can be concluded that the $1$-dim. quadratic SK geometry
determined by the prepotential (\ref{quadr-1}) admits \textit{all}
stable critical points of $V_{BH}$.

\textit{ii}) the $n=-2$ element of the reducible rank-$3$ infinite sequence $%
\frac{SU(1,1)}{U(1)}\otimes \frac{SO(2,2+n)}{SO(2)\otimes SO(2+n)}$ (with $%
n_{V}=n+rank=n+3$). endowed with a cubic holomorphic prepotential function
reading (in a suitable projective special coordinate, with K\"{a}hler gauge
fixed such that $X^{0}=1$; see \cite{BFGM1} and Refs. therein; see also
observation below Eqs. (\ref{SUSY-breaking})-(\ref{SUSY-breaking2}))
\begin{equation}
\mathcal{F}(z)=\varrho z^{3},~\varrho \in \mathbb{C},  \label{cub-1}
\end{equation}
constrained by the condition $Im\left( z\right) <0$. By recalling the first
of Eqs. (\ref{C}), such a prepotential yields $C=6\varrho e^{K}$ (and thus
Eq. (\ref{DC=0-1})), and consequently it also yields the constant scalar
curvature
\begin{equation}
\mathcal{R}\equiv g^{-2}R=g^{-2}\left( -2g^{2}+g^{-1}\left| C\right|
^{2}\right) =-\frac{2}{3},  \label{curv-cub-1}
\end{equation}
where the SKG constraints (i.e. the third of Eqs. (\ref{C})) and the global
value\footnote{%
The global value $\left| C\right| ^{2}g^{-3}=\frac{4}{3}$ for homogeneous
symmetric cubic $n_{V}=1$ SK geometries actually is nothing but the $n_{V}=1$
case of the general global relation holding in a generic homogeneous
symmetric cubic $n_{V}$-dimensional SK geometry \cite{CVP}:
\begin{equation*}
C_{p(kl}C_{ij)n}g^{n\overline{n}}g^{p\overline{p}}\overline{C}_{\overline{n}%
\overline{p}\overline{m}}=C_{\left( p\right| (kl}C_{ij)\left| n\right) }g^{n%
\overline{n}}g^{p\overline{p}}\overline{C}_{\overline{n}\overline{p}%
\overline{m}}=\frac{4}{3}g_{\left( l\right| \overline{m}}C_{\left|
ijk\right) }.
\end{equation*}
} $\left| C\right| ^{2}g^{-3}=\frac{4}{3}$ have been used. As it can be
computed (see \textit{e.g.} \cite{Saraikin-Vafa-1}), the $1$-dim. SK
geometry determined by the prepotential (\ref{cub-1}) admits, beside the
(stable) $\frac{1}{2}$-BPS ones, stable non-BPS $Z\neq 0$ critical points of
$V_{BH}$. Thus, it is another example in which \textit{all} critical points
of $V_{BH}$ actually are attractors in a strict sense.\smallskip

Clearly, the quadratic and cubic homogeneous symmetric $1$-modulus SK
geometries (respectively determined by holomorphic prepotentials (\ref
{quadr-1}) and (\ref{cub-1})) are not the only ones (with $n_{V}=1$)
admitting stable non-BPS critical points of $V_{BH}$. As we will check in
Sects. \ref{quintic} and \ref{eightic}, for instance also the $1$-modulus SK
geometries of the moduli space of the (mirror) Fermat $CY_{3}$ \textit{%
quintic} $\mathcal{M}_{5}^{\prime }$ and \textit{octic} $\mathcal{M}%
_{8}^{\prime }$ admit, in a suitable neighbourhood of the LG point,
stable non-BPS ($Z\neq 0$) critical points of $V_{BH}$.
\setcounter{equation}0 \setcounter{equation}0
\def\theequation{\arabic{section}.\arabic{equation}}

\section{Holomorphic Geometry and Picard-Fuchs Equations\label{HG}}

In this and in the next Section we will present a summary of the holomorphic
geometry embedded in the SK geometry of the scalar manifolds of $N=2$, $d=4$
MESGTs. The main references for such an issue are \cite{Ferrara-Louis1} and
\cite{Ferrara-Lerche1}, to which we will refer at the relevant points of the
treatment.

The PF Equations, satisfied in SK geometry by the holomorphic period
vector (in a suitable basis, named \textit{PF basis}) are a
consequence of SK geometry and of the underlying symplectic
structure of the \textit{flat symplectic bundle }\cite{Strominger1}
(see also next Section), which encodes the differential relations
obeyed by the covariantly holomorphic sections and their covariant
derivatives. \setcounter{equation}0
\def\theequation{4.\arabic{subsection}.\arabic{equation}}

\subsection{\label{HG-n}$n_{V}$-Moduli}

Let us start by considering the K\"{a}hler-covariantly holomorphic,
symplectic $1\times \left( 2n_{V}+2\right) $ vector\footnote{%
In order to make the contact with the relevant literature easier, in this
Section, as well in the next one, we will change some notations with respect
to the previous treatment.
\par
Firstly, we will consider row (\textit{i.e.} $1\times \left( 2n_{V}+2\right)
$), instead of column (\textit{i.e.} $\left( 2n_{V}+2\right) \times 1$),
period vectors.
\par
Secondly, we will use lowercase Greek indices to denote homogeneous
coordinates (instead of lowercase Latin indices, as done in the previous
Section). Lowercase Latin indices will rather be used to denote indices
pertaining to the so-called \textit{holomorphic geometry} we are going to
discuss.}
\begin{equation}
V\left( z,\overline{z}\right) \equiv \left( L^{\Lambda }\left( z,\overline{z}%
\right) ,M_{\Lambda }\left( z,\overline{z}\right) \right) =e^{\frac{1}{2}%
K\left( z,\overline{z}\right) }\Pi ^{T}\left( z\right) ,  \label{V}
\end{equation}
with $\Pi ^{T}\left( z\right) $ defined by Eq. (\ref{PI-PI}).

\textit{Flatness} of the symplectic connection entails the following
relations \cite{Strominger1}:
\begin{equation}
\left\{
\begin{array}{l}
D_{\alpha }V=U_{\alpha }; \\
\\
D_{\alpha }U_{\beta }=iC_{\alpha \beta \gamma }g^{\gamma \overline{\gamma }}%
\overline{D}_{\overline{\gamma }}\overline{V}=iC_{\alpha \beta \gamma
}g^{\gamma \overline{\gamma }}\overline{U}_{\overline{\gamma }}; \\
\\
D_{\alpha }\overline{D}_{\overline{\beta }}\overline{V}=D_{\alpha }\overline{%
U}_{\overline{\beta }}=g_{\alpha \overline{\beta }}\overline{V}; \\
\\
D_{\alpha }\overline{V}=0.
\end{array}
\right.  \label{SKG-rels2}
\end{equation}
Notice that, by the definition (\ref{V}), the $\mathcal{N}=2$ central charge
function (defined by Eq. (\ref{Z})) can be rewritten (in the notation for
period vectors used in the present Section) as $Z=\widetilde{\Gamma }\Omega
V^{T}$, and the defining relations (\ref{SKG-rels1}) of SK geometry can thus
be obtained by transposing the relations (\ref{SKG-rels2}) and by further
left-multiplying them by the $1\times \left( 2n_{V}+2\right) $ vector $%
\widetilde{\Gamma }\Omega $.

Let us now consider a new $1\times \left( 2n_{V}+2\right) $ vector of
holomorphic sections\footnote{%
The subscript ``$h$'' stands for ``holomorphic''.} ($a=1,...,n_{V}$)
\begin{equation}
V_{h}\left( X(z)\right) \equiv \left( X^{0}\left( z\right) ,X^{a}\left(
z\right) ,F_{a}\left( X(z)\right) ,-F_{0}\left( X(z)\right) \right) .
\label{Vh}
\end{equation}

We notice that, while $V\left( z,\overline{z}\right) $ defined in Eq. (\ref
{V}) is symplectic with respect to the symplectic metric $\Omega $, this
does not hold for $V_{h}\left( X(z)\right) $ defined in Eq. (\ref{Vh}),
which is instead symplectic with respect to a newly defined anti-diagonal
symplectic metric ($Q^{T}=-Q$, $Q^{2}=-\mathbb{I}_{2n_{V}+2}$):
\begin{equation}
Q\equiv \left(
\begin{array}{cccc}
&  &  & 1 \\
&  & -\mathbb{I}_{n_{V}} &  \\
& \mathbb{I}_{n_{V}} &  &  \\
-1 &  &  &
\end{array}
\right) ,  \label{Q}
\end{equation}
where unwritten elements vanish.

In the treatment which follows we will assume the existence of an
holomorphic prepotential $F\left( X\left( z\right) \right) $ of $\mathcal{N}%
=2$, $d=4$ vector multiplet couplings such that
\begin{equation}
F_{\Lambda }\left( z\right) \equiv \frac{\partial F\left( X\left( z\right)
\right) }{\partial X^{\Lambda }},  \label{21march-afternoon-1}
\end{equation}
which is in turn implied by the assumption that the holomorphic square
matrix
\begin{equation}
e_{\alpha }^{a}\left( z\right) \equiv \frac{\partial \left[ \frac{%
X^{a}\left( z\right) }{X^{0}\left( z\right) }\right] }{\partial z^{\alpha }}%
\equiv \frac{\partial t^{a}\left( z\right) }{\partial z^{\alpha }}=\frac{%
\partial _{\alpha }X^{a}\left( z\right) }{X^{0}\left( z\right) }-\frac{%
X^{a}\left( z\right) }{\left( X^{0}\left( z\right) \right) ^{2}}\partial
_{\alpha }X^{0}(z)  \label{e}
\end{equation}
is invertible (non-singular), where in the last step we introduced the
\textit{homogeneous} (K\"{a}hler-invariant) coordinates (see \textit{e.g.}
\cite{CDF})
\begin{equation}
t^{a}\left( z\right) \equiv \frac{X^{a}\left( z\right) }{X^{0}\left(
z\right) }.  \label{t}
\end{equation}
The matrix $e_{\alpha }^{a}\left( z\right) $ is nothing but the Jacobian of
the change of basis between the $t^{a}\left( z\right) $s and the $z^{\alpha
} $s. \textit{Special (symplectic) coordinates} correspond to the case $%
e_{\alpha }^{a}\left( z\right) =\delta _{\alpha }^{a}$, implying that $%
t^{a}\left( z\right) \equiv \frac{X^{a}\left( z\right) }{X^{0}\left(
z\right) }=z^{a}$ (in such a case $a$-indices and $\alpha $-indices do
coincide). By further fixing the K\"{a}hler gauge such that $X^{0}=1$, one
finally gets $t^{a}\left( z\right) =X^{a}\left( z\right) =z^{a}$ and $%
X^{0}=1 $, which is the usual definition of special coordinates (yielding $%
\partial _{\alpha }X^{\Lambda }=\delta _{\alpha }^{a}$).

In special coordinates, the holomorphic period vector $V_{h}\left(
X(z)\right) $ introduced in Eq. (\ref{Vh}) reads as follows (K\"{a}hler
gauge $X^{0}=1$ fixed understood throughout, unless otherwise noted; here
and below the subscript ``$sp.$'' stands for ``special''):
\begin{equation}
V_{h,sp.}\left( z\right) \equiv \left( 1,z^{a},\partial _{a}\mathcal{F}%
\left( z\right) ,-\mathcal{F}_{0}\left( z\right) \right) =\left(
1,z^{a},\partial _{a}\mathcal{F}\left( z\right) ,z^{a}\partial _{a}\mathcal{F%
}\left( z\right) -2\mathcal{F}\left( z\right) \right) ,  \label{Vh-special}
\end{equation}
where
\begin{equation}
\mathcal{F}\left( z\right) \equiv F\left( \frac{X(z)}{X^{0}(z)}\right)
=\left( X^{0}(z)\right) ^{-2}F\left( X(z)\right)  \label{F-call}
\end{equation}
is the K\"{a}hler gauge-invariant holomorphic prepotential (here in special
coordinates and for $X^{0}=1$). In the second step the homogeneity of degree
2 of the prepotential was used; for general symplectic and special ($X^{0}=1$%
) coordinates it respectively reads
\begin{equation}
\begin{array}{l}
X^{\Lambda }\partial _{\Lambda }F(X)=X^{0}\partial _{0}F\left( X\right)
+X^{a}\partial _{a}F\left( X\right) =2F(X); \\
\\
\mathcal{F}_{0}\left( z\right) +z^{a}\partial _{a}\mathcal{F}\left( z\right)
=2\mathcal{F}(z).
\end{array}
\label{hom-deg-2-F}
\end{equation}
By starting from Eq. (\ref{Vh-special}) and by differentiating once and
twice $V_{h,sp.}\left( z\right) $, one respectively achieves
\begin{eqnarray}
&&
\begin{array}{l}
\partial _{b}V_{h,sp.}\left( z\right) =\left( 0,\delta _{b}^{a},\partial
_{a}\partial _{b}\mathcal{F}\left( z\right) ,-\partial _{b}\mathcal{F}\left(
z\right) +z^{a}\partial _{a}\partial _{b}\mathcal{F}\left( z\right) \right) ;
\end{array}
\label{dbVh-special} \\
&&  \notag \\
&&
\begin{array}{l}
\partial _{b}\partial _{c}V_{h,sp.}\left( z\right) =\left( 0,0,\partial
_{a}\partial _{b}\partial _{c}\mathcal{F}\left( z\right) ,z^{a}\partial
_{a}\partial _{b}\partial _{c}\mathcal{F}\left( z\right) \right) ,
\end{array}
\end{eqnarray}
implying that
\begin{equation}
\begin{array}{l}
\partial _{b}\partial _{c}V_{h,sp.}\left( z\right) =W_{abc,sp.}\left(
z\right) V_{h,sp.}^{a}\left( z\right) ; \\
\\
\partial _{a}V_{h,sp.}^{b}\left( z\right) =\delta _{a}^{b}V_{h,sp.}^{0},
\end{array}
\label{res1}
\end{equation}
where $W_{abc,sp.}\left( z\right) \equiv \partial _{a}\partial _{b}\partial
_{c}\mathcal{F}\left( z\right) $ is the holomorphic part of $C_{\alpha \beta
\gamma }$ in special coordinates and for $X^{0}=1$ (see the second row of
the first of relations (\ref{C})) and
\begin{eqnarray}
&&
\begin{array}{l}
V_{h,sp.}^{a}(z)\equiv \left( 0,0,\delta _{d}^{a},z^{a}\right) ,
\end{array}
\label{Vah-special} \\
&&  \notag \\
&&
\begin{array}{l}
V_{h,sp.}^{0}(z)\equiv \left( 0,0,0,1\right) .
\end{array}
\label{V0h-special}
\end{eqnarray}
By adding the definition $V_{h,sp.,a}\left( z\right) \equiv \partial
_{a}V_{h,sp.}(z)$ and the\ trivial result $\partial _{a}V_{h,sp.}^{0}(z)=0$
to Eqs. (\ref{res1}), one finally gets the set of differential relations
\cite{Ferrara-Lerche1}
\begin{equation}
\begin{array}{l}
\partial _{a}V_{h,sp.}(z)=V_{h,sp.,a}\left( z\right) , \\
\\
\partial _{a}\partial _{b}V_{h,sp.}(z)=\partial _{a}V_{h,sp.,b}\left(
z\right) =W_{abc,sp.}\left( z\right) V_{h,sp.}^{c}\left( z\right) , \\
\\
\partial _{a}V_{h,sp.}^{b}\left( z\right) =\delta _{a}^{b}V_{h,sp.}^{0}, \\
\\
\partial _{a}V_{h,sp.}^{0}(z)=0,
\end{array}
\label{holomorphic-rels-special}
\end{equation}
which are the holomorphic counterparts of SK relations (\ref{SKG-rels2}),
written in special coordinates and for $X^{0}=1$.

By \textit{``holomorphically covariantizing''} the relations (\ref
{holomorphic-rels-special}), \textit{i.e.} by writing them in a generic
system of homogeneous coordinates, one obtains (notice that here $a$-indices
and $\alpha $-indices in general do not coincide) \cite{Ferrara-Lerche1}
\begin{equation}
\begin{array}{l}
\widehat{D}_{\alpha }V_{h}(z)=V_{h,\alpha }(z), \\
\\
\widehat{D}_{\alpha }\widehat{D}_{\beta }V_{h}(z)=\widehat{D}_{\alpha
}V_{h,\beta }(z)=W_{\alpha \beta \gamma }\left( z\right) V_{h}^{\gamma }(z),
\\
\\
\widehat{D}_{\alpha }V_{h}^{\beta }(z)=\delta _{\alpha }^{\beta
}V_{h}^{0}\left( z\right) , \\
\\
\widehat{D}_{\alpha }V_{h}^{0}(z)=0,
\end{array}
\label{holomorphic-rels-homogeneous}
\end{equation}
where $V_{h}(z)$, $V_{h,\beta }(z)$, $V_{h}^{\beta }(z)$ and $V_{h}^{0}(z)$
are respectively given by the following formul\ae \footnote{%
The first of relations (\ref{jazz}) corresponds to Eq. (\ref{Vh}) by using
the definition of $t^{a}\left( z\right) $s and the homogeneity of degree 2
of the prepotential $F$.} \cite{Ferrara-Lerche1}:
\begin{eqnarray}
&&
\begin{array}{l}
V_{h}(z)=\left( X^{0}(z),X^{a}(z),X^{0}(z)e_{a}^{\alpha }(z)\partial
_{\alpha }\mathcal{F}(z),X^{a}(z)e_{a}^{\alpha }(z)\partial _{\alpha }%
\mathcal{F}(z)-2X^{0}(z)\mathcal{F}(z)\right) ; \\
\\
V_{h,\beta }(z)=\widehat{D}_{\beta }V_{h}(z)=\left( 0,X^{0}(z)e_{\beta
}^{a}(z),X^{0}(z)e_{a}^{\alpha }(z)\widehat{D}_{\alpha }\partial _{\beta }%
\mathcal{F}(z),-X^{0}(z)\partial _{\beta }\mathcal{F}(z)+X^{a}(z)e_{a}^{%
\alpha }(z)\widehat{D}_{\alpha }\partial _{\beta }\mathcal{F}(z)\right) ; \\
\\
V_{h}^{\beta }(z)=\left( 0,0,\left( X^{0}(z)\right) ^{-1}e_{a}^{\beta
}(z),\left( X^{0}(z)\right) ^{-2}X^{a}(z)e_{a}^{\beta }(z)\right) ; \\
\\
V_{h}^{0}(z)=\left( 0,0,0,\left( X^{0}(z)\right) ^{-1}\right) ,
\end{array}
\notag \\
&&  \label{jazz}
\end{eqnarray}
which correspond to the \textit{``holomorphically covariantized''}
counterparts of Eqs. (\ref{Vh-special}), (\ref{dbVh-special}), (\ref
{Vah-special}) and (\ref{V0h-special}), respectively. $e_{a}^{\alpha
}(z)\equiv e_{a}^{\alpha }(t\left( z\right) )$ is the inverse of $e_{\alpha
}^{a}(z)$ (defined by Eq. (\ref{e})):
\begin{equation}
e_{a}^{\alpha }(t\left( z\right) )\frac{\partial t^{a}\left( z\right) }{%
\partial z^{\beta }}\equiv \delta _{\beta }^{\alpha },~e_{a}^{\alpha
}(t\left( z\right) )\frac{\partial t^{b}\left( z\right) }{\partial z^{\alpha
}}\equiv \delta _{a}^{b}\Longleftrightarrow e_{a}^{\alpha }(t\left( z\right)
)\equiv \left. \frac{\partial z^{\alpha }\left( t\right) }{\partial t^{a}}%
\right| _{t=t(z)}.
\end{equation}

$W_{\alpha \beta \gamma }\left( z\right) $ (completely symmetric, with
K\"{a}hler weights $\left( 4,0\right) $, and defined in the second row of
the first of relations (\ref{C})) is the covariant part of $C_{\alpha \beta
\gamma }\left( z,\overline{z}\right) $ in generic homogeneous coordinates.
It can be further elaborated as follows:
\begin{eqnarray}
W_{\alpha \beta \gamma }\left( z\right) &\equiv &\left( \partial _{\alpha
}X^{\Lambda }\left( z\right) \right) \left( \partial _{\beta }X^{\Sigma
}\left( z\right) \right) \left( \partial _{\gamma }X^{\Xi }\left( z\right)
\right) \left. \partial _{\Xi }\partial _{\Sigma }\partial _{\Lambda
}F\left( X\right) \right| _{X=X(z)}=  \label{W-1} \\
&=&\left( X^{0}\left( z\right) \right) ^{2}e_{\alpha }^{a}\left( z\right)
e_{\beta }^{b}\left( z\right) e_{\gamma }^{c}\left( z\right) \left. \partial
_{c}\partial _{b}\partial _{a}\mathcal{F}\left( t\right) \right| _{t=t(z)}=
\label{W-2} \\
&&  \notag \\
&=&\left( X^{0}\left( z\right) \right) ^{2}\left[
\begin{array}{l}
\partial _{\gamma }\partial _{\beta }\partial _{\alpha }\mathcal{F}\left(
z\right) + \\
\\
-\left( \partial _{\gamma }e_{\beta }^{b}\left( z\right) \right)
e_{b}^{\beta ^{\prime }}\left( t\left( z\right) \right) \partial _{\beta
^{\prime }}\partial _{\alpha }\mathcal{F}\left( z\right) -\left( \partial
_{\gamma }e_{\alpha }^{a}\left( z\right) \right) e_{a}^{\alpha ^{\prime
}}\left( t\left( z\right) \right) \partial _{\beta }\partial _{\alpha
^{\prime }}\mathcal{F}\left( z\right) + \\
\\
-\left( \partial _{\beta }e_{\alpha }^{a}\left( z\right) \right)
e_{a}^{\alpha ^{\prime }}\left( t\left( z\right) \right) \partial _{\gamma
}\partial _{\alpha ^{\prime }}\mathcal{F}\left( z\right) +e_{\alpha
}^{a}\left( z\right) \left( \partial _{\gamma }\partial _{\beta
}e_{a}^{\alpha ^{\prime }}\left( t\left( z\right) \right) \right) \partial
_{\alpha ^{\prime }}\mathcal{F}\left( z\right) + \\
\\
+\left( \partial _{\beta ^{\prime }}e_{\alpha }^{a}\left( z\right) \right)
\left( \partial _{\gamma }e_{\beta }^{b}\left( z\right) \right) e_{b}^{\beta
^{\prime }}\left( t\left( z\right) \right) e_{a}^{\alpha ^{\prime }}\left(
t\left( z\right) \right) \partial _{\alpha ^{\prime }}\mathcal{F}\left(
z\right)
\end{array}
\right] .  \label{W-3}
\end{eqnarray}

Notice that a new holomorphic covariant derivative $\widehat{D}_{\alpha }$
has been introduced. In analogy with the usual covariant derivative in
K\"{a}hler-Hodge manifold, the action of $\widehat{D}_{\alpha }$ on a vector
$\phi _{\beta }$ with K\"{a}hler weight $p$ reads \cite
{Ferrara-Louis1,Ferrara-Lerche1}
\begin{equation}
\widehat{D}_{\alpha }\phi _{\beta }\left( z,\overline{z}\right) =\left(
\partial _{\alpha }+\frac{p}{2}\widehat{K}_{\alpha }\left( z\right) \right)
\phi _{\beta }\left( z,\overline{z}\right) +\widehat{\Gamma }_{\alpha \beta
}^{~~~\gamma }\left( z\right) \phi _{\gamma }\left( z,\overline{z}\right) ,
\end{equation}
where $\widehat{\Gamma }_{\alpha \beta }^{~~~\gamma }\left( z\right) $ is
the holomorphic part of the Christoffel connection $\Gamma _{\alpha \beta
}^{~~~\gamma }\left( z,\overline{z}\right) $ of the SK manifold being
considered \cite{Ferrara-Louis1,Ferrara-Lerche1} ($e_{\alpha }^{a}\left(
z\right) e_{a}^{\gamma }(z)=\delta _{\alpha }^{\gamma }$, $e_{\alpha
}^{a}\left( z\right) e_{b}^{\alpha }(z)=\delta _{b}^{a}$; also recall
definition (\ref{Christoffel-n})):
\begin{eqnarray}
\widehat{\Gamma }_{\alpha \beta }^{~~~\gamma }\left( z\right) &\equiv
&-\left( \partial _{\beta }e_{\alpha }^{a}\left( z\right) \right)
e_{a}^{\gamma }(z)=\Gamma _{\alpha \beta }^{~~~\gamma }\left( z,\overline{z}%
\right) -T_{\alpha \beta }^{~~~\gamma }\left( z,\overline{z}\right) \equiv
\label{Gamma-hat} \\
&&  \notag \\
&\equiv &-g^{\gamma \overline{\gamma }}\left( z,\overline{z}\right) \partial
_{\alpha }\partial _{\beta }\overline{\partial }_{\overline{\gamma }}K\left(
z,\overline{z}\right) +e_{\alpha }^{a}(z)e_{\beta }^{b}(z)\left[ \frac{%
\partial ^{3}K\left( t\left( z\right) ,\overline{t}\left( \overline{z}%
\right) \right) }{\partial t^{b}\partial t^{a}\overline{\partial }\overline{t%
}^{\overline{d}}}\right] g^{c\overline{d}}\left( z,\overline{z}\right)
e_{c}^{\gamma }\left( z\right) .  \label{Gamma-hat2}
\end{eqnarray}
It can be checked that $\widehat{\Gamma }_{\alpha \beta }^{~~~\gamma }\left(
z\right) $ transforms as a connection under holomorphic reparametrizations. $%
\widehat{\Gamma }_{\alpha \beta }^{~~~\gamma }\left( z\right) $ is usually
called the \textit{Christoffel holomorphic connection}. By its very
definition (\ref{Gamma-hat}), it follows that $e_{\alpha }^{a}\left(
z\right) $ can be seen as the \textit{Christoffel holomorphic }$n_{V}$%
\textit{-bein}.

It is worth pointing out that the $\widehat{\Gamma }_{\alpha \beta
}^{~~~\gamma }$s are the Christoffel symbols of the second kind of an
holomorphic Riemann metric
\begin{equation}
\widehat{g}_{\alpha \beta }\left( z\right) \equiv e_{\alpha }^{a}(z)e_{\beta
}^{b}(z)\eta _{ab}=\frac{\partial \left[ \frac{X^{a}\left( z\right) }{%
X^{0}\left( z\right) }\right] }{\partial z^{\alpha }}\frac{\partial \left[
\frac{X^{b}\left( z\right) }{X^{0}\left( z\right) }\right] }{\partial
z^{\beta }}\eta _{ab},  \label{g-hat}
\end{equation}
where $\eta _{ab}$ is constant (invertible) symmetric matrix (note that $%
\widehat{g}_{\alpha \beta }\left( z\right) $ has two holomorphic indices, in
contrast to the K\"{a}hler metric $g_{\alpha \overline{\beta }}\left( z,%
\overline{z}\right) =\partial _{\alpha }\overline{\partial }_{\overline{%
\beta }}K(z,\overline{z})$). $\widehat{g}_{\alpha \beta }\left( z\right) $
is the metric tensor of the so-called \textit{holomorphic geometry}
``embedded'' in the considered SK geometry. Due to Eq. (\ref{e}), it can be
checked that $\widehat{\Gamma }_{\alpha \beta }^{~~~\gamma }\left( z\right) $
is actually a Riemann-flat connection, since it holds that
\begin{equation}
\widehat{R}_{\delta \alpha \beta }^{~~~~~~\gamma }\left( z\right) \equiv
\partial _{\delta }\widehat{\Gamma }_{\alpha \beta }^{~~~\gamma }\left(
z\right) -\partial _{\alpha }\widehat{\Gamma }_{\delta \beta }^{~~~\gamma
}\left( z\right) -\widehat{\Gamma }_{\alpha \beta }^{~~~\zeta }\left(
z\right) \widehat{\Gamma }_{\zeta \delta }^{~~~\gamma }\left( z\right) +%
\widehat{\Gamma }_{\delta \beta }^{~~~\zeta }\left( z\right) \widehat{\Gamma
}_{\zeta \alpha }^{~~~\gamma }\left( z\right) =0.
\end{equation}

Moreover, since $X^{0}\left( z\right) $ has K\"{a}hler weights $\left(
2,0\right) $, the quantity
\begin{equation}
\widehat{K}_{\alpha }\left( z\right) \equiv -\partial _{\alpha }\left[
ln\left( X^{0}\left( z\right) \right) \right]  \label{Kappa-hat}
\end{equation}
transforms as a connection under K\"{a}hler gauge transformations:
\begin{equation}
K\left( z,\overline{z}\right) \longrightarrow K\left( z,\overline{z}\right)
+f(z)+\overline{f}(\overline{z})\Longrightarrow \widehat{K}_{\alpha }\left(
z\right) \longrightarrow \widehat{K}_{\alpha }\left( z\right) +\partial
_{\alpha }f(z).
\end{equation}
$\widehat{K}_{\alpha }\left( z\right) $ is usually called the \textit{%
K\"{a}hler holomorphic connection}. By its very definition (\ref{Kappa-hat}%
), it thus follows that $X^{0}\left( z\right) $ can be seen as the \textit{%
K\"{a}hler holomorphic }$1$\textit{-bein}.

In the formalism of \textit{holomorphic geometry }introduced above, it can
be easily checked that the quantity $e_{a}^{\alpha }\left( z\right) \widehat{%
D}_{\alpha }\partial _{\beta }\mathcal{F}(z)$, appearing in the vector $%
V_{h,\beta }(z)$ introduced in the second of relations (\ref{jazz}), can be
further elaborated as follows:
\begin{equation}
e_{a}^{\alpha }\left( z\right) \widehat{D}_{\alpha }\partial _{\beta }%
\mathcal{F}(z)=e_{a}^{\alpha }\left( z\right) \widehat{D}_{\alpha }\widehat{D%
}_{\beta }\mathcal{F}(z)=\partial _{\beta }\left[ e_{a}^{\alpha }\left(
z\right) \partial _{\alpha }\mathcal{F}(z)\right] =e_{a}^{\alpha }\left(
z\right) \widehat{D}_{\beta }\partial _{\alpha }\mathcal{F}(z)=e_{a}^{\alpha
}\left( z\right) \widehat{D}_{\beta }\widehat{D}_{\alpha }\mathcal{F}(z).
\label{wow1}
\end{equation}

Finally, it should observed that special coordinates are \textit{flat}
coordinates for the holomorphic geometry, because for special coordinates ($%
e_{\alpha }^{a}\left( z\right) =\delta _{\alpha }^{a}$) (in the K\"{a}hler
gauge $X^{0}=1$) Eqs. (\ref{Gamma-hat}), (\ref{Kappa-hat}) and (\ref{g-hat})
respectively reduce to
\begin{eqnarray}
&&
\begin{array}{l}
\widehat{\Gamma }_{\alpha \beta }^{~~~\gamma }\left( z\right) =0;
\end{array}
\\
&&
\begin{array}{l}
\widehat{K}_{\alpha }\left( z\right) =0;
\end{array}
\\
&&
\begin{array}{l}
\widehat{g}_{\alpha \beta }\left( z\right) =\eta _{\alpha \beta }.
\end{array}
\end{eqnarray}

It is worth pointing out that the system (\ref{holomorphic-rels-homogeneous}%
) is the holomorphic counterpart of the system (\ref{SKG-rels2}), and it is
manifestly covariant with respect to the holomorphic geometry defined by $%
\widehat{g}_{\alpha \beta }\left( z\right) $ and $\widehat{K}_{\alpha
}\left( z\right) $. By breaking the \textit{``holomorphic covariance''} and
choosing special coordinates (and fixing K\"{a}hler gauge such that $X^{0}=1$%
), the system (\ref{holomorphic-rels-homogeneous}) reduces to the system (%
\ref{holomorphic-rels-special}). The system of holomorphic
differential relations (\ref{holomorphic-rels-homogeneous}) is
usually referred to as the (holomorphic) \textit{Picard-Fuchs (PF)
system}. \setcounter{equation}0
\def\theequation{4.\arabic{subsection}.\arabic{equation}}

\subsection{\label{HG-1}$1$-Modulus}

Let us now specialize the treatment to the 1-modulus case. Once again, such
a case is peculiarly simple, since for $n_{V}=1$ the Christoffel holomorphic
$n_{V}$-bein is nothing but an holomorphic function ($z^{1}\equiv \psi $, $%
t^{1}\equiv t$, $\partial _{\psi }\equiv \partial $):
\begin{equation}
e_{\psi }^{1}\left( \psi \right) \equiv e\left( \psi \right) \equiv \frac{%
\partial \left[ \frac{X^{1}\left( \psi \right) }{X^{0}\left( \psi \right) }%
\right] }{\partial \psi }\equiv \frac{\partial t\left( \psi \right) }{%
\partial \psi }=\frac{X^{0}\left( \psi \right) \partial X^{1}\left( \psi
\right) -X^{1}\left( \psi \right) \partial X^{0}\left( \psi \right) }{\left[
X^{0}\left( \psi \right) \right] ^{2}},  \label{e-1-modulus}
\end{equation}
and the connections and metric of holomorphic geometry reduce to
\begin{equation}
\begin{array}{l}
\widehat{\Gamma }_{\psi \psi }^{~~~\psi }\left( \psi \right) \equiv \widehat{%
\Gamma }\left( \psi \right) =-\partial \left[ ln\left( e\left( \psi \right)
\right) \right] ; \\
\\
\widehat{K}_{\psi }\left( \psi \right) \equiv \widehat{K}\left( \psi \right)
=-\partial \left[ ln\left( X^{0}\left( \psi \right) \right) \right] ; \\
\\
\widehat{g}_{\psi \psi }\left( \psi \right) \equiv \widehat{g}\left( \psi
\right) =\left[ e\left( \psi \right) \right] ^{2}\eta ,~\eta \in \mathbb{C}%
_{0}.
\end{array}
\label{jazz2}
\end{equation}
Consequently, the action of $\widehat{D}_{\psi }\equiv \widehat{D}$ on a
1-vector (function) $\phi \left( \psi ,\overline{\psi }\right) $ with
K\"{a}hler weights $\left( p,\overline{p}\right) $ reads
\begin{eqnarray}
\widehat{D}\phi \left( \psi ,\overline{\psi }\right) &=&\left( \partial +%
\frac{p}{2}\widehat{K}\left( \psi \right) +\widehat{\Gamma }\left( \psi
\right) \right) \phi \left( \psi ,\overline{\psi }\right) =  \notag \\
&&  \notag \\
&=&\left\{ \partial -\frac{p}{2}\partial \left[ ln\left( X^{0}\left( \psi
\right) \right) \right] -\partial \left[ ln\left( e\left( \psi \right)
\right) \right] \right\} \phi \left( \psi ,\overline{\psi }\right) .
\label{1-modulus-cov-der}
\end{eqnarray}

It can be shown \cite{Ferrara-Louis1,Ferrara-Lerche1} that in the 1-modulus
case the PF system (\ref{holomorphic-rels-homogeneous}) is equivalent to the
following complex differential relation:
\begin{equation}
\widehat{D}^{2}\left\{ \left[ \mathcal{W}\left( \psi \right) \right] ^{-1}%
\widehat{D}^{2}V_{h}\left( \psi \right) \right\} =0,  \label{PF1}
\end{equation}
where (see Eqs. (\ref{W-1})-(\ref{W-3}), and the second row of the first of
relations (\ref{C}))
\begin{equation}
\begin{array}{l}
\mathcal{W}\left( \psi \right) \equiv W_{\psi \psi \psi }\left( \psi \right)
=e^{-K\left( \psi ,\overline{\psi }\right) }C\left( \psi ,\overline{\psi }%
\right) = \\
\\
=\left[ \partial X^{0}\left( \psi \right) \right] ^{3}\left. \frac{\partial
^{3}F\left( X\right) }{\left( \partial X^{0}\right) ^{3}}\right| _{X=X\left(
\psi \right) }+3\left[ \partial X^{0}\left( \psi \right) \right] ^{2}\left[
\partial X^{1}\left( \psi \right) \right] \left. \frac{\partial ^{3}F\left(
X\right) }{\left( \partial X^{0}\right) ^{2}\partial X^{1}}\right|
_{X=X\left( \psi \right) }+ \\
\\
+3\left[ \partial X^{1}\left( \psi \right) \right] ^{2}\left[ \partial
X^{0}\left( \psi \right) \right] \left. \frac{\partial ^{3}F\left( X\right)
}{\left( \partial X^{1}\right) ^{2}\partial X^{0}}\right| _{X=X\left( \psi
\right) }+\left[ \partial X^{1}\left( \psi \right) \right] ^{3}\left. \frac{%
\partial ^{3}F\left( X\right) }{\left( \partial X^{1}\right) ^{3}}\right|
_{X=X\left( \psi \right) }= \\
\\
=\left( X^{0}\left( \psi \right) \right) ^{2}\left[ e\left( \psi \right)
\right] ^{3}\left. \frac{\partial ^{3}\mathcal{F}\left( t\right) }{\left(
\partial t\right) ^{3}}\right| _{t=t(\psi )}= \\
\\
=\left( X^{0}\left( \psi \right) \right) \left[
\begin{array}{l}
3\left[ e\left( \psi \right) \right] ^{-2}\left[ \partial e\left( \psi
\right) \right] ^{2}\partial \mathcal{F}\left( \psi \right) -\left[ e\left(
\psi \right) \right] ^{-1}\left[ \partial ^{2}e\left( \psi \right) \right]
\partial \mathcal{F}\left( \psi \right) + \\
\\
-3\left[ e\left( \psi \right) \right] ^{-1}\left[ \partial e\left( \psi
\right) \right] \partial ^{2}\mathcal{F}\left( \psi \right) +\partial ^{3}%
\mathcal{F}\left( \psi \right)
\end{array}
\right] ,
\end{array}
\label{W-1-modulus}
\end{equation}
and (see Eq. (\ref{Vh}) and the first of relations (\ref{jazz}))
\begin{eqnarray}
V_{h}\left( \psi \right) &=&\left( X^{0}\left( \psi \right) ,X^{1}\left(
\psi \right) ,F_{1}\left( X(\psi )\right) ,-F_{0}\left( X(\psi )\right)
\right) =  \notag \\
&=&\left( X^{0}(\psi ),X^{1}(\psi ),X^{0}(\psi )\left[ e(\psi )\right]
^{-1}\partial \mathcal{F}(\psi ),X^{1}(\psi )\left[ e(\psi )\right]
^{-1}\partial \mathcal{F}(\psi )-2X^{0}(\psi )\mathcal{F}(\psi )\right) .
\label{jazzit}
\end{eqnarray}

Eq. (\ref{PF1}) can also be rewritten as a fourth order linear ordinary
differential equation for $V_{h}\left( \psi \right) $ (\textit{1-modulus PF
Eq.}) \cite{Ferrara-Louis1,Ferrara-Lerche1}:
\begin{equation}
\sum_{n=0}^{4}a_{n}\left( \psi \right) \partial ^{n}V_{h}\left( \psi \right)
=0,  \label{PF2}
\end{equation}
where $\partial ^{n}\equiv \frac{\partial ^{n}}{\left( \partial \psi \right)
^{n}}$ ($n=0$ corresponds to the identity operator)\footnote{%
For a general treatment of the $n_{V}$-moduli case, see \cite
{Ferrara-Lerche1}.}. The functions $a_{n}\left( \psi \right) $ can be
obtained by comparing Eqs. (\ref{PF1}) and (\ref{PF2}) \cite{Ferrara-Louis1}
(notation of dependence on $\psi $ omitted throughout):
\begin{eqnarray}
&&
\begin{array}{l}
a_{4}\equiv \mathcal{W}^{-1}; \\
\\
a_{3}\equiv 2\partial \left( \mathcal{W}^{-1}\right) ; \\
\\
a_{2}\equiv \mathcal{W}^{-1}\left( \partial \widehat{\Lambda }-\widehat{%
\Lambda }^{2}+2\widehat{\Sigma }\right) +\left[ \partial \left( \mathcal{W}%
^{-1}\right) \right] \widehat{\Lambda }+\partial ^{2}\left( \mathcal{W}%
^{-1}\right) ; \\
\\
a_{1}\equiv \mathcal{W}^{-1}\left( \partial ^{2}\widehat{\Lambda }+2\partial
\widehat{\Sigma }-2\widehat{\Lambda }\partial \widehat{\Lambda }\right) +%
\left[ \partial \left( \mathcal{W}^{-1}\right) \right] \left( 2\widehat{%
\Sigma }+2\partial \widehat{\Lambda }-\widehat{\Lambda }^{2}\right) +\left[
\partial ^{2}\left( \mathcal{W}^{-1}\right) \right] \widehat{\Lambda }; \\
\\
a_{0}\equiv \mathcal{W}^{-1}\left( \widehat{\Sigma }^{2}-\widehat{\Sigma }%
\partial \widehat{\Lambda }-\widehat{\Lambda }\partial \widehat{\Sigma }%
+\partial ^{2}\widehat{\Sigma }\right) +\left[ \partial \left( \mathcal{W}%
^{-1}\right) \right] \left( 2\partial \widehat{\Sigma }-\widehat{\Lambda }%
\widehat{\Sigma }\right) +\left[ \partial ^{2}\left( \mathcal{W}^{-1}\right)
\right] \widehat{\Sigma },
\end{array}
\notag \\
&&  \label{a}
\end{eqnarray}
where the following holomorphic functions have been introduced (recall the
first two Eqs. of (\ref{jazz2})):
\begin{eqnarray}
&&\widehat{\Lambda }\equiv 2\widehat{K}+\widehat{\Gamma }=-2\partial \left[
ln\left( X^{0}\right) \right] -\partial \left[ ln\left( e\right) \right]
=-\partial ln\left( X^{0}\partial X^{1}-X^{1}\partial X^{0}\right) =-\frac{%
X^{0}\partial ^{2}X^{1}-X^{1}\partial ^{2}X^{0}}{X^{0}\partial
X^{1}-X^{1}\partial X^{0}};~  \label{Lambda-hat} \\
&&  \notag \\
&&
\begin{array}{l}
\widehat{\Sigma }\equiv \partial \widehat{K}+\widehat{K}^{2}+\widehat{\Gamma
}\widehat{K}=-\partial ^{2}\left[ ln\left( X^{0}\right) \right] +\left[
\partial \left[ ln\left( X^{0}\right) \right] \right] ^{2}+\left[ \partial
\left[ ln\left( X^{0}\right) \right] \right] \left[ \partial \left[ ln\left(
e\right) \right] \right] = \\
\\
=\frac{\left( \partial X^{0}\right) \partial ^{2}X^{1}-\left( \partial
X^{1}\right) \partial ^{2}X^{0}}{X^{0}\partial X^{1}-X^{1}\partial X^{0}}=%
\widehat{\Lambda }+\frac{\left( X^{0}+\partial X^{0}\right) \partial
^{2}X^{1}-\left( X^{1}+\partial X^{1}\right) \partial ^{2}X^{0}}{%
X^{0}\partial X^{1}-X^{1}\partial X^{0}}.
\end{array}
\notag \\
&&  \label{Sigma-hat}
\end{eqnarray}

The definitions (\ref{a}) entail the following differential relations
between the functional coeffiecients of 1-modulus PF Eq. (\ref{PF2}):
\begin{equation}
\begin{array}{l}
a_{3}=2\partial a_{4}; \\
\\
a_{1}=\partial \left( a_{2}-\frac{1}{2}\partial a_{3}\right) .
\end{array}
\label{diff-rels-a}
\end{equation}
Thus, the holomorphic functions $a_{n}$ only depend on the (inverse of the)
holomorphic Yukawa coupling function $\mathcal{W}$, on the holomorphic
connections $\widehat{K}$ and $\widehat{\Gamma }$ (or, equivalently, on the
holomorphic $1$-beins $X^{0}$ and $e$). If $\mathcal{W}$, $\widehat{K}$ and $%
\widehat{\Gamma }$ (or $\mathcal{W}$, $X^{0}$ and $e$) are given as input,
Eq. (\ref{PF2}) is a fourth order linear ordinary differential equation for $%
V_{h}\left( \psi \right) $. Also, for given $a_{n}$ the definitions (\ref{a}%
) turn into non-linear ordinary differential equations for $X^{0}$ and $e$.

In special coordinates (with K\"{a}hler gauge fixed such that $X^{0}=1$) one
has $X^{1}\left( \psi \right) =t\left( \psi \right) =\psi $ and $e\left(
\psi \right) =1$, and the $a_{n}\left( \psi \right) $s simplify drastically%
\footnote{%
However, it is here worth pointing out that in the following general
treatment of (mirror) Fermat $CY_{3}$s $t\left( \psi \right) $ is \textit{not%
} a special coordinate, \textit{i.e.} $t\left( \psi \right) \neq \psi $, and
moreover the K\"{a}hler gauge is left unfixed (in particular, $X^{0}\left(
\psi \right) \neq 1$).}:
\begin{equation}
\begin{array}{l}
a_{4,sp.}\equiv \mathcal{W}_{sp.}^{-1}; \\
\\
a_{3,sp.}\equiv 2\partial \left( \mathcal{W}_{sp.}^{-1}\right) =-2\mathcal{W}%
_{sp.}^{-1}\frac{\partial \mathcal{W}_{sp.}}{\mathcal{W}_{sp.}}; \\
\\
a_{2,sp.}\equiv \partial ^{2}\left( \mathcal{W}_{sp.}^{-1}\right) =\mathcal{W%
}_{sp.}^{-1}\left[ 2\left( \frac{\partial \mathcal{W}_{sp.}}{\mathcal{W}%
_{sp.}}\right) ^{2}-\frac{\partial ^{2}\mathcal{W}_{sp.}}{\mathcal{W}_{sp.}}%
\right] ; \\
\\
a_{1,sp.}=0=a_{0,sp.},
\end{array}
\end{equation}
and thus Eq. (\ref{PF2}) can be rewritten as
\begin{equation}
\sum_{n=0}^{4}a_{n,sp.}\left( \psi \right) \partial ^{n}V_{h,sp.}\left( \psi
\right) =\left\{ \partial ^{4}-2\left( \partial ln\left( \mathcal{W}%
_{sp.}\right) \right) \partial ^{3}+\left[ 2\left( \partial ln\left(
\mathcal{W}_{sp.}\right) \right) ^{2}-\frac{\partial ^{2}\mathcal{W}_{sp.}}{%
\mathcal{W}_{sp.}}\right] \partial ^{2}\right\} V_{h,sp.}=0,
\label{PF2-special}
\end{equation}
where, by recalling Eq. (\ref{jazzit}), $V_{h,sp.}$ reads
\begin{equation}
V_{h,sp.}\left( \psi \right) =\left( 1,\psi ,\partial \mathcal{F}\left( \psi
\right) ,\psi \partial \mathcal{F}\left( \psi \right) -2\mathcal{F}\left(
\psi \right) \right) ,  \label{jazzit-special}
\end{equation}
which is nothing but the case $n_{V}=1$ of Eq. (\ref{Vh-special}). Moreover,
since in special coordinates (with $X^{0}=1$) the holomorphic Yukawa
coupling function $\mathcal{W}$ defined by Eq. (\ref{W-1-modulus}) reads
\begin{equation}
\mathcal{W}_{sp.}=\partial ^{3}\mathcal{F},  \label{W-1-modulus-special}
\end{equation}
Eq. (\ref{PF2-special}) can be also rewritten as
\begin{equation}
\left\{ \partial ^{4}-2\frac{\partial ^{4}\mathcal{F}}{\partial ^{3}\mathcal{%
F}}\partial ^{3}+\left[ 2\left( \frac{\partial ^{4}\mathcal{F}}{\partial ^{3}%
\mathcal{F}}\right) ^{2}-\frac{\partial ^{5}\mathcal{F}}{\partial ^{3}%
\mathcal{F}}\right] \partial ^{2}\right\} V_{h,sp.}=0.
\end{equation}

The 1-modulus PF Eq. (\ref{PF2}) can be further elaborated, by recalling
some basic facts about fourth order linear ordinary differential equations
\cite{Forsyth}. It is interesting to notice that not all the $a_{n}\left(
\psi \right) $s are actually relevant. Firstly, one can scale $a_{4}$ out
from the 1-modulus PF Eq. (\ref{PF2}), and furthermore drop the coefficient
of the third derivative by performing the following rescaling redefinition
of $V_{h}\left( \psi \right) $ \cite{Ferrara-Lerche1}:
\begin{equation}
V_{h}\left( \psi \right) \longrightarrow V_{h}\left( \psi \right) exp\left[ -%
\frac{1}{4}\int^{\psi }d\psi ^{\prime }\frac{a_{3}\left( \psi ^{\prime
}\right) }{a_{4}\left( \psi ^{\prime }\right) }\right] =V_{h}\left( \psi
\right) exp\left[ -\frac{1}{2}\int^{\psi }d\psi ^{\prime }\partial ln\left(
a_{4}\left( \psi ^{\prime }\right) \right) \right] =\left[ a_{4}\left( \psi
\right) \right] ^{-1/2}V_{h}\left( \psi \right) ,  \label{redef-Vh}
\end{equation}
where the first of differential relations (\ref{diff-rels-a}) was used. By
doing this, the PF Eq. (\ref{PF2}) can be recast in the following form:
\begin{equation}
\sum_{n=0}^{4}c_{n}\left( \psi \right) \partial ^{n}V_{h}\left( \psi \right)
=0,  \label{PF3}
\end{equation}
where
\begin{equation}
\begin{array}{l}
c_{4}=1; \\
\\
c_{3}=0; \\
\\
c_{2}=a_{4}^{-1}\left( \frac{3}{8}a_{4}^{-1}a_{3}^{2}-\frac{3}{2}\partial
a_{3}+a_{2}\right) ; \\
\\
c_{1}=-\frac{1}{2}a_{4}^{-1}\left( \frac{3}{4}%
a_{4}^{-2}a_{3}^{3}-3a_{4}^{-1}a_{3}\partial a_{3}+3\partial
^{2}a_{3}+a_{4}^{-1}a_{2}a_{3}-2\partial a_{2}\right) ; \\
\\
c_{0}=-\frac{1}{16}a_{4}^{-1}\left[
\begin{array}{l}
-\frac{45}{16}a_{4}^{-3}a_{3}^{4}+\frac{27}{2}a_{4}^{-2}a_{3}^{2}\partial
a_{3}-9a_{4}^{-1}\left( \partial a_{3}\right) ^{2}+ \\
\\
-10a_{4}^{-1}a_{3}\partial ^{2}a_{3}+4\partial
^{3}a_{3}-3a_{4}^{-2}a_{3}^{2}a_{2}+ \\
\\
+4a_{4}^{-1}a_{2}\partial a_{3}+4a_{4}^{-1}a_{3}\partial a_{2}-16a_{0}
\end{array}
\right] .
\end{array}
\end{equation}

Furthermore, it is possible to remove the dependence of Eq. (\ref{PF2}) on
the holomorphic K\"{a}hler connection by introducing the K\"{a}hler
invariant holomorphic vector \cite{Ferrara-Louis1}
\begin{equation}
\widetilde{V}_{h}\left( \psi \right) \equiv \left[ X^{0}\left( \psi \right) %
\right] ^{-1}V_{h}\left( \psi \right) .
\end{equation}
By doing this, the PF Eq. (\ref{PF2}) can be recast in the following form:
\begin{equation}
\sum_{n=0}^{4}d_{n}\left( \psi \right) \partial ^{n}\widetilde{V}_{h}\left(
\psi \right) =0,  \label{PF4}
\end{equation}
where
\begin{equation}
\begin{array}{l}
d_{4}=1; \\
\\
d_{3}=4\frac{\partial X^{0}}{X^{0}}+\frac{a_{3}}{a_{4}}=-2\Theta ; \\
\\
d_{2}=6\frac{\partial ^{2}X^{0}}{X^{0}}+3\frac{a_{3}}{a_{4}}\frac{\partial
X^{0}}{X^{0}}+\frac{a_{2}}{a_{4}}=\Theta ^{2}-\partial \Theta -\widehat{%
\Gamma }\Theta -\widehat{\Gamma }^{2}+\partial \widehat{\Gamma }=\frac{1}{4}%
d_{3}^{2}+\frac{1}{2}\partial d_{3}+\frac{1}{2}\widehat{\Gamma }d_{3}-%
\widehat{\Gamma }^{2}+\partial \widehat{\Gamma }; \\
\\
d_{1}=4\frac{\partial ^{3}X^{0}}{X^{0}}+3\frac{a_{3}}{a_{4}}\frac{\partial
^{2}X^{0}}{X^{0}}+2\frac{a_{2}}{a_{4}}\frac{\partial X^{0}}{X^{0}}+\frac{%
a_{1}}{a_{4}}=\left( \Theta ^{2}-\partial \Theta \right) \widehat{\Gamma }+%
\widehat{\Gamma }^{2}\Theta -2\left( \partial \widehat{\Gamma }\right)
\Theta +\partial ^{2}\widehat{\Gamma }-2\widehat{\Gamma }\partial \widehat{%
\Gamma }= \\
\\
=\frac{1}{4}\widehat{\Gamma }d_{3}^{2}-\left( \frac{1}{2}\widehat{\Gamma }%
^{2}-\partial \widehat{\Gamma }\right) d_{3}+\frac{1}{2}\widehat{\Gamma }%
\partial d_{3}+\partial ^{2}\widehat{\Gamma }-2\widehat{\Gamma }\partial
\widehat{\Gamma }=\widehat{\Gamma }d_{2}-\left( \widehat{\Gamma }%
^{2}-\partial \widehat{\Gamma }\right) d_{3}+\partial ^{2}\widehat{\Gamma }-3%
\widehat{\Gamma }\partial \widehat{\Gamma }+\widehat{\Gamma }^{3}; \\
\\
d_{0}=\frac{1}{X^{0}}\left( \partial ^{4}+\frac{a_{3}}{a_{4}}\partial ^{3}+%
\frac{a_{2}}{a_{4}}\partial ^{2}+\frac{a_{1}}{a_{4}}\partial +\frac{a_{0}}{%
a_{4}}\right) X^{0}=0,
\end{array}
\label{ds}
\end{equation}
where the K\"{a}hler gauge-invariant holomorphic function
\begin{equation}
\Theta \equiv \partial ln\left[ \frac{\mathcal{W}}{\left( X^{0}\right) ^{2}}%
\right] =\frac{\partial \mathcal{W}}{\mathcal{W}}-2\frac{\partial X^{0}}{%
X^{0}}=-\frac{1}{2}\frac{a_{3}}{a_{4}}+2\widehat{K}
\end{equation}
has been introduced. Eqs. (\ref{ds}) imply the $d_{n}$ to be formally
independent on $\widehat{K}$, and all their dependence on $X^{0}$ appears
through the function $\Theta $. Since $X^{0}$ is the first component of the
holomorphic period vector $V_{h}$ given by Eq. (\ref{jazzit}), the vanishing
of $d_{0}$ directly follows from the PF Eq. (\ref{PF2}) (with $a_{4}\neq 0$
and $X^{0}\neq 0$).

\section{\label{HG2}Holomorphic Geometry and Flat Holomorphic Matrix
Connection}

\setcounter{equation}0
\def\theequation{5.\arabic{subsection}.\arabic{equation}}

\subsection{\label{HG2-n}$n_{V}$-Moduli}

As shown in \cite{Strominger1}, the basic, defining differential relations (%
\ref{SKG-rels2}) of SK geometry can be recast as a vanishing condition for a
suitably defined flat symplectic non-holomorphic connection. Analogously,
the holomorphic differential Eqs. (\ref{holomorphic-rels-homogeneous}) can
be rewritten as a vanishing condition for a suitably defined flat
holomorphic connection, \textit{i.e.} as \cite{Ferrara-Lerche1}
\begin{equation}
\left( \mathbb{I}_{2n_{V}+2}\partial _{\alpha }-\mathbf{A}_{\alpha }\left(
z\right) \right) \mathbf{V}_{h}\left( z\right) =0,  \label{PF-sys}
\end{equation}
where $\mathbf{V}_{h}\left( z\right) $ is a $\left( 2n_{V}+2\right) \times
\left( 2n_{V}+2\right) $ holomorphic matrix ($\left( 2n_{V}+2\right) \times
1 $ vector with $1\times \left( 2n_{V}+2\right) $ vector entries) defined as
follows \cite{Ferrara-Lerche1}:
\begin{equation}
\mathbf{V}_{h}\left( z\right) \equiv \left(
\begin{array}{l}
V_{h}\left( z\right) \\
~ \\
V_{h,\beta }\left( z\right) \\
~ \\
V_{h}^{\beta }\left( z\right) \\
~ \\
V_{h}^{0}\left( z\right)
\end{array}
\right) ,  \label{Vh-matrix}
\end{equation}
where the entries are defined in Eqs. (\ref{jazz}). On the other hand, $%
\mathbf{A}_{\alpha }\left( z\right) $ is the following $\left(
2n_{V}+2\right) \times \left( 2n_{V}+2\right) $ holomorphic connection
matrix:
\begin{equation}
\mathbf{A}_{\alpha }\left( z\right) \equiv \left(
\begin{array}{ccccccc}
-\widehat{K}_{\alpha }\left( z\right) &  & \delta _{\alpha }^{\gamma } &  & 0
&  & 0 \\
&  &  &  &  &  &  \\
0 &  & -\left( \widehat{\Gamma }_{\alpha }\left( z\right) +\widehat{K}%
_{\alpha }\left( z\right) \mathbb{I}_{n_{V}}\right) _{\beta }^{~\gamma } &
& \left( W_{\alpha }\right) _{\gamma \beta }\left( z\right) &  & 0 \\
&  &  &  &  &  &  \\
0 &  & 0 &  & \left( \widehat{\Gamma }_{\alpha }\left( z\right) +\widehat{K}%
_{\alpha }\left( z\right) \mathbb{I}_{n_{V}}\right) _{\gamma }^{~\beta } &
& \delta _{\alpha }^{\beta } \\
&  &  &  &  &  &  \\
0 &  & 0 &  & 0 &  & \widehat{K}_{\alpha }\left( z\right)
\end{array}
\right) .  \label{A}
\end{equation}
It should be noticed that $\mathbf{A}_{\alpha }\left( z\right) $ is
Lie-algebra valued in $\mathfrak{sp}\left( 2n_{V}+2\right) $, \textit{i.e.}
it satisfies the infinitesimal symplecticity condition \cite{Ferrara-Lerche1}
\begin{equation}
\mathbf{A}_{\alpha }^{T}\left( z\right) Q+Q\mathbf{A}_{\alpha }\left(
z\right) =0,
\end{equation}
where $Q$ is the symplectic metric defined in Eq. (\ref{Q}).

Put another way, it can be stated that the PF Eqs. (\ref
{holomorphic-rels-homogeneous}) are equivalent to the matrix system (\ref
{PF-sys}), with $\mathbf{A}_{\alpha }\left( z\right) $ defined by Eq. (\ref
{A}). The general solution of such an holomorphic matrix system is given by
Eqs. (\ref{jazz}) arranged as a vector as given by Eq. (\ref{Vh-matrix}).

As expected, by specializing the holomorphic matrix system (\ref{PF-sys}) in
special coordinates and choosing the K\"{a}hler gauge to be such that $%
X^{0}=1$, one gets the following holomorphic matrix system:
\begin{equation}
\left( \mathbb{I}_{2n_{V}+2}\partial _{a}-\mathbf{A}_{a,sp.}\left( z\right)
\right) \mathbf{V}_{h,sp.}\left( z\right) =0,  \label{PF-sys-special}
\end{equation}
which is equivalent to the holomorphic system (\ref{holomorphic-rels-special}%
).

$\mathbf{V}_{h,sp.}\left( z\right) $ is a $\left( 2n_{V}+2\right) \times
\left( 2n_{V}+2\right) $ holomorphic matrix ($\left( 2n_{V}+2\right) \times
1 $ vector with $1\times \left( 2n_{V}+2\right) $ vector entries) defined as
follows \cite{Ferrara-Lerche1}:
\begin{equation}
\mathbf{V}_{h,sp.}\left( z\right) \equiv \left(
\begin{array}{l}
V_{h,sp.}\left( z\right) \\
~ \\
V_{h,sp.,b}\left( z\right) \\
~ \\
V_{h,sp.}^{b}\left( z\right) \\
~ \\
V_{h,sp.}^{0}(z)
\end{array}
\right) ,  \label{Vh-matrix-special}
\end{equation}
where the entries are defined in Eqs. (\ref{Vh-special}), (\ref{dbVh-special}%
), (\ref{Vah-special}) and (\ref{V0h-special}). It is worth mentioning that
the matrices $\mathbf{V}_{h,sp.}\left( z\right) $ and $\mathbf{V}_{h}\left(
z\right) $ have a symplectic structure with respect to the symplectic metric
relevant for holomorphic geometry, \textit{i.e.} with respect to $Q$ defined
in Eq. (\ref{Q}):
\begin{equation}
\begin{array}{l}
\mathbf{V}_{h,sp.}^{T}\left( z\right) Q\mathbf{V}_{h,sp.}\left( z\right) =Q;
\\
\\
\mathbf{V}_{h}^{T}\left( z\right) Q\mathbf{V}_{h}\left( z\right) =Q.
\end{array}
\end{equation}
$\mathbf{A}_{a,sp.}\left( z\right) $ (named $\mathbb{C}_{a}$ in Eq. (3.6) of
the first Ref. of \cite{Ferrara-Lerche1}) is the $\left( 2n_{V}+2\right)
\times \left( 2n_{V}+2\right) $ holomorphic connection matrix obtained by $%
\mathbf{A}_{\alpha }\left( z\right) $ (given by Eq. (\ref{A})) by putting $%
\widehat{\Gamma }_{\alpha }\left( z\right) =0=\widehat{K}_{\alpha }\left(
z\right) $ (also recalling that in special coordinates $a$-indices and $%
\alpha $-indices coincide). Clearly, as its ``holomorphically covariant''
counterpart $\mathbf{A}_{\alpha }\left( z\right) $, also $\mathbf{A}%
_{a,sp.}\left( z\right) $ is Lie-algebra valued in $\mathfrak{sp}\left(
2n_{V}+2\right) $, and therefore it satisfies a corresponding infinitesimal
symplecticity condition.

In other words, it can be stated that the the holomorphic system
(\ref {holomorphic-rels-special}) can be recast in the matrix form
(\ref {PF-sys-special}), with $\mathbf{A}_{a,sp.}\left( z\right) $
defined by Eq. (3.6) of the first Ref. of \cite{Ferrara-Lerche1}.
The general solution of such an holomorphic matrix system is given
by Eqs. (\ref{Vh-special}), (\ref {dbVh-special}),
(\ref{Vah-special}) and (\ref{V0h-special}) arranged as a vector as
given by Eq. (\ref{Vh-matrix-special}). \setcounter{equation}0
\def\theequation{5.\arabic{subsection}.\arabic{equation}}

\subsection{\label{HG2-1}$1$-Modulus}

Once again, by considering the 1-modulus case more in detail, one obtains a
major simplification. The 1-modulus PF Eq. (\ref{PF2}) can be rewritten in
matrix form as follows:
\begin{equation}
\left( \mathbb{I}_{4}\partial -\mathbf{A}\left( \psi \right) \right) \mathbf{%
V}_{h}\left( \psi \right) =0,  \label{PF-sys-1-modulus}
\end{equation}
where $\mathbf{V}_{h}\left( \psi \right) $ is a $4\times 4$ holomorphic
matrix, obtained by putting $n_{V}=1$ in Eq. (\ref{Vh-matrix}):
\begin{eqnarray}
\mathbf{V}_{h}\left( \psi \right) &\equiv &\left(
\begin{array}{l}
V_{h}\left( \psi \right) \\
~ \\
V_{h,\psi }\left( \psi \right) \\
~ \\
V_{h}^{\psi }\psi \\
~ \\
V_{h}^{0}\left( \psi \right)
\end{array}
\right) =  \label{Vh-matrix-1-1-modulus} \\
&&  \notag \\
&=&\left(
\begin{array}{ccccccc}
X^{0} &  & X^{1} &  & X^{0}e^{-1}\partial \mathcal{F} &  &
X^{1}e^{-1}\partial \mathcal{F}-2X^{0}\mathcal{F} \\
&  &  &  &  &  &  \\
0 &  & X^{0}e &  & X^{0}e^{-1}\widehat{D}\partial \mathcal{F} &  &
-X^{0}\partial \mathcal{F}+X^{1}e^{-1}\widehat{D}\partial \mathcal{F} \\
&  &  &  &  &  &  \\
0 &  & 0 &  & \left( X^{0}\right) ^{-1}e^{-1} &  & \left( X^{0}\right)
^{-2}X^{1}e^{-1} \\
&  &  &  &  &  &  \\
0 &  & 0 &  & 0 &  & \left( X^{0}\right) ^{-1}
\end{array}
\right) ,  \notag \\
&&  \label{Vh-matrix-2-1-modulus}
\end{eqnarray}
where the first row is given by Eq. (\ref{jazzit}), and (recall Eq. (\ref
{F-call}))
\begin{equation}
\mathcal{F}\left( \psi \right) \equiv F\left( \frac{X^{1}(\psi )}{X^{0}(\psi
)}\right) =\left( X^{0}\left( \psi \right) \right) ^{-2}F\left( X(\psi
)\right) \ref{F-call}
\end{equation}
is the K\"{a}hler gauge-invariant holomorphic prepotential. Moreover, the
second row of $\mathbf{V}_{h}$ can be further elaborated by recalling Eq. (%
\ref{wow1}) and the first of definitions (\ref{jazz2}):
\begin{equation}
\widehat{D}\partial \mathcal{F}=\widehat{D}^{2}\mathcal{F}=e\partial \left[
e^{-1}\partial \mathcal{F}\right] =\left[ \partial ^{2}-e^{-1}\left(
\partial e\right) \partial \right] \mathcal{F}=\left[ \partial ^{2}+\widehat{%
\Gamma }\partial \right] \mathcal{F}.  \label{ddF-1-modulus}
\end{equation}

On the other hand, $\mathbf{A}\left( \psi \right) \equiv \mathbf{A}_{\psi
}\left( \psi \right) $ is a $4\times 4$ holomorphic connection matrix, which
is Lie-algebra valued in $\mathfrak{sp}\left( 4\right) $ and corresponds to $%
n_{V}=1$ in Eq. (\ref{A}):
\begin{eqnarray}
&&
\begin{array}{l}
\mathbf{A}\left( \psi \right) = \\
\\
=\left(
\begin{array}{ccccccc}
\partial ln\left( X^{0}\right) &  & 1 &  & 0 &  & 0 \\
&  &  &  &  &  &  \\
0 &  & \partial ln\left( e\right) +\partial ln\left( X^{0}\right) &  &
\mathcal{W} &  & 0 \\
&  &  &  &  &  &  \\
0 &  & 0 &  & -\partial ln\left( e\right) -\partial ln\left( X^{0}\right) &
& 1 \\
&  &  &  &  &  &  \\
0 &  & 0 &  & 0 &  & -\partial ln\left( X^{0}\right)
\end{array}
\right) .
\end{array}
\notag \\
&&  \label{A-1-modulus}
\end{eqnarray}
where use of Eqs. (\ref{jazz2}) has been made (see also Eq. (\ref
{W-1-modulus})). Eq. (\ref{A-1-modulus}) can be further elaborated; indeed,
by using Eqs. (\ref{e-1-modulus}), (\ref{jazz2}), (\ref{Lambda-hat}) and (%
\ref{Sigma-hat}), one easily gets that
\begin{eqnarray}
&&
\begin{array}{l}
\partial ln\left( e\right) +\partial ln\left( X^{0}\right) =-\left( \widehat{%
\Gamma }+\widehat{K}\right) =-\widehat{\Lambda }-\frac{\partial X^{0}}{X^{0}}%
=-\widehat{\Lambda }+\widehat{K}= \\
\\
=-\widehat{\Sigma }+\frac{\left[ \left( X^{0}\right) ^{2}+X^{0}\partial X^{0}%
\right] \partial ^{2}X^{1}-\left( X^{1}+\partial X^{1}\right) X^{0}\partial
^{2}X^{0}-X^{0}\partial X^{0}\partial X^{1}-X^{1}\left( \partial
X^{0}\right) ^{2}}{\left( X^{0}\right) ^{2}\partial X^{1}-X^{0}X^{1}\partial
X^{0}}= \\
\\
=\partial ln\left( \frac{X^{0}\partial X^{1}-X^{1}\partial X^{0}}{X^{0}}%
\right) =\partial ln\left( \partial X^{1}-X^{1}\partial lnX^{0}\right) =%
\frac{\left( X^{0}\right) ^{2}\partial ^{2}X^{1}-X^{0}X^{1}\partial
^{2}X^{0}-X^{0}\partial X^{0}\partial X^{1}-X^{1}\left( \partial
X^{0}\right) ^{2}}{\left( X^{0}\right) ^{2}\partial X^{1}-X^{0}X^{1}\partial
X^{0}}.
\end{array}
\notag \\
&&
\end{eqnarray}

It can be stated that the 1-modulus PF Eq. (\ref{PF2}) is equivalent to the
matrix system (\ref{PF-sys-1-modulus}), with $\mathbf{A}\left( \psi \right) $
defined by Eq. (\ref{A-1-modulus}). The general solution of such an
holomorphic matrix system (which corresponds to the most general solution of
the fourth order linear PF Eq. (\ref{PF2})) is given by Eqs. (\ref
{Vh-matrix-1-1-modulus})-(\ref{Vh-matrix-2-1-modulus}) (implemented by Eq. (%
\ref{ddF-1-modulus})).\medskip

Let us now further specialize our treatment to the 1-modulus SK geometries
endowing the moduli space of Fermat $CY_{3}$s. As previously mentioned, the
fourth order linear PF ordinary differential equation for each of the four
threefolds (classified by the index $k=5,6,8,10$: see next Section) of such
a class of $CY_{3}$s has been obtained for $k=5$ in \cite{CDLOGP1,CDLOGP2}
(see in particular Eq. (3.9) of \cite{CDLOGP1}, where $z\equiv \psi ^{-5}$
and $\varpi _{0}\sim \frac{1}{\psi }V_{h}$; see also \cite{Cadavid-Ferrara}%
), and for $k=6,8,10$ in \cite{KT} (see Eq. (3.1) of such a Ref., with
notation $\alpha \equiv \psi $), where a unified, $k$-parametrized treatment
was exploited (see also \cite{Font}).

In order to recast the 1-modulus PF Eqs. given by Eq. (3.1) of \cite{KT} in
the form (\ref{PF2}) with the differential relations (\ref{diff-rels-a})
between the $a_{n}\left( \psi \right) $s holding, one must multiply them by
the function $\psi ^{-\xi _{k}}$, with $\xi _{k}=0,3,6,8$ for $k=5,6,8,10$
respectively. By doing this, one achieves the result that for Fermat $CY_{3}$%
s the fourth order linear PF Eqs. (\ref{PF2}) can be recast in the following
$k$-parametrized form\footnote{%
As we will see in Sect. \ref{const-norm}, for self-consistency reasons the
1-modulus PF Eqs. (\ref{PF2-corr}) (with Tables 1 and 2) (which are
``corrected'' by an overall factor $\psi ^{-\xi _{k}}$ with respect to the
ones given in Eq. (3.1) of \cite{KT}) need also to be further multiplied by
a suitable ``normalization'' constant (see Eq. (\ref{PF2-corr-corr})).}:
\begin{equation}
\begin{array}{l}
\sum_{n=0}^{4}a_{n,k}\left( \psi \right) \partial ^{n}V_{h}\left( \psi
\right) =0, \\
\\
a_{n,k}\left( \psi \right) \equiv -\sigma _{n}\psi ^{n+1}+\left( -1\right)
^{n}\tau _{n,k}\psi ^{n+1-k},
\end{array}
\label{PF2-corr}
\end{equation}
with the constants $\sigma _{n}$s and $\tau _{n,k}$s given by the following
Tables:
\begin{table}[h]
\begin{center}
\begin{tabular}{|c||c|}
\hline
$n$ & $\sigma _{n}$ \\ \hline\hline
$0$ & $1$ \\ \hline
$1$ & $15$ \\ \hline
$2$ & $25$ \\ \hline
$3$ & $10$ \\ \hline
$4$ & $1$ \\ \hline
\end{tabular}
\end{center}
\caption{{}Values of the integer constants $\protect\sigma _{n}$}
\end{table}
\begin{table}[h]
\begin{center}
\begin{tabular}{|c||c|c|c|c|}
\hline
$k\longrightarrow $ & $5$ & $6$ & $8$ & $10$ \\ \hline\hline
$\tau _{0,k}$ & $0$ & $0$ & $0$ & $0$ \\ \hline
$\tau _{1,k}$ & $0$ & $0$ & $15$ & $35$ \\ \hline
$\tau _{2,k}$ & $0$ & $2$ & $15$ & $35$ \\ \hline
$\tau _{3,k}$ & $0$ & $2$ & $6$ & $10$ \\ \hline
$\tau _{4,k}$ & $1$ & $1$ & $1$ & $1$ \\ \hline
\end{tabular}
\end{center}
\caption{{}Values of the integer constants $\protect\tau _{n,k}$}
\end{table}
\setcounter{equation}0
\def\theequation{\arabic{section}.\arabic{equation}}

\section{General Analysis of (Mirror) Fermat Calabi-Yau Threefolds\newline
near the Landau-Ginzburg Point\label{GA}}

In the present Section we will briefly present the formalism of one-modulus
(mirror) Fermat Calabi-Yau threefolds ($CY_{3}$s), focussing our analysis in
a suitable neighbourhood of the LG point $\psi =0$. We will mainly follow
\cite{KT}, and cite where appropriate other relevant works. We also derive
original formal results, which will be then used in the case-by-case
analysis of extremal BH LG attractors performed in next Sections.

Fermat $CY_{3}$s can be defined as the vanishing \textit{locus} of
quasi-homogeneous polynomials in 5 complex variables, of the general form:
\begin{equation}
\mathcal{P}_{0}=\sum_{i=0}^{4}\nu _{i}\left( x^{i}\right) ^{n_{i}}=0;
\end{equation}
such a \textit{locus} gives the embedding of the considered $CY_{3}$ in a
suitably weighted complex projective space $\mathbb{W}\mathbb{CP}_{\nu
_{0},\nu _{1},\nu _{2},\nu _{3},\nu _{4}}^{4}$. By imposing the defining
conditions of vanishing first Chern class and of absence of singularities,
it is possible to show that only four possible sets of $\left\{ \nu
_{i},n_{i}\right\} _{i=0,1,2,3,4}$ exist, \textit{all} corresponding to $%
CY_{3}$s with $h\left( 1,1\right) =dim\left( H^{1,1}\left( CY_{3}\right)
\right) =1$ (\textit{i.e.} only one K\"{a}hler modulus). The four existing
Fermat $CY_{3}$s can be classified by introducing the \textit{Fermat
parameter }$k$, defined as the smallest common multiple of the $n_{i}$s%
\footnote{$k$ can equivalently be defined as the degree of $\mathcal{W}_{0}$%
. It turns out that $k=n_{i}\nu _{i}$ (no summation on $i$) $\forall
i=0,1,2,3,4$. Moreover, it also holds that $k=\sum_{i=0}^{4}\nu _{i}$.}; the
only allowed values of $k$ turn out to be $k=5,6,8,10$. Thus, the four
existing Fermat $CY_{3}$s $\mathcal{M}_{k}$ are given by the following
geometrical \textit{loci}\footnote{%
Here and below, we give a name to the Fermat $CY_{3}$s corresponding to the
various possible values of the Fermat parameter.
\par
The Fermat $CY_{3}$ with $k=5$ has been named \textit{quintic }some time ago
(see \textit{e.g. } \textit{\ }\cite{Strom-Witten,Cadavid-Ferrara,
CDLOGP1,CDLOGP2,KT}).
\par
In a similar fashion, by using the corresponding Latin cardinal adjectives,
we name \textit{sextic}, \textit{octic}, and \textit{dectic} the Fermat $%
CY_{3}$s with $k=6,8,10$, respectively.} \cite
{Strom-Witten,CDLOGP1,CDLOGP2,KT,Font}:
\begin{eqnarray}
&&
\begin{array}{l}
k=5:\text{\textit{quintic} }\mathcal{M}_{5}=\left\{ x^{i}\in \left( \mathbb{W%
}\right) \mathbb{CP}_{1,1,1,1,1}^{4}:\mathcal{P}_{0,5}=\sum_{i=0}^{4}\left(
x^{i}\right) ^{5}=0\right\} ; \\
\\
k=6:\text{\textit{sextic} }\mathcal{M}_{6}=\left\{ x^{i}\in \mathbb{WCP}%
_{2,1,1,1,1}^{4}:\mathcal{P}_{0,6}=2\left( x^{0}\right)
^{3}+\sum_{i=1}^{4}\left( x^{i}\right) ^{6}=0\right\} ; \\
\\
k=8:\text{\textit{octic} }\mathcal{M}_{8}=\left\{ x^{i}\in \mathbb{WCP}%
_{4,1,1,1,1}^{4}:\mathcal{P}_{0,8}=4\left( x^{0}\right)
^{2}+\sum_{i=1}^{4}\left( x^{i}\right) ^{8}=0\right\} ; \\
\\
k=10:\text{\textit{dectic} }\mathcal{M}_{10}=\left\{ x^{i}\in \mathbb{WCP}%
_{5,2,1,1,1}^{4}:\mathcal{P}_{0,10}=5\left( x^{0}\right) ^{2}+2\left(
x^{1}\right) ^{5}+\sum_{i=2}^{4}\left( x^{i}\right) ^{10}=0\right\} .
\end{array}
\notag \\
&&  \label{14mar}
\end{eqnarray}

By orbifolding the $\mathcal{M}_{k}$s and quotienting by the full phase
symmetry group $G$ (see \cite{KT} and Refs. therein), one obtains a pair of
Fermat $CY_{3}$s\footnote{%
For simplicity's sake, we denote in the same way the starting Fermat $CY_{3}$
and the one obtained by orbifolding and then quotienting by $G$.} $\left(
\mathcal{M}_{k},\mathcal{M}_{k}^{\prime }\right) $ related by the so-called
\textit{mirror symmetry }\cite{mirror1,mirror2,mirror3,CDLOGP1,CDLOGP2},
with $h(1,1)$ and $h(2,1)=dim\left( H^{2,1}\left( CY_{3}\right) \right) $
interchanged (and therefore opposite Euler number $\varkappa $).
Correspondingly, the defining vanishing geometrical \textit{loci} will be
``deformed'' as follows:
\begin{equation}
\mathcal{P}_{0}\longrightarrow \mathcal{P}\equiv \mathcal{P}_{0}-k\psi
\prod_{i=0}^{4}x^{i}.
\end{equation}
All the relevant topological data of Fermat $CY_{3}$s $\mathcal{M}_{k}$s are
given in Table 3.
\begin{table}[tbp]
\begin{center}
\begin{tabular}{|c||c|c|c|c|c|}
\hline
$
\begin{array}{c}
\\
k \\
\downarrow
\end{array}
$ & $
\begin{array}{c}
\\
G \\
~
\end{array}
$ & $
\begin{array}{c}
\\
Ord\left( G\right) \\
~
\end{array}
$ & $
\begin{array}{c}
\\
h\left( 1,1\right) \\
~
\end{array}
$ & $
\begin{array}{c}
\\
h\left( 2,1\right) \\
~
\end{array}
$ & $
\begin{array}{c}
\\
\varkappa \equiv 2\left[ h\left( 1,1\right) -h\left( 2,1\right) \right] \\
~
\end{array}
$ \\ \hline\hline
$
\begin{array}{c}
\\
5 \\
~
\end{array}
$ & $
\begin{array}{c}
\\
\left( \mathbb{Z}_{5}\right) ^{3} \\
~
\end{array}
$ & $
\begin{array}{c}
\\
5^{3} \\
~
\end{array}
$ & $
\begin{array}{c}
\\
1 \\
~
\end{array}
$ & $
\begin{array}{c}
\\
101 \\
~
\end{array}
$ & $
\begin{array}{c}
\\
-200 \\
~
\end{array}
$ \\ \hline
$
\begin{array}{c}
\\
6 \\
~
\end{array}
$ & $
\begin{array}{c}
\\
\mathbb{Z}_{3}\otimes \left( \mathbb{Z}_{6}\right) ^{2} \\
~
\end{array}
$ & $
\begin{array}{c}
\\
3\cdot 6^{2} \\
~
\end{array}
$ & $
\begin{array}{c}
\\
1 \\
~
\end{array}
$ & $
\begin{array}{c}
\\
103 \\
~
\end{array}
$ & $
\begin{array}{c}
\\
-204 \\
~
\end{array}
$ \\ \hline
$
\begin{array}{c}
\\
8 \\
~
\end{array}
$ & $
\begin{array}{c}
\\
\left( \mathbb{Z}_{8}\right) ^{2}\otimes \mathbb{Z}_{2} \\
~
\end{array}
$ & $
\begin{array}{c}
\\
2\cdot 8^{2} \\
~
\end{array}
$ & $
\begin{array}{c}
\\
1 \\
~
\end{array}
$ & $
\begin{array}{c}
\\
149 \\
~
\end{array}
$ & $
\begin{array}{c}
\\
-296 \\
~
\end{array}
$ \\ \hline
$
\begin{array}{c}
\\
10 \\
~
\end{array}
$ & $
\begin{array}{c}
\\
\left( \mathbb{Z}_{10}\right) ^{2} \\
~
\end{array}
$ & $
\begin{array}{c}
\\
1\cdot 10^{2} \\
~
\end{array}
$ & $
\begin{array}{c}
\\
1 \\
~
\end{array}
$ & $
\begin{array}{c}
\\
145 \\
~
\end{array}
$ & $
\begin{array}{c}
\\
-288 \\
~
\end{array}
$ \\ \hline
\end{tabular}
\end{center}
\caption{\textbf{Basic topological data of Fermat }$CY_{3}$\textbf{s } $%
\mathcal{M}_{k}$\textbf{s}}
\end{table}

$\psi $ is the \textit{K\"{a}hler deformation modulus} for Fermat $CY_{3}$s $%
\mathcal{M}_{k}$s (all having $h\left( 1,1\right) =1$) and the \textit{%
complex structure deformation modulus} for the corresponding \textit{mirror}
Fermat $CY_{3}$s $\mathcal{M}_{k}^{\prime }$s (all having $h\left(
2,1\right) =1$). Since in the treatment and computations performed below we
will consider $\psi $ as a complex structure modulus, we will be actually
working in the \textit{mirror description} of the considered $CY_{3} $s,
\textit{i.e.} we will be considering the \textit{mirror} Fermat $CY_{3}$s $%
\mathcal{M}_{k}^{\prime }$s ($k=5,6,8,10$).

In such a framework, the relevant quantities for the $d=4$ low-energy
effective Lagrangian of the $d=10$ superstring theory compactified on $%
\mathcal{M}_{k}$ are given (within the complex structure moduli space ($dim_{%
\mathbb{C}}=1$)) by the K\"{a}hler metric and Yukawa couplings on $\mathcal{M%
}_{k}^{\prime }$ (related to $\mathcal{M}_{k}$ by mirror symmetry). All such
quantities will be obtained by the solutions of the fourth order linear PF
ordinary differential Eqs. (\ref{PF2}).

Near the LG point $\psi =0$, the $4\times 1$ period vector\footnote{%
Once again, in order to make the contact with the relevant literature
easier, in this Section as well as in the next ones, we will reconsider
column (\textit{i.e.} $4\times 1$), rather than row (\textit{i.e.} $1\times
4 $), period vectors. Moreover, the holomorphic period vector in the
symplectic basis (hitherto named $V_{h}$) will be henceforth denoted by $\Pi
$.} in the PF basis $\varpi _{k}\left( \psi \right) $ is obtained by solving
the PF Eqs.\footnote{%
When comparing Eq. (\ref{PF2-PF}) to Eq. (\ref{PF2}) (and, more in general,
considering the treatment given in Sects. \ref{HG} and \ref{HG2}), the $%
4\times 1$ symplectic holomorphic period vector $V_{h}\equiv \Pi $ and the $%
4\times 1$ PF holomorphic period vector $\varpi $ turn out to satisfy the
same fourth order linear ordinary differential equation.
\par
Consequently, they necessarily have to be related by a \textit{global} (%
\textit{i.e.} $\psi $\textit{-independent}) ``rotation'' in the moduli
space. This is precisely what happens, with such a ``rotation'' in the
moduli space expressed by the $4\times 4$ real matrices $M_{k}$s given in
Eqs. (\ref{M1})-(\ref{M2}) (see Sect. 4 of \cite{KT}).}
\begin{equation}
\sum_{n=0}^{4}a_{n,k}\left( \psi \right) \partial ^{n}\varpi _{k}\left( \psi
\right) =0.  \label{PF2-PF}
\end{equation}
Here we choose the normalization and the gauge of the holomorphic 3-form
defined on $\mathcal{M}_{k}^{\prime }$ such that\footnote{%
The normalization of $\varpi _{k}\left( \psi \right) $ adopted in the
present work is the same of \cite{CDLOGP1,CDLOGP2,KT}, and it differs from
the one adopted in (a part of the) literature on flux compactifications (see
\textit{e.g.} Subsect. 3.2 of \cite{G2}) by a factor $\frac{1}{Ord\left(
G_{k}\right) }$ (the reason is that we are interested in the mirror
manifolds $\mathcal{M}_{k}^{\prime }$s, not in $\mathcal{M}_{k}$s).
\par
On the other hand, it is easy to realize that the gauge of the holomorphic
3-form $\Xi $ adopted in \cite{CDLOGP1,CDLOGP2,KT} is mostly convenient in
order to study the \textit{large complex structure modulus limit }$\psi
\longrightarrow \infty $. Since we will investigate the \textit{LG limit }$%
\psi \longrightarrow 0$, for our purposes it is better to adopt the gauge of
\cite{G2}, which amounts to redefining the holomorphic 3-form (with respect
to the holomorphic 3-form of \cite{CDLOGP1,CDLOGP2,KT}) as follows:
\begin{equation*}
\Xi \longrightarrow \frac{1}{\psi }\Xi .
\end{equation*}
Such a redefinition yields the following K\"{a}hler gauge transformation:
\begin{equation*}
K\left( \psi ,\overline{\psi }\right) \longrightarrow K\left( \psi ,%
\overline{\psi }\right) +f\left( \psi \right) +\overline{f}\left( \overline{%
\psi }\right) ,~~f\left( \psi \right) \equiv 2ln\left( \psi \right) .
\end{equation*}
\par
Such a gauge transformation will affect the K\"{a}hler-covariant quantities (%
\textit{i.e.} the ones with non-vanishing K\"{a}hler weights), such as $%
exp\left( K\right) $, $\mathcal{W}$ and $C$, but clearly not the
K\"{a}hler-invariant ones, such as the metric $g$. In the \textit{LG limit},
for the K\"{a}hler-invariant quantities we obtain a matching with the
results of \cite{CDLOGP1,CDLOGP2,KT} (see also \cite{Font}); such a matching
can be checked to hold also for K\"{a}hler-covariant quantities, up to the
different overall normalization and K\"{a}hler gauge discussed above.}
\begin{equation}
\varpi _{k}\left( \psi \right) \equiv -\frac{1}{\psi }\frac{(2\pi i)^{3}}{%
Ord(G_{k})}\left(
\begin{array}{l}
\omega _{2,k}(\psi ) \\
~ \\
\omega _{1,k}(\psi ) \\
~ \\
\omega _{0,k}(\psi ) \\
~ \\
\omega _{k-1,k}(\psi )
\end{array}
\right) ,  \label{omega-PF}
\end{equation}
with
\begin{equation}
\begin{array}{l}
\omega _{j,k}(\psi )\equiv \omega _{0,k}(\beta _{k}^{2j}\psi ), \\
\\
\beta _{k}\equiv exp(\frac{\pi i}{k})
\end{array}
\label{rel-j}
\end{equation}
all being solutions of Eq. (\ref{PF2-PF}) ($j=0,1,...,k-1$). However, since
Eq. (\ref{PF2-PF}) is a fourth order (linear) differential equation, only 4
linearly independent solutions $\omega _{j,k}(\psi )$s exist. Thus, $\forall
k=5,6,8,10$, $k-4$ linear relations between the $\omega _{j,k}(\psi )$s
hold. One possible choice is the following one \cite{CDLOGP1,CDLOGP2,KT}:
\begin{eqnarray}
&&
\begin{array}{l}
k=5:
\begin{array}{l}
\sum_{j=0}^{4}\omega _{j,5}(\psi )=0.
\end{array}
\\
~
\end{array}
\\
&&
\begin{array}{l}
k=6:\left\{
\begin{array}{l}
\omega _{0,6}(\psi )+\omega _{2,6}(\psi )+\omega _{4,6}(\psi )=0; \\
\\
\omega _{1,6}(\psi )+\omega _{3,6}(\psi )+\omega _{5,6}(\psi )=0.
\end{array}
\right. \\
~
\end{array}
\\
&&
\begin{array}{l}
k=8:\omega _{i,8}(\psi )+\omega _{i+4,8}(\psi )=0,~~~i=0,1,2,3. \\
~
\end{array}
\\
&&
\begin{array}{l}
k=10:\left\{
\begin{array}{l}
\omega _{i,10}(\psi )+\omega _{i+5,10}(\psi )=0,~~~i=0,1,2,3,4; \\
\\
\omega _{0,10}(\psi )+\omega _{2,10}(\psi )+\omega _{3,10}(\psi )+\omega
_{4,10}(\psi )+\omega _{5,10}(\psi )=0.
\end{array}
\right. \\
~
\end{array}
\end{eqnarray}
The defining Eq. (\ref{omega-PF}) expresses the usual conventions, in which
one takes $\omega _{0,k}(\psi )$, $\omega _{1,k}(\psi )$, $\omega
_{2,k}(\psi )$ and $\omega _{k-1,k}(\psi )$ as basis for $\varpi _{k}\left(
\psi \right) $. Therefore, due to relations (\ref{rel-j}), the key quantity
turns out to be the holomorphic function $\omega _{0,k}$, whose series
expansion (convergent for $\left| \psi \right| <1$, with the \textit{%
fundamental region} \cite{CDLOGP1,CDLOGP2,KT} selected by $0\leqslant
arg\left( \psi \right) <\frac{2\pi }{k}$) reads \cite{KT}
\begin{equation}
\omega _{0,k}(\psi )=-\sum_{m=1}^{\infty }C_{k,m-1}\beta _{k}^{(k-1)m}\psi
^{m},  \label{omega0-1}
\end{equation}
with
\begin{equation}
\begin{array}{l}
C_{k,m-1}\equiv \frac{\Gamma (\frac{m}{k})\Gamma (1-\frac{m}{k})k^{m-1}}{%
\Gamma (m)\Pi _{i=0}^{4}\Gamma (1-\frac{m}{k}\nu _{i,k})}\gamma _{k}^{m}\in
\mathbb{R}, \\
~ \\
\gamma _{k}\equiv \Pi _{i=0}^{4}(\nu _{i,k})^{-\nu _{i,k}/k}\in \mathbb{R}%
_{0},
\end{array}
\label{omega0-2}
\end{equation}
where in $C_{k,m-1}$ $\Gamma $ denotes the Euler gamma function
\begin{equation}
\Gamma \left( s\right) \equiv \int_{0}^{\infty }t^{s-1}e^{-t}dt,~~~Re\left(
s\right) >0.  \label{Euler-Gamma}
\end{equation}
By using Eqs. (\ref{omega0-1})-(\ref{omega0-2}), the series expansion
(convergent for $\left| \psi \right| <1$, $0\leqslant arg\left( \psi \right)
<\frac{2\pi }{k}$) of $\varpi _{k}\left( \psi \right) $ can be written as
follows:
\begin{equation}
\varpi _{k}\left( \psi \right) =-\frac{(2\pi i)^{3}}{Ord(G_{k})}%
\sum_{m=1}^{\infty }(-1)^{m}C_{k,m-1}\psi ^{m-1}\left(
\begin{array}{l}
\beta _{k}^{3m} \\
~ \\
\beta _{k}^{m} \\
~ \\
\beta _{k}^{-m} \\
~ \\
\beta _{k}^{-3m}
\end{array}
\right) .  \label{omega-PF-series}
\end{equation}
The change between the PF basis and the symplectic basis for holomorphic $%
4\times 1$ period vector is given by:
\begin{equation}
\Pi _{k}(\psi )=M_{k}\varpi _{k}\left( \psi \right) .  \label{pi}
\end{equation}
where the $4\times 4$ constant matrices $M_{k}$ read \cite
{CDLOGP1,CDLOGP2,KT}
\begin{eqnarray}
M_{5} &=&\left(
\begin{array}{cccc}
-\frac{3}{5} & -\frac{1}{5} & \frac{21}{5} & \frac{8}{5} \\
0 & 0 & -1 & 0 \\
-1 & 0 & 8 & 3 \\
0 & 1 & -1 & 0
\end{array}
\right) \ ,\quad M_{6}=\left(
\begin{array}{cccc}
-\frac{1}{3} & -\frac{1}{3} & \frac{1}{3} & \frac{1}{3} \\
0 & 0 & -1 & 0 \\
-1 & 0 & 3 & 2 \\
0 & 1 & -1 & 0
\end{array}
\right) \ ,  \label{M1} \\
&&  \notag \\
&&  \notag \\
M_{8} &=&\left(
\begin{array}{cccc}
-\frac{1}{2} & -\frac{1}{2} & \frac{1}{2} & \frac{1}{2} \\
0 & 0 & -1 & 0 \\
-1 & 0 & 3 & 2 \\
0 & 1 & -1 & 0
\end{array}
\right) \ ,\quad M_{10}=\left(
\begin{array}{cccc}
0 & 1 & 1 & 1 \\
0 & 0 & -1 & 0 \\
1 & 0 & 0 & -1 \\
0 & 1 & -1 & 0
\end{array}
\right) \ .  \label{M2}
\end{eqnarray}

In SKG the holomorphic period vector $\Pi $ is usually normalized such that
(see \textit{e.g.} the report \cite{CDF}; recall Eq. (\ref{Omega}))
\begin{equation}
K\left( z,\overline{z}\right) =-ln\left[ i\Pi ^{T}\left( z\right) \Omega
\overline{\Pi }\left( \overline{z}\right) \right] =-ln\left\{ i\left[
\overline{X}^{\Lambda }\left( \overline{z}\right) F_{\Lambda }\left(
z\right) -X^{\Lambda }\left( z\right) \overline{F}_{\Lambda }\left(
\overline{z}\right) \right] \right\} .  \label{norm-PI}
\end{equation}
Specializing such a formula for the class of (mirror) $CY_{3}$s being
treated, one obtains that the K\"{a}hler potential is given by:
\begin{eqnarray}
K_{k}\left( \psi ,\overline{\psi }\right) &=&-ln\left[ i\;\Pi _{k}^{T}\left(
\psi \right) \Omega \overline{\Pi }_{k}\left( \overline{\psi }\right) \right]
=-ln\left[ -i\;\Pi _{k}^{\dag }\left( \overline{\psi }\right) \Omega \Pi
_{k}\left( \psi \right) \right] =  \notag \\
&=&-ln\left\{ i\left[ \overline{X}_{k}^{\Lambda }\left( \overline{\psi }%
\right) F_{k,\Lambda }\left( \psi \right) -X_{k}^{\Lambda }\left( \psi
\right) \overline{F}_{k,\Lambda }\left( \overline{\psi }\right) \right]
\right\} = \\
&=&-ln\left[ i\;\Pi _{k}^{\dag }\left( \overline{\psi }\right) \Sigma \Pi
_{k}\left( \psi \right) \right] =-ln\left[ -i\;\Pi _{k}^{T}\left( \psi
\right) \Sigma \overline{\Pi }_{k}\left( \overline{\psi }\right) \right] =
\notag \\
&=&-ln\left[ -i\;\varpi _{k}^{\dag }\left( \overline{\psi }\right)
m_{k}\varpi _{k}\left( \psi \right) \right] =-ln\left[ i\;\varpi
_{k}^{T}\left( \psi \right) m_{k}\overline{\varpi }_{k}\left( \overline{\psi
}\right) \right] ,  \label{kahlerpot}
\end{eqnarray}
where\footnote{%
Note the change of convention with respect to (the case $n_{V}=1$ of) the
defining Eq. (\ref{Omega}): $\Sigma =-\left. \Omega \right|
_{n_{V}=1}=\left. \Omega ^{T}\right| _{n_{V}=1}$. On the other hand, Eq. (%
\ref{PI-1-CY}) is nothing but the ($k$-indexed) case $n_{V}=1$ of the
definition (\ref{PI-PI}).}
\begin{eqnarray}
&&
\begin{array}{l}
\Sigma \equiv \left(
\begin{array}{ccc}
0_{2} &  & \mathbb{I}_{2} \\
&  &  \\
-\mathbb{I}_{2} &  & 0_{2}
\end{array}
\right) ;
\end{array}
\\
&&
\begin{array}{l}
\Pi _{k}\left( \psi \right) \equiv \left(
\begin{array}{l}
X_{k}^{0}\left( \psi \right) \\
X_{k}^{1}\left( \psi \right) \\
F_{k,0}\left( X\left( \psi \right) \right) \\
F_{k,1}\left( X\left( \psi \right) \right)
\end{array}
\right) ;
\end{array}
\label{PI-1-CY} \\
&&
\begin{array}{l}
m_{k}\equiv M_{k}^{\dag }\Sigma M_{k}=M_{k}^{T}\Sigma M_{k}=\frac{1}{\chi
_{k}}\left(
\begin{array}{cccc}
0 & -1 & -\lambda _{k} & -1 \\
1 & 0 & -\varsigma _{k} & -\lambda _{k} \\
\lambda _{k} & \varsigma _{k} & 0 & -1 \\
1 & \lambda _{k} & 1 & 0
\end{array}
\right) ,
\end{array}
\label{mk}
\end{eqnarray}
with the values of $\chi _{k}$, $\lambda _{k}$ and $\varsigma _{k}$ given in
Table 4. By recalling the third column from the left of Table 3, one can
observe that $Ord\left( G_{k}\right) =$ $\chi _{k}k^{2}$.
\begin{table}[tbp]
\begin{center}
\begin{tabular}{|c||c|c|c|}
\hline
$
\begin{array}{c}
\\
k \\
\downarrow
\end{array}
$ & $
\begin{array}{c}
\\
\chi _{k} \\
~
\end{array}
$ & $
\begin{array}{c}
\\
\lambda _{k} \\
~
\end{array}
$ & $
\begin{array}{c}
\\
\varsigma _{k} \\
~
\end{array}
$ \\ \hline\hline
$
\begin{array}{c}
\\
5 \\
~
\end{array}
$ & $
\begin{array}{c}
\\
5 \\
~
\end{array}
$ & $
\begin{array}{c}
\\
3 \\
~
\end{array}
$ & $
\begin{array}{c}
\\
3 \\
~
\end{array}
$ \\ \hline
$
\begin{array}{c}
\\
6 \\
~
\end{array}
$ & $
\begin{array}{c}
\\
3 \\
~
\end{array}
$ & $
\begin{array}{c}
\\
2 \\
~
\end{array}
$ & $
\begin{array}{c}
\\
0 \\
~
\end{array}
$ \\ \hline
$
\begin{array}{c}
\\
8 \\
~
\end{array}
$ & $
\begin{array}{c}
\\
2 \\
~
\end{array}
$ & $
\begin{array}{c}
\\
2 \\
~
\end{array}
$ & $
\begin{array}{c}
\\
1 \\
~
\end{array}
$ \\ \hline
$
\begin{array}{c}
\\
10 \\
~
\end{array}
$ & $
\begin{array}{c}
\\
1 \\
~
\end{array}
$ & $
\begin{array}{c}
\\
1 \\
~
\end{array}
$ & $
\begin{array}{c}
\\
-1 \\
~
\end{array}
$ \\ \hline
\end{tabular}
\end{center}
\caption{\textbf{Values of the integer constants }$\protect\chi _{k}$\textbf{%
, }$\protect\lambda _{k}$\textbf{\ and }$\protect\varsigma _{k}$}
\end{table}
Substituting Eq. (\ref{omega-PF-series}) and definition (\ref{mk}) into Eq. (%
\ref{kahlerpot}), one obtains the series expansion (converging for $\left|
\psi \right| <1$, $0\leqslant arg\left( \psi \right) <\frac{2\pi }{k}$) of
the K\"{a}hler potential:
\begin{equation}
K_{k}\left( \psi ,\overline{\psi }\right) =-ln\left( \frac{(2\pi )^{6}}{%
(Ord(G_{k}))^{2}}\sum_{m,n=1}^{\infty }C_{k,m-1}C_{k,n-1}\psi ^{m-1}%
\overline{\psi }^{n-1}F_{k,mn}\right) ,  \label{kahlerpot1}
\end{equation}
where the following infinite rank-2 tensor has been introduced\footnote{%
In the present paper $F$ will be used to denote two different entities:
\par
1) the holomorphic prepotential (with K\"{a}hler weights $\left( 4,0\right) $%
) of SK geometry, as defined by Eq. (\ref{21march-afternoon-1});
\par
2) the infinite rank-2 tensor, defined by Eq. (\ref{F}), introduced in the
study of the LG limit in the moduli space of (mirror) Fermat $CY_{3}$s.
\par
Attention should be paid to carefully identify the meaning of $F$, which
however can be understood at a glance in the various frameworks. It should
also be recalled that $\mathcal{F}$ denotes the K\"{a}hler-invariant
holomorphic prepotential of SK geometry, as defined by Eq. (\ref{F-call}).}:
\begin{eqnarray}
&&
\begin{array}{l}
F_{k,m\,n}\equiv i(-1)^{m+n+1}\left(
\begin{array}{cccc}
\beta _{k}^{-3n}, & \beta _{k}^{-n}, & \beta _{k}^{n}, & \beta _{k}^{3n}
\end{array}
\right) m_{k}\left(
\begin{array}{l}
\beta _{k}^{3m} \\
\beta _{k}^{m} \\
\beta _{k}^{-m} \\
\beta _{k}^{-3m}
\end{array}
\right) = \\
\\
=\frac{2}{\chi _{k}}e^{i(m+n)\pi }\left\{ sin(\frac{3n-m}{k}\pi )+sin(\frac{%
3m-n}{k}\pi )+sin(\frac{3(n+m)}{k}\pi )+\right. \\
\\
\left. +\varsigma _{k}sin(\frac{n+m}{k}\pi )+\lambda _{k}\left[ sin(\frac{%
3n+m}{k}\pi )+sin(\frac{3m+n}{k}\pi )\right] \right\} .
\end{array}
\notag \\
&&  \label{F}
\end{eqnarray}
From such a definition, $F_{k,mn}$ turns out to have the following relevant
properties:
\begin{equation}
\begin{array}{l}
\overline{F_{k,m\,n}}=F_{k,m\,n}; \\
\\
F_{k,m\,n}=F_{k,n\,m}; \\
\\
F_{k,m+k~\,n}=F_{k,m\,~n+k}=(-1)^{k+1}F_{k,m\,n}; \\
\\
F_{k,m\,n}=0\text{ if }n+m=k; \\
\\
F_{k,kk}=0.
\end{array}
\label{prop-F}
\end{equation}
Consequently, at most only $\frac{k(k+1)}{2}-\left[ \frac{k}{2}\right] -1$
(real) non-vanishing independent elements of $F_{k,m\,n}$ exist (where $%
\left[ \frac{k}{2}\right] $ denotes the integer part of $\frac{k}{2}$), even
though, as evident from Eqs. (\ref{k=5-Fmn}), (\ref{k=6-Fmn}), (\ref{k=8-Fmn}%
) and (\ref{k=10-Fmn}) below, actually such an upper bound is never reached
for the allowed values of the Fermat parameter $k=5,6,8,10$.

The holomorphic superpotential (also named $\mathcal{N}=2$ holomorphic
central charge function) is given by:
\begin{equation}
W_{k}\left( \psi ;q,p\right) =\Gamma \Pi _{k}\left( \psi \right) ,
\label{sup}
\end{equation}
where the $1\times 4$ BH charge vector in the symplectic basis is here
defined as\footnote{%
Notice the change in the notation of the symplectic charge vectors with
respect to the notation used in Sects. \ref{Intro} and \ref{SKG-gen}. $%
\widetilde{\Gamma }$ defined in the second row of (\ref{Gamma-Fermat}) has
the electric and magnetic charges interchanged with respect to the (case $%
n_{V}=1$ of the) symplectic vector defined by Eq. (\ref{Gamma-tilde}).}
\begin{equation}
\begin{array}{l}
\Gamma \equiv (-p^{0},\,-p^{1},\,q_{0},\,q_{1})=\widetilde{\Gamma }\Sigma ,
\\
\\
\widetilde{\Gamma }\equiv (q_{0},\,q_{1},p^{0},\,p^{1}).
\end{array}
\label{Gamma-Fermat}
\end{equation}
Using Eqs. (\ref{omega-PF-series}), (\ref{pi}) and (\ref{mk}), one can
obtain the following series expansion (convergent for $\left| \psi \right|
<1 $, $0\leqslant arg\left( \psi \right) <\frac{2\pi }{k}$) of the
holomorphic superpotential:
\begin{equation}
W_{k}\left( \psi ;q,p\right) =A_{k}\sum_{m=1}^{\infty }C_{k,m-1}\psi
^{m-1}N_{k,m}\left( q,p\right) ,  \label{sup2}
\end{equation}
where the following quantities have been introduced:
\begin{eqnarray}
&&
\begin{array}{l}
A_{k}\equiv -\frac{1}{\chi _{k}}\frac{(2\pi i)^{3}}{Ord(G_{k})};
\end{array}
\\
&&  \notag \\
&&
\begin{array}{l}
N_{k,m}\left( q,p\right) \equiv (-1)^{m}\left[ n_{k,1}\left( q,p\right)
\beta _{k}^{3m}+n_{k,2}\left( q,p\right) \beta _{k}^{m}+n_{k,3}\left(
q,p\right) \beta _{k}^{-m}+n_{k,4}\left( q,p\right) \beta _{k}^{-3m}\right] ,
\end{array}
\label{N-def}
\end{eqnarray}
where
\begin{equation}
n_{k}\left( q,p\right) \equiv \chi _{k}\Gamma M_{k}\in \mathbb{Z}^{4}
\label{PF-charge}
\end{equation}
is the $1\times 4$ BH charge vector in the PF basis.

By inverting the definition (\ref{PF-charge}), one obtains
\begin{equation}
\Gamma =\frac{1}{\chi _{k}}n_{k}\left( q,p\right) M_{k}^{-1}\in \mathbb{Z}%
^{4},  \label{PF-charge-inv}
\end{equation}
where $\Gamma $ is defined in Eq. (\ref{Gamma-Fermat}). Eqs. (\ref{PF-charge}%
)-(\ref{PF-charge-inv}) express the change between the symplectic and PF
basis of BH charges.

By recalling Eq. (\ref{VBH1-1-modulus}) and using Eqs. (\ref{kahlerpot1})
and (\ref{sup2}), the general form of the ``effective BH potential''
function $V_{BH,k}\left( \psi ,\overline{\psi };q,p\right) $ for the
Calabi-Yau threefolds $\mathcal{M}_{k}^{\prime }$s reads
\begin{eqnarray}
V_{BH,k}\left( \psi ,\overline{\psi };q,p\right) &=&\frac{1}{\chi
_{k}^{2}F_{k,11}}\left[ exp\left[ \widetilde{K}_{k}\left( \psi ,\overline{%
\psi }\right) \right] \right] \left[ \left| \widetilde{W}_{k}\right|
^{2}\left( \psi ,\overline{\psi };q,p\right) +(g_{\psi \bar{\psi},k}\left(
\psi ,\overline{\psi }\right) )^{-1}\left| D_{\psi }\widetilde{W}_{k}\right|
^{2}\left( \psi ,\overline{\psi };q,p\right) \right] \equiv  \notag \\
&&  \notag \\
&\equiv &\frac{1}{\chi _{k}^{2}F_{k,11}}\widetilde{V}_{BH,k}\left( \psi ,%
\overline{\psi };q,p\right) ,  \label{pot}
\end{eqnarray}
where
\begin{equation}
\begin{array}{l}
\widetilde{K}_{k}\left( \psi ,\overline{\psi }\right) \equiv K_{k}\left(
\psi ,\overline{\psi }\right) +ln\left[ \frac{(2\pi )^{6}}{(Ord(G_{k}))^{2}}%
C_{k,0}^{2}F_{k,11}\right] ; \\
\\
\widetilde{W}_{k}\left( \psi ;q,p\right) \equiv \frac{W_{k}\left( \psi
;q,p\right) }{A_{k}C_{k,0}}.
\end{array}
\label{W-tilde-def}
\end{equation}
\bigskip

\subsection*{Remark}

It is worth spending a few words concerning the holomorphic prepotential $%
F\left( X\left( \psi \right) \right) $. In the treatment of 1-modulus SK
geometry underlying the moduli space of Fermat $CY_{3}$-compactifications,
we will assume it to exist. This corresponds to the choice of a symplectic
gauge such that the holomorphic function (see Eq. (\ref{e-1-modulus}))
\begin{equation}
e_{k}\equiv \frac{\partial \left( \frac{X_{k}^{1}}{X_{k}^{0}}\right) }{%
\partial \psi }=\frac{X_{k}^{0}\partial X_{k}^{1}-X_{k}^{1}\partial X_{k}^{0}%
}{\left[ X_{k}^{0}\right] ^{2}}
\end{equation}
does not vanish, at least for $\psi $ belonging to a suitable neighbourhood
of the LG point $\psi =0$ (see Subsect. 4.5 of the second Ref. of \cite{DFF}%
). By specializing Eq. (\ref{hom-deg-2-F}) for $n_{V}=1$, one achieves:
\begin{equation}
F\left( X\left( \psi \right) \right) =\frac{1}{2}\left[ F_{0}\left( \psi
\right) X^{0}\left( \psi \right) +F_{1}\left( \psi \right) X^{1}\left( \psi
\right) \right] =\frac{1}{2}\left[ \Pi ^{1}\left( \psi \right) \Pi
^{3}\left( \psi \right) +\Pi ^{2}\left( \psi \right) \Pi ^{4}\left( \psi
\right) \right] ,
\end{equation}
where $\Pi ^{i}\left( \psi \right) $ denotes the $i$-th component ($%
i=1,2,3,4 $) of the $4\times 1$ symplectic holomorphic period vector $\Pi
_{k}\left( \psi \right) $ given by Eq. (\ref{PI-1-CY}).

Consequently, by recalling Eqs. (\ref{omega-PF-series}) and (\ref{pi}), $%
F\left( X\left( \psi \right) \right) $ can be explicitly computed in power
series expansion (convergent for $\left| \psi \right| <1$, $0\leqslant
arg\left( \psi \right) <\frac{2\pi }{k}$) for the $k$-parametrized class of
Fermat $CY_{3}$s. By introducing the $4\times 1$ vector
\begin{equation}
\xi _{k,m}=\left(
\begin{array}{l}
\xi _{k,m}^{1} \\
\xi _{k,m}^{2} \\
\xi _{k,m}^{3} \\
\xi _{k,m}^{4}
\end{array}
\right) \equiv M_{k}\left(
\begin{array}{l}
\beta _{k}^{3m} \\
\beta _{k}^{m} \\
\beta _{k}^{-m} \\
\beta _{k}^{-3m}
\end{array}
\right) ,  \label{csi-csi}
\end{equation}
one obtains that (in the third line round brackets denote symmetrization
with respect to enclosed indices)
\begin{eqnarray}
F_{k}\left( X_{k}\left( \psi \right) \right) &=&\frac{1}{2}\left[ \Pi
_{k}^{1}\left( \psi \right) \Pi _{k}^{3}\left( \psi \right) +\Pi
_{k}^{2}\left( \psi \right) \Pi _{k}^{4}\left( \psi \right) \right] =  \notag
\\
&=&\frac{1}{2}\frac{(2\pi i)^{6}}{\left( Ord(G_{k})\right) ^{2}}%
\sum_{m,n=1}^{\infty }(-1)^{m+n}C_{k,m-1}C_{k,n-1}\left( \xi _{k,m}^{1}\xi
_{k,n}^{3}+\xi _{k,m}^{2}\xi _{k,n}^{4}\right) \psi ^{m+n-2}=  \notag \\
&=&-\frac{32\pi ^{6}}{\left( Ord(G_{k})\right) ^{2}}\sum_{m,n=1}^{\infty
}(-1)^{m+n}C_{k,m-1}C_{k,n-1}\Phi _{k,\left( m,n\right) }\psi
^{m+n-2},\smallskip  \label{hol-prep-F-CY} \\
&&  \notag \\
\Phi _{k,m,n} &\equiv &\left( \xi _{k,m}^{1}\xi _{k,n}^{3}+\xi _{k,m}^{2}\xi
_{k,n}^{4}\right) .  \label{PHIZ}
\end{eqnarray}
Thence, one can symmetrically truncate with respect to the dummy indices $m$
and $n$, both ranging up to $l$. The resulting series expansion is complete
up to $\mathcal{O}\left( \psi ^{l-1}\right) $ included; we will denote such
a symmetric truncation as
\begin{equation}
F_{k,l}\left( X_{k}\left( \psi \right) \right) =-\frac{32\pi ^{6}}{\left(
Ord(G_{k})\right) ^{2}}\sum_{m=1}^{l}%
\sum_{n=1}^{l}(-1)^{m+n}C_{k,m-1}C_{k,n-1}\psi ^{m+n-2}\Phi _{k,\left(
m,n\right) }.  \label{hol-prep-F-l-CY}
\end{equation}

On the other hand, by recalling Eq. (\ref{F-call}) and Eq. (\ref{PI-1-CY}),
the K\"{a}hler gauge-invariant holomorphic prepotential reads
\begin{equation}
\mathcal{F}_{k}\left( \psi \right) \equiv F_{k}\left( \frac{X_{k}^{1}(\psi )%
}{X_{k}^{0}(\psi )}\right) =\frac{F_{k}\left( X_{k}^{0}(\psi
),X_{k}^{1}(\psi )\right) }{\left( X_{k}^{0}(\psi )\right) ^{2}}=\frac{1}{2}%
\frac{\sum_{m,n=1}^{\infty }(-1)^{m+n}C_{k,m-1}C_{k,n-1}\Phi _{k,\left(
m,n\right) }\psi ^{m+n-2}}{\left[ \sum_{r=1}^{\infty }(-1)^{r}C_{k,r-1}\xi
_{k,r}^{1}\psi ^{r-1}\right] ^{2}}.  \label{F-call-CY}
\end{equation}

The vector $\xi _{k,m}$, as well as the infinite rank-2 tensor $\Phi
_{k,m,n} $ and the holomorphic prepotentials $F_{k,l}\left( X_{k}\left( \psi
\right) \right) $ and $\mathcal{F}_{k}\left( \psi \right) $ will be computed
for each value of the Fermat parameter $k=5,6,8,10$ (classifying the mirror
Fermat $CY_{3}$s $\mathcal{M}_{k}^{\prime }$s) in Subsubsects. \ref{prep-5},
\ref{prep-6}, \ref{prep-8} and \ref{prep-10}, respectively.

As previously pointed out, it is here worth remarking that in the general
treatment of (mirror) Fermat $CY_{3}$s given above (and explicited the next
Sects. \ref{quintic}-\ref{tentic}) $t_{k}\left( \psi \right) \equiv
X_{k}^{1}\left( \psi \right) /X_{k}^{0}\left( \psi \right) $ is \textit{not}
a special coordinate, \textit{i.e.} $t_{k}\left( \psi \right) \neq \psi $,
and moreover the K\"{a}hler gauge is left unfixed (in particular, $%
X_{k}^{0}\left( \psi \right) \neq 1$).\bigskip

The next Sects. \ref{quintic}-\ref{tentic} will be devoted to consistently
solving the 1-modulus AEs (\ref{AEs-1-modulus-W}) (with covariant
derivatives given by Eqs. (\ref{DW})-(\ref{DDW})) near the LG point $\psi =0$%
, using all the formal machinery elaborated above in the framework of
1-modulus SK geometry underlying the moduli space of Fermat $CY_{3}$%
-compactifications. In other words, we will solve the criticality condition
for the ``effective BH potential'' (\ref{pot}) near the LG point $\psi =0$,
obtaining the constraints which define the BH charge configurations
supporting the LG point to be a critical point of $V_{BH,k}$ given by Eq. (%
\ref{pot}). Furthermore, we will address the issue of the stability, by
inspecting the real form of the $2\times 2$ Hessian matrix, and checking the
(un)fulfilling of the relevant stability conditions treated in Subsect. \ref
{Sect4-1}, as well.

We will find that (beside the $\frac{1}{2}$-BPS solutions, existing and
stable in all cases) for $k=5,8$ only non-BPS, $Z\neq 0$ solutions exist,
and they are attractors (local \textit{minima} of $V_{BH}$), whereas for $%
k=6,10$ only non-BPS, $Z=0$ solutions exist, and they are \textit{not}
attractors in a strict sense (since they are \textit{saddle points} of $%
V_{BH}$).

\section{$k=5$ : Mirror \textit{Quintic} near LG Point\label{quintic}}

\setcounter{equation}0
\def\theequation{7.\arabic{subsection}.\arabic{equation}}

\subsection{Geometric Setup}

In the case of mirror \textit{quintic} $\mathcal{M}_{5}^{\prime }$ it is
easy to realize that one has to consider the ``effective BH potential'' (\ref
{pot}) (at least) up to $\mathcal{O}\left( \psi ^{3}\right) $ (or, as always
understood below, $\mathcal{O}\left( \overline{\psi }^{3}\right) $). As a
consequence, the AEs (\ref{AEs-1-modulus-W}) and the Hessian matrix will be
known up to $\mathcal{O}\left( \psi \right) $.

For $k=5$ the definitions (\ref{omega0-2}) yield
\begin{equation}
C_{5,5l-1}=0,\quad l\in \mathbb{N;}
\end{equation}
moreover, since $F_{5,m+5,\,n}=F_{5,m\,,n+5}=F_{5,m\,n}$ (see the third of
properties (\ref{prop-F})), the only independent elements of the rank-2
tensor $F_{5}$ belong to the $5\times 5$ matrix
\begin{eqnarray}
F_{5,m\,n} &=&\left(
\begin{array}{cccccc}
\sqrt{5+2\sqrt{5}} & 0 & 0 & 0 & -\sqrt{5+2\sqrt{5}} &  \\
0 & -\sqrt{5-2\sqrt{5}} & 0 & 0 & \sqrt{5-2\sqrt{5}} &  \\
0 & 0 & \sqrt{5-2\sqrt{5}} & 0 & -\sqrt{5-2\sqrt{5}} &  \\
0 & 0 & 0 & -\sqrt{5+2\sqrt{5}} & \sqrt{5+2\sqrt{5}} &  \\
-\sqrt{5+2\sqrt{5}} & \sqrt{5-2\sqrt{5}} & -\sqrt{5-2\sqrt{5}} & \sqrt{5+2%
\sqrt{5}} & 0 &
\end{array}
\right) .  \notag \\
&&  \label{k=5-Fmn}
\end{eqnarray}

Let us now write down all the relevant quantities up to the needed order
(here and below, unless otherwise specified, we omit the Fermat parameter $%
k=5$):
\begin{eqnarray}
&&
\begin{array}{l}
\widetilde{K}\approx (\sqrt{5}-2)\frac{C_{1}^{2}}{C_{0}^{2}}\left[ \psi \bar{%
\psi}-\left( \frac{C_{2}^{2}}{C_{1}^{2}}-\frac{(\sqrt{5}-2)}{2}\frac{%
C_{1}^{2}}{C_{0}^{2}}\right) (\psi \bar{\psi})^{2}+\frac{C_{5}C_{0}}{%
C_{1}^{2}}(\sqrt{5}+2)(\psi ^{5}+\bar{\psi}^{5})\right] +\mathcal{O}(\psi
^{6}); \\
~
\end{array}
\\
&&
\begin{array}{l}
g\approx (\sqrt{5}-2)\frac{C_{1}^{2}}{C_{0}^{2}}\left[ 1-4\left( \frac{%
C_{2}^{2}}{C_{1}^{2}}-\frac{(\sqrt{5}-2)}{2}\frac{C_{1}^{2}}{C_{0}^{2}}%
\right) \psi \bar{\psi}\right] +\mathcal{O}(\psi ^{4}); \\
~
\end{array}
\label{k=5-metric} \\
&&
\begin{array}{l}
\widetilde{W}\approx N_{1}+\frac{C_{1}}{C_{0}}N_{2}\psi +\frac{C_{2}}{C_{0}}%
\bar{N}_{2}\psi ^{2}+\frac{C_{3}}{C_{0}}\bar{N}_{1}\psi ^{3}+\mathcal{O}%
(\psi ^{5}). \\
~
\end{array}
\end{eqnarray}
Now, by using the formul\ae\ of the general analysis exploited in Sect. \ref
{GA}, we can get the ``effective BH potential'' and the holomorphic
superpotential, as well as their (covariant) derivatives, up to $\mathcal{O}%
\left( \psi \right) $ (notice that in all the treatments of Sects. \ref
{quintic}-\ref{tentic} we are interested only in ordinary derivatives of $%
\widetilde{V}_{BH}$, since they coincide with the covariant ones at the
critical points of $\widetilde{V}_{BH}$):
\begin{eqnarray}
&&
\begin{array}{l}
\widetilde{W}=N_{1}+\frac{C_{1}}{C_{0}}N_{2}\psi ;\label{W-k=5} \\
~
\end{array}
\\
&&
\begin{array}{l}
D\widetilde{W}=\frac{C_{1}}{C_{0}}\left[ N_{2}+2\frac{C_{2}}{C_{1}}\bar{N}%
_{2}\psi +\frac{C_{1}}{C_{0}}(\sqrt{5}-2)N_{1}\bar{\psi}\right] ; \\
~
\end{array}
\label{covder} \\
&&
\begin{array}{l}
D^{2}\widetilde{W}=2\frac{C_{2}}{C_{0}}\bar{N}_{2}+6\frac{C_{3}}{C_{0}}\bar{N%
}_{1}\psi +4\frac{C_{2}^{2}}{C_{0}C_{1}}N_{2}\bar{\psi}, \\
~
\end{array}
\\
&&
\begin{array}{l}
\widetilde{V}_{BH}=|N_{1}|^{2}+(\sqrt{5}+2)|N_{2}|^{2}+2\frac{C_{1}}{C_{0}}%
\left[ N_{2}\bar{N}_{1}+(\sqrt{5}+2)\frac{C_{2}C_{0}}{C_{1}^{2}}(\bar{N}%
_{2})^{2}\right] \psi + \\
~ \\
\qquad \quad +2\frac{C_{1}}{C_{0}}\left[ \bar{N}_{2}N_{1}+(\sqrt{5}+2)\frac{%
C_{2}C_{0}}{C_{1}^{2}}(N_{2})^{2}\right] \bar{\psi}; \\
~
\end{array}
\label{Vtilde-k=5} \\
&&
\begin{array}{l}
\partial \widetilde{V}_{BH}=2\frac{C_{1}}{C_{0}}\left[ N_{2}\bar{N}_{1}+(%
\sqrt{5}+2)\frac{C_{2}C_{0}}{C_{1}^{2}}(\bar{N}_{2})^{2}\right] +6\frac{C_{2}%
}{C_{0}}\left[ 1+(\sqrt{5}+2)\frac{C_{3}C_{0}}{C_{2}C_{1}}\right] \bar{N}_{1}%
\bar{N}_{2}\psi + \\
~ \\
\qquad +2\frac{C_{1}^{2}}{C_{0}^{2}}\left[ |N_{1}|^{2}(\sqrt{5}%
-2)+|N_{2}|^{2}\left( 1+4(\sqrt{5}+2)\frac{C_{0}^{2}C_{2}^{2}}{C_{1}^{4}}%
\right) \right] \bar{\psi}; \\
~
\end{array}
\label{AE-k=5}
\end{eqnarray}
\begin{eqnarray}
&&
\begin{array}{l}
\partial ^{2}\widetilde{V}_{BH}=6\frac{C_{2}}{C_{0}}\left[ 1+(\sqrt{5}+2)%
\frac{C_{3}C_{0}}{C_{2}C_{1}}\right] \bar{N}_{1}\bar{N}_{2}+ \\
~ \\
\quad +24\frac{C_{3}}{C_{0}}(\bar{N}_{1})^{2}\psi +4\frac{C_{1}}{C_{0}}\left[
\frac{C_{2}}{C_{0}}\left( 1+4(\sqrt{5}+2)\frac{C_{0}^{2}C_{2}^{2}}{C_{1}^{4}}%
\right) (\bar{N}_{2})^{2}\right. + \\
~ \\
\quad +\left. (\sqrt{5}-2)\frac{C_{1}^{2}}{C_{0}^{2}}\left( 1+(\sqrt{5}+2)%
\frac{C_{0}^{2}C_{2}^{2}}{C_{1}^{4}}+3(\sqrt{5}+2)^{2}\frac{%
C_{0}^{3}C_{2}C_{3}}{C_{1}^{5}}\right) \bar{N}_{1}N_{2}\right] \bar{\psi};
\\
~ \\
~
\end{array}
\label{ddV-k=5} \\
&&
\begin{array}{l}
\partial \overline{\partial }\widetilde{V}_{BH}=2\frac{C_{1}^{2}}{C_{0}^{2}}%
\left[ |N_{1}|^{2}(\sqrt{5}-2)+|N_{2}|^{2}\left( 1+4(\sqrt{5}+2)\frac{%
C_{0}^{2}C_{2}^{2}}{C_{1}^{4}}\right) \right] + \\
~ \\
\quad +4\frac{C_{1}}{C_{0}}\left[ \frac{C_{2}}{C_{0}}\left( 1+4(\sqrt{5}+2)%
\frac{C_{0}^{2}C_{2}^{2}}{C_{1}^{4}}\right) (\bar{N}_{2})^{2}\right. + \\
~ \\
\quad +\left. (\sqrt{5}-2)\frac{C_{1}^{2}}{C_{0}^{2}}\left( 1+(\sqrt{5}+2)%
\frac{C_{0}^{2}C_{2}^{2}}{C_{1}^{4}}+3(\sqrt{5}+2)^{2}\frac{%
C_{0}^{3}C_{2}C_{3}}{C_{1}^{5}}\right) \bar{N}_{1}N_{2}\right] \psi + \\
~ \\
\quad +4\frac{C_{1}}{C_{0}}\left[ \frac{C_{2}}{C_{0}}\left( 1+4(\sqrt{5}+2)%
\frac{C_{0}^{2}C_{2}^{2}}{C_{1}^{4}}\right) (N_{2})^{2}\right. + \\
~ \\
+\left. (\sqrt{5}-2)\frac{C_{1}^{2}}{C_{0}^{2}}\left( 1+(\sqrt{5}+2)\frac{%
C_{0}^{2}C_{2}^{2}}{C_{1}^{4}}+3(\sqrt{5}+2)^{2}\frac{C_{0}^{3}C_{2}C_{3}}{%
C_{1}^{5}}\right) N_{1}\bar{N}_{2}\right] \bar{\psi}.
\end{array}
\label{ddV-k=5-2}
\end{eqnarray}
\setcounter{equation}0
\def\theequation{7.1.\arabic{subsubsection}.\arabic{equation}}

\subsubsection{\label{prep-5}The Holomorphic Prepotentials $F$ and $\mathcal{%
F}$}

In order to compute the holomorphic prepotential $F\left( X\left( \psi
\right) \right) $ and its ``K\"{a}hler-invariant counterpart'' $\mathcal{F}%
\left( \psi \right) $ defined by Eq. (\ref{F-call}), we have to recall the
general formul\ae\ for (mirror) Fermat $CY_{3}$s introduced in Sect. \ref{GA}%
, and specialize them for $k=5$. By recalling the definitions (\ref{csi-csi}%
), (\ref{PHIZ}), (\ref{hol-prep-F-l-CY}), (\ref{F-call-CY}), one can
compute:
\begin{eqnarray}
&&
\begin{array}{l}
\xi _{5,m}=\left(
\begin{array}{l}
\frac{1}{5}\left[ 3+2cos\left( \frac{2\pi m}{5}\right) \right] \left[
5cos\left( \frac{\pi m}{5}\right) -11isin\left( \frac{\pi m}{5}\right)
\right] \\
\\
-cos\left( \frac{\pi m}{5}\right) +isin\left( \frac{\pi m}{5}\right) \\
\\
2\left[ 3+2cos\left( \frac{2\pi m}{5}\right) \right] \left[ cos\left( \frac{%
\pi m}{5}\right) -2isin\left( \frac{\pi m}{5}\right) \right] \\
\\
2isin\left( \frac{\pi m}{5}\right)
\end{array}
\right) ;
\end{array}
\label{csi-csi-5} \\
&&  \notag \\
&&  \notag \\
&&
\begin{array}{l}
\Phi _{5,m,n}= \\
\\
=\frac{1}{5}\left\{
\begin{array}{l}
\left[
\begin{array}{l}
-24cos\left( \frac{\pi \left( n-3m\right) }{5}\right) -24cos\left( \frac{\pi
\left( 3n-m\right) }{5}\right) +64cos\left( \frac{\pi \left( 3n+m\right) }{5}%
\right) +64cos\left( \frac{\pi \left( n+3m\right) }{5}\right) + \\
\\
-13cos\left( \frac{\pi \left( n-m\right) }{5}\right) -17cos\left( \frac{3\pi
\left( n-m\right) }{5}\right) +173cos\left( \frac{\pi \left( n+m\right) }{5}%
\right) +27cos\left( \frac{3\pi \left( n+m\right) }{5}\right)
\end{array}
\right] + \\
\\
-i\left[
\begin{array}{l}
-21sin\left( \frac{\pi \left( n-3m\right) }{5}\right) +21sin\left( \frac{\pi
\left( 3n-m\right) }{5}\right) +63sin\left( \frac{\pi \left( 3n+m\right) }{5}%
\right) +63sin\left( \frac{\pi \left( n+3m\right) }{5}\right) + \\
\\
+173sin\left( \frac{\pi \left( n+m\right) }{5}\right) +21sin\left( \frac{%
3\pi \left( n+m\right) }{5}\right)
\end{array}
\right]
\end{array}
\right\} =\Phi _{5,\left( m,n\right) };
\end{array}
\notag \\
&&  \label{PHIZ-5}
\end{eqnarray}
\begin{eqnarray}
&&
\begin{array}{l}
F_{k=5,l=5}\left( X_{5}\left( \psi \right) \right) =-\frac{32\pi ^{6}}{5^{6}}%
\sum_{m=1}^{5}\sum_{n=1}^{5}(-1)^{m+n}C_{5,m-1}C_{5,n-1}\psi ^{m+n-2}\Phi
_{5,m,n}= \\
\\
\\
=-\frac{32\pi ^{8}}{5^{6}}\left[
\begin{array}{l}
-\frac{43+11\sqrt{5}+i\sqrt{11890+5302\sqrt{5}}}{\Gamma ^{10}\left( \frac{4}{%
5}\right) }+\frac{4\left( 115-5\sqrt{5}+11i\sqrt{10\left( 5+\sqrt{5}\right) }%
\right) }{\Gamma ^{5}\left( \frac{3}{5}\right) \Gamma ^{5}\left( \frac{4}{5}%
\right) }\psi + \\
\\
-5\left[ \frac{20\left( -40+3\sqrt{5}-i\sqrt{5725-2510\sqrt{5}}\right) }{%
\left( 5+\sqrt{5}\right) \Gamma ^{10}\left( \frac{3}{5}\right) }-\frac{2%
\sqrt{5}\left( 5+23\sqrt{5}+11i\sqrt{10-2\sqrt{5}}\right) }{\Gamma
^{5}\left( \frac{2}{5}\right) \Gamma ^{5}\left( \frac{4}{5}\right) }\right]
\psi ^{2}+ \\
\\
+\frac{625\sqrt{2050+710\sqrt{5}}}{16\pi ^{5}}\psi ^{3}+ \\
\\
+125\left[ \frac{5\left( -40+3\sqrt{5}+i\sqrt{5725-2510\sqrt{5}}\right) }{%
\left( 5+\sqrt{5}\right) \Gamma ^{10}\left( \frac{2}{5}\right) }-\frac{2%
\sqrt{5}\left( 5+23\sqrt{5}-11i\sqrt{10-2\sqrt{5}}\right) }{3\Gamma
^{5}\left( \frac{1}{5}\right) \Gamma ^{5}\left( \frac{3}{5}\right) }\right]
\psi ^{4}
\end{array}
\right] +\mathcal{O}\left( \psi ^{5}\right) ;
\end{array}
\notag \\
&&  \label{F55-CY}
\end{eqnarray}
\begin{eqnarray}
&&
\begin{array}{l}
\mathcal{F}_{5}\left( \psi \right) \equiv F_{5}\left( \frac{X_{5}^{1}(\psi )%
}{X_{5}^{0}(\psi )}\right) =\frac{F_{5}\left( X_{5}^{0}(\psi
),X_{5}^{1}(\psi )\right) }{\left( X_{5}^{0}(\psi )\right) ^{2}}=\frac{1}{2}%
\frac{\sum_{m,n=1}^{\infty }(-1)^{m+n}C_{5,m-1}C_{5,n-1}\Phi _{5,m,n}\psi
^{m+n-2}}{\left[ \sum_{r=1}^{\infty }(-1)^{r}C_{5,r-1}\xi _{5,r}^{1}\psi
^{r-1}\right] ^{2}}\approx \\
\\
\approx \frac{\Phi _{5,1,1}}{2\left( \xi _{5,1}^{1}\right) ^{2}}\left[ 1-2%
\frac{C_{5,1}\Phi _{5,2,1}}{C_{5,0}\Phi _{5,1,1}}\psi +\mathcal{O}\left(
\psi ^{2}\right) \right] \left[ 1+2\frac{C_{5,1}\xi _{5,2}^{1}}{C_{5,0}\xi
_{5,1}^{1}}\psi +\mathcal{O}\left( \psi ^{2}\right) \right] \approx \\
\\
\approx \frac{\Phi _{5,1,1}}{2\left( \xi _{5,1}^{1}\right) ^{2}}\left[
1+2\left( \frac{\xi _{5,2}^{1}}{\xi _{5,1}^{1}}-\frac{\Phi _{5,2,1}}{\Phi
_{5,1,1}}\right) \frac{C_{5,1}}{C_{5,0}}\psi \right] +\mathcal{O}\left( \psi
^{2}\right) ,
\end{array}
\notag \\
&&  \label{F-call-CY-5}
\end{eqnarray}
where in the formula (\ref{F-call-CY-5}) for the K\"{a}hler-invariant
holomorphic prepotential $\mathcal{F}_{5}$ only the leading terms in the
\textit{LG limit }$\psi \longrightarrow 0$ have been retained. Eq. (\ref
{F55-CY}) is the exact expression (complete up to $\mathcal{O}\left( \psi
^{4}\right) $ included) of the holomorphic prepotential $F_{5}$ in a
suitable neighbourhood of the LG point of the moduli space of the (mirror)%
\textit{\ quintic} $\mathcal{M}_{5}^{\prime }$.
\setcounter{equation}0
\def\theequation{7.\arabic{subsection}.\arabic{equation}}

\subsection{\textit{``Criticality Condition''} Approach and Attractor
Equation}

Let us now find the solutions of the AE $\partial \widetilde{V}_{BH}\left(
\psi ,\overline{\psi };q,p\right) =0$, and check their stability. Since we
are working (very) near the LG point, by using Eq. (\ref{Vtilde-k=5}) we can
rewrite the AE for $\mathcal{M}_{5}^{\prime }$ as follows:
\begin{equation}
N_{2}\bar{N}_{1}+(\sqrt{5}+2)\frac{C_{2}C_{0}}{C_{1}^{2}}(\bar{N}%
_{2})^{2}\approx 0.  \label{attreq}
\end{equation}
Here we simply put $\psi =0$ in Eq. (\ref{AE-k=5}). Solving Eq. (\ref{attreq}%
), we will find one (or more) set(s) of BH charges supporting $\psi
\approx 0 $ to be a critical point of $V_{BH}$. As understood
throughout all our treatment of Sects. \ref{quintic}-\ref{tentic},
\textit{\c{c}a va sans dire} that actual BH charges are very close
to the found one, and also that the critical value of $\psi $ may
not be zero, rather it may belong to a suitable neighbourhood of the
LG point. \setcounter{equation}0
\def\theequation{7.\arabic{subsection}.\arabic{equation}}

\subsection{\label{21march-1}Critical Hessian of $V_{BH}$}

The stability of the critical point $\psi \approx 0$ of $V_{BH}$ is governed
by the real form of the symmetric Hessian $2\times 2$ matrix of $V_{BH}$
evaluated at the considered extremum; in general, it reads (see Eq. (\ref
{Hessian-real}))
\begin{equation}
H_{\text{real form}}^{V_{BH}}\equiv \left(
\begin{array}{cc}
\mathcal{A} & \mathcal{C} \\
\mathcal{C} & \mathcal{B}
\end{array}
\right) ,  \label{Hessian-real-form}
\end{equation}
where $\mathcal{A}$, $\mathcal{B}$, $\mathcal{C}\in \mathbb{R}$ are given in
terms of $\partial ^{2}V_{BH}$, $\partial \overline{\partial }V_{BH}$, $%
\overline{\partial }^{2}V_{BH}\in \mathbb{C}$ by (the $n_{V}=1$ case of) Eq.
(\ref{4jan2}). By using such an Eq., and also Eqs. (\ref{ddV-k=5})-(\ref
{ddV-k=5-2}) evaluated along the criticality condition (\ref{attreq}), it
can be computed that the components of $H_{\text{real form}}^{\widetilde{V}%
_{BH}}$ constrained by the AE (\ref{attreq}) read as follows:
\begin{eqnarray}
&&
\begin{array}{l}
\mathcal{A}\approx 2\frac{C_{1}^{2}}{C_{0}^{2}}\left[ |N_{1}|^{2}(\sqrt{5}%
-2)+|N_{2}|^{2}\left( 1+4(\sqrt{5}+2)\frac{C_{0}^{2}C_{2}^{2}}{C_{1}^{4}}%
\right) \right] + \\
~ \\
+3\frac{C_{2}}{C_{0}}\left[ 1+(\sqrt{5}+2)\frac{C_{3}C_{0}}{C_{2}C_{1}}%
\right] (\bar{N}_{1}\bar{N}_{2}+N_{1}N_{2}); \\
~
\end{array}
\\
&&  \notag \\
&&
\begin{array}{l}
\mathcal{B}\approx 2\frac{C_{1}^{2}}{C_{0}^{2}}\left[ |N_{1}|^{2}(\sqrt{5}%
-2)+|N_{2}|^{2}\left( 1+4(\sqrt{5}+2)\frac{C_{0}^{2}C_{2}^{2}}{C_{1}^{4}}%
\right) \right] + \\
~ \\
-3\frac{C_{2}}{C_{0}}\left[ 1+(\sqrt{5}+2)\frac{C_{3}C_{0}}{C_{2}C_{1}}%
\right] (\bar{N}_{1}\bar{N}_{2}+N_{1}N_{2}); \\
~
\end{array}
\qquad \\
&&\qquad \qquad  \notag \\
&&
\begin{array}{l}
\mathcal{C}\approx -3i\frac{C_{2}}{C_{0}}\left[ 1+(\sqrt{5}+2)\frac{%
C_{3}C_{0}}{C_{2}C_{1}}\right] (\bar{N}_{1}\bar{N}_{2}-N_{1}N_{2}).
\end{array}
\end{eqnarray}
The resulting real eigenvalues of $H_{\text{real form}}^{\widetilde{V}_{BH}}$
constrained by the AE (\ref{attreq}) read:
\begin{eqnarray}
\lambda _{\pm } &\approx &2\frac{C_{1}^{2}}{C_{0}^{2}}\left[ |N_{1}|^{2}(%
\sqrt{5}-2)+|N_{2}|^{2}\left( 1+4(\sqrt{5}+2)\frac{C_{0}^{2}C_{2}^{2}}{%
C_{1}^{4}}\right) \right] +  \notag \\
&&\pm 6\frac{C_{2}}{C_{0}}\left[ 1+(\sqrt{5}+2)\frac{C_{3}C_{0}}{C_{2}C_{1}}%
\right] |N_{1}||N_{2}|.  \label{k=5-gen-eigenvalues}
\end{eqnarray}
By recalling Eq. (\ref{pot}) and using Eq. (\ref{Vtilde-k=5}) with $\psi
\approx 0$ and constrained by the AE (\ref{attreq}), one obtains that the
purely charge-dependent LG critical values of the ``effective BH potential''
for the mirror \textit{quintic} $\mathcal{M}_{5}^{\prime }$ are
\begin{equation}
V_{BH,LG-critical,k=5}\approx \frac{1}{25\sqrt{5+2\sqrt{5}}}\left[
|N_{1}|^{2}+(\sqrt{5}+2)|N_{2}|^{2}\right] ;  \label{k=5-VBH-crit}
\end{equation}
by recalling formula (\ref{BHEA}), this directly yields the following purely
charge-dependent values of the BH entropy at the LG critical points of $%
V_{BH,5}$ in the moduli space of $\mathcal{M}_{5}^{\prime }$:
\begin{equation}
S_{BH,LG-critical,k=5}\approx \frac{\pi }{25\sqrt{5+2\sqrt{5}}}\left[
|N_{1}|^{2}+(\sqrt{5}+2)|N_{2}|^{2}\right] \approx 0.013\pi \left[
|N_{1}|^{2}+(\sqrt{5}+2)|N_{2}|^{2}\right] .  \label{k=5-SBH}
\end{equation}
By recalling Eqs. (\ref{14mar}) and (\ref{omega0-2}), it is helpful to write
down here the numerical values of constants relevant to our treatment:
\begin{equation}
C_{0}\approx 2.5,\quad C_{1}\approx 2.25,\quad C_{2}\approx 0.77,\quad
C_{3}\approx 0.054.  \label{con5}
\end{equation}
\medskip \setcounter{equation}0
\def\theequation{7.\arabic{subsection}.\arabic{equation}}

\subsection{\label{21march-2}Solutions to Attractor Equation}

Let us now analyze more in depth the species of LG attractor points arising
from the AE (\ref{attreq}). As it can be easily seen, the AE (\ref{attreq})
has two \textit{non-degenerate} solutions (\textit{i.e.} with non-vanishing $%
V_{BH}$ and therefore with non-vanishing BH entropy, see Eq. (\ref{BHEA}%
)):\medskip

\textbf{I.} The first non-degenerate solution to AE (\ref{attreq}) is
\begin{equation}
N_{2}\approx 0.  \label{k=5-1/2BPS}
\end{equation}
As one can see from Eqs. (\ref{W-k=5})-(\ref{covder}), such a solution
corresponds to a $\frac{1}{2}$-BPS LG critical point of $V_{BH}$ ($%
\widetilde{W}\neq 0$, $D\widetilde{W}=0$). From the definition (\ref{N-def})
with $\left( k,m\right) =\left( 5,2\right) $, in order to get the solution (%
\ref{k=5-1/2BPS}), one has to fine-tune two PF BH charges out of four in the
following way:
\begin{equation}
n_{3}\approx \frac{1}{2}(1+\sqrt{5})(n_{2}-n_{1}),\quad n_{4}\approx -\frac{1%
}{2}(1+\sqrt{5})n_{1}+n_{2}.  \label{chcon1/2}
\end{equation}
The charges $n_{1}$, $n_{2}$ are not fixed; they only satisfy the
non-degeneration condition $N_{1}\neq 0$. The real eigenvalues (\ref
{k=5-gen-eigenvalues}) for the $\frac{1}{2}$-BPS critical solution coincide
and, as it is well known \cite{FGK,BFM,AoB}, are strictly positive:
\begin{equation}
\lambda _{+,\frac{1}{2}-BPS}=\lambda _{-,\frac{1}{2}-BPS}\approx 2(\sqrt{5}%
-2)\frac{C_{1}^{2}}{C_{0}^{2}}|N_{1}|_{N_{2}\approx 0}^{2}>0.
\end{equation}
Consequently, the $\frac{1}{2}$-BPS LG critical point $\psi \approx 0$
supported by the PF BH charge configuration (\ref{chcon1/2}) is a stable
extremum, since it is a minimum of $V_{BH}$, and it is therefore an
attractor in a strict sense. The classical (Bekenstein-Hawking) BH entropy
of such a (class of) $\frac{1}{2}$-BPS LG attractor(s) takes the value
\begin{equation}
S_{BH,\frac{1}{2}-BPS}=\pi V_{BH,\frac{1}{2}-BPS}\approx 0.013\pi
|N_{1}|_{N_{2}\approx 0}^{2},
\end{equation}
where $\left. N_{1}\right| _{N_{2}\approx 0}$ is given by the general
formula (\ref{N-def}) with $\left( k,m\right) =\left( 5,1\right) $
constrained by Eq. (\ref{chcon1/2}):
\begin{equation}
\left. N_{1}\right| _{N_{2}\approx 0}\approx \frac{n_{1}}{8}\left[ 4\sqrt{5}%
-i\sqrt{2(5+\sqrt{5})^{3}}\right] -\frac{n_{2}}{4}\left[ 5+\sqrt{5}-i\sqrt{%
10(5-\sqrt{5})}\right] .
\end{equation}
\medskip \medskip

\textbf{II}. The second non-degenerate solution to AE (\ref{attreq}) is
\begin{equation}
\begin{array}{l}
|N_{1}|\approx \xi |N_{2}|, \\
~ \\
\xi \equiv (\sqrt{5}+2)\frac{C_{0}C_{2}}{C_{1}^{2}}\approx 1,6; \\
~ \\
arg{(N_{1})}-3arg{(N_{2})}\approx \pi ,
\end{array}
\quad  \label{chnon}
\end{equation}
where $N_{1}$ and $N_{2}$ are given by the general formula (\ref{N-def})
with $k=5$ and $m=1,2$, respectively:
\begin{eqnarray}
N_{1} &=&\frac{\sqrt{5}-1}{4}\left( n_{1}+n_{4}-\frac{(3+\sqrt{5})}{2}%
(n_{2}+n_{3})\right) +\frac{i}{2}\sqrt{\frac{(5+\sqrt{5})}{2}}\left(
n_{4}-n_{1}+\frac{(\sqrt{5}-1)}{2}(n_{3}-n_{2})\right) ,  \label{N1} \\
&&  \notag \\
N_{2} &=&-\frac{\sqrt{5}+1}{4}\left( n_{1}+n_{4}-\frac{(3-\sqrt{5})}{2}%
(n_{2}+n_{3})\right) +\frac{i}{2}\sqrt{\frac{(5-\sqrt{5})}{2}}\left(
n_{4}-n_{1}-\frac{(\sqrt{5}+1)}{2}(n_{3}-n_{2})\right) .  \label{N2}
\end{eqnarray}
The first of fine-tuning conditions (\ref{chnon}) substituted in Eq. (\ref
{k=5-SBH}) yields
\begin{equation}
S_{BH,non-BPS,Z\neq 0}\approx \frac{\pi }{25\sqrt{5+2\sqrt{5}}}\left( \xi
^{2}+\sqrt{5}+2\right) |N_{2}|^{2}\approx 0.088\pi |N_{2}|^{2}.
\label{k=5-SBH-non-BPS}
\end{equation}
As one can see from Eqs. (\ref{W-k=5})-(\ref{covder}), the solution (\ref
{chnon}) corresponds to a non-BPS, $Z\neq 0$ LG critical point of $V_{BH}$ ($%
\widetilde{W}\neq 0$, $D\widetilde{W}\neq 0$). The real eigenvalues (\ref
{k=5-gen-eigenvalues}) for such a non-BPS, $Z\neq 0$ critical solution read
\begin{equation}
\lambda _{\pm ,non-BPS,Z\neq 0}\approx 2\frac{C_{1}^{2}}{C_{0}^{2}}%
|N_{2}|^{2}\left[ 1+5(\sqrt{5}-2)\xi ^{2}\pm 3\left( \xi (\sqrt{5}-2)+(\sqrt{%
5}+2)\frac{C_{0}^{2}C_{3}}{C_{1}^{3}}\right) \xi \right] .
\label{k=5-non-BPS-eigenvalues}
\end{equation}
Substituting the numerical values (\ref{con5}) of the involved constants in
Eq. (\ref{k=5-non-BPS-eigenvalues}), one reaches the conclusion that both $%
\lambda _{\pm ,non-BPS,Z\neq 0}$ are strictly positive:
\begin{equation}
\lambda _{\pm ,non-BPS,Z\neq 0}\approx 2|N_{2}|^{2}\left[ 3.277\pm 1.97%
\right] >0.
\end{equation}
Thus, the non-BPS, $Z\neq 0$ LG critical point $\psi \approx 0$ supported by
the PF BH charge configuration (\ref{chnon})-(\ref{N2}) is a minimum of $%
V_{BH}$ and consequently an attractor in a strict sense.\medskip

Let us now find the fine-tuning conditions for PF BH charges supporting the
considered non-BPS, $Z\neq 0$ LG attractor for the mirror \textit{quintic} $%
\mathcal{M}_{5}^{\prime }$. This amounts to solving Eqs. (\ref{chnon})-(\ref
{N2}) by recalling the definitions (\ref{N-def}) and (\ref{PF-charge}). By
doing so, one gets the following three different sets of constraining
relations on the PF BH charges:
\begin{eqnarray}
&&
\begin{array}{l}
\mathbf{II.1}:\left\{
\begin{array}{l}
~\frac{n_{2,\pm }}{n_{1}}=\frac{a_{2}-a_{1,\pm }-1}{a_{1,\pm }+2}\approx
\left\{
\begin{array}{l}
``+":0.342 \\
\multicolumn{1}{c}{``-":-138.3}
\end{array}
\right. , \\
\\
\frac{n_{3,\pm }}{n_{1}}=-\frac{a_{2}+a_{1,\pm }-1}{a_{1,\pm }+2}\approx
\left\{
\begin{array}{l}
``+":-1.352 \\
``-":35
\end{array}
\right. , \\
\\
\frac{n_{4,\pm }}{n_{1}}=\frac{a_{1,\pm }-2}{a_{1,\pm }+2}\approx \left\{
\begin{array}{l}
``+":0.009 \\
``-":102.3
\end{array}
\right. , \\
\\
a_{1,\pm }\left( \xi \right) \equiv \pm 2\sqrt{\frac{\sqrt{5}(1+3\xi )}{2-%
\sqrt{5}+\xi \left[ -\sqrt{5}-4\xi +2(1+\sqrt{5})\xi ^{2}\right] }}\approx
\pm 2.04, \\
\\
a_{2}\left( \xi \right) \equiv \frac{\sqrt{5\sqrt{5}(2+\sqrt{5})}\left( 1+%
\sqrt{2}\xi \right) }{\sqrt{\sqrt{5}\left( \sqrt{5}-2\right) }+\sqrt{2\sqrt{5%
}(\sqrt{5}+2)}\xi }\approx 2.92, \\
\\
n_{1}\neq 0;
\end{array}
\right.
\end{array}
\label{fc51} \\
&&  \notag \\
&&
\begin{array}{l}
\mathbf{II.2}:\left\{
\begin{array}{l}
n_{2}=n_{1}\frac{\left( \sqrt{5}-1\right) -\xi (\sqrt{5}+1)}{\left( \sqrt{5}%
+1\right) -\xi (\sqrt{5}-1)}\approx -3.13n_{1}, \\
\\
n_{3}=n_{2}, \\
\\
n_{4}=n_{1};
\end{array}
\right.
\end{array}
\label{fc52}
\end{eqnarray}
\begin{eqnarray}
&&
\begin{array}{l}
\mathbf{II.3}:\left\{
\begin{array}{l}
\qquad \left\{
\begin{array}{l}
n_{2,\pm }+n_{3,\pm }-n_{1,\pm }-n_{4,\pm }=a_{\pm }, \\
\\
n_{1,\pm }+n_{2,\pm }+n_{3,\pm }+n_{4,\pm }=b_{\pm }, \\
\\
2n_{3,\pm }-2n_{2,\pm }+d=c, \\
\\
n_{4,\pm }-n_{1,\pm }=d,
\end{array}
\right. \Longleftrightarrow \left\{
\begin{array}{l}
n_{1,\pm }=\frac{1}{4}\left( -a_{\pm }+b_{\pm }-2d\right) ,~ \\
\\
n_{2,\pm }=\frac{1}{4}\left( a_{\pm }+b_{\pm }-c+d\right) , \\
\\
~n_{3,\pm }=\frac{1}{4}\left( a_{\pm }+b_{\pm }+c-d\right) ,~ \\
\\
n_{4,\pm }=\frac{1}{4}\left( -a_{\pm }+b_{\pm }+2d\right) ,
\end{array}
\right. \\
\\
\\
a_{\pm }\left( \xi ;b_{\pm },c,d\right) \equiv -\frac{\sqrt{5+2\sqrt{5}}c-%
\sqrt{5(5-2\sqrt{5})}d+\sqrt{2(5+\sqrt{5})}(-c+\sqrt{5}d)\xi }{\sqrt{5(5-2%
\sqrt{5})}c-5\sqrt{(5+2\sqrt{5})}d+\sqrt{2(5+\sqrt{5})}(\sqrt{5}c-5d)\xi }%
b_{\pm }\left( \xi ;c,d\right) \approx \\
\\
\approx \left( 0.26157+\frac{0.213c}{-0.3325c+d}\right) b_{\pm }\left( \xi
;c,d\right) \approx \pm 0.49\sqrt{\frac{c-0.98995d}{c-2.89698d}}\frac{\left(
c-3.0102d\right) \left( c-3.9798d\right) }{c-3.0077d}, \\
\\
b_{\pm }\left( \xi ;c,d\right) \equiv \pm \frac{1}{2}\sqrt{\frac{\left[ 5-2%
\sqrt{5}+2\xi (5-\sqrt{5}+(5+\sqrt{5})\xi )\right] \left[ 2\sqrt{5}+\sqrt{5}%
\left( 1+\sqrt{5}\right) \xi \right] }{\left[ -2\sqrt{5}+3\sqrt{5}\left( 1+%
\sqrt{5}\right) \xi \right] }}\left[ c-\frac{\sqrt{5}d\left( 1+\sqrt{5}+4\xi
\right) }{3-\sqrt{5}+4\xi }\right] \sqrt{\frac{\left[ c-\frac{\sqrt{5}%
d\left( -2+\left( \sqrt{5}+1\right) \xi \right) }{2+\left( \sqrt{5}+1\right)
\xi }\right] }{\left[ c-\frac{\sqrt{5}d\left( 2+3\left( \sqrt{5}+1\right)
\xi \right) }{-2+3\left( \sqrt{5}+1\right) \xi }\right] }}\approx \\
\\
\approx \pm 2.481\left( c-3.00769d\right) \sqrt{\frac{c-0.98995d}{c-2.89698d}%
}.
\end{array}
\right.
\end{array}
\notag \\
&&  \label{fc53}
\end{eqnarray}
By using the definitions (\ref{PF-charge}) and (\ref{PF-charge-inv}) for the
case $k=5$, one obtains (recall the first of definitions (\ref{Gamma-Fermat}%
))
\begin{equation}
\begin{array}{c}
\left( n_{1},n_{2},n_{3},n_{4}\right) =\left(
3p^{0}-5q_{0},p^{0}+5q_{1},-21p^{0}+5p^{1}+40q_{0}-5q_{1},-8p^{0}+15q_{0}%
\right) \\
\\
\Updownarrow \\
\\
\Gamma =\left( -3n_{1}-n_{4},\frac{1}{5}\left(
4n_{1}-n_{2}-n_{3}+4n_{4}\right) ,\frac{1}{5}(8n_{1}+3n_{4}),\frac{1}{5}%
(-3n_{1}+n_{2}-n_{4})\right) .
\end{array}
\label{rel-sympl-PF-ch-k=5}
\end{equation}
\setcounter{equation}0
\def\theequation{7.4.\arabic{subsubsection}.\arabic{equation}}

\subsubsection{Analysis of non-BPS, $Z\neq 0$ supporting BH charge
configurations}

Let us analyze more in depth the obtained supporting BH charge
configurations.\medskip

\textbf{II.1}. This configuration is characterized by a physically
meaningful ``$\pm $'' split, affecting the charge-dependent expression of
the BH entropy. By using Eqs. (\ref{N2}), (\ref{k=5-SBH-non-BPS}) and (\ref
{rel-sympl-PF-ch-k=5}), Eqs. (\ref{fc51}) yield
\begin{eqnarray}
&&
\begin{array}{l}
N_{2,II.1,\pm }=\left[ -\frac{\sqrt{5}}{2}\left( 3+\sqrt{5}\right) \frac{%
a_{1,\pm }}{a_{1,\pm }+2}+\frac{i}{2}\left( 1+\sqrt{5}\right) \sqrt{\frac{%
\sqrt{5}\left( \sqrt{5}-1\right) }{2}}\frac{a_{2}-\sqrt{5}}{a_{1,\pm }+2}%
\right] n_{1}\approx \\
\\
~~~~~~~~~~\approx \left[ \left\{
\begin{array}{l}
``+":-2.95603 \\
\\
``-":-298.5592
\end{array}
\right. +i\left\{
\begin{array}{l}
``+":0.322 \\
\\
``-":-32.5229
\end{array}
\right. \right] \left( 3p^{0}-5q_{0}\right) ; \\
\\
\\
S_{BH,non-BPS,Z\neq 0,II.1,\pm }\approx 0.088\pi |N_{2,II.1,\pm
}|^{2}\approx \left\{
\begin{array}{l}
``+":0.121\pi \left( 3p^{0}-5q_{0}\right) ^{2}. \\
\\
``-":1192.1\pi \left( 3p^{0}-5q_{0}\right) ^{2}.
\end{array}
\right.
\end{array}
\notag \\
&&  \label{SBH-II1}
\end{eqnarray}
In order to give an example of the range of the symplectic
(electric/magnetic) charges in the considered BH charge configuration, by
recalling that PF BH charges are integers as are the symplectic BH charges
(see definitions (\ref{PF-charge})-(\ref{PF-charge-inv})), one can fulfill
the fine-tuning conditions (\ref{fc51}) by taking \textit{e.g.}
\begin{equation}
\begin{array}{c}
n_{1}=10 \\
\Downarrow \\
n_{2,\pm }\approx \left\{
\begin{array}{l}
``+":3 \\
``-":-1383
\end{array}
\right. ,n_{3,\pm }\approx \left\{
\begin{array}{l}
``+":-14 \\
``-":350
\end{array}
\right. ,n_{4,\pm }\approx \left\{
\begin{array}{l}
``+":0 \\
``-":1023
\end{array}
\right. .
\end{array}
\end{equation}
Using Eq. (\ref{rel-sympl-PF-ch-k=5}), one finally gets
\begin{equation}
\Gamma _{\pm }\approx \left\{
\begin{array}{l}
``+":\left( -30,51,16,-5\right) ; \\
\\
``-":\left( -1053,1033,630,-487\right) .
\end{array}
\right. .  \label{porgy1}
\end{equation}
\medskip

\textbf{II.2}. This configuration is the only one not having any kind of
split, and having a purely real $N_{2}$. Also, it is the one adopted in \cite
{TT} (see in particular Sect. 4 and App. C of such a Ref.).\textbf{\ }By
using Eqs. (\ref{N2}), (\ref{k=5-SBH-non-BPS}) and (\ref{rel-sympl-PF-ch-k=5}%
), Eqs. (\ref{fc52}) yield
\begin{eqnarray}
&&
\begin{array}{l}
N_{2,II.2}=-\frac{2\sqrt{5}}{1+\sqrt{5}+\left( 1-\sqrt{5}\right) \xi }%
n_{1}\approx -3.554\left( 3p^{0}-5q_{0}\right) ; \\
\\
S_{BH,non-BPS,Z\neq 0,II.2}\approx 0.088\pi \left( N_{2,II.2}\right)
^{2}\approx 1.111\pi \left( 3p^{0}-5q_{0}\right) ^{2}.
\end{array}
\notag \\
&&  \label{SBH-II2}
\end{eqnarray}
Concerning the range of the symplectic (electric/magnetic) charges in the
considered BH charge configuration, by repeating the same procedure
exploited above for the case II.1, one achieves the same results obtained at
the end of Sect. 4 of \cite{TT}.\medskip

\textbf{II.3}. This case, beside being split in ``$\pm $'' (as the case
II.1), is also parameterized by the couple $\left( c,d\right) \in \mathbb{Z}%
^{2}$ ($\mathbb{R}^{2}$ in the -semi- classical limit of large charges),
further constrained by the request $a_{\pm }\left( b_{\pm },c,d\right) $, $%
b_{\pm }\left( c,d\right) \in \mathbb{Z}$ ($\mathbb{R}$ for large charges),
implying that
\begin{equation}
c\neq 3.077d,\text{ ~and~~}\left\{
\begin{array}{l}
\left\{
\begin{array}{l}
c\geqslant 0.98995d, \\
\\
c>2.89698d;
\end{array}
\right. \\
\multicolumn{1}{c}{\text{or}} \\
\left\{
\begin{array}{l}
c\leqslant 0.98995d, \\
\\
c<2.89698d.
\end{array}
\right.
\end{array}
\right. \text{ ~}\text{\ \ ~}  \label{yield1}
\end{equation}
By using Eq. (\ref{N2}), Eqs. (\ref{fc53}) yield
\begin{eqnarray}
&&
\begin{array}{l}
N_{2,II.3,\pm }(c,d)=-\frac{1}{4}\left[ -\sqrt{5}a_{\pm }\left( c,d\right)
+b_{\pm }\left( c,d\right) \right] -\frac{i}{4}\sqrt{\frac{5+\sqrt{5}}{2}}%
\left[ c-\sqrt{5}d\right] = \\
\\
=-\frac{1}{4}\left[ \frac{\sqrt{5+2\sqrt{5}}+\sqrt{5-2\sqrt{5}}}{\sqrt{5-2%
\sqrt{5}}+\sqrt{\sqrt{5}(2+2\sqrt{5})}\xi }\frac{\left( c-\sqrt{5}d\right) }{%
\left[ c-\frac{\left( \sqrt{5+2\sqrt{5}}d+\sqrt{\sqrt{5}(2+2\sqrt{5})}\xi
\right) }{\left( \sqrt{5-2\sqrt{5}}+\sqrt{\sqrt{5}(2+2\sqrt{5})}\xi \right) }%
\sqrt{5}d\right] }b_{\pm }\left( c,d\right) \right] + \\
\\
-\frac{i}{4}\sqrt{\frac{5+\sqrt{5}}{2}}\left( c-\sqrt{5}d\right) = \\
\\
=-\frac{1}{4\sqrt{2}}\left\{ \pm \left\{
\begin{array}{c}
\sqrt{\frac{\left( 1+\sqrt{5}\right) \left[ 5-2\sqrt{5}+2\xi (5-\sqrt{5}+(5+%
\sqrt{5})\xi )\right] \left[ 2\sqrt{5}+\sqrt{5}\left( 1+\sqrt{5}\right) \xi
\right] }{\left[ -2\sqrt{5}+3\sqrt{5}\left( 1+\sqrt{5}\right) \xi \right]
\left[ \left( \sqrt{5}-2\right) +2\sqrt{2}\sqrt{3-\sqrt{5}}\xi +2(1+\sqrt{5}%
)\xi ^{2}\right] }}\cdot \\
\\
\cdot \sqrt{\frac{c-\frac{-2+\left( \sqrt{5}+1\right) \xi }{2+\left( \sqrt{5}%
+1\right) \xi }\sqrt{5}d}{c-\frac{2+3\left( \sqrt{5}+1\right) \xi }{%
-2+3\left( \sqrt{5}+1\right) \xi }\sqrt{5}d}}\frac{c-\frac{1+\sqrt{5}+4\xi }{%
3-\sqrt{5}+4\xi }\sqrt{5}d}{c-\frac{\left( \sqrt{5+2\sqrt{5}}d+\sqrt{\sqrt{5}%
(2+2\sqrt{5})}\xi \right) }{\left( \sqrt{5-2\sqrt{5}}+\sqrt{\sqrt{5}(2+2%
\sqrt{5})}\xi \right) }\sqrt{5}d}
\end{array}
\right\} +i\sqrt{5+\sqrt{5}}\right\} \left( c-\sqrt{5}d\right) \approx \\
\\
\approx \left[ \mp 0.34631\sqrt{\frac{c-0.98995d}{c-2.89698d}}-0.47552i%
\right] \left( c-2.23607d\right) ;
\end{array}
\notag \\
&&
\end{eqnarray}
\begin{eqnarray}
&&
\begin{array}{l}
\left| N_{2,II.3,\pm }(c,d)\right| ^{2}= \\
\\
=\frac{1}{32}\left\{ \left\{
\begin{array}{c}
\frac{\left( 1+\sqrt{5}\right) \left[ 5-2\sqrt{5}+2\xi (5-\sqrt{5}+(5+\sqrt{5%
})\xi )\right] \left[ 2\sqrt{5}+\sqrt{5}\left( 1+\sqrt{5}\right) \xi \right]
}{\left[ -2\sqrt{5}+3\sqrt{5}\left( 1+\sqrt{5}\right) \xi \right] \left[
\left( \sqrt{5}-2\right) +2\sqrt{2}\sqrt{3-\sqrt{5}}\xi +2(1+\sqrt{5})\xi
^{2}\right] }\cdot \\
\\
\cdot \frac{c-\frac{-2+\left( \sqrt{5}+1\right) \xi }{2+\left( \sqrt{5}%
+1\right) \xi }\sqrt{5}d}{c-\frac{2+3\left( \sqrt{5}+1\right) \xi }{%
-2+3\left( \sqrt{5}+1\right) \xi }\sqrt{5}d}\frac{\left[ c-\frac{1+\sqrt{5}%
+4\xi }{3-\sqrt{5}+4\xi }\sqrt{5}d\right] ^{2}}{\left[ c-\frac{\left( \sqrt{%
5+2\sqrt{5}}d+\sqrt{\sqrt{5}(2+2\sqrt{5})}\xi \right) }{\left( \sqrt{5-2%
\sqrt{5}}+\sqrt{\sqrt{5}(2+2\sqrt{5})}\xi \right) }\sqrt{5}d\right] ^{2}}
\end{array}
\right\} +\left( 5+\sqrt{5}\right) \right\} \left( c-\sqrt{5}d\right)
^{2}\approx \\
\\
\approx \left[ 0.11993\left( c-0.98995d\right) +0.22613\left(
c-2.89698d\right) \right] \frac{\left( c-2.23607d\right) ^{2}}{c-2.89698d}%
\approx \\
\\
\approx 0.34606\frac{\left( c-2.23607d\right) ^{3}}{c-2.89698d}.
\end{array}
\notag \\
&&
\end{eqnarray}
By recalling Eqs. (\ref{k=5-SBH-non-BPS}) and (\ref{rel-sympl-PF-ch-k=5}),
one finally gets (notice that in this case the ``$\pm $'' split is
unphysical, because it does not affect the charge-dependent expression of
the BH entropy) :
\begin{equation}
\begin{array}{l}
S_{BH,non-BPS,Z\neq 0,II.3,\pm }(c,d)\approx 0.088\pi \left| N_{2,II.3,\pm
}(c,d)\right| ^{2}\approx 0.03045\pi \frac{\left( c-2.23607d\right) ^{3}}{%
c-2.89698d}\approx \\
\\
\approx 0.03045\pi \frac{\left(
-30.40325p^{0}+10p^{1}+55.27864q_{0}-20q_{1}\right) ^{3}}{%
-23.13316p^{0}+10p^{1}+42.0603q_{0}-20q_{1}}= \\
\\
=S_{BH,non-BPS,Z\neq 0,II.3,\pm }(p^{0},p^{1},q_{0},q_{1}).
\end{array}
\label{SBH-II3}
\end{equation}
\medskip

Thus, by comparing Eqs. (\ref{SBH-II1}), (\ref{SBH-II2}) and (\ref{SBH-II3}%
), one can conclude that \textit{all} three distinct sets of fine-tuning
conditions for PF BH charges (\ref{fc51})-(\ref{fc53}) do support a non-BPS,
$Z\neq 0$ LG attractor in a strict sense, but they yield different
charge-dependent expressions for the classical BH entropy $%
S_{BH,non-BPS,Z\neq 0}$. As non-BPS, $Z\neq 0$ attractor, the LG point of
the moduli space of the Fermat mirror \textit{quintic} $\mathcal{M}%
_{5}^{\prime }$ turns out to originate a \textit{threefold splitting} of the
supporting BH charges, and correspondingly of the purely charge-dependent
classical (Bekenstein-Hawking) BH entropy. This fact confirms the results of
the recent investigation in \cite{Saraikin-Vafa-1}.\medskip

Finally, by recalling the definition (\ref{SUSY-breaking}), one can compute
the supersymmetry-breaking order parameter for the non-BPS, $Z\neq 0$ LG
attractor in the mirror \textit{quintic} $\mathcal{M}_{5}^{\prime }$; by
using Eqs. (\ref{k=5-metric}), (\ref{W-k=5}), (\ref{covder}) and (\ref{chnon}%
), one gets
\begin{equation}
\mathcal{O}_{non-BPS,Z\neq 0}\equiv \left[ \frac{g^{-1}|DW|^{2}}{|W|^{2}}%
\right] _{non-BPS,Z\neq 0}=\left( g^{-1}|Dln\widetilde{W}|^{2}\right)
_{non-BPS,Z\neq 0}\approx \frac{(\sqrt{5}+2)}{\xi ^{2}}\approx 1.65,
\end{equation}
which is consistent with the result obtained at the end of Sect. 4 of \cite
{TT}.

\section{$k=6$ : Mirror \textit{Sextic} near LG Point\label{sixtic}}

\setcounter{equation}0
\def\theequation{8.\arabic{subsection}.\arabic{equation}}

\subsection{Geometric Setup}

For the mirror \textit{sextic} $\mathcal{M}_{6}^{\prime }$ the computations (%
\textit{but not the results!}) go the same way as for the mirror \textit{%
quintic} $\mathcal{M}_{5}^{\prime }$. Thus, also for $k=6$ it is easy to
realize that one has to consider the ``effective BH potential'' (\ref{pot})
(at least) up to $\mathcal{O}\left( \psi ^{3}\right) $ in order to get the
AE (\ref{AEs-1-modulus-W}) and the Hessian matrix up to $\mathcal{O}\left(
\psi \right) $.

For $k=6$ the definitions (\ref{omega0-2}) yield
\begin{equation}
C_{6,3l-1}=0,\quad l\in \mathbb{N}
\end{equation}
moreover, since $F_{6,m+6,\,n}=F_{6,m\,,n+6}=-F_{6,m\,n}$ (see the third of
properties (\ref{prop-F})), the only independent elements of the rank-2
tensor $F_{6}$ belong to the $6\times 6$ matrix
\begin{equation}
F_{6,m\,n}=\left(
\begin{array}{cccccc}
2\sqrt{3} & 0 & -\sqrt{3} & 0 & 0 & 3 \\
0 & -\frac{2}{\sqrt{3}} & 1 & 0 & 0 & -\frac{1}{\sqrt{3}} \\
-\sqrt{3} & 1 & 0 & -1 & \sqrt{3} & -2 \\
0 & 0 & -1 & \frac{2}{\sqrt{3}} & 0 & -\frac{1}{\sqrt{3}} \\
0 & 0 & \sqrt{3} & 0 & -2\sqrt{3} & 3 \\
3 & -\frac{1}{\sqrt{3}} & -2 & -\frac{1}{\sqrt{3}} & 3 & 0
\end{array}
\right) .  \label{k=6-Fmn}
\end{equation}
Let us now write down all the relevant quantities up to the needed order
(here and below, unless otherwise specified, we omit the Fermat parameter $%
k=6$):
\begin{eqnarray}
&&
\begin{array}{l}
\widetilde{K}\approx \frac{1}{3}\frac{C_{1}^{2}}{C_{0}^{2}}\left[ \psi \bar{%
\psi}+\frac{1}{6}\frac{C_{1}^{2}}{C_{0}^{2}}(\psi \bar{\psi})^{2}\right] +%
\mathcal{O}(\psi ^{6}); \\
~
\end{array}
\\
&&
\begin{array}{l}
g\approx \frac{1}{3}\frac{C_{1}^{2}}{C_{0}^{2}}\left[ 1+\frac{2}{3}\frac{%
C_{1}^{2}}{C_{0}^{2}}\psi \bar{\psi}\right] +\mathcal{O}(\psi ^{4}); \\
~
\end{array}
\label{k=6-metric} \\
&&
\begin{array}{l}
\widetilde{W}\approx N_{1}+\frac{C_{1}}{C_{0}}N_{2}\psi -\frac{C_{3}}{C_{0}}%
\bar{N}_{2}\psi ^{3}-\frac{C_{4}}{C_{0}}\bar{N}_{1}\psi ^{4}+\mathcal{O}%
(\psi ^{6}). \\
~
\end{array}
\end{eqnarray}
Now, by using the formul\ae\ of the general analysis exploited in Sect. \ref
{GA}, we can get the ``effective BH potential'' and the holomorphic
superpotential, as well as their (covariant) derivatives, up to $\mathcal{O}%
\left( \psi \right) $:
\begin{eqnarray}
&&
\begin{array}{l}
\widetilde{W}=N_{1}+\frac{C_{1}}{C_{0}}N_{2}\psi ; \\
~
\end{array}
\label{W-k=6} \\
&&
\begin{array}{l}
D\widetilde{W}=\frac{C_{1}}{C_{0}}\left[ N_{2}+\frac{1}{3}\frac{C_{1}}{C_{0}}%
N_{1}\bar{\psi}\right] ; \\
~
\end{array}
\label{covder6} \\
&&
\begin{array}{l}
D^{2}\widetilde{W}=-6\frac{C_{3}}{C_{0}}\bar{N}_{2}\psi -12\frac{C_{4}}{C_{0}%
}\bar{N}_{1}\psi ^{2}; \\
~
\end{array}
\label{DDW-k=6} \\
&&
\begin{array}{l}
\widetilde{V}_{BH}=|N_{1}|^{2}+3|N_{2}|^{2}+2\frac{C_{1}}{C_{0}}N_{2}\bar{N}%
_{1}\psi +2\frac{C_{1}}{C_{0}}\bar{N}_{2}N_{1}\bar{\psi}; \\
~
\end{array}
\label{pot6}
\end{eqnarray}
\begin{eqnarray}
&&
\begin{array}{l}
\partial \widetilde{V}_{BH}=2\frac{C_{1}}{C_{0}}N_{2}\bar{N}_{1}-18\frac{%
C_{3}}{C_{1}}(\bar{N}_{2})^{2}\psi +\frac{2}{3}\frac{C_{1}^{2}}{C_{0}^{2}}%
\left( |N_{1}|^{2}+3|N_{2}|^{2}\right) \bar{\psi}; \\
~
\end{array}
\\
&&
\begin{array}{l}
\partial ^{2}\widetilde{V}_{BH}=-18\frac{C_{3}}{C_{1}}(\bar{N}_{2})^{2}-24%
\frac{C_{3}}{C_{0}}\left( 1+\frac{C_{4}C_{0}}{C_{3}C_{1}}\right) \bar{N}_{1}%
\bar{N}_{2}\psi +\frac{4}{3}\frac{C_{1}^{3}}{C_{0}^{3}}\bar{N}_{1}N_{2}\bar{%
\psi}; \\
~
\end{array}
\label{ddV-k=6} \\
&&
\begin{array}{l}
\partial \overline{\partial }\widetilde{V}_{BH}=\frac{2}{3}\frac{C_{1}^{2}}{%
C_{0}^{2}}\left( |N_{1}|^{2}+3|N_{2}|^{2}\right) +\frac{4}{3}\frac{C_{1}^{3}%
}{C_{0}^{3}}\bar{N}_{1}N_{2}\psi +\frac{4}{3}\frac{C_{1}^{3}}{C_{0}^{3}}N_{1}%
\bar{N}_{2}\bar{\psi}. \\
~
\end{array}
\label{ddV-k=6-2}
\end{eqnarray}
From the definition (\ref{N-def}) with $k=6$ and $m=1,2$, one gets
respectively:
\begin{eqnarray}
N_{1} &=&-\frac{\sqrt{3}}{2}(n_{2}+n_{3})+\frac{i}{2}%
(2n_{4}-2n_{1}+n_{3}-n_{2});  \label{N1-6} \\
&&  \notag \\
N_{2} &=&\frac{1}{2}(n_{3}+n_{2}-2n_{4}-2n_{1})-\frac{\sqrt{3}i}{2}%
(n_{3}-n_{2}).  \label{N2-6}
\end{eqnarray}
\setcounter{equation}0
\def\theequation{8.1.\arabic{subsubsection}.\arabic{equation}}

\subsubsection{\label{prep-6}The Holomorphic Prepotentials $F$ and $\mathcal{%
F}$}

As done for $k=5$ in Subsubsect. \ref{prep-5}, in order to compute the
holomorphic prepotential $F\left( X\left( \psi \right) \right) $ and its
``K\"{a}hler-invariant counterpart'' $\mathcal{F}\left( \psi \right) $
defined by Eq. (\ref{F-call}), we have to recall the general formul\ae\ for
(mirror) Fermat $CY_{3}$s introduced in Sect. \ref{GA}, and specialize them
for $k=6$. By recalling the definitions (\ref{csi-csi}), (\ref{PHIZ}), (\ref
{hol-prep-F-l-CY}), (\ref{F-call-CY}), one can compute:
\begin{eqnarray}
&&
\begin{array}{l}
\xi _{6,m}=\left(
\begin{array}{l}
-\frac{2}{3}i\left[ sin\left( \frac{\pi m}{6}\right) +sin\left( \frac{\pi m}{%
2}\right) \right] \\
\\
-cos\left( \frac{\pi m}{6}\right) +isin\left( \frac{\pi m}{6}\right) \\
\\
4cos^{2}\left( \frac{\pi m}{6}\right) \left[ cos\left( \frac{\pi m}{6}%
\right) -3isin\left( \frac{\pi m}{6}\right) \right] \\
\\
2isin\left( \frac{\pi m}{6}\right)
\end{array}
\right) ;
\end{array}
\notag \\
&&  \label{csi-csi-6}
\end{eqnarray}
\begin{eqnarray}
&&
\begin{array}{l}
\Phi _{6,m,n}= \\
\\
\begin{array}{l}
=\left[
\begin{array}{l}
-cos\left( \frac{\pi \left( n-3m\right) }{6}\right) -cos\left( \frac{\pi
\left( 3n-m\right) }{6}\right) -2cos\left( \frac{\pi \left( n-m\right) }{6}%
\right) -cos\left( \frac{\pi \left( n-m\right) }{2}\right) + \\
\\
+cos\left( \frac{\pi \left( n+3m\right) }{6}\right) +cos\left( \frac{\pi
\left( 3n+m\right) }{5}\right) +2cos\left( \frac{\pi \left( n+m\right) }{6}%
\right) +cos\left( \frac{\pi \left( n+m\right) }{2}\right)
\end{array}
\right] + \\
\\
-\frac{2}{3}icos\left( \frac{\pi m}{6}\right) \left[ 3sin\left( \frac{\pi n}{%
6}\right) +8cos^{3}\left( \frac{\pi n}{6}\right) sin\left( \frac{\pi m}{3}%
\right) \right] \neq \Phi _{6,\left( m,n\right) };
\end{array}
\end{array}
\notag \\
&&  \label{PHIZ-6} \\
&&  \notag \\
&&
\begin{array}{l}
F_{k=6,l=5}\left( X_{6}\left( \psi \right) \right) =-\frac{2\pi ^{6}}{3^{6}}%
\sum_{m=1}^{5}\sum_{n=1}^{5}(-1)^{m+n}C_{6,m-1}C_{6,n-1}\psi ^{m+n-2}\Phi
_{6,\left( m,n\right) }= \\
\\
\\
=-\frac{2\pi ^{6}}{3^{6}}\left[ -\frac{2^{4/3}\left( 5+2i\sqrt{3}\right) \pi
^{2}}{\Gamma ^{2}\left( \frac{2}{3}\right) \Gamma ^{8}\left( \frac{5}{6}%
\right) }+\frac{24\left( i+\sqrt{3}\right) \pi }{\Gamma ^{4}\left( \frac{2}{3%
}\right) \Gamma ^{4}\left( \frac{5}{6}\right) }\psi +\frac{3i2^{2/3}\left( 2%
\sqrt{3}-9i\right) }{\Gamma ^{6}\left( \frac{2}{3}\right) }\psi ^{2}-\frac{%
32^{2/3}\left( 6+i\sqrt{3}\right) }{\pi \Gamma ^{2}\left( \frac{2}{3}\right)
\Gamma ^{2}\left( \frac{5}{6}\right) }\psi ^{3}+\frac{207\sqrt{3}}{4\pi ^{3}}%
\psi ^{4}\right] +\mathcal{O}\left( \psi ^{5}\right) ;
\end{array}
\notag \\
&&  \label{F65-CY} \\
&&  \notag \\
&&
\begin{array}{l}
\mathcal{F}_{6}\left( \psi \right) \equiv F_{6}\left( \frac{X_{6}^{1}(\psi )%
}{X_{6}^{0}(\psi )}\right) =\frac{F_{6}\left( X_{6}^{0}(\psi
),X_{6}^{1}(\psi )\right) }{\left( X_{6}^{0}(\psi )\right) ^{2}}=\frac{1}{2}%
\frac{\sum_{m,n=1}^{\infty }(-1)^{m+n}C_{6,m-1}C_{6,n-1}\Phi _{6,\left(
m,n\right) }\psi ^{m+n-2}}{\left[ \sum_{r=1}^{\infty }(-1)^{r}C_{6,r-1}\xi
_{6,r}^{1}\psi ^{r-1}\right] ^{2}}\approx \\
\\
\approx \frac{\Phi _{6,1,1}}{2\left( \xi _{6,1}^{1}\right) ^{2}}\left[ 1-%
\frac{C_{6,1}\left( \Phi _{6,2,1}+\Phi _{6,1,2}\right) }{C_{6,0}\Phi _{6,1,1}%
}\psi +\mathcal{O}\left( \psi ^{2}\right) \right] \left[ 1+2\frac{C_{6,1}\xi
_{6,2}^{1}}{C_{6,0}\xi _{6,1}^{1}}\psi +\mathcal{O}\left( \psi ^{2}\right)
\right] \approx \\
\\
\approx \frac{\Phi _{6,1,1}}{2\left( \xi _{6,1}^{1}\right) ^{2}}\left[
1+2\left( \frac{\xi _{6,2}^{1}}{\xi _{6,1}^{1}}-\frac{\Phi _{6,2,1}+\Phi
_{6,1,2}}{2\Phi _{6,1,1}}\right) \frac{C_{6,1}}{C_{6,0}}\psi \right] +%
\mathcal{O}\left( \psi ^{2}\right) ,
\end{array}
\notag \\
&&  \label{F-call-CY-6}
\end{eqnarray}
where in the formula (\ref{F-call-CY-6}) for the K\"{a}hler-invariant
holomorphic prepotential $\mathcal{F}_{6}$ only the leading terms in the
\textit{LG limit }$\psi \longrightarrow 0$ have been retained. Eq. (\ref
{F65-CY}) is the exact expression (complete up to $\mathcal{O}\left( \psi
^{4}\right) $ included) of the holomorphic prepotential $F_{6}$ in a
suitable neighbourhood of the LG point of the moduli space of the (mirror)%
\textit{\ sextic} $\mathcal{M}_{6}^{\prime }$.
\setcounter{equation}0
\def\theequation{8.\arabic{subsection}.\arabic{equation}}

\subsection{\textit{``Criticality Condition''} Approach and Attractor
Equation}

Let us now find the solutions of the AE $\partial \widetilde{V}_{BH}\left(
\psi ,\overline{\psi };q,p\right) =0$, and check their stability. Since we
are working near the LG point, by using Eq. (\ref{pot6}) we can rewrite the
AE for $\mathcal{M}_{6}^{\prime }$ as follows:
\begin{equation}
N_{2}\bar{N}_{1}\approx 0.  \label{attreq6}
\end{equation}
Here we simply put $\psi =0$. Solving Eq. (\ref{attreq6}), we will
find one (or more) set(s) of BH charges supporting $\psi \approx 0$
to be a critical point of $V_{BH}$. \setcounter{equation}0
\def\theequation{8.\arabic{subsection}.\arabic{equation}}

\subsection{\label{21march-5}Critical Hessian of $V_{BH}$}

By using (the $n_{V}=1$ case of) Eq. (\ref{4jan2}) and Eqs. (\ref{ddV-k=6})-(%
\ref{ddV-k=6-2}) evaluated along the criticality condition (\ref{attreq6}),
it can be computed that the components of $H_{\text{real form}}^{\widetilde{V%
}_{BH}}$ (given by Eq. (\ref{Hessian-real-form})) constrained by the AE (\ref
{attreq6}) read as follows:
\begin{eqnarray}
&&\mathcal{A}=\frac{2}{3}\frac{C_{1}^{2}}{C_{0}^{2}}\left(
|N_{1}|^{2}+3|N_{2}|^{2}\right) -9\frac{C_{3}}{C_{1}}\left( (\bar{N}%
_{2})^{2}+(N_{2})^{2}\right) ; \\
&&  \notag \\
&&\mathcal{B}=\frac{2}{3}\frac{C_{1}^{2}}{C_{0}^{2}}\left(
|N_{1}|^{2}+3|N_{2}|^{2}\right) +9\frac{C_{3}}{C_{1}}\left( (\bar{N}%
_{2})^{2}+(N_{2})^{2}\right) ; \\
&&  \notag \\
&&\mathcal{C}=9i\frac{C_{3}}{C_{1}}\left( (\bar{N}_{2})^{2}-(N_{2})^{2}%
\right) .
\end{eqnarray}
The resulting real eigenvalues of $H_{\text{real form}}^{\widetilde{V}_{BH}}$
constrained by the AE (\ref{attreq6}) read:
\begin{equation}
\lambda _{\pm }\approx 2\frac{C_{1}^{2}}{C_{0}^{2}}\left[ \frac{1}{3}%
|N_{1}|^{2}+|N_{2}|^{2}\left( 1\pm 9\frac{C_{0}^{2}C_{3}}{C_{1}^{3}}\right) %
\right] .  \label{k=6-gen-eigenvalues}
\end{equation}
By recalling Eq. (\ref{pot}) and using Eq. (\ref{pot6}) with $\psi \approx 0$
and constrained by the AE (\ref{attreq6}), one obtains that the purely
charge-dependent LG critical values of the ``effective BH potential'' for
the mirror \textit{sextic} $\mathcal{M}_{6}^{\prime }$ are
\begin{equation}
V_{BH,LG-critical,k=6}\approx \frac{1}{18\sqrt{3}}\left[
|N_{1}|^{2}+3|N_{2}|^{2}\right] ;
\end{equation}
by recalling formula (\ref{BHEA}), this directly yields the following purely
charge-dependent values of the BH entropy at the LG critical points of $%
V_{BH,6}$ in the moduli space of $\mathcal{M}_{6}^{\prime }$:
\begin{equation}
S_{BH,LG-critical,k=6}\approx \frac{\pi }{18\sqrt{3}}\left[
|N_{1}|^{2}+3|N_{2}|^{2}\right] .
\end{equation}

Let us write down here the numerical values of constants relevant to our
treatment:
\begin{equation}
C_{0}\approx 2.27,\quad C_{1}\approx 1.52,\quad C_{3}\approx -0.247,\quad
C_{4}\approx 0.054;\quad 9\frac{C_{0}^{2}C_{3}}{C_{1}^{3}}\approx -3.25.
\label{con6}
\end{equation}
\medskip \setcounter{equation}0
\def\theequation{8.\arabic{subsection}.\arabic{equation}}

\subsection{\label{21march-6}Solutions to Attractor Equations}

Let us now analyze more in depth the species of LG attractor points arising
from the AE (\ref{attreq6}). As it can be easily seen, the AE (\ref{attreq6}%
) has two \textit{non-degenerate} solutions:\medskip

\textbf{I.} The first non-degenerate solution to AE (\ref{attreq6}) is
\begin{equation}
N_{2}\approx 0.  \label{k=6-1/2BPS}
\end{equation}
This is nothing but the $k=5$ solution (\ref{k=5-1/2BPS}). As one can see
from Eq. (\ref{W-k=6})-(\ref{covder6}), also for $k=6$ such a solution
corresponds to a $\frac{1}{2}$-BPS LG critical point of $V_{BH}$ ($%
\widetilde{W}\neq 0$, $D\widetilde{W}=0$). From Eq. (\ref{N2-6}), in order
to get the solution (\ref{k=6-1/2BPS}), we have to fine-tune two PF BH
charges out of four in the following way:
\begin{equation}
n_{3}\approx n_{2},\quad n_{4}\approx n_{2}-n_{1}.  \label{k=6-chcon1/2}
\end{equation}
The charges $n_{1}$, $n_{2}$ are not fixed; they only satisfy the
non-degeneration condition
\begin{equation}
\left. N_{1}\right| _{N_{2}\approx 0}\approx -\sqrt{3}n_{2}+i(n_{2}-2n_{1})%
\neq 0\Longleftrightarrow n_{2}\neq 0,n_{1}\neq \frac{n_{2}}{2}.
\end{equation}
As it was for $k=5$, also the real eigenvalues (\ref{k=6-gen-eigenvalues})
for the $\frac{1}{2}$-BPS critical solution coincide and, as it is well
known \cite{FGK,BFM,AoB}, are strictly positive:
\begin{equation}
\lambda _{+,\frac{1}{2}-BPS}=\lambda _{-,\frac{1}{2}-BPS}\approx \frac{2}{3}%
\frac{C_{1}^{2}}{C_{0}^{2}}|N_{1}|_{N_{2}\approx 0}^{2}>0.
\end{equation}
Consequently, the $\frac{1}{2}$-BPS LG critical point $\psi \approx 0$
supported by the PF BH charge configuration (\ref{k=6-chcon1/2}) is a stable
extremum, since it is a minimum of $V_{BH}$, and it is therefore an
attractor in a strict sense. The classical (Bekenstein-Hawking) BH entropy
at such a (class of) $\frac{1}{2}$-BPS LG attractor(s) takes the value
\begin{equation}
S_{BH,\frac{1}{2}-BPS}=\pi V_{BH,\frac{1}{2}-BPS}\approx 0.032\pi
|N_{1}|_{N_{2}\approx 0}^{2}=0.128\pi \left(
n_{1}^{2}+n_{2}^{2}-n_{1}n_{2}\right) .
\end{equation}
\medskip \medskip

\textbf{II}. The second non-degenerate solution to AE (\ref{attreq6}) is
\begin{equation}
N_{1}\approx 0.  \label{k=6-non-BPS}
\end{equation}
As one can see from Eq. (\ref{W-k=6})-(\ref{covder6}), such a solution
corresponds to a non-BPS, $Z=0$ LG critical point of $V_{BH}$ ($\widetilde{W}%
=0$, $D\widetilde{W}\neq 0$). The real eigenvalues (\ref{k=6-gen-eigenvalues}%
) for such a non-BPS, $Z=0$ critical solution read
\begin{equation}
\lambda _{\pm ,non-BPS,Z=0}\approx 2\frac{C_{1}^{2}}{C_{0}^{2}}\left( 1\pm 9%
\frac{C_{0}^{2}C_{3}}{C_{1}^{3}}\right) |N_{2}|_{N_{1}\approx 0}^{2}.
\label{k=6-non-BPS-eigenvalues}
\end{equation}
Substituting the numerical values (\ref{con6}) of the involved constants in
Eq. (\ref{k=6-non-BPS-eigenvalues}), one reaches the conclusion that one
eigenvalue is positive and the other one is negative:
\begin{equation}
\lambda _{\pm ,non-BPS,Z=0}\approx 2\left( 0.45\mp 1.46\right)
|N_{2}|_{N_{1}\approx 0}^{2}\lessgtr 0.
\end{equation}
Let us now find the fine-tuning conditions for PF BH charges supporting the
considered non-BPS, $Z=0$ LG attractor for the mirror \textit{sextic} $%
\mathcal{M}_{6}^{\prime }$. This amounts to solving Eq. (\ref{k=6-non-BPS})
by recalling Eq. (\ref{N1-6}). By doing so, one gets the following unique
set of constraining relations on PF BH charges:
\begin{equation}
n_{3}\approx -n_{2},\quad n_{4}\approx n_{2}+n_{1}.  \label{ft6}
\end{equation}
The charges $n_{1},n_{2}$ are not fixed; they only satisfy the
non-degeneration condition
\begin{equation}
\left. N_{2}\right| _{N_{1}\approx 0}\approx -\left( 2n_{1}+n_{2}\right) +%
\sqrt{3}in_{2}\neq 0\Longleftrightarrow n_{2}\neq 0,n_{1}\neq -\frac{n_{2}}{2%
}.  \label{k=6-N2-non-BPS}
\end{equation}
Thus, the non-BPS, $Z=0$ LG critical point $\psi \approx 0$ supported by the
PF BH charge configuration (\ref{ft6}) is a saddle point of $V_{BH}$ and
consequently it is \textit{not} an attractor in a strict sense.

The classical (Bekenstein-Hawking) BH entropy at such a (class of) non-BPS, $%
Z=0$ LG saddle point(s) takes the value
\begin{equation}
S_{BH,non-BPS,Z=0}=\pi V_{BH,non-BPS,Z=0}\approx 0.096\pi
|N_{2}|_{N_{1}\approx 0}^{2}=0.384\pi \left(
n_{1}^{2}+n_{2}^{2}+n_{1}n_{2}\right) .  \label{k=6-V-non-BPS}
\end{equation}
\medskip

By using the definitions (\ref{PF-charge}) and (\ref{PF-charge-inv}) for the
case $k=6$, one obtains (recall the first of definitions (\ref{Gamma-Fermat}%
))
\begin{equation}
\begin{array}{c}
\left( n_{1},n_{2},n_{3},n_{4}\right) =\left(
p^{0}-3q_{0},p^{0}+3q_{1},-p^{0}+3p^{1}+9q_{0}-3q_{1},-p^{0}+6q_{0}\right)
\\
\\
\Updownarrow \\
\\
\Gamma =\left( -2n_{1}-n_{4},\frac{1}{3}\left(
3n_{1}-n_{2}-n_{3}+3n_{4}\right) ,\frac{1}{3}(n_{1}+n_{4}),\frac{1}{3}%
(-2n_{1}+n_{2}-n_{4})\right) .
\end{array}
\label{k=6-rel-n-sympl}
\end{equation}
Thus, the fine-tuning conditions (\ref{k=6-chcon1/2}) and (\ref{ft6}) can be
rewritten in terms of symplectic BH charges respectively as follows:
\begin{eqnarray}
&&
\begin{array}{l}
\frac{1}{2}-BPS:\left\{
\begin{array}{l}
-2p^{0}+3p^{1}+9q_{0}-6q_{1}\approx 0,\quad 3q_{0}\approx p^{0}+3q_{1}, \\
\\
p^{0}\neq -3q_{1},p^{0}\neq 6q_{0}+3q_{1};
\end{array}
\right.
\end{array}
\label{k=6-chcon1/2-sympl} \\
&&  \notag \\
&&  \notag \\
&&
\begin{array}{l}
non-BPS,Z=0:\left\{
\begin{array}{l}
p^{0}\approx 3q_{0}-q_{1},\quad p^{1}\approx -3q_{0}, \\
\\
p^{0}\neq -3q_{1},p^{0}\neq 6q_{0}-3q_{1}.
\end{array}
\right.
\end{array}
\label{ft6-symplectic}
\end{eqnarray}
\medskip

\subsection*{Remark}

The non-BPS, $Z=0$ critical solution for the mirror \textit{sextic} $%
\mathcal{M}_{6}^{\prime }$ had been previously investigated in Sect. 7 of
\cite{G}. Up to irrelevant changes of notation, Eq. (\ref{ft6-symplectic})
coincides with Eq. (7.8) of \cite{G}. By considering the second derivatives (%
\ref{ddV-k=6})-(\ref{ddV-k=6-2}) of the ``effective BH potential''
constrained by Eq. (\ref{k=6-non-BPS}) and comparing them with Eq. (7.9) of
\cite{G}, one can state that the crucial difference between the results of
\cite{G} and ours lies in the critical value of the second holomorphic
derivative of $V_{BH}$. Indeed, Eq. (7.9) of \cite{G} reads
\begin{equation}
\left( \partial ^{2}V_{BH}\right) _{non-BPS,Z=0}=0.  \label{Gi}
\end{equation}
From our previous computations, consistently taking into account the needed
orders in $\psi $ to get the series expansion for $V_{BH}$ up to $\mathcal{O}%
\left( \psi ^{2}\right) $ (or $\mathcal{O}\left( \psi ^{3}\right) $), we
disagree with the critical value of the second holomorphic derivative of $%
V_{BH}$ at the considered non-BPS, $Z=0$ critical point given by Eq. (\ref
{Gi}). According to our results, the statement made in \cite{G} that the
considered non-BPS, $Z=0$ LG critical point of $V_{BH}$ is actually an
attractor in a strict sense for all possible supporting symplectic BH charge
configurations (\ref{ft6-symplectic}) does \textit{not} hold. Instead, as
correctly stated above, the non-BPS, $Z=0$ LG critical point $\psi \approx 0$
supported by the BH charge configurations (\ref{ft6}) (PF) and (\ref
{ft6-symplectic}) (symplectic) is a saddle point of $V_{BH}$ and
consequently it is \textit{not} an attractor in a strict sense.

Also, by recalling Eq. (\ref{k=6-V-non-BPS}) and using Eqs. (\ref
{k=6-N2-non-BPS}) and (\ref{k=6-rel-n-sympl})-(\ref{ft6-symplectic}), our
analysis yields that the ``effective BH potential'' and BH entropy at the
considered (class of) non-BPS, $Z=0$ LG saddle point(s) take the purely
charge-dependent values:
\begin{equation}
V_{BH,non-BPS,Z=0}=\frac{1}{2\sqrt{3}}%
(3q_{0}^{2}+q_{1}^{2}),~~~S_{BH,non-BPS,Z=0}=\frac{\pi }{2\sqrt{3}}%
(3q_{0}^{2}+q_{1}^{2}).  \label{k=6-VBH-SBH-non-BPS-Z=0}
\end{equation}
As one can see, the value of $V_{BH,non-BPS,Z=0}$ given by Eq. (\ref
{k=6-VBH-SBH-non-BPS-Z=0}) does \textit{not} coincide with the one given by
Eq. (7.10) of \cite{G}.

\section{$k=8$ : Mirror \textit{Octic} near LG Point\label{eightic}}

\setcounter{equation}0
\def\theequation{9.\arabic{subsection}.\arabic{equation}}

\subsection{Geometric Setup}

The case of mirror \textit{octic} $\mathcal{M}_{8}^{\prime }$ (as well as
the one of mirror \textit{dectic} $\mathcal{M}_{10}^{\prime }$ treated in
Sect. \ref{tentic}) needs a different approach with respect to the cases of
mirror \textit{quintic} $\mathcal{M}_{5}^{\prime }$ and mirror sextic $%
\mathcal{M}_{6}^{\prime }$, respectively treated in Sects. \ref{quintic} and
\ref{sixtic}.

Indeed, contrary to what happens for $k=5,6$ (see Eqs. (\ref{k=5-metric})
and (\ref{k=6-metric}), respectively), for $k=8,10$ the series expansion of
the K\"{a}hler metric $g\left( \psi ,\overline{\psi }\right) $ near the LG
point starts with \textit{no} constant term (namely, it is \textit{not}
regular at $\psi =0$). As a consequence, one has to consider the series
expansion of the ``effective BH potential'' $V_{BH}$ up to $\mathcal{O}%
\left( \psi ^{4}\right) $ (rather than up to $\mathcal{O}\left( \psi
^{3}\right) $, as it is for $k=5,6$), in order to obtain all the relevant
quantities up to $\mathcal{O}\left( \psi ^{2}\right) $ (rather than up to $%
\mathcal{O}\left( \psi \right) $, as it is for $k=5,6$).

For $k=8$ the definitions (\ref{omega0-2}) yield
\begin{equation}
C_{8,2l-1}=0,\quad l\in \mathbb{N;}
\end{equation}
moreover, since $F_{8,m+8,\,n}=F_{8,m\,,n+8}=-F_{8,m\,n}$ (see the third of
properties (\ref{prop-F})), the only independent elements of the rank-2
tensor $F_{8}$ belong to the $8\times 8$ matrix
\begin{eqnarray}
F_{8,mn} &=&{\tiny \left(
\begin{array}{cccccccc}
2(2+\sqrt{2}) & -\sqrt{10+7\sqrt{2}} & 0 & \sqrt{2+\sqrt{2}} & 0 & -\sqrt{2+%
\sqrt{2}} & 0 & \sqrt{10+7\sqrt{2}} \\
{\rule{0pt}{12pt}}-\sqrt{10+7\sqrt{2}} & 2 & \sqrt{2-\sqrt{2}} & -\sqrt{2} &
\sqrt{10-7\sqrt{2}} & 0 & \sqrt{2+\sqrt{2}} & -3\sqrt{2} \\
{\rule{0pt}{12pt}}0 & \sqrt{2-\sqrt{2}} & 2(-2+\sqrt{2}) & \sqrt{2-\sqrt{2}}
& 0 & -\sqrt{10-7\sqrt{2}} & 0 & \sqrt{10-7\sqrt{2}} \\
{\rule{0pt}{12pt}}\sqrt{2+\sqrt{2}} & -\sqrt{2} & \sqrt{2-\sqrt{2}} & 0 & -%
\sqrt{2-\sqrt{2}} & \sqrt{2} & -\sqrt{2+\sqrt{2}} & 2 \\
{\rule{0pt}{12pt}}0 & \sqrt{10-7\sqrt{2}} & 0 & -\sqrt{2-\sqrt{2}} & 4-2%
\sqrt{2} & -\sqrt{2-\sqrt{2}} & 0 & \sqrt{10-7\sqrt{2}} \\
{\rule{0pt}{12pt}}-\sqrt{2+\sqrt{2}} & 0 & -\sqrt{10-7\sqrt{2}} & \sqrt{2} &
-\sqrt{2-\sqrt{2}} & -2 & \sqrt{10+7\sqrt{2}} & -3\sqrt{2} \\
{\rule{0pt}{12pt}}0 & \sqrt{2+\sqrt{2}} & 0 & -\sqrt{2+\sqrt{2}} & 0 & \sqrt{%
10+7\sqrt{2}} & -4-2\sqrt{2} & \sqrt{10+7\sqrt{2}} \\
{\rule{0pt}{12pt}}\sqrt{10+7\sqrt{2}} & -3\sqrt{2} & \sqrt{10-7\sqrt{2}} & 2
& \sqrt{10-7\sqrt{2}} & -3\sqrt{2} & \sqrt{10+7\sqrt{2}} & 0
\end{array}
\right) .}  \notag \\
&&  \label{k=8-Fmn}
\end{eqnarray}
Let us now write down all the relevant quantities up to the needed order
(here and below, unless otherwise specified, we omit the Fermat parameter $%
k=8$):
\begin{eqnarray}
&&
\begin{array}{l}
\widetilde{K}\approx (3-2\sqrt{2})\frac{C_{2}^{2}}{C_{0}^{2}}(\psi \bar{\psi}%
)^{2}\left[ 1-\left( \frac{C_{4}^{2}}{C_{2}^{2}}-\frac{1}{2}(3-2\sqrt{2})%
\frac{C_{2}^{2}}{C_{0}^{2}}\right) (\psi \bar{\psi})^{2}\right] +\frac{C_{8}%
}{C_{0}}(\psi ^{8}+\bar{\psi}^{8})+\mathcal{O}(\psi ^{9}); \\
~
\end{array}
\\
&&
\begin{array}{l}
g\approx 4(3-2\sqrt{2})\frac{C_{2}^{2}}{C_{0}^{2}}\psi \bar{\psi}\left[
1-4\left( \frac{C_{4}^{2}}{C_{2}^{2}}-\frac{1}{2}(3-2\sqrt{2})\frac{C_{2}^{2}%
}{C_{0}^{2}}\right) (\psi \bar{\psi})^{2}\right] +\mathcal{O}(\psi ^{7}); \\
~
\end{array}
\label{k=8-metric} \\
&&
\begin{array}{l}
\widetilde{W}\approx N_{1}+\frac{C_{2}}{C_{0}}N_{3}\psi ^{2}-\frac{C_{4}}{%
C_{0}}\bar{N}_{3}\psi ^{4}-\frac{C_{6}}{C_{0}}\bar{N}_{1}\psi ^{6}+\mathcal{O%
}(\psi ^{8}). \\
~
\end{array}
\end{eqnarray}
Now, by using the formul\ae\ of the general analysis exploited in Sect. \ref
{GA}, we can get the ``effective BH potential'' and the holomorphic
superpotential, as well as their (covariant) derivatives, up to $\mathcal{O}%
\left( \psi ^{2}\right) $:
\begin{eqnarray}
&&
\begin{array}{l}
\widetilde{W}=N_{1}+\frac{C_{2}}{C_{0}}N_{3}\psi ^{2}; \\
~
\end{array}
\label{W-k=8} \\
&&
\begin{array}{l}
D\widetilde{W}=2\frac{C_{2}}{C_{0}}N_{3}\psi -4\frac{C_{4}}{C_{0}}\bar{N}%
_{3}\psi ^{3}+2(3-2\sqrt{2})\frac{C_{2}^{2}}{C_{0}^{2}}N_{1}\psi \bar{\psi}%
^{2}; \\
~
\end{array}
\label{covder8} \\
&&
\begin{array}{l}
D^{2}\widetilde{W}=-8\frac{C_{4}}{C_{0}}\bar{N}_{3}\psi ^{2}; \\
~
\end{array}
\\
&&
\begin{array}{l}
\widetilde{V}_{BH}=|N_{1}|^{2}+(3+2\sqrt{2})|N_{3}|^{2}+2\frac{C_{2}}{C_{0}}%
\left( N_{3}\bar{N}_{1}-(3+2\sqrt{2})\frac{C_{4}C_{0}}{C_{2}^{2}}(\bar{N}%
_{3})^{2}\right) \psi ^{2}+ \\
~ \\
\qquad \quad +2\frac{C_{2}}{C_{0}}\left( \bar{N}_{3}N_{1}-(3+2\sqrt{2})\frac{%
C_{4}C_{0}}{C_{2}^{2}}(N_{3})^{2}\right) \bar{\psi}^{2};
\end{array}
\label{pot8}
\end{eqnarray}
\begin{eqnarray}
&&
\begin{array}{l}
\partial \widetilde{V}_{BH}=4\psi \left[ \frac{C_{2}}{C_{0}}\left( N_{3}\bar{%
N}_{1}-(3+2\sqrt{2})\frac{C_{4}C_{0}}{C_{2}^{2}}(\bar{N}_{3})^{2}\right) -3%
\frac{C_{4}}{C_{0}}\left( 1+(3+2\sqrt{2})\frac{C_{6}C_{0}}{C_{4}C_{2}}%
\right) \bar{N}_{1}\bar{N}_{3}\psi ^{2}+\right. \\
~ \\
\qquad \qquad \left. +\frac{C_{2}^{2}}{C_{0}^{2}}\left( |N_{1}|^{2}(3-2\sqrt{%
2})+|N_{3}|^{2}\left( 1+4(3+2\sqrt{2})\frac{C_{0}^{2}C_{4}^{2}}{C_{2}^{4}}%
\right) \right) \bar{\psi}^{2}\right] ; \\
~
\end{array}
\\
&&
\begin{array}{l}
\partial ^{2}\widetilde{V}_{BH}=4\frac{C_{2}}{C_{0}}\left( N_{3}\bar{N}%
_{1}-(3+2\sqrt{2})\frac{C_{4}C_{0}}{C_{2}^{2}}(\bar{N}_{3})^{2}\right) -36%
\frac{C_{4}}{C_{0}}\left( 1+(3+2\sqrt{2})\frac{C_{6}C_{0}}{C_{4}C_{2}}%
\right) \bar{N}_{1}\bar{N}_{3}\psi ^{2}+ \\
~ \\
~~~~~~~~~~~~+4\frac{C_{2}^{2}}{C_{0}^{2}}\left( |N_{1}|^{2}(3-2\sqrt{2}%
)+|N_{3}|^{2}\left( 1+4(3+2\sqrt{2})\frac{C_{0}^{2}C_{4}^{2}}{C_{2}^{4}}%
\right) \right) \bar{\psi}^{2}; \\
~
\end{array}
\label{ddV-k=8} \\
&&
\begin{array}{l}
\partial \overline{\partial }\widetilde{V}_{BH}=8\frac{C_{2}^{2}}{C_{0}^{2}}%
\left( |N_{1}|^{2}(3-2\sqrt{2})+|N_{3}|^{2}\left( 1+4(3+2\sqrt{2})\frac{%
C_{0}^{2}C_{4}^{2}}{C_{2}^{4}}\right) \right) \psi \bar{\psi}. \\
~
\end{array}
\label{ddV-k=8-2}
\end{eqnarray}
Let us stress once again that, contrary to the treatment of Sects. \ref
{quintic} and \ref{sixtic}, and as evident from Eqs. (\ref{pot8})-(\ref
{ddV-k=8-2}), for the case of mirror \textit{octic} we truncate the series
expansion of the ``effective BH potential'' and of its second derivatives
around the LG point up to $\mathcal{O}\left( \psi ^{2}\right) $ included,
and the series expansion of its first derivative around the LG point up to $%
\mathcal{O}\left( \psi ^{3}\right) $ included. This is due to the absence of
an $\mathcal{O}\left( \psi \right) $ term in expression of $\widetilde{V}%
_{BH}$ given by Eq. (\ref{pot8}). As mentioned at the start of the present
Section, such a fact can be traced back to the non-regularity of $g$ at $%
\psi =0$ (see Eq. (\ref{k=8-metric})). \setcounter{equation}0
\def\theequation{9.1.\arabic{subsubsection}.\arabic{equation}}

\subsubsection{\label{prep-8}The Holomorphic Prepotentials $F$ and $\mathcal{%
F}$}

As done for $k=5,6$ respectively in Subsubsects. \ref{prep-5} and \ref
{prep-6}, in order to compute the holomorphic prepotential $F\left( X\left(
\psi \right) \right) $ and its ``K\"{a}hler-invariant counterpart'' $%
\mathcal{F}\left( \psi \right) $ defined by Eq. (\ref{F-call}), we have to
recall the general formul\ae\ for (mirror) Fermat $CY_{3}$s introduced in
Sect. \ref{GA}, and specialize them for $k=8$. By recalling the definitions (%
\ref{csi-csi}), (\ref{PHIZ}), (\ref{hol-prep-F-l-CY}), (\ref{F-call-CY}),
one can compute:
\begin{eqnarray}
&&
\begin{array}{l}
\xi _{8,m}=\left(
\begin{array}{l}
-i\left[ sin\left( \frac{\pi m}{8}\right) +sin\left( \frac{3\pi m}{8}\right)
\right] \\
\\
-cos\left( \frac{\pi m}{8}\right) +isin\left( \frac{\pi m}{8}\right) \\
\\
4cos^{2}\left( \frac{\pi m}{8}\right) \left[ cos\left( \frac{\pi m}{8}%
\right) -3isin\left( \frac{\pi m}{8}\right) \right] \\
\\
2isin\left( \frac{\pi m}{8}\right)
\end{array}
\right) ;
\end{array}
\label{csi-csi-8} \\
&&  \notag \\
&&  \notag \\
&&
\begin{array}{l}
\Phi _{8,m,n}= \\
\\
=\frac{1}{2}\left\{
\begin{array}{l}
\left[
\begin{array}{l}
-3cos\left( \frac{\pi \left( n-3m\right) }{8}\right) -3cos\left( \frac{\pi
\left( 3n-m\right) }{8}\right) -5cos\left( \frac{\pi \left( n-m\right) }{8}%
\right) -3cos\left( \frac{3\pi \left( n-m\right) }{8}\right) + \\
\\
+5cos\left( \frac{\pi \left( n+m\right) }{8}\right) +3cos\left( \frac{3\pi
\left( n+m\right) }{8}\right) +3cos\left( \frac{\pi \left( 3n+m\right) }{8}%
\right) +3cos\left( \frac{\pi \left( n+3m\right) }{8}\right)
\end{array}
\right] + \\
\\
+i\left[
\begin{array}{l}
3sin\left( \frac{\pi \left( n-3m\right) }{8}\right) +sin\left( \frac{\pi
\left( 3n-m\right) }{8}\right) +sin\left( \frac{\pi \left( n-m\right) }{8}%
\right) +sin\left( \frac{3\pi \left( n-m\right) }{8}\right) + \\
\\
-5sin\left( \frac{\pi \left( n+m\right) }{8}\right) -sin\left( \frac{3\pi
\left( n+m\right) }{8}\right) -sin\left( \frac{\pi \left( 3n+m\right) }{8}%
\right) -3sin\left( \frac{\pi \left( n+3m\right) }{8}\right)
\end{array}
\right]
\end{array}
\right\} \neq \Phi _{8,\left( m,n\right) };
\end{array}
\notag \\
&&  \label{PHIZ-8} \\
&&  \notag \\
&&
\begin{array}{l}
F_{k=8,l=7}\left( X_{8}\left( \psi \right) \right) =-\frac{\pi ^{6}}{2^{9}}%
\sum_{m=1}^{7}\sum_{n=1}^{7}(-1)^{m+n}C_{8,m-1}C_{8,n-1}\psi ^{m+n-2}\Phi
_{8,\left( m,n\right) }= \\
\\
\\
=-\frac{\pi ^{6}}{2^{9}}\left[
\begin{array}{l}
-\frac{\left( 4+\sqrt{2}+2i\left( 1+\sqrt{2}\right) \right) \pi }{\Gamma
^{2}\left( \frac{5}{6}\right) \Gamma ^{6}\left( \frac{7}{8}\right) }+\frac{%
8\left( i+\sqrt{2}\right) \sqrt{\pi }}{\Gamma ^{3}\left( \frac{5}{8}\right)
\Gamma \left( \frac{5}{6}\right) \Gamma ^{3}\left( \frac{7}{8}\right) }\psi
^{2}+ \\
\\
+\left[ \frac{-16+4\sqrt{2}+8i\left( 1-\sqrt{2}\right) }{\Gamma ^{6}\left(
\frac{5}{8}\right) }-\frac{4\left( 2+i\right) 2^{3/4}}{\pi ^{3/2}\Gamma
^{3}\left( \frac{3}{4}\right) }\right] \psi ^{4}+ \\
\\
+\frac{1}{\sqrt{2}\pi ^{4}}\left[ \frac{\left( 4-\sqrt{2}\right) \sqrt{\pi }%
\Gamma \left( \frac{5}{6}\right) }{sin^{3}\left( \frac{7\pi }{8}\right) }+%
\frac{2\left( 4+\sqrt{2}\right) \Gamma ^{2}\left( \frac{7}{6}\right) }{%
sin^{3}\left( \frac{5\pi }{8}\right) }\right] \psi ^{6}
\end{array}
\right] +\mathcal{O}\left( \psi ^{7}\right) =F_{k=8,l=7}\left( \psi
^{2}\right) ;
\end{array}
\notag \\
&&  \label{F87-CY}
\end{eqnarray}
\begin{eqnarray}
&&
\begin{array}{l}
\mathcal{F}_{8}\left( \psi \right) \equiv F_{8}\left( \frac{X_{8}^{1}(\psi )%
}{X_{8}^{0}(\psi )}\right) =\frac{F_{8}\left( X_{8}^{0}(\psi
),X_{8}^{1}(\psi )\right) }{\left( X_{8}^{0}(\psi )\right) ^{2}}=\frac{1}{2}%
\frac{\sum_{m,n=1}^{\infty }(-1)^{m+n}C_{8,m-1}C_{8,n-1}\Phi _{8,\left(
m,n\right) }\psi ^{m+n-2}}{\left[ \sum_{r=1}^{\infty }(-1)^{r}C_{8,r-1}\xi
_{8,r}^{1}\psi ^{r-1}\right] ^{2}}\approx \\
\\
\approx \frac{\Phi _{8,1,1}}{2\left( \xi _{8,1}^{1}\right) ^{2}}\left[ 1+%
\frac{C_{8,2}\left( \Phi _{8,1,3}+\Phi _{8,3,1}\right) }{C_{8,0}\Phi _{8,1,1}%
}\psi ^{2}+\mathcal{O}\left( \psi ^{3}\right) \right] \left[ 1-2\frac{%
C_{8,2}\xi _{8,3}^{1}}{C_{8,0}\xi _{8,1}^{1}}\psi ^{2}+\mathcal{O}\left(
\psi ^{3}\right) \right] \approx \\
\\
\approx \frac{\Phi _{8,1,1}}{2\left( \xi _{8,1}^{1}\right) ^{2}}\left[
1-2\left( \frac{\xi _{8,3}^{1}}{\xi _{8,1}^{1}}-\frac{\Phi _{8,1,3}+\Phi
_{8,3,1}}{2\Phi _{8,1,1}}\right) \frac{C_{8,2}}{C_{8,0}}\psi ^{2}\right] +%
\mathcal{O}\left( \psi ^{3}\right) ,
\end{array}
\notag \\
&&  \label{F-call-CY-8}
\end{eqnarray}
where in the formula (\ref{F-call-CY-8}) for the K\"{a}hler-invariant
holomorphic prepotential $\mathcal{F}_{8}$ only the leading terms in the
\textit{LG limit }$\psi \longrightarrow 0$ have been retained. Eq. (\ref
{F87-CY}) is the exact expression (complete up to $\mathcal{O}\left( \psi
^{6}\right) $ included) of the holomorphic prepotential $F_{8}$ in a
suitable neighbourhood of the LG point of the moduli space of the (mirror)%
\textit{\ octic} $\mathcal{M}_{8}^{\prime }$. \setcounter{equation}0
\def\theequation{9.\arabic{subsection}.\arabic{equation}}

\subsection{\textit{``Criticality Condition''} Approach and Attractor
Equation}

Let us now find the solutions of the AE $\partial \widetilde{V}_{BH}\left(
\psi ,\overline{\psi };q,p\right) =0$, and check their stability. Since we
are working near the LG point, by using Eq. (\ref{pot8}) we can rewrite the
AE for $\mathcal{M}_{8}^{\prime }$ as follows:
\begin{equation}
\begin{array}{l}
\frac{C_{2}}{C_{0}}\left( N_{3}\bar{N}_{1}-(3+2\sqrt{2})\frac{C_{4}C_{0}}{%
C_{2}^{2}}(\bar{N}_{3})^{2}\right) +\frac{C_{2}^{2}}{C_{0}^{2}}\left(
|N_{1}|^{2}(3-2\sqrt{2})+|N_{3}|^{2}\left( 1+4(3+2\sqrt{2})\frac{%
C_{0}^{2}C_{4}^{2}}{C_{2}^{4}}\right) \right) \bar{\psi}^{2}\approx \\
~ \\
\approx 3\frac{C_{4}}{C_{0}}\left( 1+(3+2\sqrt{2})\frac{C_{6}C_{0}}{%
C_{4}C_{2}}\right) \bar{N}_{1}\bar{N}_{3}\psi ^{2}.
\end{array}
\qquad \quad  \label{attreq8}
\end{equation}
Solving Eq. (\ref{attreq8}), we will find one (or more) set(s) of BH charges
supporting $\psi \approx 0$ to be a critical point of $V_{BH}$. Since we are
working near the LG point, it is clear that the first term in the left-hand
side (l.h.s.) of Eq. (\ref{attreq8}) must be small enough. This implies the
following fine-tuning condition:
\begin{equation}
\begin{array}{l}
N_{3}\bar{N}_{1}-\vartheta (\bar{N}_{3})^{2}\approx 0, \\
\\
\vartheta \equiv (3+2\sqrt{2})\frac{C_{4}C_{0}}{C_{2}^{2}}.
\end{array}
\label{atr8}
\end{equation}
\setcounter{equation}0
\def\theequation{9.\arabic{subsection}.\arabic{equation}}

\subsection{\label{21march-3}Critical Hessian of $V_{BH}$}

By using (the $n_{V}=1$ case of) Eq. (\ref{4jan2}) and Eqs. (\ref{ddV-k=8})-(%
\ref{ddV-k=8-2}) evaluated along the criticality condition (\ref{attreq8})-(%
\ref{atr8}), it can be computed that the components of $H_{\text{real form}%
}^{\widetilde{V}_{BH}}$ (given by Eq. (\ref{Hessian-real-form})) constrained
by Eqs. (\ref{attreq8})-(\ref{atr8}) read as follows:
\begin{eqnarray}
&&
\begin{array}{l}
\mathcal{A}=-12\frac{C_{4}}{C_{0}}\left( 1+(3+2\sqrt{2})\frac{C_{6}C_{0}}{%
C_{4}C_{2}}\right) \left( \bar{N}_{1}\bar{N}_{3}\psi ^{2}+N_{1}N_{3}\bar{\psi%
}^{2}\right) + \\
~ \\
\quad +8\frac{C_{2}^{2}}{C_{0}^{2}}\psi \bar{\psi}\left( |N_{1}|^{2}(3-2%
\sqrt{2})+|N_{3}|^{2}\left( 1+4(3+2\sqrt{2})\frac{C_{0}^{2}C_{4}^{2}}{%
C_{2}^{4}}\right) \right) ; \\
~
\end{array}
\\
&&  \notag \\
&&
\begin{array}{l}
\mathcal{B}=12\frac{C_{4}}{C_{0}}\left( 1+(3+2\sqrt{2})\frac{C_{6}C_{0}}{%
C_{4}C_{2}}\right) \left( \bar{N}_{1}\bar{N}_{3}\psi ^{2}+N_{1}N_{3}\bar{\psi%
}^{2}\right) + \\
~ \\
+8\frac{C_{2}^{2}}{C_{0}^{2}}\psi \bar{\psi}\left( |N_{1}|^{2}(3-2\sqrt{2}%
)+|N_{3}|^{2}\left( 1+4(3+2\sqrt{2})\frac{C_{0}^{2}C_{4}^{2}}{C_{2}^{4}}%
\right) \right) ; \\
~
\end{array}
\\
&&  \notag \\
&&
\begin{array}{l}
\mathcal{C}=-12i\frac{C_{4}}{C_{0}}\left( 1+(3+2\sqrt{2})\frac{C_{6}C_{0}}{%
C_{4}C_{2}}\right) \left( N_{1}N_{3}\bar{\psi}^{2}-\bar{N}_{1}\bar{N}%
_{3}\psi ^{2}\right) . \\
~
\end{array}
\end{eqnarray}
The resulting real eigenvalues of $H_{\text{real form}}^{\widetilde{V}_{BH}}$
constrained by Eqs. (\ref{attreq8})-(\ref{atr8}) read:
\begin{equation}
\lambda _{\pm }\approx 8\psi \bar{\psi}\left[
\begin{array}{l}
\frac{C_{2}^{2}}{C_{0}^{2}}\left( |N_{1}|^{2}(3-2\sqrt{2})+|N_{3}|^{2}\left(
1+4(3+2\sqrt{2})\frac{C_{0}^{2}C_{4}^{2}}{C_{2}^{4}}\right) \right) + \\
\\
\pm 3\frac{C_{4}}{C_{0}}\left( 1+(3+2\sqrt{2})\frac{C_{6}C_{0}}{C_{4}C_{2}}%
\right) |N_{1}||N_{3}|
\end{array}
\right] .  \label{k=8-gen-eigenvalues}
\end{equation}

By recalling Eq. (\ref{pot}) and using Eq. (\ref{pot8}) with $\psi \approx 0$
and constrained by Eqs. (\ref{attreq8})-(\ref{atr8}), one obtains that the
purely charge-dependent LG critical values of the ``effective BH potential''
for the mirror \textit{octic} $\mathcal{M}_{8}^{\prime }$ are
\begin{equation}
V_{BH,LG-critical,k=8}\approx \frac{1}{8(2+\sqrt{2})}\left[ |N_{1}|^{2}+(3+2%
\sqrt{2})|N_{3}|^{2}\right] ;
\end{equation}
by recalling formula (\ref{BHEA}), this directly yields the following purely
charge-dependent values of the BH entropy at the LG critical points of $%
V_{BH,8}$ in the moduli space of $\mathcal{M}_{8}^{\prime }$:
\begin{equation}
S_{BH,LG-critical,k=8}\approx \frac{\pi }{8(2+\sqrt{2})}\left[
|N_{1}|^{2}+(3+2\sqrt{2})|N_{3}|^{2}\right] .  \label{k=8-SBH}
\end{equation}

Let us write down here the numerical values of constants relevant to our
treatment:
\begin{equation}
C_{0}\approx 1.64,\;C_{2}\approx -0.9,\;C_{4}\approx 0.24,\;C_{6}\approx
-0.007,\;C_{8}\approx -0.004;\quad \vartheta \approx 2.83.  \label{con8}
\end{equation}
\medskip \setcounter{equation}0
\def\theequation{9.\arabic{subsection}.\arabic{equation}}

\subsection{\label{21march-4}Solutions to Attractor Equations}

Let us now analyze more in depth the species of LG attractor points arising
from the AE (\ref{atr8}). As it can be easily seen, the AE (\ref{atr8}) has
two \textit{non-degenerate} solutions:\smallskip \medskip

\textbf{I.} The first non-degenerate solution to AE (\ref{atr8}) is
\begin{equation}
N_{3}\approx 0.  \label{k=8-1/2BPS}
\end{equation}
As one can see from Eq. (\ref{W-k=8})-(\ref{covder8}), such a solution
corresponds to a $\frac{1}{2}$-BPS LG critical point of $V_{BH}$ ($%
\widetilde{W}\neq 0$, $D\widetilde{W}=0$). From the definition (\ref{N-def})
with $\left( k,m\right) =\left( 8,3\right) $, in order to get the solution (%
\ref{k=8-1/2BPS}), one has to fine-tune two PF BH charges out of four in the
following way:
\begin{equation}
n_{3}\approx -n_{1}+\sqrt{2}n_{2},\quad n_{4}\approx -\sqrt{2}n_{1}+n_{2},
\label{k=8-chcon1/2}
\end{equation}
The charges $n_{1}$, $n_{2}$ are not fixed; they only satisfy the
non-degeneration condition $N_{1}\neq 0$. The real eigenvalues (\ref
{k=8-gen-eigenvalues}) for the $\frac{1}{2}$-BPS critical solution coincide
and, as it is well known \cite{FGK,BFM,AoB}, are strictly positive:
\begin{equation*}
\lambda _{+,\frac{1}{2}-BPS}=\lambda _{-,\frac{1}{2}-BPS}\approx 8\psi \bar{%
\psi}\frac{C_{2}^{2}}{C_{0}^{2}}(3-2\sqrt{2})|N_{1}|_{N_{3}\approx 0}^{2}>0.
\end{equation*}
Consequently, the $\frac{1}{2}$-BPS LG critical point $\psi \approx 0$
supported by the PF BH charge configuration (\ref{k=8-chcon1/2}) is a stable
extremum, since it is a (local) minimum of $V_{BH}$, and it is therefore an
attractor in a strict sense. The classical (Bekenstein-Hawking) BH entropy
at such a (class of) $\frac{1}{2}$-BPS LG attractor(s) takes the value
\begin{equation}
S_{BH,\frac{1}{2}-BPS}=\pi V_{BH,\frac{1}{2}-BPS}\approx 0.0366\pi
|N_{1}|_{N_{3}\approx 0}^{2},
\end{equation}
where $\left. N_{1}\right| _{N_{3}\approx 0}$ is given by the general
formula (\ref{N-def}) with $\left( k,m\right) =\left( 8,1\right) $
constrained by Eq. (\ref{k=8-chcon1/2}):
\begin{equation}
\left. N_{1}\right| _{N_{3}\approx 0}\approx \sqrt{4-2\sqrt{2}}[n_{1}(1-i(1+%
\sqrt{2}))-n_{2}(1+\sqrt{2}-i)].
\end{equation}
\medskip \medskip

\textbf{II}. The second non-degenerate solution to AE (\ref{atr8}) reads
(from Eqs. (\ref{atr8}) and (\ref{con8}): $\vartheta \equiv (3+2\sqrt{2})%
\frac{C_{4}C_{0}}{C_{2}^{2}}\approx 2.83$):
\begin{equation}
\begin{array}{l}
|N_{1}|\approx \vartheta |N_{3}|, \\
\\
arg{(N_{1})}\approx 3arg{(N_{3}),}
\end{array}
\quad  \label{ft8}
\end{equation}
where $N_{1}$ and $N_{3}$ are given by the general formula (\ref{N-def})
with $k=8$ and $m=1,3$, respectively:
\begin{eqnarray}
N_{1} &=&-\frac{\sqrt{2-\sqrt{2}}}{2}\left( n_{1}+n_{4}+(1+\sqrt{2}%
)(n_{2}+n_{3})-i((n_{4}-n_{1})(1+\sqrt{2})+n_{3}-n_{2})\right) ;  \label{8N1}
\\
&&  \notag \\
N_{3} &=&\frac{\sqrt{2+\sqrt{2}}}{2}\left( n_{1}+n_{4}+(1-\sqrt{2}%
)(n_{2}+n_{3})+i((n_{4}-n_{1})(1-\sqrt{2})+n_{3}-n_{2})\right) .  \label{8N3}
\end{eqnarray}
The first of fine-tuning conditions (\ref{ft8}) substituted in Eq. (\ref
{k=8-SBH}) yields
\begin{equation}
S_{BH,non-BPS,Z\neq 0}\approx \frac{\pi }{8(2+\sqrt{2})}\left( \vartheta
^{2}+3+2\sqrt{2}\right) |N_{3}|^{2}\approx 0.5066\pi |N_{3}|^{2}.
\label{k=8-SBH-non-BPS}
\end{equation}
As one can see from Eq. (\ref{W-k=8})-(\ref{covder8}), such a solution
corresponds to a non-BPS, $Z\neq 0$ LG critical point of $V_{BH}$ ($%
\widetilde{W}\neq 0$, $D\widetilde{W}\neq 0$). The real eigenvalues (\ref
{k=8-gen-eigenvalues}) for such a non-BPS, $Z\neq 0$ critical solution read
\begin{eqnarray}
&&\lambda _{\pm ,non-BPS,Z\neq 0}\approx 8\psi \bar{\psi}|N_{3}|^{2}\frac{%
C_{2}^{2}}{C_{0}^{2}}\left[ (1+5(3-2\sqrt{2})\vartheta ^{2})\pm 3(3-2\sqrt{2}%
)\vartheta ^{2}\left( 1+(3+2\sqrt{2})\frac{C_{0}C_{6}}{C_{2}C_{4}}\right)
\right] \approx  \notag \\
&&  \notag \\
&&\qquad \qquad \qquad \approx 8\psi \bar{\psi}|N_{3}|^{2}\frac{C_{2}^{2}}{%
C_{0}^{2}}\left[ 7.9\pm 5.5\right] >0,
\end{eqnarray}
where in the second line we replaced some constants with their numerical
values by using Eq. (\ref{con8}).

Thus, the non-BPS, $Z\neq 0$ LG critical point $\psi \approx 0$ supported by
the PF BH charge configuration (\ref{ft8})-(\ref{8N3}) is a (local) minimum
of $V_{BH}$ and consequently an attractor in a strict sense.\medskip

Let us now find the fine-tuning conditions for PF BH charges supporting the
considered non-BPS, $Z\neq 0$ LG attractor for the mirror \textit{octic} $%
\mathcal{M}_{8}^{\prime }$. This amounts to solving Eqs. (\ref{ft8})-(\ref
{8N3}) by recalling the definitions (\ref{N-def}) and (\ref{PF-charge}). By
doing so, one obtains the following three different sets of constraining
relations on PF BH charges:
\begin{eqnarray}
&&
\begin{array}{c}
\mathbf{II.1}:\left\{
\begin{array}{l}
n_{2,\pm }=(a_{1,\pm }+a_{2})n_{1}\approx \left\{
\begin{array}{l}
``+":2.4581n_{1} \\
\multicolumn{1}{c}{``-":-1.32378n_{1}}
\end{array}
\right. , \\
\\
n_{3,\pm }=(a_{1,\pm }-a_{2})n_{1}=-n_{2,\mp }, \\
\\
n_{4}=-n_{1}, \\
\\
a_{1,\pm }\left( \vartheta \right) \equiv \pm \frac{\sqrt{2}}{\vartheta +1}%
\sqrt{\frac{3(2-\sqrt{2})\vartheta +\sqrt{2}}{(2-\sqrt{2})\vartheta -\sqrt{2}%
}}\approx \pm 1.89094, \\
\\
a_{2}\left( \vartheta \right) \equiv \frac{\sqrt{2}+\vartheta (2-\sqrt{2})}{%
\sqrt{2}(1+\vartheta )}\approx 0.56716, \\
\\
n_{1}\neq 0;
\end{array}
\right.
\end{array}
\label{ch81} \\
&&  \notag \\
&&
\begin{array}[t]{c}
\mathbf{II.2}:\left\{
\begin{array}{l}
n_{2}=\frac{\sqrt{2}-1+\vartheta }{-1+(\sqrt{2}-1)\vartheta }n_{1}\approx
18.83713n_{1}, \\
\\
n_{3}=n_{2}, \\
\\
n_{4}=n_{1}, \\
\\
n_{1}\neq 0;
\end{array}
\right.
\end{array}
\label{ch82} \\
&&
\begin{array}{l}
\mathbf{II.3}:\left\{
\begin{array}{l}
\left\{
\begin{array}{l}
n_{2,\pm }+n_{3,\pm }=a_{\pm }, \\
\\
n_{1,\pm }+n_{4,\pm }=b_{\pm }, \\
\\
n_{2,\pm }-n_{3,\pm }=c, \\
\\
n_{1,\pm }-n_{4,\pm }=d,
\end{array}
\right. \Longleftrightarrow \left\{
\begin{array}{l}
n_{1,\pm }=\frac{b_{\pm }+d}{2}, \\
\\
n_{2,\pm }=\frac{a_{\pm }+c}{2}, \\
\\
n_{3,\pm }=\frac{a_{\pm }-c}{2}, \\
\\
n_{4,\pm }=\frac{b_{\pm }-d}{2},
\end{array}
\right. \quad \quad \quad \\
\\
a_{\pm }\left( \vartheta ;b_{\pm },c,d\right) \equiv -\frac{c+d-(\sqrt{2}+1)%
\left[ c-(\sqrt{2}-1)d\right] \vartheta }{c-d+\left[ c-(\sqrt{2}-1)d\right]
\vartheta }b_{\pm }\left( \vartheta ;c,d\right) \approx \left( 1.76316+\frac{%
c}{-4.1598c+d}\right) b_{\pm }\left( \vartheta ;c,d\right) \approx \\
\\
\approx \pm 2.29892\sqrt{\frac{c-0.053086d}{c-0.55928d}}\frac{\left(
c-0.31647d\right) (c-0.56716d)}{c-0.24039d}, \\
\\
b_{\pm }\left( \vartheta ;c,d\right) \equiv \pm \left( 1+\vartheta \right)
\sqrt{\frac{1+\left( \sqrt{2}+1\right) \vartheta }{2\left( -1+3\left( \sqrt{2%
}+1\right) \vartheta \right) }}\left[ c-\frac{1+\left( \sqrt{2}-1\right)
\vartheta }{1+\vartheta }d\right] \sqrt{\frac{c-\left( \sqrt{2}+1\right)
\frac{-1+\left( \sqrt{2}-1\right) \vartheta }{1+\left( \sqrt{2}+1\right)
\vartheta }d}{c-\left( \sqrt{2}+1\right) \frac{-1-3\left( \sqrt{2}-1\right)
\vartheta }{1-3\left( \sqrt{2}+1\right) \vartheta }d}}\approx \\
\\
\approx \pm 1.71651(c-0.56716d)\sqrt{\frac{c-0.053086d}{c-0.55928d}}.
\end{array}
\right.
\end{array}
\label{ch83}
\end{eqnarray}
\smallskip

By using the definitions (\ref{PF-charge}) and (\ref{PF-charge-inv}) for the
case $k=8$, one obtains (recall the first of definitions (\ref{Gamma-Fermat}%
))
\begin{equation}
\begin{array}{c}
\left( n_{1},n_{2},n_{3},n_{4}\right) =\left(
p^{0}-2q_{0},p^{0}+2q_{1},-p^{0}+2p^{1}+6q_{0}-2q_{1},-p^{0}+4q_{0}\right)
\\
\\
\Updownarrow \\
\\
\Gamma =\left( -2n_{1}-n_{4},\frac{1}{2}\left(
3n_{1}-n_{2}-n_{3}+3n_{4}\right) ,\frac{1}{2}(n_{1}+n_{4}),\frac{1}{2}%
(-2n_{1}+n_{2}-n_{4})\right) .
\end{array}
\label{rel-sympl-PF-ch-k=8}
\end{equation}
\ \setcounter{equation}0
\def\theequation{9.4.\arabic{subsubsection}.\arabic{equation}}

\subsubsection{Analysis of non-BPS, $Z\neq 0$ supporting BH charge
configurations}

Let us analyze more in depth the obtained supporting BH charge
configurations.\medskip

\textbf{II.1}. As for $k=5$, this configuration is characterized by a
physically meaningful ``$\pm $'' split, affecting the charge-dependent
expression of the BH entropy. By using Eqs. (\ref{8N3}), (\ref
{k=8-SBH-non-BPS}) and (\ref{rel-sympl-PF-ch-k=8}), Eqs. (\ref{ch81}) yield
\begin{equation}
\begin{array}{l}
N_{3,II.1,\pm }=\frac{\sqrt{2}\sqrt{2-\sqrt{2}}}{\vartheta +1}\left[ \pm
\sqrt{\frac{3(\sqrt{2}-1)\vartheta +1}{(\sqrt{2}-1)\vartheta -1}}+i\right]
n_{1}\approx 0.2826\left( \pm 5.12109+i\right) n_{1}; \\
\\
S_{BH,non-BPS,Z\neq 0,II.1,\pm }\approx 0.5066\pi |N_{3,II.1,\pm
}|^{2}\approx \left\{
\begin{array}{l}
``+":1.10157\pi \left( p^{0}-2q_{0}\right) ^{2}. \\
\\
``-":-1.02065\pi \left( p^{0}-2q_{0}\right) ^{2}.
\end{array}
\right.
\end{array}
\label{SBH-II1-k=8}
\end{equation}
In order to give an example of the range of the symplectic
(electric/magnetic) charges in the considered BH charge configuration, by
recalling that PF BH charges are integers as are the symplectic BH charges
(see definitions (\ref{PF-charge})-(\ref{PF-charge-inv})), one can fulfill
the fine-tuning conditions (\ref{ch81}) by taking \textit{e.g.}
\begin{equation}
\begin{array}{c}
n_{1}=10 \\
\Downarrow \\
n_{2,\pm }\approx \left\{
\begin{array}{l}
``+":25 \\
``-":-13
\end{array}
\right. ,n_{3,\pm }\approx \left\{
\begin{array}{l}
``+":13 \\
``-":-25
\end{array}
\right. ,n_{4}=-10.
\end{array}
\end{equation}
Using Eq. (\ref{rel-sympl-PF-ch-k=8}), one finally gets
\begin{equation}
\Gamma _{\pm }\approx \left\{
\begin{array}{l}
``+":\left( -10,-19,0,7\right) \\
\\
``-":\left( -10,19,0,-11\right)
\end{array}
\right. .  \label{porgy2}
\end{equation}
\smallskip \medskip

\textbf{II.2}.\textbf{\ }As for $k=5$, this configuration is the only one
not having any kind of split, and having a purely real $N_{3}$. It is worth
remarking that Eq. (\ref{ch82}) is the analogue for the mirror \textit{octic}
$\mathcal{M}_{8}^{\prime }$ of the fine-tuning condition (\ref{fc52})
adopted for the mirror \textit{quintic} $\mathcal{M}_{5}^{\prime }$ in \cite
{TT} (in particular, see Sect. 4 and App. C of such a Ref.).\textbf{\ }By
using Eqs. (\ref{8N3}), (\ref{k=8-SBH-non-BPS}) and (\ref
{rel-sympl-PF-ch-k=8}), Eqs. (\ref{ch82}) yield
\begin{equation}
\begin{array}{l}
N_{3,II.2}=\frac{2\sqrt{2}\sqrt{2-\sqrt{2}}}{-1+(\sqrt{2}-1)\vartheta }%
n_{1}\approx 12.56955\left( p^{0}-2q_{0}\right) ; \\
\\
S_{BH,non-BPS,Z\neq 0,II.2}\approx 0.5066\pi |N_{3,II.2}|^{2}\approx
80.03965\pi \left( p^{0}-2q_{0}\right) ^{2}.
\end{array}
\label{SBH-II2-k=8}
\end{equation}
In order to give an example of the range of the symplectic
(electric/magnetic) charges in the considered BH charge configuration, by
recalling that PF BH charges are integers as are the symplectic BH charges
(see definitions (\ref{PF-charge})-(\ref{PF-charge-inv})), one can fulfill
the fine-tuning conditions (\ref{ch82}) by taking \textit{e.g.}
\begin{equation}
\begin{array}{c}
n_{1}=10 \\
\Downarrow \\
n_{2}\approx 188,~~n_{3}\approx 188,~~n_{4}=10.
\end{array}
\end{equation}
Using Eq. (\ref{rel-sympl-PF-ch-k=8}), one finally gets
\begin{equation}
\Gamma =\left( -30,-158,10,79\right) .
\end{equation}
\medskip

\textbf{II.3}. As for $k=5$, this case, beside being split in ``$\pm $'' (as
the case II.1), is also parameterized by the couple $\left( c,d\right) \in
\mathbb{Z}^{2}$ ($\mathbb{R}^{2}$ in the -semi- classical limit of large
charges), further constrained by the request $a_{\pm }\left( b_{\pm
},c,d\right) $, $b_{\pm }\left( c,d\right) \in \mathbb{Z}$ ($\mathbb{R}$ for
large charges), implying that
\begin{equation}
c\neq 0.24039d,\text{ ~and~~}\left\{
\begin{array}{l}
\left\{
\begin{array}{l}
c\geqslant 0.053086d, \\
\\
c>0.55928d;
\end{array}
\right.  \\
\multicolumn{1}{c}{\text{or}} \\
\left\{
\begin{array}{l}
c\leqslant 0.053086d, \\
\\
c<0.55928d.
\end{array}
\right.
\end{array}
\right. \text{ ~}\text{\ \ ~}
\end{equation}
By using Eq. (\ref{8N3}), Eqs. (\ref{ch83}) yield
\begin{eqnarray}
&&
\begin{array}{l}
N_{3,II.3,\pm }\left( c,d\right) =\frac{\sqrt{2+\sqrt{2}}}{2}\left\{ \left[
(1-\sqrt{2})a_{\pm }\left( c,d\right) +b_{\pm }\left( c,d\right) \right] -i%
\left[ c+(1-\sqrt{2})d\right] \right\} = \\
\\
=\frac{\sqrt{2+\sqrt{2}}}{2}\left\{ \frac{\sqrt{2}c-\sqrt{2}\left( \sqrt{2}%
-1\right) d}{c-d+\left[ c-(\sqrt{2}-1)d\right] \vartheta }b_{\pm }\left(
c,d\right) -i\left[ c+(1-\sqrt{2})d\right] \right\} = \\
\\
=\frac{\sqrt{2+\sqrt{2}}}{2}\left\{ \pm \sqrt{\frac{1+\left( \sqrt{2}%
+1\right) \vartheta }{-1+3\left( \sqrt{2}+1\right) \vartheta }}\sqrt{\frac{%
c-\left( \sqrt{2}+1\right) \frac{-1+\left( \sqrt{2}-1\right) \vartheta }{%
1+\left( \sqrt{2}+1\right) \vartheta }d}{c-\left( \sqrt{2}+1\right) \frac{%
-1-3\left( \sqrt{2}-1\right) \vartheta }{1-3\left( \sqrt{2}+1\right)
\vartheta }d}}-i\right\} \left[ c+\left( 1-\sqrt{2}\right) d\right] \approx
\\
\\
\approx 0.92388\left\{ \pm 0.63381\sqrt{\frac{c-0.053086d}{c-0.55928d}}%
-i\right\} \left( c-0.41421d\right) ;
\end{array}
\notag \\
&&
\end{eqnarray}
\begin{eqnarray}
&&
\begin{array}{l}
\left| N_{3,II.3,\pm }\left( c,d\right) \right| ^{2}= \\
\\
=\frac{2+\sqrt{2}}{4}\left\{ \frac{\left[ 1+\left( \sqrt{2}+1\right)
\vartheta \right] c+\left( \sqrt{2}+1\right) \left[ 1-\left( \sqrt{2}%
-1\right) \vartheta \right] d}{\left[ -1+3\left( \sqrt{2}+1\right) \vartheta
\right] c+\left( \sqrt{2}+1\right) \left[ -1-3\left( \sqrt{2}-1\right)
\vartheta \right] d}+1\right\} \left[ c+\left( 1-\sqrt{2}\right) d\right]
^{2}\approx  \\
\\
\approx 1.19644\frac{\left( c-0.41421d\right) ^{3}}{c-0.55928d}.
\end{array}
\notag \\
&&
\end{eqnarray}
By recalling Eqs. (\ref{k=8-SBH-non-BPS}) and (\ref{rel-sympl-PF-ch-k=8}),
one finally gets (notice that in this case the ``$\pm $'' split is
unphysical, because it does not affect the charge-dependent expression of
the BH entropy) :
\begin{equation}
\begin{array}{l}
S_{BH,non-BPS,Z\neq 0,II.3,\pm }\left( c,d\right) \approx 0.5066\pi \left|
N_{2,II.3,\pm }(c,d)\right| ^{2}\approx 0.60611\pi \frac{\left(
c-0.41421d\right) ^{3}}{c-0.55928d}\approx  \\
\\
\approx 0.60611\pi \frac{\left( 2p^{0}-2p^{1}-3.51474q_{0}+4q_{1}\right) ^{3}%
}{0.88144p^{0}-2p^{1}-2.64432q_{0}+4q_{1}}= \\
\\
=S_{BH,non-BPS,Z\neq 0,II.3,\pm }\left( p^{0},p^{1},q_{0},q_{1}\right) .
\end{array}
\label{SBH-II3-k=8}
\end{equation}

Thus, by comparing Eqs. (\ref{SBH-II1-k=8}), (\ref{SBH-II2-k=8}) and (\ref
{SBH-II3-k=8}), also for $k=8$ one can conclude that \textit{all} three
distinct sets of fine-tuning conditions for PF BH charges (\ref{ch81})-(\ref
{ch83}) do support a non-BPS, $Z\neq 0$ LG attractor in a strict sense, but
they yield different charge-dependent expressions for the classical BH
entropy $S_{BH,non-BPS,Z\neq 0}$. As non-BPS, $Z\neq 0$ attractor, the LG
point of the moduli space of the Fermat mirror \textit{octic} $\mathcal{M}%
_{8}^{\prime }$ turns out to originate a \textit{threefold splitting} of the
supporting BH charges, and correspondingly of the purely charge-dependent
classical (Bekenstein-Hawking) BH entropy. Once again, this confirms the
results of the recent investigation in \cite{Saraikin-Vafa-1}.\medskip\

Finally, by recalling the definition (\ref{SUSY-breaking}), one can compute
the supersymmetry-breaking order parameter for the non-BPS, $Z\neq 0$ LG
attractor in the mirror \textit{octic} $\mathcal{M}_{8}^{\prime }$; by using
Eqs. (\ref{k=8-metric}), (\ref{W-k=8}), (\ref{covder8}) and (\ref{ft8}), one
gets
\begin{eqnarray}
\mathcal{O}_{non-BPS,Z\neq 0} &\equiv &\left[ \frac{g^{-1}|DW|^{2}}{|W|^{2}}%
\right] _{non-BPS,Z\neq 0}=\left( g^{-1}|Dln\widetilde{W}|^{2}\right)
_{non-BPS,Z\neq 0}\approx \frac{3+2\sqrt{2}}{\vartheta ^{2}}\approx 0.72.
\notag \\
&&
\end{eqnarray}

\section{$k=10$ : Mirror \textit{Dectic} near LG Point\label{tentic}}

\setcounter{equation}0
\def\theequation{10.\arabic{subsection}.\arabic{equation}}

\subsection{Geometric Setup}

For the mirror \textit{dectic} $\mathcal{M}_{10}^{\prime }$ the computations
(\textit{but not the results!}) go the same way as for the mirror \textit{%
octic} $\mathcal{M}_{8}^{\prime }$.

For $k=10$ the definitions (\ref{omega0-2}) yield
\begin{equation}
C_{10,2l-1}=0=C_{10,5l-1},\quad l\in \mathbb{N;}
\end{equation}
moreover, since $F_{10,m+10,\,n}=F_{10,m\,,n+10}=-F_{10,m\,n}$ (see
the third of properties (\ref{prop-F})), the only independent
elements of the rank-2 tensor $F_{10}$ belong to the $10\times 10$
matrix \newpage
\begin{equation}
\begin{array}{l}
F_{10,m%
\,n}=~~~~~~~~~~~~~~~~~~~~~~~~~~~~~~~~~~~~~~~~~~~~~~~~~~~~~~~~~~~~~~~~~~~~~~~~~~~~~~~~~~~~~~~~~~~~~~~~~~~~~~~~~~~~~~~~~~~~~~~~~~~~~~~~~~~~~~~~~
\end{array}
\label{k=10-Fmn}
\end{equation}
\begin{eqnarray}
\begin{array}{l}
\\
{\tiny {\left(
\begin{array}{cccccccccc}
\sqrt{5(5+2\sqrt{5})} & -3-\sqrt{5} & 0 & 3+\sqrt{5} & -\sqrt{5(5+2\sqrt{5})}
& 2+\sqrt{5} & 0 & -2 & 0 & 2+\sqrt{5} \\
{\rule{0pt}{12pt}}-3-\sqrt{5} & \sqrt{5+2\sqrt{5}} & 3-\sqrt{5} & -\sqrt{2(5+%
\sqrt{5})} & 2+\sqrt{5} & -\sqrt{10-2\sqrt{5}} & -2+\sqrt{5} & 0 & 2 & -%
\sqrt{25-2\sqrt{5}} \\
{\rule{0pt}{12pt}}0 & 3-\sqrt{5} & -\sqrt{5(5-2\sqrt{5})} & 2 & -\sqrt{5(5-2%
\sqrt{5})} & 3-\sqrt{5} & 0 & 2-\sqrt{5} & 0 & -2+\sqrt{5} \\
{\rule{0pt}{12pt}}3+\sqrt{5} & -\sqrt{2(5+\sqrt{5})} & 2 & -\sqrt{5-2\sqrt{5}%
} & -2+\sqrt{5} & 0 & -3+\sqrt{5} & \sqrt{10-2\sqrt{5}} & -2-\sqrt{5} &
\sqrt{25+2\sqrt{5}} \\
{\rule{0pt}{12pt}}-\sqrt{5(5+2\sqrt{5})} & 2+\sqrt{5} & -\sqrt{5(5-2\sqrt{5})%
} & -2+\sqrt{5} & 0 & 2-\sqrt{5} & \sqrt{5(5-2\sqrt{5})} & -2-\sqrt{5} &
\sqrt{5(5+2\sqrt{5})} & 8 \\
{\rule{0pt}{12pt}}2+\sqrt{5} & -\sqrt{10-2\sqrt{5}} & 3-\sqrt{5} & 0 & 2-%
\sqrt{5} & \sqrt{5-2\sqrt{5}} & -2 & \sqrt{2(5+\sqrt{5})} & -3-\sqrt{5} &
\sqrt{25+2\sqrt{5}} \\
{\rule{0pt}{12pt}}0 & -2+\sqrt{5} & 0 & -3+\sqrt{5} & \sqrt{5(5-2\sqrt{5})}
& -2 & \sqrt{5(5-2\sqrt{5})} & -3+\sqrt{5} & 0 & -2+\sqrt{5} \\
{\rule{0pt}{12pt}}-2 & 0 & 2-\sqrt{5} & \sqrt{10-2\sqrt{5}} & -2-\sqrt{5} &
\sqrt{2(5+\sqrt{5})} & -3+\sqrt{5} & -\sqrt{5+2\sqrt{5}} & 3+\sqrt{5} & -%
\sqrt{25-2\sqrt{5}} \\
{\rule{0pt}{12pt}}0 & 2 & 0 & 2-\sqrt{5} & \sqrt{5(5+2\sqrt{5})} & -3-\sqrt{5%
} & 0 & 3+\sqrt{5} & -\sqrt{5(5+2\sqrt{5})} & 2+\sqrt{5} \\
{\rule{0pt}{12pt}}2+\sqrt{5} & -\sqrt{25-2\sqrt{5}} & -2+\sqrt{5} & \sqrt{%
25+2\sqrt{5}} & 8 & \sqrt{25+2\sqrt{5}} & -2+\sqrt{5} & -\sqrt{25-2\sqrt{5}}
& 2+\sqrt{5} & 0
\end{array}
\right) }}
\end{array}
&&  \notag \\
&&  \notag
\end{eqnarray}
Let us now write down all the relevant quantities up to the needed order
(here and below, unless otherwise specified, we omit the Fermat parameter $%
k=10$):
\begin{eqnarray}
&&
\begin{array}{l}
\widetilde{K}\approx (5-\sqrt{2})\frac{C_{2}^{2}}{C_{0}^{2}}(\psi \bar{\psi}%
)^{2}\left[ 1+\frac{(5-\sqrt{2})}{2}\frac{C_{2}^{2}}{C_{0}^{2}}(\psi \bar{%
\psi})^{2}\right] +\mathcal{O}(\psi ^{9}); \\
~
\end{array}
\\
&&
\begin{array}{l}
g\approx 4(5-\sqrt{2})\frac{C_{2}^{2}}{C_{0}^{2}}\psi \bar{\psi}\left[ 1+2(5-%
\sqrt{2})\frac{C_{2}^{2}}{C_{0}^{2}}(\psi \bar{\psi})^{2}\right] +\mathcal{O}%
(\psi ^{7}); \\
~
\end{array}
\label{k=10-metric} \\
&&
\begin{array}{l}
\widetilde{W}\approx N_{1}+\frac{C_{2}}{C_{0}}N_{3}\psi ^{2}-\frac{C_{6}}{%
C_{0}}\bar{N}_{3}\psi ^{6}+\mathcal{O}(\psi ^{8}). \\
~
\end{array}
\end{eqnarray}
Now, by using the formul\ae\ of the general analysis exploited in Sect. \ref
{GA}, we can get the ``effective BH potential'' and the holomorphic
superpotential, as well as their (covariant) derivatives, up to $\mathcal{O}%
\left( \psi ^{2}\right) $:
\begin{eqnarray}
&&
\begin{array}{l}
\widetilde{W}=N_{1}+\frac{C_{2}}{C_{0}}N_{3}\psi ^{2}; \\
~
\end{array}
\label{W-k=10} \\
&&
\begin{array}{l}
D\widetilde{W}=2\frac{C_{2}}{C_{0}}N_{3}\psi +2(\sqrt{5}-2)\frac{C_{2}^{2}}{%
C_{0}^{2}}N_{1}\psi \bar{\psi}^{2}; \\
~
\end{array}
\label{covder10} \\
&&
\begin{array}{l}
D^{2}\widetilde{W}=-24\frac{C_{6}}{C_{0}}\bar{N}_{3}\psi ^{4}; \\
~
\end{array}
\\
&&
\begin{array}{l}
\widetilde{V}_{BH}=|N_{1}|^{2}+(\sqrt{5}+2)|N_{3}|^{2}+2\frac{C_{2}}{C_{0}}%
N_{3}\bar{N}_{1}\psi ^{2}+2\frac{C_{2}}{C_{0}}\bar{N}_{3}N_{1}\bar{\psi}^{2};
\\
~
\end{array}
\label{pot10}
\end{eqnarray}
\begin{eqnarray}
&&
\begin{array}{l}
\partial \widetilde{V}_{BH}=4\psi \left( \frac{C_{2}}{C_{0}}N_{3}\bar{N}%
_{1}-3(\sqrt{5}+2)\frac{C_{6}}{C_{2}}(\bar{N}_{3})^{2}\psi ^{2}+\frac{%
C_{2}^{2}}{C_{0}^{2}}\left( |N_{1}|^{2}(\sqrt{5}-2)+|N_{3}|^{2}\right) \bar{%
\psi}^{2}\right) ; \\
~
\end{array}
\\
&&
\begin{array}{l}
\partial ^{2}\widetilde{V}_{BH}=4\left( \frac{C_{2}}{C_{0}}N_{3}\bar{N}%
_{1}-9(\sqrt{5}+2)\frac{C_{6}}{C_{2}}(\bar{N}_{3})^{2}\psi ^{2}+\frac{%
C_{2}^{2}}{C_{0}^{2}}\left( |N_{1}|^{2}(\sqrt{5}-2)+|N_{3}|^{2}\right) \bar{%
\psi}^{2}\right) ; \\
~
\end{array}
\label{ddV-k=10} \\
&&
\begin{array}{l}
\partial \overline{\partial }\widetilde{V}_{BH}=8\frac{C_{2}^{2}}{C_{0}^{2}}%
\left( |N_{1}|^{2}(\sqrt{5}-2)+|N_{3}|^{2}\right) \psi \bar{\psi}. \\
~
\end{array}
\label{ddV-k=10-2}
\end{eqnarray}

Contrary to the cases of mirror \textit{quintic} and \textit{sextic} (see
Sects. \ref{quintic} and \ref{sixtic}, respectively), and similarly to the
case of mirror \textit{octic} (see Sect. \ref{eightic}), for the case of
mirror \textit{dectic} it is evident from Eqs. (\ref{pot10})-(\ref
{ddV-k=10-2}) that we truncate the series expansion of the ``effective BH
potential'' and of its second derivatives around the LG point up to $%
\mathcal{O}\left( \psi ^{2}\right) $ included, and the series expansion of
its first derivative around the LG point up to $\mathcal{O}\left( \psi
^{3}\right) $ included. This is due to the absence of an $\mathcal{O}\left(
\psi \right) $ term in expression of $\widetilde{V}_{BH}$ given by Eq. (\ref
{pot10}). As mentioned in Sect. \ref{eightic}, such a fact can be traced
back to the non-regularity of $g$ at $\psi =0$ (see Eq. (\ref{k=10-metric}%
)). \setcounter{equation}0
\def\theequation{10.1.\arabic{subsubsection}.\arabic{equation}}

\subsubsection{\label{prep-10}The Holomorphic Prepotentials $F$ and $%
\mathcal{F}$}

As done for $k=5,6,8$ respectively in Subsubsects. \ref{prep-5}, \ref{prep-6}
and \ref{prep-8}, in order to compute the holomorphic prepotential $F\left(
X\left( \psi \right) \right) $ and its ``K\"{a}hler-invariant counterpart'' $%
\mathcal{F}\left( \psi \right) $ defined by Eq. (\ref{F-call}), we have to
recall the general formul\ae\ for (mirror) Fermat $CY_{3}$s introduced in
Sect. \ref{GA}, and specialize them for $k=10$. By recalling the definitions
(\ref{csi-csi}), (\ref{PHIZ}), (\ref{hol-prep-F-l-CY}), (\ref{F-call-CY}),
one can compute\footnote{%
As implied by Eqs. (\ref{PHIZ-5}), (\ref{PHIZ-6}), (\ref{PHIZ-8}) and (\ref
{PHIZ-10}), it holds that
\begin{equation*}
Re\left( \Phi _{k,m,n}\right) =Re\left( \Phi _{k,(m,n)}\right) ,~\forall
k=5,6,8,10,
\end{equation*}
while the same relation for the imaginary part holds only for $k=5$. From
the definitions (\ref{csi-csi}) and (\ref{PHIZ}), this fact can be traced
back to value of $k$ itself and to the explicit form of the matrices $M_{k}$%
s, given by Eqs. (\ref{M1}) and (\ref{M2}).}:
\begin{eqnarray}
&&
\begin{array}{l}
\xi _{10,m}=\left(
\begin{array}{l}
2cos\left( \frac{\pi m}{10}\right) +cos\left( \frac{3\pi m}{10}\right)
-isin\left( \frac{3\pi m}{10}\right) \\
\\
-cos\left( \frac{\pi m}{10}\right) +isin\left( \frac{\pi m}{10}\right) \\
\\
2isin\left( \frac{3\pi m}{10}\right) \\
\\
2isin\left( \frac{\pi m}{10}\right)
\end{array}
\right) ;
\end{array}
\label{csi-csi-10} \\
&&  \notag \\
&&
\begin{array}{l}
\Phi _{10,m,n}= \\
\\
\begin{array}{l}
=\left[
\begin{array}{l}
cos\left( \frac{3\pi \left( n-m\right) }{10}\right) -cos\left( \frac{\pi
\left( n-m\right) }{10}\right) +cos\left( \frac{\pi \left( n+m\right) }{10}%
\right) -cos\left( \frac{3\pi \left( n+m\right) }{10}\right) + \\
\\
+5cos\left( \frac{\pi \left( n+m\right) }{8}\right) +3cos\left( \frac{3\pi
\left( n+m\right) }{8}\right) +3cos\left( \frac{\pi \left( 3n+m\right) }{8}%
\right) +3cos\left( \frac{\pi \left( n+3m\right) }{8}\right)
\end{array}
\right] + \\
\\
+i\left[
\begin{array}{l}
sin\left( \frac{3\pi \left( n-m\right) }{10}\right) -sin\left( \frac{\pi
\left( n-m\right) }{10}\right) +2sin\left( \frac{\pi \left( 3n-m\right) }{10}%
\right) -sin\left( \frac{\pi \left( n+m\right) }{10}\right) + \\
\\
+2sin\left( \frac{\pi \left( 3n+m\right) }{10}\right) +sin\left( \frac{3\pi
\left( n+m\right) }{10}\right)
\end{array}
\right] \neq \Phi _{10,\left( m,n\right) };
\end{array}
\end{array}
\notag \\
&&  \label{PHIZ-10}
\end{eqnarray}

\begin{eqnarray}
&&
\begin{array}{l}
F_{k=10,l=9}\left( X_{10}\left( \psi \right) \right) =-\frac{2\pi ^{6}}{5^{4}%
}\sum_{m=1}^{9}\sum_{n=1}^{9}(-1)^{m+n}C_{10,m-1}C_{10,n-1}\psi ^{m+n-2}\Phi
_{10,\left( m,n\right) }= \\
\\
\\
=-\frac{2\pi ^{6}}{5^{4}}\left[
\begin{array}{l}
\frac{\left( 3+\sqrt{5}\right) \left( 2\sqrt{10}+i\left( 5+\sqrt{5}\right)
^{3/2}\right) \pi }{10\Gamma ^{2}\left( \frac{4}{5}\right) \Gamma ^{6}\left(
\frac{9}{10}\right) 2^{9/10}}+ \\
\\
-\frac{5\left( 5-3\sqrt{5}\right) \left( 1+i\sqrt{5-2\sqrt{5}}\right) \pi }{%
\Gamma ^{2}\left( \frac{2}{5}\right) \Gamma ^{6}\left( \frac{7}{10}\right)
2^{6/5}}\psi ^{4}+ \\
\\
+\frac{25\sqrt{5\left( 25+2\sqrt{5}\right) }}{96\pi ^{3}}\psi ^{8}
\end{array}
\right] +\mathcal{O}\left( \psi ^{9}\right) =F_{k=10,l=9}\left( \psi
^{4}\right) ;
\end{array}
\notag \\
&&  \label{F109-CY} \\
&&
\begin{array}{l}
\mathcal{F}_{10}\left( \psi \right) \equiv F_{10}\left( \frac{%
X_{10}^{1}(\psi )}{X_{10}^{0}(\psi )}\right) =\frac{F_{10}\left(
X_{10}^{0}(\psi ),X_{10}^{1}(\psi )\right) }{\left( X_{10}^{0}(\psi )\right)
^{2}}=\frac{1}{2}\frac{\sum_{m,n=1}^{\infty
}(-1)^{m+n}C_{10,m-1}C_{10,n-1}\Phi _{10,\left( m,n\right) }\psi ^{m+n-2}}{%
\left[ \sum_{r=1}^{\infty }(-1)^{r}C_{10,r-1}\xi _{10,r}^{1}\psi ^{r-1}%
\right] ^{2}}\approx \\
\\
\approx \frac{\Phi _{10,1,1}}{2\left( \xi _{10,1}^{1}\right) ^{2}}\left[ 1+%
\frac{C_{10,2}\left( \Phi _{10,1,3}+\Phi _{10,3,1}\right) }{C_{10,0}\Phi
_{10,1,1}}\psi ^{2}+\mathcal{O}\left( \psi ^{3}\right) \right] \left[ 1-2%
\frac{C_{10,2}\xi _{10,3}^{1}}{C_{10,0}\xi _{10,1}^{1}}\psi ^{2}+\mathcal{O}%
\left( \psi ^{3}\right) \right] \approx \\
\\
\approx \frac{\Phi _{10,1,1}}{2\left( \xi _{10,1}^{1}\right) ^{2}}\left[
1-2\left( \frac{\xi _{10,3}^{1}}{\xi _{10,1}^{1}}-\frac{\Phi _{10,1,3}+\Phi
_{10,3,1}}{2\Phi _{10,1,1}}\right) \frac{C_{10,2}}{C_{10,0}}\psi ^{2}\right]
+\mathcal{O}\left( \psi ^{3}\right) ,
\end{array}
\notag \\
&&  \label{F-call-CY-10}
\end{eqnarray}
where in the formula (\ref{F-call-CY-10}) for the K\"{a}hler-invariant
holomorphic prepotential $\mathcal{F}_{10}$ only the leading terms in the
\textit{LG limit }$\psi \longrightarrow 0$ have been retained. Eq. (\ref
{F109-CY}) is the exact expression (complete up to $\mathcal{O}\left( \psi
^{8}\right) $ included) of the holomorphic prepotential\footnote{%
It is worth noticing that $F_{8}$ and $F_{10}$ (respectively given by Eqs. (%
\ref{F87-CY}) and (\ref{F109-CY})) are in turn actually functions of $\psi
^{2}$ (as explicitly checked up to $\mathcal{O}\left( \psi ^{10}\right) $
included) and $\psi ^{4}$ (as explicitly checked up to $\mathcal{O}\left(
\psi ^{16}\right) $ included). This fact distinguishes the cases $k=8,10$
from the cases $k=5,6$ (respectively given by Eqs. (\ref{F55-CY}) and (\ref
{F65-CY})), and it can be traced back to the explicit form of the matrices $%
M_{k}$s (given by Eqs. (\ref{M1}) and (\ref{M2})) and of the constants $%
C_{k,m-1}$ (defined in the first of Eqs. (\ref{omega0-2})).} $F_{10}$ in a
suitable neighbourhood of the LG point of the moduli space of the (mirror)%
\textit{\ dectic} $\mathcal{M}_{10}^{\prime }$.
\setcounter{equation}0
\def\theequation{10.\arabic{subsection}.\arabic{equation}}

\subsection{\textit{``Criticality Condition''} Approach and Attractor
Equation}

Let us now find the solutions of the AE $\partial \widetilde{V}_{BH}\left(
\psi ,\overline{\psi };q,p\right) =0$, and check their stability. Since we
are working near the LG point, by using Eq. (\ref{pot10}) we can rewrite the
AE for $\mathcal{M}_{10}^{\prime }$ as follows:
\begin{equation}
\frac{C_{2}}{C_{0}}N_{3}\bar{N}_{1}+\frac{C_{2}^{2}}{C_{0}^{2}}\left(
|N_{1}|^{2}(\sqrt{5}-2)+|N_{3}|^{2}\right) \bar{\psi}^{2}\approx 3(\sqrt{5}%
+2)\frac{C_{6}}{C_{2}}(\bar{N}_{3})^{2}\psi ^{2}  \label{attreq10}
\end{equation}
Solving Eq. (\ref{attreq8}), we will find one (or more) set(s) of BH charges
supporting $\psi \approx 0$ to be a critical point of $V_{BH}$. Since we are
working near the LG point, it is clear that the first term in the l.h.s. of
Eq. (\ref{attreq10}) must be small enough. This implies the following
fine-tuning condition:
\begin{equation}
N_{3}\bar{N}_{1}\approx 0.  \label{atr10}
\end{equation}
\setcounter{equation}0
\def\theequation{10.\arabic{subsection}.\arabic{equation}}

\subsection{\label{21march-7}Critical Hessian of $V_{BH}$}

By using (the $n_{V}=1$ case of) Eq. (\ref{4jan2}) and Eqs. (\ref{ddV-k=10}%
)-(\ref{ddV-k=10-2}) evaluated along the criticality condition (\ref
{attreq10})-(\ref{atr10}), it can be computed that the components of $H_{%
\text{real form}}^{\widetilde{V}_{BH}}$ (given by Eq. (\ref
{Hessian-real-form})) constrained by the Eqs. (\ref{attreq10})-(\ref{atr10})
read as follows:
\begin{eqnarray}
&&\mathcal{A}=-12(\sqrt{5}+2)\frac{C_{6}}{C_{2}}\left( (\bar{N}_{3})^{2}\psi
^{2}+(N_{3})^{2}\bar{\psi}^{2}\right) +8\frac{C_{2}^{2}}{C_{0}^{2}}\psi \bar{%
\psi}\left( |N_{1}|^{2}(\sqrt{5}-2)+|N_{3}|^{2}\right) ; \\
&&  \notag \\
&&\mathcal{B}=12(\sqrt{5}+2)\frac{C_{6}}{C_{2}}\left( (\bar{N}_{3})^{2}\psi
^{2}+(N_{3})^{2}\bar{\psi}^{2}\right) +8\frac{C_{2}^{2}}{C_{0}^{2}}\psi \bar{%
\psi}\left( |N_{1}|^{2}(\sqrt{5}-2)+|N_{3}|^{2}\right) ; \\
&&  \notag \\
&&\mathcal{C}=-12i(\sqrt{5}+2)\frac{C_{6}}{C_{2}}\left( (N_{3})^{2}\bar{\psi}%
^{2}-(\bar{N}_{3})^{2}\psi ^{2}\right) .
\end{eqnarray}
The resulting real eigenvalues of $H_{\text{real form}}^{\widetilde{V}_{BH}}$
constrained by Eqs. (\ref{attreq10})-(\ref{atr10}) read:
\begin{equation}
\lambda _{\pm }\approx 8\psi \bar{\psi}\left[ \frac{C_{2}^{2}}{C_{0}^{2}}%
\left( |N_{1}|^{2}(\sqrt{5}-2)+|N_{3}|^{2}\right) \pm 3(\sqrt{5}+2)\frac{%
C_{6}}{C_{2}}|N_{3}|^{2}\right] .  \label{k=10-gen-eigenvalues}
\end{equation}

By recalling Eq. (\ref{pot}) and using Eq. (\ref{pot10}) with $\psi \approx
0 $ and constrained by by Eqs. (\ref{attreq10})-(\ref{atr10}), one obtains
that the purely charge-dependent LG critical values of the ``effective BH
potential'' for the mirror \textit{dectic} $\mathcal{M}_{10}^{\prime }$ are
\begin{equation}
V_{BH,LG-critical,k=10}\approx \frac{1}{\sqrt{5(5+2\sqrt{5})}}\left[
|N_{1}|^{2}+(\sqrt{5}+2)|N_{3}|^{2}\right] ;  \label{k=10-VBH-crit}
\end{equation}
by recalling formula (\ref{BHEA}), this directly yields the following purely
charge-dependent values of the BH entropy at the LG critical points of $%
V_{BH,10}$ in the moduli space of $\mathcal{M}_{10}^{\prime }$:
\begin{equation}
S_{BH,LG-critical,k=10}\approx \frac{\pi }{\sqrt{5(5+2\sqrt{5})}}\left[
|N_{1}|^{2}+(\sqrt{5}+2)|N_{3}|^{2}\right] .  \label{k=10-SBH}
\end{equation}
From the definition (\ref{N-def}), for $k=10$ and $m=1,3$ one respectively
gets that
\begin{eqnarray}
N_{1} &=&-\frac{1}{2}\sqrt{\frac{5-\sqrt{5}}{2}}\left( n_{1}+n_{4}+\frac{(1+%
\sqrt{5})}{2}(n_{2}+n_{3})\right) +i\frac{\sqrt{5}+1}{4}\left( n_{4}-n_{1}+%
\frac{3-\sqrt{5}}{2}(n_{3}-n_{2})\right) ;  \label{10N1} \\
&&  \notag \\
N_{3} &=&\frac{1}{2}\sqrt{\frac{(5-\sqrt{5})}{2}}\left( n_{1}+n_{4}-\frac{%
\sqrt{5}-1}{2}(n_{2}+n_{3})\right) +i\frac{\sqrt{5}-1}{4}\left( n_{4}-n_{1}+%
\frac{3+\sqrt{5}}{2}(n_{3}-n_{2})\right) .  \label{10N3}
\end{eqnarray}

Let us write down here the numerical values of constants relevant to our
treatment:
\begin{equation}
C_{0}\approx 1.57,\quad C_{2}\approx -0.66,\quad C_{6}\approx 0.077;\quad 3(%
\sqrt{5}+2)\frac{C_{6}}{C_{2}}\approx -1.48.  \label{con10}
\end{equation}
\medskip \setcounter{equation}0
\def\theequation{10.\arabic{subsection}.\arabic{equation}}

\subsection{\label{21march-8}Solutions to Attractor Equations}

Let us now analyze more in depth the species of LG attractor points arising
from the AE (\ref{atr10}). As it can be easily seen, the AE (\ref{atr10})
has two \textit{non-degenerate} solutions:\medskip

\textbf{I.} The first non-degenerate solution to AE (\ref{atr10}) is
\begin{equation}
N_{3}\approx 0.  \label{k=10-1/2BPS}
\end{equation}
This is nothing but the $k=8$ solution (\ref{k=8-1/2BPS}). As one can see
from Eq. (\ref{W-k=10})-(\ref{covder10}), also for $k=10$ such a solution
corresponds to a $\frac{1}{2}$-BPS LG critical point of $V_{BH}$ ($%
\widetilde{W}\neq 0$, $D\widetilde{W}=0$). From Eq. (\ref{10N3}), in order
to get the solution (\ref{k=10-1/2BPS}), one has to fine-tune two PF BH
charges out of four in the following way:
\begin{equation}
n_{3}\approx \frac{1}{2}(\sqrt{5}-1)(n_{1}+n_{2}),\quad n_{4}\approx -\frac{1%
}{2}(\sqrt{5}-1)n_{1}+n_{2}.  \label{k=10-chcon1/2}
\end{equation}
The charges $n_{1}$, $n_{2}$ are not fixed; they only satisfy the
non-degeneration condition $N_{1}\neq 0$. As it was for $k=5,6,8$, also the
real eigenvalues (\ref{k=10-gen-eigenvalues}) for the $\frac{1}{2}$-BPS
critical solution coincide and, as it is well known \cite{FGK,BFM,AoB}, are
strictly positive:
\begin{equation}
\lambda _{+,\frac{1}{2}-BPS}=\lambda _{-,\frac{1}{2}-BPS}\approx 8(\sqrt{5}%
-2)\psi \bar{\psi}\frac{C_{2}^{2}}{C_{0}^{2}}|N_{1}|_{N_{3}\approx 0}^{2}>0.
\end{equation}
Consequently, the $\frac{1}{2}$-BPS LG critical point $\psi \approx 0$
supported by the PF BH charge configuration (\ref{k=10-chcon1/2}) is a
stable extremum, since it is a (local) minimum of $V_{BH}$, and it is
therefore an attractor in a strict sense. The classical (Bekenstein-Hawking)
BH entropy at such a (class of) $\frac{1}{2}$-BPS LG attractor(s) takes the
value
\begin{equation}
S_{BH,\frac{1}{2}-BPS}=\pi V_{BH,\frac{1}{2}-BPS}\approx 0.166\pi
|N_{1}|_{N_{3}\approx 0}^{2},
\end{equation}
where $\left. N_{1}\right| _{N_{3}\approx 0}$ is given by Eq. (\ref{10N1})
constrained by Eq. (\ref{k=10-chcon1/2}):
\begin{equation}
\left. N_{1}\right| _{N_{3}\approx 0}\approx -\frac{\sqrt{5}}{2}\left( \sqrt{%
5-2\sqrt{5}}n_{1}+\sqrt{\frac{5+\sqrt{5}}{2}}n_{2}+i\left( n_{1}-\frac{\sqrt{%
5}-1}{2}n_{2}\right) \right) .
\end{equation}
\medskip \medskip

\textbf{II}. The second non-degenerate solution to AE (\ref{atr10}) is
\begin{equation}
N_{1}\approx 0.  \label{k=10-non-BPS}
\end{equation}
Interestingly, this is nothing but the $k=6$ solution (\ref{k=6-non-BPS}).
As one can see from Eq. (\ref{W-k=10})-(\ref{covder10}), also for $k=10$
such a solution corresponds to a non-BPS, $Z=0$ LG critical point of $V_{BH}$
($\widetilde{W}=0$, $D\widetilde{W}\neq 0$). The real eigenvalues (\ref
{k=10-gen-eigenvalues}) for such a non-BPS, $Z=0$ critical solution read
\begin{equation}
\lambda _{\pm ,non-BPS,Z=0}\approx 8\psi \bar{\psi}\left[ \frac{C_{2}^{2}}{%
C_{0}^{2}}\pm 3(\sqrt{5}+2)\frac{C_{6}}{C_{2}}\right] |N_{3}|_{N_{1}\approx
0}^{2}.  \label{k=10-non-BPS-eigenvalues}
\end{equation}
Substituting the numerical values (\ref{con10}) of the involved constants in
Eq. (\ref{k=10-non-BPS-eigenvalues}), one reaches the conclusion that one
eigenvalue is positive and the other one is negative:
\begin{equation}
\lambda _{\pm ,non-BPS,Z=0}\approx 8\psi \bar{\psi}\left[ 0.18\mp 1.48\right]
|N_{3}|_{N_{1}\approx 0}^{2}\lessgtr 0.
\end{equation}
Let us now find the fine-tuning conditions for PF BH charges supporting the
considered non-BPS, $Z=0$ LG attractor for the mirror \textit{dectic} $%
\mathcal{M}_{10}^{\prime }$. This amounts to solving Eq. (\ref{k=10-non-BPS}%
) by recalling Eq. (\ref{10N1}). By doing so, one gets the following unique
set of constraining relations on PF BH charges:
\begin{equation}
n_{3}\approx -\frac{1}{2}(1+\sqrt{5})(n_{1}+n_{2}),\quad n_{4}\approx \frac{1%
}{2}(1+\sqrt{5})n_{1}+n_{2}.  \label{ft10}
\end{equation}
The charges $n_{1},n_{2}$ are not fixed; they only satisfy the
non-degeneration condition $N_{3}\neq 0$. Thus, the non-BPS, $Z=0$ LG
critical point $\psi \approx 0$ supported by the PF BH charge configuration (%
\ref{ft10}) is a saddle point of $V_{BH}$ and consequently it is \textit{not}
an attractor in a strict sense.

The classical (Bekenstein-Hawking) BH entropy at such a (class of) non-BPS, $%
Z=0$ LG saddle point(s) takes the value
\begin{equation}
S_{BH,non-BPS,Z=0}=\pi V_{BH,non-BPS,Z=0}\approx 0.7\pi
|N_{3}|_{N_{1}\approx 0}^{2},
\end{equation}
where $\left. N_{3}\right| _{N_{1}\approx 0}$ is given by Eq. (\ref{10N3})
constrained \ by Eq. (\ref{ft10}):
\begin{equation}
\left. N_{3}\right| _{N_{1}\approx 0}\approx \frac{\sqrt{5}}{2}\left( \sqrt{%
5+2\sqrt{5}}n_{1}+\sqrt{\frac{5-\sqrt{5}}{2}}n_{2}-i\left( n_{1}+\frac{\sqrt{%
5}+1}{2}n_{2}\right) \right) .
\end{equation}

Similarly to the treatment of Sects. \ref{quintic}-\ref{eightic}, one can
also switch to the symplectic (electric/magnetic) basis for BH charges by
using Eq. (\ref{PF-charge-inv}) for $k=10$, and thus re-express the
fine-tuning conditions (\ref{k=10-chcon1/2}) and (\ref{ft10}) in terms of
the (components of the) symplectic BH charge vector $\Gamma $ defined by Eq.
(\ref{Gamma-Fermat}).

\section{Special K\"{a}hler Geometrical Identities and Attractors\label%
{SKG-id-approach}}

\setcounter{equation}0
\def\theequation{11.\arabic{subsection}.\arabic{equation}}

\subsection{\label{SKG-id-approach-gen}$n_{V}=1$ Identities near the LG Point%
}

Let us consider the real part of the $n_{V}=1$ case of SK geometry
identities (\ref{SKG-identities1}) \cite{FBC,K1,K2,BFM,AoB,K3}; by taking
into account the change in the notation of the symplectic charge vectors
pointed out in the Footnote before Eq. (\ref{Gamma-Fermat}), one achieves:
\begin{equation}
\widetilde{\Gamma }^{T}=2e^{K}Im\left[ W\overline{\Pi }+g^{-1}\left( \bar{D}%
\overline{W}\right) D\Pi \right] ,  \label{SKG-Identities-1-modulus}
\end{equation}
where the $1\times 4$ BH symplectic charge vector $\widetilde{\Gamma }$ is
defined in Eq. (\ref{Gamma-Fermat}).

Next, let us switch to more convenient variables for the treatment of
1-modulus SK geometries endowing the moduli space of Fermat $CY_{3}$s. By
recalling the definition (\ref{PF-charge}) of the $1\times 4$ PF BH charge
vector $n$, we can rewrite Eq. (\ref{SKG-Identities-1-modulus}) as follows
(here and below, unless otherwise specified, we omit the classifying Fermat
parameter $k=5,6,8,10$):
\begin{equation}
n^{T}=\frac{2}{\chi F_{1\,1}}e^{\widetilde{K}}Im\left[ \widetilde{W}%
\overline{\widetilde{\Upsilon }}+g^{-1}\left( \bar{D}\overline{\widetilde{W}}%
\right) D\widetilde{\Upsilon }\right] ,  \label{new}
\end{equation}
where the notations introduced in Sect. (\ref{GA}) have been used.
Furthermore, we defined the $4\times 1$ holomorphic vector
\begin{equation}
\widetilde{\Upsilon }\left( \psi \right) \equiv \frac{1}{AC_{0}}M^{T}\Sigma
^{T}\Pi \left( \psi \right) =\frac{1}{AC_{0}}m^{T}\varpi \left( \psi \right)
,  \label{PHI}
\end{equation}
where Eqs. (\ref{pi}) and (\ref{mk}) have been used.

Let us investigate Eq.(\ref{new}) in a certain neighbourhood of the LG point
$\psi =0$. The treatment given in Sects. \ref{quintic}-\ref{tentic} yields
that, by its very definition (\ref{PHI}), $\widetilde{\Upsilon }$ has the
following series expansion near the LG point\footnote{%
Notice that, consistently with the approach to the truncation of series
expansions near the LG point performed in Sects. \ref{quintic}-\ref{tentic},
for $k=5,6$ we truncate up to $\mathcal{O}\left( \psi \right) $ included,
whereas for $k=8,10$ we truncate up to $\mathcal{O}\left( \psi ^{2}\right) $
included.}:
\begin{eqnarray}
&&
\begin{array}{l}
k=5:\left\{
\begin{array}{l}
\widetilde{\Upsilon }=\varphi _{1}+\frac{C_{1}}{C_{0}}\varphi _{2}\psi , \\
\\
D\widetilde{\Upsilon }=\frac{C_{1}}{C_{0}}\left[ \varphi _{2}+2\frac{C_{2}}{%
C_{1}}\overline{\varphi }_{2}\psi +\frac{C_{1}}{C_{0}}(\sqrt{5}-2)\varphi
_{1}\bar{\psi}\right] ;
\end{array}
\right. \\
~
\end{array}
\label{k=5-PHI} \\
&&
\begin{array}{l}
k=6:\left\{
\begin{array}{l}
\widetilde{\Upsilon }=\varphi _{1}+\frac{C_{1}}{C_{0}}\varphi _{2}\psi , \\
\\
D\widetilde{\Upsilon }=\frac{C_{1}}{C_{0}}\left[ \varphi _{2}+\frac{1}{3}%
\frac{C_{1}}{C_{0}}\varphi _{1}\bar{\psi}\right] ;
\end{array}
\right. \\
~
\end{array}
\\
&&
\begin{array}{l}
k=8:\left\{
\begin{array}{l}
\widetilde{\Upsilon }=\varphi _{1}+\frac{C_{2}}{C_{0}}\varphi _{3}\psi ^{2},
\\
\\
D\widetilde{\Upsilon }=2\frac{C_{2}}{C_{0}}\psi \left[ \varphi _{3}-2\frac{%
C_{4}}{C_{2}}\bar{\varphi}_{3}\psi ^{2}+\frac{C_{2}}{C_{0}}(3-2\sqrt{2}%
)\varphi _{1}\bar{\psi}^{2}\right] ;
\end{array}
\right. \\
~
\end{array}
\\
&&
\begin{array}{l}
k=10:\left\{
\begin{array}{l}
\widetilde{\Upsilon }=\varphi _{1}+\frac{C_{2}}{C_{0}}\varphi _{3}\psi ^{2},
\\
\\
D\widetilde{\Upsilon }=2\frac{C_{2}}{C_{0}}\psi \left[ \varphi _{3}+\frac{%
C_{2}}{C_{0}}(\sqrt{5}-2)\varphi _{1}\bar{\psi}^{2}\right] .
\end{array}
\right. \\
~
\end{array}
\label{k=10-PHI}
\end{eqnarray}
We defined the complex $4\times 1$ vector\footnote{%
By recalling the definitions (\ref{mk}) and (\ref{csi-csi}), it is worth
noticing that
\begin{equation*}
\varphi _{k,m}=M_{k}^{T}\Sigma \mathbf{\xi }_{k,m}.
\end{equation*}
}
\begin{equation}
\varphi _{k,m}\equiv m_{k}\left(
\begin{array}{c}
\beta _{k}^{3m} \\
~ \\
\beta _{k}^{m} \\
~ \\
\beta _{k}^{-m} \\
~ \\
\beta _{k}^{-3m}
\end{array}
\right) ,  \label{phi-phi}
\end{equation}
whose explicit forms (for $m=1,2,3$, the only values relevant for the series
expansion (\ref{k=5-PHI})-(\ref{k=10-PHI})) read as follows:
\begin{eqnarray}
&&
\begin{array}{l}
k=5:\left\{
\begin{array}{c}
\varphi _{1}=\frac{1}{2}\sqrt{\frac{5+\sqrt{5}}{2}}\left(
\begin{array}{l}
-\sqrt{5+2\sqrt{5}}+i\sqrt{5} \\
-\sqrt{\frac{5+\sqrt{5}}{2}}+\frac{1}{2}i(5+3\sqrt{5}) \\
\sqrt{\frac{5+\sqrt{5}}{2}}+\frac{1}{2}i(5+3\sqrt{5}) \\
\sqrt{5+2\sqrt{5}}+i\sqrt{5}
\end{array}
\right) , \\
\\
\varphi _{2}=\frac{1}{2}\sqrt{\frac{5-\sqrt{5}}{2}}\left(
\begin{array}{l}
\sqrt{5-2\sqrt{5}}-i\sqrt{5} \\
-\sqrt{\frac{5-\sqrt{5}}{2}}+\frac{1}{2}i(5-3\sqrt{5}) \\
\sqrt{\frac{5-\sqrt{5}}{2}}+\frac{1}{2}i(5-3\sqrt{5}) \\
-\sqrt{5-2\sqrt{5}}-i\sqrt{5}
\end{array}
\right) ;
\end{array}
\right. \\
~
\end{array}
\label{k=5-phi-phi} \\
&&
\begin{array}{l}
k=6:\left\{
\begin{array}{l}
\varphi _{1}=\frac{3}{2}\left(
\begin{array}{l}
-\sqrt{3}+i \\
2i \\
2i \\
\sqrt{3}+i
\end{array}
\right) , \\
\\
\varphi _{2}=\frac{1}{2}\left(
\begin{array}{l}
1-i\sqrt{3} \\
-1 \\
1 \\
-1-i\sqrt{3}
\end{array}
\right) ;
\end{array}
\right. \\
~
\end{array}
\label{k=6-phi-phi} \\
&&
\begin{array}{l}
k=8:\left\{
\begin{array}{l}
\varphi _{1}=\frac{1}{2}\sqrt{2+\sqrt{2}}\left(
\begin{array}{l}
-2-\sqrt{2}+i\sqrt{2} \\
-\sqrt{2}+i(2+\sqrt{2}) \\
\sqrt{2}+i(2+\sqrt{2}) \\
2+\sqrt{2}+i\sqrt{2}
\end{array}
\right) , \\
\\
\varphi _{3}=\frac{1}{2}\sqrt{2-\sqrt{2}}\left(
\begin{array}{l}
-2+\sqrt{2}+i\sqrt{2} \\
\sqrt{2}+i(-2+\sqrt{2}) \\
-\sqrt{2}+i(-2+\sqrt{2}) \\
2-\sqrt{2}+i\sqrt{2}
\end{array}
\right) ;
\end{array}
\right. \\
~
\end{array}
\\
&&
\begin{array}{l}
k=10:\left\{
\begin{array}{l}
\varphi _{1}=\frac{1}{2}\sqrt{\frac{5+\sqrt{5}}{2}}\left(
\begin{array}{l}
-\frac{1}{2}\sqrt{3+\sqrt{5}}+i\sqrt{\frac{5+\sqrt{5}}{10}} \\
1+i\sqrt{1+\frac{2}{\sqrt{5}}} \\
-1+i\sqrt{1+\frac{2}{\sqrt{5}}} \\
\frac{1}{2}\sqrt{3+\sqrt{5}}+i\sqrt{\frac{5+\sqrt{5}}{10}}
\end{array}
\right) , \\
\\
\varphi _{3}=\frac{1}{2}\sqrt{\frac{5-\sqrt{5}}{2}}\left(
\begin{array}{l}
\frac{1}{2}\sqrt{-3+\sqrt{5}}+i\sqrt{\frac{5-\sqrt{5}}{10}} \\
1-i\sqrt{1-\frac{2}{\sqrt{5}}} \\
-1-i\sqrt{1-\frac{2}{\sqrt{5}}} \\
\frac{1}{2}\sqrt{3-\sqrt{5}}+i\sqrt{\frac{5-\sqrt{5}}{10}}
\end{array}
\right) .
\end{array}
\right.
\end{array}
\label{k=10-phi-phi}
\end{eqnarray}

Using Eqs. (\ref{PHI})-(\ref{k=10-phi-phi}), one obtains that Eq. (\ref{new}%
) near the LG point of the moduli space of Fermat $CY_{3}$-compactifications
(when consistently truncated up to the order in $\psi $ considered above)
reads
\begin{eqnarray}
k &=&5,6:\quad n^{T}=\frac{2}{\chi F_{1\,1}}Im\left[ N_{1}\bar{\varphi}_{1}-%
\frac{F_{1\,1}}{F_{2\,2}}\bar{N}_{2}\varphi _{2}\right] ;  \label{id-1} \\
&&  \notag \\
k &=&8,10:\quad n^{T}=\frac{2}{\chi F_{1\,1}}Im\left[ N_{1}\bar{\varphi}_{1}-%
\frac{F_{1\,1}}{F_{3\,3}}\bar{N}_{3}\varphi _{3}\right] .  \label{id-2}
\end{eqnarray}
Thence, it is easy to check that, substituting the explicit forms of $%
N_{k,m}\left( q,p\right) $, $\varphi _{k,m}$ and $F_{k,m\,n}$ (see Eqs. (\ref
{N-def}), (\ref{phi-phi}) and (\ref{F}), respectively) in Eqs. (\ref{id-1})-(%
\ref{id-2}), they become trivial identities, yielding nothing but
\begin{equation}
n_{1}=n_{1},\quad n_{2}=n_{2},\quad n_{3}=n_{3},\quad n_{4}=n_{4}.
\label{identical}
\end{equation}
In other words, the 4 real Eqs. (\ref{new}) are not equations, but
rather they are identities. Therefore, they are satisfied at every
point in the moduli space and for every BH charge configuration.
Therefore, it is no surprise if, when evaluating them in a certain
neighbourhood of the LG point as we did, one finds the identical
relations (\ref{identical}). Thus, we found nothing new but another
confirmation of a well known fact of SK geometry
\cite{FBC,K1,K2,BFM,AoB,K3}. \setcounter{equation}0
\def\theequation{11.\arabic{subsection}.\arabic{equation}}

\subsection{\label{SKG-id-approach-apply}\textit{``Special K\"{a}hler
Geometrical Identities''} Approach\newline
to LG Attractors in Fermat $CY_{3}$s}

However, the 4 real identities (\ref{new}) can still be used to find
extremal BH attractors, when properly evaluated along the constraints
defining the various species of such attractors satisfying the criticality
condition of the ``effective BH potential'' $V_{BH}$. Put another way, in
the 1-modulus case with which we are concerned, when evaluated at the
geometrical \textit{loci} in the moduli space defining the various
tipologies (\textit{i.e.} $\frac{1}{2}$-BPS, non-BPS $Z\neq 0$ and non-BPS $%
Z=0$) of attractors, the 4 real identities (\ref{new}) become 4 real
equations. These are equivalent to the 2 real equations given by the real
and imaginary part of the criticality condition $\partial V_{BH}=0$. This
approach has recently been used in \cite{K1} for the general $n_{V}$-moduli
case, and then further investigated in \cite{K2}. The SK geometrical
identities in the general $n_{V}$-moduli case had previously been formulated
in \cite{FBC} in terms of the decomposition of the third real cohomology $%
H^{3}\left( CY_{3};\mathbb{R}\right) $ of the $CY_{3}$ in the Dolbeaut
cohomology basis (see \cite{FBC}, and \cite{CDF} for further Refs.).

We will now focus on the 1-modulus case related to Fermat $CY_{3}$%
-compactifications, and we will evaluate the 4 real identities (\ref{new})
at the geometrical \textit{loci} in the moduli space defining the various
classes of extremal BH attractors. We will consequently show that solving
the obtained 4 real equations is equivalent to solving the 2 real equations
corresponding to the real and imaginary parts of the criticality condition (%
\ref{AEs-1-modulus-W}). Thus, it follows that only 2 equations are
independent out of the starting 4 ones. From a computational point of view,
one can realize that, at least in the framework we are considering, the
\textit{``criticality condition''} approach is simpler than the \textit{``SK
geometrical identities''} approach, at least for the non-BPS, $Z\neq 0$ case.

Let us now evaluate the 4 real SK identities (\ref{new}) along the 3
geometrical \textit{loci} defining the 3 species of critical points of $%
V_{BH}$ arising in SK geometry. \setcounter{equation}0
\def\theequation{11.2.\arabic{subsubsection}.\arabic{equation}}

\subsubsection{\label{SKG-id-approach-BPS}$\frac{1}{2}$-BPS}

The corresponding geometrical \textit{locus} in the moduli space is given by
the constraints $\widetilde{W}\neq 0$, $D\widetilde{W}=0$, which directly
solve the criticality condition (\textit{i.e.} the 1-modulus AE) (\ref
{AEs-1-modulus-W}). By evaluating the 4 real SK identities (\ref{new}) along
such critical constraints, one gets
\begin{equation}
n^{T}=\frac{2}{\chi F_{1\,1}}\left[ e^{\widetilde{K}}Im\left( \widetilde{W}%
\overline{\widetilde{\Upsilon }}\right) \right] _{\frac{1}{2}-BPS}.
\label{1/2}
\end{equation}
Such 4 real equations constrain the PF BH charge configurations along the
\textit{locus} \smallskip $\widetilde{W}\neq 0$, $D\widetilde{W}=0$ of $%
\frac{1}{2}$-BPS critical points of $V_{BH}$ in the moduli space ($dim_{%
\mathbb{C}}=1$)\ of Fermat $CY_{3}$s. One can explicitly check that
for all Fermat $CY_{3}$s the solutions of the 4 real Eqs.
(\ref{1/2}) near the LG point give nothing but the
$\frac{1}{2}$-BPS-supporting PF BH charge configurations previously
computed in Sects. \ref{quintic}-\ref{tentic} exploiting the
so-called \textit{``criticality condition''} approach.
\setcounter{equation}0
\def\theequation{11.2.\arabic{subsubsection}.\arabic{equation}}

\subsubsection{\label{SKG-id-approach-non-BPS-Z<>0}Non-BPS, $Z\neq 0$}

The corresponding geometrical \textit{locus} in the moduli space is given by
the conditions $\widetilde{W}\neq 0$, $D\widetilde{W}\neq 0$, further
constrained by the criticality condition (\ref{AEs-1-modulus-W}); by also
recalling the definition (\ref{W-tilde-def}), one obtains
\begin{equation}
\left( \overline{D}\overline{\widetilde{W}}\right) _{non-BPS,Z\neq 0}=-\left[
\frac{g^{-1}\left( \overline{D}^{2}\overline{\widetilde{W}}\right) D%
\widetilde{W}}{2\widetilde{W}}\right] _{non-BPS,Z\neq 0}.  \label{non-BPS1}
\end{equation}
By inserting Eq. (\ref{non-BPS1}) in the 4 real SK identities (\ref{new}),
one gets \cite{K1,K2,AoB}
\begin{eqnarray}
n^{T} &=&\frac{2}{\chi F_{1\,1}}\left\{ e^{\widetilde{K}}Im\left[ \widetilde{%
W}\overline{\widetilde{\Upsilon }}-\frac{g^{-2}\left( \overline{D}^{2}%
\overline{\widetilde{W}}\right) D\widetilde{W}}{2\widetilde{W}}D\widetilde{%
\Upsilon }\right] \right\} _{non-BPS,Z\neq 0}=  \label{zneq0} \\
&=&\frac{2}{\chi F_{1\,1}}\left\{ e^{\widetilde{K}}Im\left[ \widetilde{W}%
\overline{\widetilde{\Upsilon }}+i\frac{g^{-3}C\left( \overline{D}\overline{%
\widetilde{W}}\right) ^{2}}{2\overline{\widetilde{W}}}\overline{D}\overline{%
\widetilde{\Upsilon }}\right] \right\} _{non-BPS,Z\neq 0},
\label{zneq0-other}
\end{eqnarray}
where, in the second line, we used the $n_{V}=1$ case of the second SK
differential relation of (\ref{SKG-rels1}), yielding globally
\begin{equation}
D^{2}\widetilde{W}=iCg^{-1}\overline{D}\overline{\widetilde{W}}.
\label{SKG-rel-1-modulus-C}
\end{equation}
The 4 real Eqs. (\ref{zneq0})-(\ref{zneq0-other}) constrain the PF BH charge
configurations along the \textit{locus} (\ref{non-BPS1}) of non-BPS $Z\neq 0$
critical points of $V_{BH}$ in the moduli space of Fermat $CY_{3}$s.

Let us for example consider the mirror \textit{quintic} $\mathcal{M}%
_{5}^{\prime }$. From the treatment given in Sect. \ref{quintic} and above,
the 4 real Eqs. (\ref{zneq0}) take the following form near the LG point:
\begin{equation}
n^{T}=\frac{2}{\chi F_{1\,1}}Im\left[ N_{1}\bar{\varphi}_{1}-(\sqrt{5}+2)\xi
\frac{N_{2}N_{2}}{N_{1}}\varphi _{2}\right] ,  \label{zneq0-1}
\end{equation}
where $\xi $ is defined in Eq. (\ref{chnon}). Substituting into Eqs. (\ref
{zneq0-1}) the explicit expressions for the $N$s (see Sect. \ref{quintic})
and the $\varphi $s (see Eq. (\ref{k=5-phi-phi})) and performing long but
straightforward computations, it can be shown that one generally recovers
all the three distinct sets of BH charge configurations (\ref{fc51})-(\ref
{fc53}) supporting the considered non-BPS $Z\neq 0$ LG attractor.\smallskip\
The same can be explicitly checked for the mirror \textit{octic} $\mathcal{M}%
_{8}^{\prime }$. \setcounter{equation}0
\def\theequation{11.2.\arabic{subsubsection}.\arabic{equation}}

\subsubsection{\label{SKG-id-approach-non-BPS-Z=0}Non-BPS, $Z=0$}

The corresponding geometrical \textit{locus} in the moduli space is given by
the conditions $\widetilde{W}=0$, $D\widetilde{W}\neq 0$, further
constrained by the criticality condition (\ref{AEs-1-modulus-W}); by also
recalling and the definition (\ref{W-tilde-def}), one obtains
\begin{equation}
\left. D^{2}\widetilde{W}\right| _{non-BPS,Z=0}=0.  \label{at}
\end{equation}
By recalling Eq. (\ref{DDW}), the replacement of $\widetilde{W}=0$ and of
the condition (\ref{at}) into the 4 real SK geometrical identities (\ref{new}%
) yields the following 4 real equations:
\begin{equation}
n^{T}=-\frac{2}{\chi F_{1\,1}}\left\{ g^{-1}e^{\widetilde{K}}Im\left[ \frac{%
\partial ^{2}\widetilde{W}+\left( \partial {\widetilde{K}}\right) \partial
\widetilde{W}}{\partial \left[ ln\left( {g}\right) {-\widetilde{K}}\right] }%
\overline{D}\overline{\widetilde{\Upsilon }}\right] \right\} _{non-BPS,Z=0}.
\label{non-BPS-Z=0-constr}
\end{equation}
Such 4 real equations constrain the PF BH charge configurations along the
\textit{locus} (\ref{at}) of non-BPS $Z=0$ critical points of $V_{BH}$ in
the moduli space of Fermat $CY_{3}$s.

Let us for example consider the mirror \textit{sextic} $\mathcal{M}%
_{6}^{\prime }$. As one can easily check by using Eqs. (\ref{DDW}) and (\ref
{W-k=6})-(\ref{DDW-k=6}), in this case $\widetilde{W}=0$ directly satisfies
the criticality condition (\ref{AEs-1-modulus-W}). Consequently, rather than
Eqs. (\ref{non-BPS-Z=0-constr}), in order to exploit the so-called \textit{%
``SK geometrical identities'' }approach, one can consider the 4 real
equations
\begin{equation}
n^{T}=\frac{2}{\chi F_{1\,1}}\left\{ g^{-1}e^{\widetilde{K}}Im\left[ \left(
\overline{\partial }\overline{\widetilde{W}}\right) D\widetilde{\Upsilon }%
\right] \right\} _{non-BPS,Z=0},  \label{non-BPS-Z=0-constr-2}
\end{equation}
obtained from identities (\ref{new}) by simply putting $\widetilde{W}=0$ and
by replacing $\overline{D}\overline{\widetilde{W}}$ with $\overline{\partial
}\overline{\widetilde{W}}$, as implied by $\widetilde{W}=0$. From the
treatment of Sect. \ref{sixtic} and above, the 4 real Eqs. (\ref
{non-BPS-Z=0-constr-2}) take the following form near the LG point:
\begin{equation}
n^{T}=-\frac{2}{\chi F_{2\,2}}Im\left[ \bar{N}_{2}\varphi _{2}\right] .
\label{non-BPS-Z=0-constr-3}
\end{equation}
Substituting the explicit expressions for $N_{2}$ (see Eq. (\ref{N2-6})) and
for $\varphi _{2}$ (see Eq. (\ref{k=6-phi-phi})) into Eqs. (\ref
{non-BPS-Z=0-constr-3}), it can be shown that one obtains nothing but the
fine-tuning conditions (\ref{ft6}) for PF BH charges supporting the
considered non-BPS $Z=0$ LG attractor.\smallskip\ The same can be explicitly
checked for the mirror \textit{dectic} $\mathcal{M}_{10}^{\prime }$.

\section{The (LG Limit of) Yukawa Coupling Function of Fermat $CY_{3}$s\label%
{const-norm}}

\setcounter{equation}0
\def\theequation{12.\arabic{subsection}.\arabic{equation}}

\subsection{\label{const-norm-1}Different Approaches to Computation and LG
Limit}

The \textit{Yukawa coupling function} $C$ is nothing but the case $n_{V}=1$
of the fundamental tensor $C_{ijk}$ of SK geometry, defined by Eqs. (\ref{C}%
). Due to its covariant holomorphicity, $C$ can be written as (see the first
of Eqs. (\ref{C})):
\begin{equation}
C=e^{K}\mathcal{W},~~\overline{\partial }\mathcal{W}=0.  \label{CC}
\end{equation}
Because of $C$ has K\"{a}hler weights $\left( 2,-2\right) $, the function $%
\mathcal{W}$, usually called the \textit{holomorphic Yukawa coupling function%
}, has K\"{a}hler weights $\left( 4,0\right) $. The determination of $C$ is
far from being academic; as evident from the treatments exploited in
previous Sects., $C$ and its (covariant) derivative are key ingredients in
the study of extremal BH attractors in $n_{V}=1$, $\mathcal{N}=2$, $d=4$
MESGTs.

Thus, let us now face the question of how to determine $C$ in the framework
of the $1$-dim. SK geometry (endowing the moduli space of (mirror) Fermat $%
CY_{3}$s).

There exist at least three independent (and equivalent, as we will discuss
further below) ways to compute $C$:\medskip

I) Use Eqs. (\ref{W-1-modulus}) and (\ref{CC}), obtaining
\begin{eqnarray}
&&
\begin{array}{l}
C\left( \psi ,\overline{\psi }\right) =e^{K\left( \psi ,\overline{\psi }%
\right) }\mathcal{W}\left( \psi \right) = \\
\\
=e^{K}\left[
\begin{array}{l}
\left[ \partial X^{0}\left( \psi \right) \right] ^{3}\left. \frac{\partial
^{3}F\left( X\right) }{\left( \partial X^{0}\right) ^{3}}\right| _{X=X\left(
\psi \right) }+3\left[ \partial X^{0}\left( \psi \right) \right] ^{2}\left[
\partial X^{1}\left( \psi \right) \right] \left. \frac{\partial ^{3}F\left(
X\right) }{\left( \partial X^{0}\right) ^{2}\partial X^{1}}\right|
_{X=X\left( \psi \right) }+ \\
\\
+3\left[ \partial X^{1}\left( \psi \right) \right] ^{2}\left[ \partial
X^{0}\left( \psi \right) \right] \left. \frac{\partial ^{3}F\left( X\right)
}{\left( \partial X^{1}\right) ^{2}\partial X^{0}}\right| _{X=X\left( \psi
\right) }+\left[ \partial X^{1}\left( \psi \right) \right] ^{3}\left. \frac{%
\partial ^{3}F\left( X\right) }{\left( \partial X^{1}\right) ^{3}}\right|
_{X=X\left( \psi \right) }
\end{array}
\right] = \\
\\
=e^{K}\left( X^{0}\left( \psi \right) \right) ^{2}\left[ e\left( \psi
\right) \right] ^{3}\left. \frac{\partial ^{3}\mathcal{F}\left( t\right) }{%
\left( \partial t\right) ^{3}}\right| _{t=t(\psi )}= \\
\\
=e^{K}\left( X^{0}\left( \psi \right) \right) \left[
\begin{array}{l}
3\left[ e\left( \psi \right) \right] ^{-2}\left[ \partial e\left( \psi
\right) \right] ^{2}\partial \mathcal{F}\left( \psi \right) -\left[ e\left(
\psi \right) \right] ^{-1}\left[ \partial ^{2}e\left( \psi \right) \right]
\partial \mathcal{F}\left( \psi \right) + \\
\\
-3\left[ e\left( \psi \right) \right] ^{-1}\left[ \partial e\left( \psi
\right) \right] \partial ^{2}\mathcal{F}\left( \psi \right) +\partial ^{3}%
\mathcal{F}\left( \psi \right)
\end{array}
\right] .
\end{array}
\notag \\
&&  \label{C-1-modulus}
\end{eqnarray}
On the other hand, by recalling Eqs. (\ref{e}), (\ref{t}) (\ref{F-call}) and
(\ref{hom-deg-2-F}), Eq. (\ref{norm-PI}) can be further elaborated as
follows:
\begin{gather}
K\left( z,\overline{z}\right) =-ln\left\{ i\left| X^{0}\left( z\right)
\right| ^{2}\left[ 2\left( \mathcal{F}\left( z\right) -\overline{\mathcal{F}}%
\left( \overline{z}\right) \right) -\left( \frac{X^{a}\left( z\right) }{%
X^{0}\left( z\right) }-\frac{\overline{X}^{a}\left( \overline{z}\right) }{%
\overline{X}^{0}\left( \overline{z}\right) }\right) \left( e_{a}^{i}\left(
z\right) \frac{\partial \mathcal{F}\left( z\right) }{\partial z^{i}}+%
\overline{e}_{a}^{\overline{i}}\left( \overline{z}\right) \frac{\partial
\overline{\mathcal{F}}\left( \overline{z}\right) }{\partial \overline{z}^{%
\overline{i}}}\right) \right] \right\} ;  \label{norm-PI2} \\
\Downarrow  \notag \\
e^{K\left( z,\overline{z}\right) }=-i\left| X^{0}\left( z\right) \right|
^{-2}\left[ 2\left( \mathcal{F}\left( z\right) -\overline{\mathcal{F}}\left(
\overline{z}\right) \right) -\left( \frac{X^{a}\left( z\right) }{X^{0}\left(
z\right) }-\frac{\overline{X}^{a}\left( \overline{z}\right) }{\overline{X}%
^{0}\left( \overline{z}\right) }\right) \left( e_{a}^{i}\left( z\right)
\frac{\partial \mathcal{F}\left( z\right) }{\partial z^{i}}+\overline{e}%
_{a}^{\overline{i}}\left( \overline{z}\right) \frac{\partial \overline{%
\mathcal{F}}\left( \overline{z}\right) }{\partial \overline{z}^{\overline{i}}%
}\right) \right] ^{-1},  \label{norm-PI3}
\end{gather}
where
\begin{equation}
\begin{array}{l}
\mathcal{F}\left( z\right) \equiv \mathcal{F}\left( \frac{X\left( z\right) }{%
X^{0}\left( z\right) }\right) =\left( X^{0}\left( z\right) \right)
^{-2}F\left( X\left( z\right) \right) ; \\
\\
e_{a}^{i}\left( z\right) \equiv e_{a}^{i}\left( \frac{X\left( z\right) }{%
X^{0}\left( z\right) }\right) ,~~\overline{e}_{a}^{\overline{i}}\left(
\overline{z}\right) =\overline{e_{a}^{i}\left( z\right) }.
\end{array}
\end{equation}
Thus, Eq. (\ref{C-1-modulus}) can be rewritten as follows (for simplicity's
sake, we omit notation of dependence on $\left( \psi ,\overline{\psi }%
\right) $):
\begin{equation}
C=e^{K}\mathcal{W}=-i\frac{\left[ 3e^{-2}\left( \partial e\right)
^{2}\partial \mathcal{F}-e^{-1}\left( \partial ^{2}e\right) \partial
\mathcal{F}-3e^{-1}\left( \partial e\right) \partial ^{2}\mathcal{F}%
+\partial ^{3}\mathcal{F}\right] }{\overline{X}^{0}\left[ 2\left( \mathcal{F}%
-\overline{\mathcal{F}}\right) -\left( \frac{X^{1}}{X^{0}}-\frac{\overline{X}%
^{1}}{\overline{X}^{0}}\right) \left( e^{-1}\partial \mathcal{F}+\overline{e}%
^{-1}\overline{\partial }\overline{\mathcal{F}}\right) \right] },
\label{C-1-modulus2}
\end{equation}
which can be also further elaborated by recalling Eq. (\ref{e-1-modulus}).

Eq. (\ref{C-1-modulus2}) holds \textit{globally} for a generic $n_{V}=1$ SK
geometry, with no K\"{a}hler gauge-fixing, with the only condition of $e\neq
0$ (and $X^{0}\neq 0$). By specializing it for the $1$-dim. SK geometry
endowing the moduli space of (mirror) Fermat $CY_{3}$s, one has to replace
the ($k$-indexed) quantities $X^{0}$, $X^{1}$, $e$ and $\mathcal{F}$ with
the first and second component of Eq. (\ref{PI-1-CY}), with consequent Eq. (%
\ref{e-1-modulus}), and with Eq. (\ref{F-call-CY}), respectively. Thence,
one has to use the results obtained for $k=5,6,8,10$ in Sects. \ref{quintic}-%
\ref{tentic}, which however converge only for $\left| \psi \right| <1$, $%
0\leqslant arg\left( \psi \right) <\frac{2\pi }{k}$, and have mainly
considered in the \textit{LG limit} $\psi \longrightarrow 0$.\medskip

II) Use Eq. (\ref{CC}) and the relation $a_{4}=\mathcal{W}^{-1}$ (given by
the first of definitions (\ref{a})), where $a_{4}$ is the coefficient of the
fourth derivative in the PF Eq. (\ref{PF2}). It is important to notice that
the relation $a_{4}=\mathcal{W}^{-1}$ was obtained in \cite{Ferrara-Louis1},
where the relation
\begin{equation}
D_{\alpha }U_{\beta }=C_{\alpha \beta \gamma }g^{\gamma \overline{\gamma }}%
\overline{U}_{\overline{\gamma }}
\end{equation}
holds (see the second of Eqs. (8) of \cite{Ferrara-Louis1}), rather than the
second of Eqs. (\ref{SKG-rels2}), consistent with the conventions we use.
This means that the $C_{ijk}$ used in the present paper is nothing but $-i$
times the $C_{ijk}$ used in \cite{Ferrara-Louis1}. In the case $n_{V}=1$, by
taking into account such a shift of phase and using (the $n_{V}=1$ case of)
Eq. (\ref{norm-PI3}), one finally gets the relation
\begin{equation}
C=-ie^{K}a_{4}^{-1}=-\left\{ a_{4}\left| X^{0}\right| ^{2}\left[ 2\left(
\mathcal{F}-\overline{\mathcal{F}}\right) -\left( \frac{X^{1}}{X^{0}}-\frac{%
\overline{X}^{1}}{\overline{X}^{0}}\right) \left( e^{-1}\partial \mathcal{F}+%
\overline{e}^{-1}\overline{\partial }\overline{\mathcal{F}}\right) \right]
\right\} ^{-1}.  \label{C-1-modulus3}
\end{equation}

As Eq. (\ref{C-1-modulus2}), Eq. (\ref{C-1-modulus3}) holds \textit{globally}
for a generic $n_{V}=1$ SK geometry, with no K\"{a}hler gauge-fixing, with
the only condition of $e\neq 0$ (and $X^{0}\neq 0$). Let us specialize it
for $1$-dim. SK geometry endowing the moduli space of (mirror) Fermat $%
CY_{3} $s. In order to do this, we need to introduce the Fermat classifying
parameter $k$, and recall Eq. (\ref{kahlerpot1}) (converging for $\left|
\psi \right| <1$, $0\leqslant arg\left( \psi \right) <\frac{2\pi }{k}$), the
first of Eqs. (\ref{W-tilde-def}) and the second of Eqs. (\ref{PF2-corr}),
thus obtaining (with the use of Tables 1 and 2):
\begin{eqnarray}
C_{k} &=&-ie^{K_{k}}\frac{1}{\psi ^{5-k}-\psi ^{5}}=-\frac{i}{\left( 2\pi
\right) ^{6}}\frac{\left( Ord\left( G_{k}\right) \right) ^{2}}{%
C_{k,0}^{2}F_{k,11}}\frac{e^{\widetilde{K}_{k}}}{1-\psi ^{k}}\psi ^{k-5}=
\label{C-1-modulus3-CY} \\
&&  \notag \\
&=&-\frac{i}{\left( 2\pi \right) ^{6}}\frac{\left( Ord\left( G_{k}\right)
\right) ^{2}}{C_{k,0}^{2}F_{k,11}}\frac{\psi ^{k-5}}{1-\psi ^{k}}\left\{ 1+%
\frac{1}{C_{k,0}^{2}F_{k,11}}\left[
\begin{array}{l}
2C_{k,0}\sum_{n=2}^{\infty }C_{k,n-1}F_{k,1n}Re\left( \psi ^{n-1}\right) +
\\
\\
+\sum_{m,n=2}^{\infty }C_{k,m-1}C_{k,n-1}F_{k,mn}\psi ^{m-1}\overline{\psi }%
^{n-1}
\end{array}
\right] \right\} ^{-1};  \notag \\
&&  \label{C-1-modulus3-CY2}
\end{eqnarray}
notice that Eq. (\ref{C-1-modulus3-CY}) holds globally, whereas Eq. (\ref
{C-1-modulus3-CY2}) converges for $\left| \psi \right| <1$, $0\leqslant
arg\left( \psi \right) <\frac{2\pi }{k}$ (as given by Eq. (\ref{kahlerpot1}%
)). By recalling that $k=5,6,8,10$, in the \textit{LG limit} $\psi
\longrightarrow 0$ such an expression can be approximated as follows:
\begin{eqnarray}
&&
\begin{array}{l}
lim_{\psi \rightarrow 0}C_{k}\approx -\frac{i}{\left( 2\pi \right) ^{6}}%
\frac{\left( Ord\left( G_{k}\right) \right) ^{2}}{C_{k,0}^{2}F_{k,11}}\frac{%
\psi ^{k-5}}{1-\psi ^{k}}\left\{ 1-\frac{1}{C_{k,0}^{2}F_{k,11}}\left[
\begin{array}{l}
2C_{k,0}\sum_{n=2}^{\infty }C_{k,n-1}F_{k,1n}Re\left( \psi ^{n-1}\right) +
\\
\\
+\sum_{m,n=2}^{\infty }C_{k,m-1}C_{k,n-1}F_{k,mn}\psi ^{m-1}\overline{\psi }%
^{n-1}
\end{array}
\right] \right\} \approx \\
\\
\approx -\frac{i}{\left( 2\pi \right) ^{6}}\frac{\left( Ord\left(
G_{k}\right) \right) ^{2}}{C_{k,0}^{2}F_{k,11}}\psi ^{k-5}\longrightarrow -%
\frac{i}{\left( 2\pi \right) ^{6}}\frac{\left( Ord\left( G_{k}\right)
\right) ^{2}}{C_{k,0}^{2}F_{k,11}}\delta _{k,5}.
\end{array}
\notag \\
&&  \label{C-1-modulus3-CY-LG}
\end{eqnarray}
Thus, in the $1$-dim. SK geometry endowing the moduli space of (mirror)
Fermat $CY_{3}$s the \textit{LG limit} of the Yukawa coupling function is
holomorphic, with the leading term $\sim \mathcal{O}$ $\left( \psi
^{k-5}\right) $, as given by Eq. (\ref{C-1-modulus3-CY-LG}).\medskip

III) Use Eq. (\ref{SKG-rel-1-modulus-C}). Clearly, it holds not only for the
rescaled holomorphic central charge $\widetilde{W}$ (introduced in the
framework of Fermat $CY_{3}$- compactifications by the second of definitions
(\ref{W-tilde-def})), but for a generic superpotential $W$ of a $1$-dim. SK
geometry.

Apparently, Eq. (\ref{SKG-rel-1-modulus-C}) yields a way to compute $C$ only
along the \textit{locus} $DW\neq 0$ of $n_{V}=1$ SK manifolds, \textit{i.e.}
only away from $\frac{1}{2}$-BPS critical points of $V_{BH}$. Instead, due
to the fact that the fundamental relations (\ref{SKG-rels2}) of SK geometry
do hold globally, the approach based on Eq. (\ref{SKG-rel-1-modulus-C})
(which is nothing but the $n_{V}=1$ case of the second of Eqs. (\ref
{SKG-rels2})) has a global validity, yielding that
\begin{equation}
C=-ig\frac{D^{2}W}{\overline{D}\overline{W}}=-i\left( \overline{\partial }%
\partial K\right) \frac{\left\{ \partial ^{2}+\partial ^{2}K+2\partial
K\partial +\left( \partial K\right) ^{2}-\left[ \partial ln\left( \partial
\overline{\partial }K\right) \right] \left( \partial +\partial K\right)
\right\} W}{\left[ \overline{\partial }+\left( \overline{\partial }K\right) %
\right] \overline{W}},  \label{SKG-rel-implied}
\end{equation}
where Eqs. (\ref{DW}) and (\ref{DDW}) have been used. Notice that the
charge-dependence of $DW$ and $D^{2}W$ vanishes in the ratio $\frac{D^{2}W}{%
\overline{D}\overline{W}}$ ($C$ is charge-independent). As Eqs. (\ref
{C-1-modulus2}) and (\ref{C-1-modulus3}), Eq. (\ref{SKG-rel-implied}) holds
\textit{globally} for a generic $n_{V}=1$ SK geometry, with \textit{a priori}
no K\"{a}hler gauge-fixing. In order to specialize it for the $1$-dim. SK
geometry endowing the moduli space of (mirror) Fermat $CY_{3}$s, we need to
introduce the Fermat classifying parameter $k$, and furthermore to recall
the general tretament given in Sect. \ref{GA}, and use the results obtained
for $k=5,6,8,10$ in Sects. \ref{quintic}-\ref{tentic}, which however
converge only for $\left| \psi \right| <1$, $0\leqslant arg\left( \psi
\right) <\frac{2\pi }{k}$, and have mainly considered in the \textit{LG limit%
} $\psi \longrightarrow 0$.\medskip\ By doing this, one finally gets
\begin{equation}
\begin{array}{l}
k=5:lim_{\psi \rightarrow 0}C_{5}\approx -2i\left( \sqrt{5}-2\right) \frac{%
C_{5,1}C_{5,2}}{C_{5,0}^{2}}; \\
\\
k=6:lim_{\psi \rightarrow 0}C_{6}\approx 2i\frac{C_{6,1}C_{6,3}}{C_{5,0}^{2}}%
\psi ; \\
\\
k=8:lim_{\psi \rightarrow 0}C_{8}\approx 16i(3-2\sqrt{2})\frac{C_{8,2}C_{8,4}%
}{C_{8,0}^{2}}\psi ^{3}; \\
\\
k=10:lim_{\psi \rightarrow 0}C_{10}=48i(\sqrt{5}-2)\frac{C_{10,2}C_{10,6}}{%
C_{10,0}^{2}}\psi ^{5}.
\end{array}
\label{C-1-modulus4-CY-LG}
\end{equation}

In order for the LG limits (\ref{C-1-modulus3-CY-LG}) and (\ref
{C-1-modulus4-CY-LG}) to coincide, it is easy to realize that $a_{4,k}$ must
be normalized as follows
\begin{equation}
\begin{array}{l}
a_{4,k}\longrightarrow \frac{\left( Ord\left( G_{k}\right) \right) ^{2}}{%
\left( 2\pi \right) ^{6}F_{k,11}\mathcal{A}_{k}}a_{4,k}; \\
\\
\begin{array}{l}
\mathcal{A}_{5}\equiv 2C_{5,1}C_{5,2}\left( \sqrt{5}-2\right) ; \\
\\
\mathcal{A}_{6}\equiv -2C_{6,1}C_{6,3}; \\
\\
\mathcal{A}_{8}\equiv -16C_{8,2}C_{8,4}\left( 3-2\sqrt{2}\right) ; \\
\\
\mathcal{A}_{10}\equiv -48C_{10,2}C_{10,6}\left( 5-\sqrt{2}\right) .
\end{array}
\end{array}
\label{a4-norm}
\end{equation}
In order to preserve the set of solutions of the PF Eq. (\ref{PF2-corr}) and
the properties of the coefficients $a_{n,k}$ (as given by Eqs. (\ref{a})-(%
\ref{diff-rels-a})), it is clear that the normalization (\ref{a4-norm})
yields the multiplication of PF Eq. (\ref{PF2-corr}) by the overall constant
$\frac{\left( Ord\left( G_{k}\right) \right) ^{2}}{\left( 2\pi \right)
^{6}F_{k,11}\mathcal{A}_{k}}$. Thus, one can state that the compatibility
between the appoaches II and III to the computation of the Yukawa coupling
function $C_{k}$ in the LG limit $\psi \longrightarrow 0$ of the ($k$%
-parameterized) $1$-dim. SK geometry of the moduli space of (mirror) Fermat $%
CY_{3}$s yields the following (\textit{irrelevant}, because overall and
constant) normalization of the PF Eq. (\ref{PF2-corr})
\begin{equation}
\begin{array}{l}
\sum_{n=0}^{4}a_{n,k}\left( \psi \right) \partial ^{n}V_{h}\left( \psi
\right) =0, \\
\\
a_{n,k}\left( \psi \right) \equiv \frac{\left( Ord\left( G_{k}\right)
\right) ^{2}}{\left( 2\pi \right) ^{6}F_{k,11}\mathcal{A}_{k}}\left[ -\sigma
_{n}\psi ^{n+1}+\left( -1\right) ^{n}\tau _{n,k}\psi ^{n+1-k}\right] ,
\end{array}
\label{PF2-corr-corr}
\end{equation}
where the $k$-dependent constants $Ord\left( G_{k}\right) $, $F_{k,11}$, $%
\mathcal{A}_{k}$, $\sigma _{n}$ and $\tau _{n,k}$ are given by Table
3, Eqs. (\ref{k=5-Fmn}), (\ref{k=6-Fmn}), (\ref{k=8-Fmn}),
(\ref{k=10-Fmn}), (\ref {a4-norm}), and Tables 1 and 2,
respectively. Finally, it can be checked that the use of approach I
does not give raise to any other problem of
consistency/compatibility with the approaches II and III in the
framework (of the LG limit) of $1$-dim. SK geometry endowing the
moduli space of (mirror) Fermat $CY_{3}$s. \setcounter{equation}0
\def\theequation{12.\arabic{subsection}.\arabic{equation}}

\subsection{\label{const-norm-1-bis}Check of Stability Conditions}

By recalling Eqs. (\ref{C-1-modulus3}) and (\ref{C-1-modulus3-CY}), and the
first of definitions (\ref{W-tilde-def}), from the correct normalization (%
\ref{a4-norm}) of $a_{n,k}$ we can achieve the exact, $k$-parametrized
global formula of the Yukawa coupling function $C_{k}$ for the ($1$-dim. SK
geometry endowing the moduli space of mirror) Fermat $CY_{3}$s:
\begin{equation}
C_{k}=-ie^{k}\left( a_{4,k}\right) ^{-1}=-i\frac{\mathcal{A}_{k}}{C_{k,0}^{2}%
}\frac{\psi ^{k-5}}{1-\psi ^{k}}e^{\widetilde{K}_{k}}.  \label{C-exact}
\end{equation}
Let us now reconsider Eq. (\ref{DC-1}); by using Eq. (\ref{C-exact}), we can
compute the exact, $k$-parametrized global formula of the covariant
derivative of the Yukawa coupling function of $1$-dim. SK geometry endowing
the moduli space of (mirror) Fermat $CY_{3}$s:
\begin{equation}
DC_{k}=\partial C_{k}+\left[ \left( \partial \widetilde{K}_{k}\right)
-3\partial ln\left( \overline{\partial }\partial \widetilde{K}_{k}\right) %
\right] C_{k}=-i\frac{\mathcal{A}_{k}}{C_{k,0}^{2}}\left[ \frac{5\psi
^{4}+\left( k-5\right) \psi ^{4-k}}{\psi ^{5-k}-\psi ^{5}}+2\left( \partial
\widetilde{K}_{k}\right) -3\left( \frac{\overline{\partial }\partial ^{2}%
\widetilde{K}_{k}}{\overline{\partial }\partial \widetilde{K}_{k}}\right) %
\right] \frac{\psi ^{k-5}}{1-\psi ^{k}}e^{\widetilde{K}_{k}}.
\label{DC-exact}
\end{equation}

We now have all the ingredients to check the stability conditions for
non-BPS $Z\neq 0$ and non-BPS $Z=0$ critical points of $V_{BH}$ in the
\textit{LG limit} of the $1$-dim. SK geometry endowing the moduli space of
(mirror) Fermat $CY_{3}$s. By recalling Eqs. (\ref{stab-1-1}) and (\ref
{stab-1-3}), using Eqs. (\ref{C-exact}) and (\ref{DC-exact}), the results
obtained for the LG limit of (the moduli space of mirror) Fermat $CY_{3}$s
for $k=5,6,8,10$ in Sects. \ref{quintic}, \ref{sixtic}, \ref{eightic} and
\ref{tentic} respectively, yield that:

1) for $k=5,8$ the condition (\ref{stab-1-1}) for $\psi \approx 0$ to be a
\textit{stable} non-BPS $Z\neq 0$ critical point of $V_{BH,k=5,8}$ (and thus
to be an \textit{attractor} in strict sense) is satisfied, consistently with
the results obtained in Subsects. \ref{21march-1}-\ref{21march-2} and \ref
{21march-3}-\ref{21march-4}, respectively;

2) for $k=6,10$ the condition (\ref{stab-1-3}) for $\psi \approx 0$
to be a \textit{stable} non-BPS $Z=0$ critical point of
$V_{BH,k=6,10}$ (and thus to be an \textit{attractor} in strict
sense) is \textit{not} satisfied, consistently with the results
obtained in Subsects. \ref{21march-5}-\ref {21march-6} and
\ref{21march-7}-\ref{21march-8}, respectively.
\setcounter{equation}0
\def\theequation{12.\arabic{subsection}.\arabic{equation}}

\subsection{\label{const-norm-2}The Exact Holomorphic Yukawa Coupling
Function of Fermat $CY_{3}$s}

Having performed a self-consistent treatment of Yukawa coupling function $C$
in the LG limit, we can now reconsider the relation $a_{4}=\mathcal{W}^{-1}$
(given by the first of definitions (\ref{a})); indeed, from the correct
normalization (\ref{a4-norm}) of $a_{n,k}$, we can achieve the exact, $k$%
-parametrized global formula of the holomorphic Yukawa coupling function $%
\mathcal{W}_{k}$ for the ($1$-dim. SK geometry endowing the moduli space of
mirror) Fermat $CY_{3}$s:
\begin{equation}
\mathcal{W}_{k}\left( \psi \right) =\left[ a_{4,k}\left( \psi \right) \right]
^{-1}=\frac{\left( 2\pi \right) ^{6}F_{k,11}\mathcal{A}_{k}}{\left(
Ord\left( G_{k}\right) \right) ^{2}}\frac{\psi ^{k-5}}{1-\psi ^{k}}.
\label{WW}
\end{equation}
By using such an exact global formula, the evaluation of $\mathcal{W}_{k}$
near the three species of regular singular of points of PF Eq. (\ref
{PF2-corr-corr}) of Fermat $CY_{3}$s yields:
\begin{eqnarray}
&&
\begin{array}{l}
\text{\textit{LG limit} : }lim_{\psi \longrightarrow 0}\mathcal{W}_{k}\left(
\psi \right) =\frac{\left( 2\pi \right) ^{6}F_{k,11}\mathcal{A}_{k}}{\left(
Ord\left( G_{k}\right) \right) ^{2}}lim_{\psi \longrightarrow 0}\frac{\psi
^{k-5}}{1-\psi ^{k}}\approx \frac{\left( 2\pi \right) ^{6}F_{k,11}\mathcal{A}%
_{k}}{\left( Ord\left( G_{k}\right) \right) ^{2}}\psi ^{k-5}\longrightarrow
\frac{\left( 2\pi \right) ^{6}F_{k,11}\mathcal{A}_{k}}{\left( Ord\left(
G_{k}\right) \right) ^{2}}\delta _{k,5};
\end{array}
\label{www} \\
&&  \notag \\
&&
\begin{array}{l}
\text{\textit{Conifold limit} : }\left| lim_{\psi ^{k}\longrightarrow 1}%
\mathcal{W}_{k}\left( \psi \right) \right| =\infty ;
\end{array}
\\
&&  \notag \\
&&
\begin{array}{l}
\text{\textit{Large complex structure modulus limit} : }lim_{\psi
\longrightarrow \infty }\mathcal{W}_{k}\left( \psi \right) \approx -\frac{%
\left( 2\pi \right) ^{6}F_{k,11}\mathcal{A}_{k}}{\left( Ord\left(
G_{k}\right) \right) ^{2}}\psi ^{-5}\longrightarrow 0.
\end{array}
\end{eqnarray}

\subsection*{Remark}

A final comment concerns the vanishing of $C_{8}$ (and $\mathcal{W}_{8}$) in
the LG limit, as yielded by Eqs. (\ref{C-1-modulus3-CY-LG}) (and (\ref{www}%
)). This might seem in contrast with the result, obtained in Sect. \ref
{eightic}, that for $k=8$ only $\frac{1}{2}$-BPS and non-BPS $Z\neq 0$ LG
critical points of $V_{BH}$ (both stable) exist, depending on the supporting
BH charge configuration(s). For non-BPS $Z\neq 0$ attractors, this implies
that $C_{8,non-BPS,Z\neq 0}\neq 0$, as given for a generic $1$-dim.\ SK
geometry by Eq. (\ref{rain2}). The contrast with the result (\ref
{C-1-modulus3-CY-LG}) can be simply explained by pointing out that for the
(mirror) octic $\mathcal{M}_{8}^{\prime }$ the exact LG point $\psi =0$
cannot be a non-BPS, $Z\neq 0$ critical point of $V_{BH}$. Rather, such a
kind of critical point turns out to be realized for $\psi \approx 0$,
\textit{i.e.} in a suitable neighbourhood of the LG point. Indeed, as
remarked below Eq. (\ref{attreq}) and as understood in all the treatment of
the present paper, the actual BH charges satisfying the $\mathcal{N}=2$, $%
d=4 $ AE in the LG limit are very close to the found one, and also
the critical value of $\psi $ may not be zero (\textit{i.e.} the
\textit{exact} LG point) , rather it may belong to a suitable
neighbourhood of the LG point. \setcounter{equation}0
\def\theequation{14.\arabic{subsection}.\arabic{equation}}

\section{Conclusions and Outlook\label{Conclusion}}

In the present work we investigated non-degenerate extremal BH attractors
near the so-called LG point $\psi =0$ (herein named \textit{LG attractors})
of the moduli space (with $dim_{\mathbb{C}}=1$) of the class of (mirror)
Fermat Calabi-Yau threefolds. We found the BH charge configurations
supporting $\psi \approx 0$ to be a critical point of the real,
positive-definite ``effective BH potential'' $V_{BH}$ defined in Eq. (\ref
{VBH1-1-modulus}).

In order to do this, we exploited two different approaches:

\textit{i}) \textit{``criticality condition'' }approach: we solved at $\psi
\approx 0$ the 2 real \textit{criticality conditions} of $V_{BH}$,
corresponding in the 1-modulus case to the real and imaginary part of the
\textit{Attractor Eq.} (\ref{AEs-1-modulus-W}) (see Sects. \ref{quintic}-\ref
{tentic});

\textit{ii}) \textit{``SK geometrical identities'' }approach: we evaluated
at $\psi \approx 0$ the 4 real fundamental identities (\ref
{SKG-Identities-1-modulus}) of 1-modulus SK geometry at the geometrical
\textit{loci} corresponding to the various species of critical points of $%
V_{BH}$ (see Sect. \ref{SKG-id-approach}).

We found that the results of two such solving approaches do coincide, in
spite of the different number of real Eqs. involved in approaches 1 and 2.
The equivalence of the above-mentioned approaches to find the critical
points of $V_{BH}$ (and the BH charge configurations supporting them) is
explicit proof of the fact that the relations (\ref{SKG-Identities-1-modulus}%
) actually are \textit{identities} and not equations, \textit{i.e.} that,
for any point of the moduli space at which we evaluate them, they do \textit{%
not} give any constraint on the charge configuration.

It is worth pointing out that the \textit{``criticality condition'' }%
approach had been previously exploited in literature only for the following
cases:

\textit{i.a}) \textit{mirror quintic} ($k=5$) in \cite{TT}, where however
peculiar \textit{Ans\"{a}tze} (on the BH charge configuration and on $\psi $
in the neighbourhood of the LG point) were used, implying a certain loss of
generality;

\textit{i.b}) \textit{mirror sextic} ($k=6$)\ in \cite{G}.

On the other hand, the \textit{``SK geometrical identities'' }approach (and
its equivalence with the \textit{``criticality condition'' }one) had been
hitherto exploited only in \cite{K2}; in such a Ref., the mirror \textit{%
quintic} was considered within the same simplifying \textit{Ans\"{a}tze}
formulated in \cite{TT}, obtaining a complete agreement with the results of
\cite{TT}.

As a by-product of our computations, we extended the results of \cite{TT}
and \cite{K2} to full generality (see Sect. \ref{quintic}). Moreover, we
found that the analysis of the stability of $\psi \approx 0$ as a non-BPS, $%
Z=0$ critical point of $V_{BH}$ in the mirror \textit{sextic}, performed in
Sect. 7 of \cite{G}, suffers from some problems of inconsistency. Indeed, in
\cite{G} it was found that the LG point (supported by a certain BH charge
configuration characterizing it as a non-BPS, $Z=0$ critical point of $%
V_{BH} $) is \textit{stable }(minimum of $V_{BH}$). Instead, our
computations (see Sect. \ref{sixtic}), which carefully took into account the
relevant orders in $\psi $ and $\overline{\psi }$ in the truncation of the
series expansion around $\psi \approx 0$, allow us to conclude that, \textit{%
for the same supporting BH charge configuration}, the LG point is \textit{%
unstable} (namely, a saddle point of $V_{BH}$).

We also checked the stability of $\psi \approx 0$ in two ways:

1) by inspecting the Hessian matrix of $V_{BH}$ in correspondence to the
various BH charge configurations supporting the LG point to be a critical
point of $V_{BH}$ (see Subsects. \ref{21march-1}-\ref{21march-2}, \ref
{21march-5}-\ref{21march-6}, \ref{21march-3}-\ref{21march-4} and \ref
{21march-7}-\ref{21march-8} for $k=5,6,8,10$, respectively);

2) by using the stability conditions for non-BPS $Z\neq 0$ and non-BPS $Z=0$
critical points of $V_{BH}$ in $1$-modulus SK geometry, respectively
obtained in Subsubsects. \ref{stab-non-BPS-Z<>0} and \ref{stab-non-BPS-Z=0}.

A sketchy summary of our results is given by the following Table:
\begin{table}[h]
\begin{center}
\begin{tabular}{|c||c|c|c|c|}
\hline
$k\longrightarrow $ & $5$ & $6$ & $8$ & $10$ \\ \hline\hline
$\frac{1}{2}$-BPS & $
\begin{array}{c}
\text{\textit{stable},} \\
\text{1 charge config.}
\end{array}
$ & $
\begin{array}{c}
\text{\textit{stable},} \\
\text{1 charge config.}
\end{array}
$ & $
\begin{array}{c}
\text{\textit{stable},} \\
\text{1 charge config.}
\end{array}
$ & $
\begin{array}{c}
\text{\textit{stable},} \\
\text{1 charge config.}
\end{array}
$ \\ \hline
non-BPS, $Z\neq 0$ & $
\begin{array}{c}
\text{\textit{stable},} \\
\text{3 charge configs.}
\end{array}
$ & $-$ & $
\begin{array}{c}
\text{\textit{stable},} \\
\text{3 charge configs.}
\end{array}
$ & $-$ \\ \hline
non-BPS, $Z=0$ & $-$ & $
\begin{array}{c}
\text{\textit{unstable},} \\
\text{1 charge config.}
\end{array}
$ & $-$ & $
\begin{array}{c}
\text{\textit{unstable},} \\
\text{1 charge config.}
\end{array}
$ \\ \hline
\end{tabular}
\end{center}
\caption{{}\textbf{Species and stability of the LG critical point }$\protect%
\psi \approx 0$\textbf{\ of }$V_{BH}$ \textbf{in the moduli space of
(mirror) Fermat }$CY_{3}$\textbf{s}}
\end{table}

The stability of $\psi \approx 0$ as a $\frac{1}{2}$-BPS attractor agrees
with the known results from general analysis of SK geometry of scalar
manifolds in $\mathcal{N}=2$, $d=4$ supergravity coupled to $n_{V}$ Abelian
vector multiplets \cite{FGK,BFM,AoB}.

For non-BPS LG critical points of $V_{BH}$ the central charge $Z$ turns out
to be a crucial quantity for stability. Indeed, regardless of the kind of BH
charge configuration supporting them, the non-BPS, $Z\neq 0$ LG attractors,
when they exist, are found to be stable (minima of $V_{BH}$). On the other
hand, the non-BPS, $Z=0$ LG attractors, when they exist, are found to be
unstable (saddle points of $V_{BH}$).

It is interesting to compare such a result to what happens in the large
volume limit of $CY_{3}$-compactifications (of Type IIA superstrings).
Indeed, in such a framework (with a generic number $n_{V}$ of complex
structure moduli) in \cite{TT} it was shown that the stability of non-BPS, $%
Z\neq 0$ critical points of $V_{BH}$ (and therefore their actual attractor
behaviour) within a certain supporting BH charge configuration, crucially
depends on the possible vanishing of $p^{0}$, \textit{i.e.} of the
asymptotical magnetic flux of the graviphoton field strength, whose
microscopical interpretation corresponds to a $D6$-brane wrapping $p^{0}$
times a 3-cycle of the considered $CY_{3}$. Nevertheless, also in such a
context in the $1$-modulus case the non-BPS, $Z\neq 0$ critical points of $%
V_{BH}$ are always stable, and therefore they are attractors in a strict
sense.

Furthermore, one can also observe that all Fermat $CY_{3}$s admit \textit{%
only one} kind of non-BPS LG attractors, either with $Z\neq 0$ or with $Z=0$%
; for the allowed values of the classifying Fermat parameter $k=5,6,8,10$,
one gets the ``pattern'' shown in Table 5 above.

Once again, such a feature is exhibited also by the large volume limit of $%
CY_{3}$-compactifications (of Type IIA superstrings), whose related SK
geometry is characterized by cubic holomorphic prepotentials; indeed, it can
be explicitly computed that the $1$-modulus prepotential (\ref{cub-1})%
\textbf{\ }corresponding to the homogeneous symmetric SK manifold $\frac{%
SU(1,1)}{U(1)}$ (see the remark at the end of Subsubsect. \ref
{stab-non-BPS-Z=0}, and \cite{BFGM1} and Refs. therein), admits $\frac{1}{2}$%
-BPS and non-BPS, $Z\neq 0$ critical points of $V_{BH}$ (both stable) only.

It is worth mentioning that the fourth order linear ordinary differential
Picard-Fuchs equations of Fermat $CY_{3}$s (\ref{PF2-corr}) (or equivalently
(\ref{PF2-corr-corr})) (specified by Tables 1 and 2) exhibit other two
species of regular singular points, namely the $k$-th roots of unity ($\psi
^{k}=1$, the so-called \textit{conifold points}) and the \textit{point at
infinity} $\psi \longrightarrow \infty $ in the moduli space, corresponding
to the so-called \textit{large complex structure modulus limit}. It would be
interesting to solve criticality conditions for $V_{BH}$ near such regular
singular points, \textit{i.e.} to investigate \textit{extremal BH conifold
attractors} and \textit{extremal BH large complex structure attractors} in
the moduli space of 1-modulus (Fermat) $CY_{3}$s, also in view of recent
investigations of extremal BH attractors in specific examples of 2-moduli $%
CY_{3}$-compactifications \cite{Misra1}.

When $CY_{3}$-compactifications with more than one complex structure
deformation modulus are considered, it is clear that interesting situations
might arise other than the ones present at 1-modulus level. Indeed,
differently from what has been studied so far \cite{Misra1}, in such
frameworks all three species of extremal BH (LG) attractors (namely $\frac{1%
}{2}$-BPS, non-BPS $Z\neq 0$ and non-BPS $Z=0$) should exist, each typology
being supported by distinct, zero-overlapping BH charge configurations.
\textit{\c{C}a va sans dire} that such an issue deserves more investigation
and analyzing efforts.

Finally, it is worth spending a few words concerning the instability of
non-BPS, $Z=0$ (LG) attractors in the $1$-modulus case. It would be
intriguing to extend to such a framework the same conjecture formulated in
\cite{K3}. In Sect. 5 of such a Ref., in the framework of (the large volume
limit of $CY_{3}$-compactifications leading to) the peculiarly symmetric
case of cubic $stu$ model, it was argued that the instability of the
considered non-BPS attractors might be only apparent, since such attractors
might correspond to multi-centre stable attractor solutions, whose stable
nature should be ``resolved'' only at sufficiently small distances. As
mentioned, it would be interesting to extend such a conjecture to the
non-BPS, $Z=0$ (LG) attractors, also in relation to the possible existence
of \textit{non-BPS lines of marginal stability} \cite{Denef1,Denef2}.

\section*{\textbf{Acknowledgments}}

It is a pleasure to acknowledge proofreading by Mrs. Suzy Vascotto and Mrs.
Helen Webster.

A. Y. would like to thank the INFN Frascati National Laboratories for the
kind hospitality extended to him during the work for the present paper.

The work of S.B. has been supported in part by the European Community Human
Potential Program under contract MRTN-CT-2004-005104 \textit{``Constituents,
fundamental forces and symmetries of the Universe''}.

The work of S.F.~has been supported in part by the European Community Human
Potential Program under contract MRTN-CT-2004-005104 \textit{``Constituents,
fundamental forces and symmetries of the Universe''}, in association with
INFN Frascati National Laboratories and by D.O.E.~grant DE-FG03-91ER40662,
Task C.

The work of A.M. has been supported by a Junior Grant of the \textit{%
``Enrico Fermi''} Centre, Rome, in association with INFN Frascati National
Laboratories.

The work of A.Y. was supported in part by the grants NFSAT-CRDF
ARPI-3328-YE-04 and INTAS-05-7928. 

\end{document}